\pdftrailerid{}
\documentclass[12pt,oneside]{book}
\usepackage{setspace}
\usepackage{sectsty}
\allsectionsfont{\singlespacing}
\usepackage{graphicx}
\usepackage{float}
\usepackage[hidelinks]{hyperref}
\usepackage[notes,backend=biber,backref,backrefstyle=none]{biblatex-chicago}
\usepackage{dialogue}
\usepackage{epigraph}
\usepackage{amsmath}

\newcommand{\version}{9539ffb3 }
\usepackage[letterpaper,left=1in,top=1in,bottom=1in,right=1in]{geometry}  
\begin{document}
\singlespace
\begin{titlepage}
   \thispagestyle{empty}
   \begin{center}
      \vspace*{1cm}
      \Huge
      \textbf{The Musical Arrow of Time}\\
      \vspace{0.5cm}
      \large
      The Role of Temporal Asymmetry in Music and Its Organicist Implications\\
      \vspace{1cm}
      \textbf{Qi Xu}\\
      \vfill
      \normalsize
      Submitted in partial fulfillment of the requirements for the Doctor of Musical Arts degree, The Juilliard School\\
      May 2022
   \end{center}
\end{titlepage}
\pagebreak
\begin{center}
   \thispagestyle{empty}
   \vspace*{\fill}
   2022 Qi Xu\\
   ALL RIGHTS RESERVED
   \vspace*{\fill}
\end{center}
\pagebreak
\thispagestyle{empty}
\vspace*{\fill}
\begin{center}
   Revision: \version
\end{center}
\vspace*{\fill}
\pagebreak

\thispagestyle{empty}
\vspace*{\fill}
\begin{center}
   \textbf{Abstract}
\end{center}

Adopting a performer-centric perspective, we frequently encounter two statements: ``music flows'', and ``music is life-like''. In this dissertation, our discussion builds on top of the two statements above, resulting in an exploration of the role of temporal asymmetry in music (generalizing ``music flows'') and its relation to the idea of organicism (generalizing ``music is life-like''). In particular, by the term ``temporal asymmetry'', we focus on its two aspects. The first aspect concerns the vastly different epistemic mechanisms with which we obtain knowledge of the past and the future. Following a discussion on epistemology, we focus on a particular musical consequence: recurrence. The epistemic difference between the past and the future shapes our experience and interpretation of recurring events in music. The second aspect concerns the arrow of time: the unambiguous ordering imposed on temporal events gives rise to the a priori pointedness of time, rendering time asymmetrical and irreversible. A discussion on thermodynamics informs us musically: the arrow of time effectuates itself in musical forms by delaying the placement of the climax.

Organicism is invited to the discussion of temporal asymmetry. Organicism functions as a mediating topic as it engages with the concept of life as in organisms. Therefore, on the one hand, organicism is related to temporal asymmetry in science via a thermodynamical interpretation of life as entropy-reducing entities. On the other hand, organicism is a topic native to music via the universally acknowledged artistic idea that music should be interpreted as a vital force possessing volitional power. With organicism as a mediator, we may better understand the role of temporal asymmetry in music. In particular, we view musical form as a process of expansion and elaboration analogous to organic growth. Finally, we present an organicist interpretation of delaying the climax: viewing musical form as the result of organic growth, the arrow of time translates to a preference for prepending structure over appending structure.

\vspace*{\fill}
\pagebreak

\thispagestyle{empty}
\vspace*{\fill}
\begin{center}
   \textbf{Autobiographical Note}
\end{center}

Qi Xu, from Shenzhen, China, is currently a DMA candidate at the Juilliard school studying piano performance. During his undergraduate years at Columbia University, Qi explored other disciplines with the intention of facilitating his musical understanding. In particular, Qi was selected in 2014 as a Columbia-Oxbridge Scholar to study abroad for one year at Cambridge University where he read mathematics. With exposure to other disciplines, Qi specializes in approaching piano performance with reason and sentiment combined. As such, during his recital tour in Germany 2015 for example, he was hailed by the German press as ``a storyteller and wild riders'' (Rheinische Post) who stands in ``between drama and charm'' (Derwesten).

\vspace*{\fill}
\pagebreak

\thispagestyle{empty}
\vspace*{\fill}
\begin{center}
   \emph{To my parents}
\end{center}

\vspace*{\fill}
\pagebreak

\thispagestyle{empty}
\vspace*{\fill}
\begin{center}
   \textbf{Acknowledgements}
\end{center}

I would like to express my gratitude to many people, whose help has made this dissertation as it stands now possible in the first place.

First and foremost, I would like to thank my advisor Dr.~Lasser for his guidance throughout the dissertation preparation process. His advice, feedback, and encouragement have been inspiring me since the brainstorming stage of the dissertation.

I would also like to thank my reviewers Dr.~Dawe and Dr.~Reynolds for their time and effort invested in reviewing the dissertation. Without their thorough and constructive feedback for revision, the dissertation would not be as polished as it is today.

Moreover, I would like to thank the entire DMA committee chaired by Prof.~Gottlieb for giving me the opportunity of being part of this lovely community and program. Through classes, seminars, performances, and conversations, I learned and grew not only as a musician and a scholar, but also as an artist, a citizen, and ultimately a well-rounded human being.

I would like to thank Dr.~Kaplinsky for her mentoring since 2009. Ever since studying with Veda, she has been more than a piano professor to me. She is my life mentor and a source of inspiration. Through her teachings in piano and life, the years of studying with her have influenced and shaped my artistic vision and worldview.

I would also like to thank Dr.~Raekallio for his piano instruction during my DMA years. The conversations with him have strengthened my appreciation of the beauty of piano playing resulting purely from the machinery of piano technique.

Finally, I want to thank my family and friends for their continuous support throughout my life. Because of them, I am constantly humbled, motivated, and uplifted to devote myself to building a better world: a world, as indicated later in this dissertation, where darkness, destruction, and chaos normally prevail as the arrow of time favors thermodynamic equilibrium.

\vspace*{\fill}
\pagebreak

\frontmatter
\tableofcontents
\pagebreak
\listoffigures
\mainmatter
\pagestyle{plain}
\part*{Prelude}
\addcontentsline{toc}{part}{Prelude}

\doublespace
\chapter{Peculiar properties of time}

\epigraph{The universe is metamorphosed into a story about running brooks, poetry, and music}

Time is a topic that fascinates many for its ability to inspire awe and curiosity. Due to its omnipresence, time is discussed from various perspectives, under different contexts, and across many disciplines. On the one hand, we observe and conceive time in everyday life for practical purposes. Musicians improve their crafts by carefully engineering time-related musical structures, thereby manipulating audience's anticipation and expectation in order to achieve desirable performance effects. For example, performers take advantage of time by using devices such as rubato to create the so-called ``magical moments'', a term listeners frequently use to describe expressive performances. On the other hand, philosophers, scientists, and theologians whose areas of study are less relevant to everyday life, inquire into the issue of time as one of the fundamental questions of our universe. Augustine, in his ``Confessions'', makes a remark concerning the puzzling and paradoxical quality of time:

\begin{quote}
\label{augustine_time}
What is time? [...] We surely know what we mean when we speak of it. We also know what is meant when we hear someone else talking about it. What then is time? provided that no one asks me, I know. If I want to explain it to an inquirer, I do not know. \autocite[230]{2009confessions}
\end{quote}

His remark suggests that time seems to possess the magical property that, upon conscious inspection, resists rational understanding and explanation. Similar phenomenon can be found in the field of art. For example, in music, listeners are astounded by outstanding performances. However, upon reflection, listeners are likely unable to explain, in technical terms, how exactly the performance they experienced is expressive. Metaphorically, musicians are like magicians who inspire awe from the audience, yet must deliberately hide the method from which the awe is derived. We attribute the term ``genius'' to performers who have the ability to present successful performances whose inner workings remain inexplicable to the general public. In his ``Critique of Judgment'', Kant even goes as far as to claim that the inner working remains ultimately unknown to the author himself/herself:

\begin{quote}
   {[}If{]} an author owes a product to his{[}/her{]} genius, he himself{[}/she herself{]} does not know how he{[}/she{]} came by the ideas for it; nor is it in his{[}/her{]} power to devise such products at his{[}/her{]} pleasure, or by following a plan, and to communicate {[}his/her procedure{]} to others in precepts that would enable them to bring about like products.\autocite[175]{kant_pluhar_1987}
\end{quote}

It is through the resistance to rational understanding that genius is valued for its uniqueness, originality, and singularity. Similarly, time is puzzling and fascinating, as it easily can be felt, but hardly understood.

After recognizing the peculiarity of time, we may indulge our curiosity further by asking: what exactly are the outstanding properties of time? What specifically makes time a fascinating topic worth discussing?

\chapter{Explanatory power of time}

By answering the questions above, we become capable of acquiring a better understanding of musical topics, as music necessarily takes place in time. The explanatory power of time is essential to understanding music. Metaphorically, as suggested by Schlesinger in his book ``Aspects of Time'', time can be interpreted as a container\label{container_metaphor_of_time}, such that ``every event occur{[}s{]} at some point in time''.\autocite[3]{schlesinger_1980} We can further extend the metaphor of the container by saying that a better understanding of the container (i.e.~time) pertaining to its characteristics will also benefit our understanding of its contained object (i.e.~music). Using the dichotomy of form and content, we may claim that our perception of the content is shaped by the form through whose medium the content is presented. Moreover, even for the sake of a musical discourse, some conclusions about music are possible to be derived only if we look at a bigger picture: time and its general properties. For example, without having the idea of temporal asymmetry in mind, we might not recognize that, as will be discussed in later sections, some microscopic musical structures such as phrase model (see section \ref{section:phrase structure}) and macroscopic structures such as musical form (see section \ref{section:bar form}) bear close resemblance. Once we use time as an overarching topic, a ``point of intersection'', we are then empowered to explain and organize various issues of interest that are commonly encountered in music. Time as a topic, has the organizational power to group commonality in music that is difficult to notice otherwise. In particular, structures occurring at different organizational levels can be explained by a common root cause: attributes of temporal processes at different scales that permeate musical phenomena.

The organizational power found in the topic of time is akin to that of music history. If we study each piece in isolation without considering the overall historical trend and stylistic characteristics as a organizing mechanism, then the repertoire of musical works would look disparate in the sense that all pieces are unrelated. One of the practical consequences is that, for performers, preparing a piece for performance becomes less effective. In such a case, one has to treat the piece as completely new, without referring to knowledge, from prior experience, of pieces similar in style (e.g.~those from the same composer or from the same time period).

\section{Time as a universal theme across topics}
\label{section:aop}

Thus, to embark on a journey inquiring the nature of time and its relations to music, amounts to crosscutting the universe (or the topoi) of musical discourse. The term ``crosscutting'' here is borrowed from the field of computer science, as appeared in the term ``cross-cutting concerns''. For a software application, we can often find ways to decompose it into different logical units bearing various possible names: functions, components, modules, features or concerns. For example, a typical music streaming application consists of several components. It should have a component that provides searching functionality, so that users can search for specific songs using a set of keywords. Meanwhile, for users without a specific search target in mind, who simply would like to explore new songs, the application should have a component that displays a collection of recommended songs to the users. Finally, the application should provide a component that plays the selected song, i.e.~a music player component. The above examples of components are, using terminology from computer science again, encapsulations of the application's functionality. Metaphorically, we can think of the components as being encapsulated into separate ``capsules'', with each capsule attaining a clearly-defined boundary. Musically, the metaphor of capsules can be illustrated using the example of sonata form: a piece of music exhibiting the formal structure of sonata form is divided into components named exposition, development, and recapitulation. Each component in sonata form can be viewed as a capsule with clearly-defined boundary. For example, the boundary between the exposition and development is often notated visually by a repeat sign. The advantage of encapsulation is that the architecture of the application's design is well structured. Practically, the team of developers can adopt a strategy called the division of labor to assign tasks concerning well-defined components to specific team members.

With each component clearly defined and encapsulated, however, there are some ``aspects'' that span across multiple components. Continuing the musical metaphor using sonata form above, certain topics are found across exposition, development, and recapitulation. For example, the topic of tension-release is an aspect that spans across exposition, development, and recapitulation. For the example of the music streaming application above, one aspect that is found in all three components is that the user must be logged-in in order to use the streaming service provided by the application. Therefore, user authentication in this case, is a cross-cutting concern of the application, because it metaphorically ``crosscuts'' three components. In computer science, there is a specific design paradigm named ``aspect oriented programming'' (abbreviated as ``AOP'') that primarily deals with cross-cutting concerns:

\begin{quote}
AOP is often defined as a technique that promotes separation of concerns in a software system. Systems are composed of several components, each responsible for a specific piece of functionality. But often these components also carry additional responsibilities beyond their core functionality. System services such as logging, transaction management, and security often find their way into components whose core responsibilities is something else. These system services are commonly referred to as cross-cutting concerns because they tend to cut across multiple components in a system.\autocite[10]{walls_2011}
\end{quote}

The concept of cross-cutting concerns can be applied to music via analogy. In addition to the specific metaphor using sonata form illustrated above, we approach music from a multitude of perspectives depending on topics of interest. As such, the field of music is partitioned into various subdisciplines. Given an arbitrary piece of music, if the interest is on the socio-historical context under which the piece is composed, then the approach takes the form of music history. If the interest is on the musical aspect of the piece, treating it as an ahistorical and autonomous object of art, then we can pick applicable analytical devices to approach the piece as we see fit. Among examples of analytical devices are harmony, counterpoint, formal analysis. As a practical consequence, these partition schemes give rise practically to standard courses commonly found in music conservatories: music theory (which can be further divided into harmony and counterpoint), music history, performance practice, etc. Within the broad categories above, there are specific theories providing unique insight and interpretations to the piece. To name a few, we encounter theories such as species counterpoint, sonata theory, Schenkerian analysis, music set theory, serialism, neo-Riemannian theory, etc. These theories may have their unique merits, weakness, and areas of focus. For example, Schenkerian analysis is effective in treating tonal music (or more specifically, a certain subset of tonal music), and its explanatory power becomes questionable once we take into account non-tonal music. However, regardless of the specific contents proposed by various theories, they necessarily share a common feature that, by definition, they are studies of music, which invariably involves arrangement of events in time. Now, we can extend the analogy of cross-cutting concerns as follows: different subdisciplines and theories about music act as encapsulated components designated to explain musical phenomena. Meanwhile, the issue of time serves as a cross-cutting concern, because it tends to ``cut across multiple components'' of musical research.

\chapter{Unique feature of time: directionality}

Among properties of time, the most distinguishable one concerns directionality of time. We seem to have a priori knowledge of the directionality of time that is independent from empirical observations. In particular, the directionality of time entails that time necessarily flows from the past to the future. A piece of music can then be figuratively described as a process of unfolding, whose direction points from the past (that is already unfolded and visible), through the present, into the future (that is still folded and invisible).

\section{Space and time: a comparison}
\label{space_vs_time}

The directionality of time becomes even more evident once we contrast it with that of space. In space, there is no absolute direction, in the sense that directions can be named arbitrarily: there is no intrinsic difference between left and right, at least for the macroscopic world we experience daily. It is up to our conventions that we define directions the way they are. For example, some countries adopt left-hand traffic while others adopt right-hand traffic. Meanwhile, both conventions are equally justified. In other words, spatial directions amount to arbitrary choice, rather than necessity. In mathematics, in order to determine the direction of cross product, the orientation of the vector space must be determined in advance. However, the choice of orientation is arbitrary. It is due to convention, not necessity, that we often pick the orientation such that the cross product abide by the right-hand rule, as opposed to the equally valid left-hand rule. Furthermore, we can consider a thought experiment as follows. A possible universe in which right-handed population is majority is not substantially different from the one in which left-handed population is majority. The arbitrariness of choice precisely refers to the fact that outcomes resulting from different choices are indistinguishable from one another.

Physicist Feynman, in his publication ``The Feynman Lectures on Physics'', delivers a vivid parable illustrating the problem of distinguishing spatial orientations:

\begin{quote}
[Imagine] that we were talking to a Martian, or someone very far away, by telephone. We are not allowed to send him any actual samples to inspect; for instance, if we could send light, we could send him right-hand circularly polarized light and say, ``That is right-hand light---just watch the way it is going.'' But we cannot give him anything, we can only talk to him. He is far away, or in some strange location, and he cannot see anything we can see. For instance, we cannot say, ``Look at Ursa major; now see how those stars are arranged. What we mean by `right' is \ldots{}'' We are only allowed to telephone him.

Now we want to tell him all about us. Of course, first we start defining numbers, and say, ``Tick, tick, two, tick, tick, tick, three, \ldots,'' so that gradually he can understand a couple of words, and so on. After a while we may become very familiar with this fellow, and he says, ``What do you guys look like?'' We start to describe ourselves, and say, ``Well, we are six feet tall.'' He says, ``Wait a minute, what is six feet?'' Is it possible to tell him what six feet is? Certainly! We say, ``You know about the diameter of hydrogen atoms---we are 17,000,000,000 hydrogen atoms high!'' That is possible because physical laws are not invariant under change of scale, and therefore we can define an absolute length. {[}\ldots{]} we start to describe the various organs on the inside, and we come to the heart, and we carefully describe the shape of it, and say, ``Now put the heart on the left side.'' He says, ``Duhhh---the left side?'' Now our problem is to describe to him which side the heart goes on without his ever seeing anything that we see, and without our ever sending any sample to him of what we mean by ``right''---no standard right-handed object. Can we do it? \autocite{feynman_leighton_sands_2013}
\end{quote}

The task involves teaching the concept of left and right in a purely verbal manner, without referring to any potentially shared experience. The solution, according to Feynman, is a convoluted yes, involving substantial use of atomic physics:

\begin{quote}
In short, we can tell a Martian where to put the heart: we say, ``Listen, build yourself a magnet, and put the coils in, and put the current on, and then take some cobalt and lower the temperature. Arrange the experiment so the electrons go from the foot to the head, then the direction in which the current goes through the coils is the direction that goes in on what we call the right and comes out on the left.'' So it is possible to define right and left, now, by doing an experiment of this kind.\autocite{feynman_leighton_sands_2013}
\end{quote}

Later on, the author proceeds to the notion of antimatter which further complicates the issue, as it makes the choice of handedness arbitrary again. Without going into further details that are beyond the scope of this paper, the sheer difficulty of answering the seemingly trivial question of distinguishing left from right demonstrates how absolute direction in space can be a tricky issue to tackle.

In contrast to that of space, orientation of time seems to be straightforward. We consider again the thought experiment of teaching the Martians. However, this time we would like to teach them about distinguishing the past from the future. Then the task is simple. By definition, every process takes place in time. Therefore, we may take any process, and mark its beginning and ending. Then the relationship between the marked beginning and the ending corresponds exactly to that between what we call the past and the future. The remaining task is about finding a strategy to mark the beginning and ending, as well as distinguishing between them. The instruction for the curious Martians is straightforward. If they are enthusiastic about music, then we may teach them the following: have a notebook ready, and start playing a piece of music. Then write down notes upon detecting change of states. Without loss of generality, we may assume that there are only two possibilities for change of states: from silence to sounding music, and from sounding music to silence. The rules of writing notes are the following: 1) if a change of state is encountered and the notebook is empty, write ``beginning'', and 2) if a change of state is encountered and the notebook already has the word ``beginning'' on it, then write ``ending''. The two rules uniquely determine the way through which the beginning and ending are marked. From the example above, we can easily notice the crux of the issue. Orientation of time is straightforward because we have ways to remember the past (by the act of note-taking in the example), whereas the idea of remembering the future leads to absurdity. In other words, we are able to potentially recall the past with the help of objects named ``records''. On the other hand, the only way we can inquire into the future is through prediction. Acts of recalling and predicting involve procedures and mechanisms that are fundamentally different. Knowing the difference between the two, as we will discuss in section \ref{section:epistemic difference between the past and the future}, is crucial for a better understanding of music, since music perception heavily depends on recall (of heard events) and anticipation (of upcoming events).

Meanwhile, orientation of time is non-arbitrary. More specifically, the choice of the direction in which time flows is not a play of words or convention in the following sense. Let us consider again the above thought experiment of imagining a possible world in which left-handed population is majority. This time, instead of substituting right-handed population for left-handed one, we choose to imagine a possible world in which time flows backwards, in the sense that cause follows effect, instead of preceding it. Consequently, such a possible world would be saliently different to our current one. In fact, it might be incomprehensible to us at all. How could we possibly understand a world in which effect precedes cause? For instance, how could we enjoy a concert whose finale comes before the opening movement? The reason for such incomprehensibility is partly that the temporal order of events is embedded in the definition of causality. Therefore, cause, by definition, comes before effect. Analogously, the opening movement, by definition, comes before the finale. To claim otherwise amounts to stating a logically contradictory statement: the finale which by definition follows the opening movement is at the same time preceding the opening movement, constituting a logical conjunction of mutually exclusive propositions.

\section{Artistic implications}

The difficulty of imagining the backward flow of time shows how the directionality of time is so ingrained and hard-coded in our mindset, that if the directionality is modified, then time is rendered incomprehensible and perplexing. In fact, it is precisely due to our ingrained intuition of the flow of time, that artists often employ strategies to counter such intuition in order to achieve surprising dramatic effects. For example, in his magnum opus ``Faust'', Goethe describes an unrealistically strange scene:

\begin{quote}
Show me the fruit that rots before it's plucked and trees that grow their greenery anew each day! \autocite[131]{1985faust}
\end{quote}

Spoken by Faust during his confrontation with Mephistopheles, the quoted passage addresses Faust's wish to experience the impossible and the transcendental. It is under such circumstance that the wager between Faust and Mephistopheles unfolds, becoming the direct cause for plots that follow. It is worth noting how the author conveys the idea of experiencing the impossible and the transcendental: through portraying time poetically, achieving the poetic effect by means of distorting our common sense about time. Our common sense indicates that fruits rot after they have grown mature, and trees turn green annually. In the quoted passage, however, the sequence of their life cycle is substantially altered by means of temporal disorientation and acceleration, respectively. While tampering with our common temporal perception may render time incomprehensible as discussed previously, which is undesirable for many practical purposes, it may be desirable for the sake of artistic effects. Besides the literary example of Faust illustrated above, manipulating temporal perception (in particular, temporal disorientation) is in general a powerful expressive tool across many artistic disciplines. Consider additionally the following examples: a broken glass is restored spontaneously from the floor onto the table; a dead person rises from the tomb. As discussed previously for similar examples, we are perplexed upon encountering such examples because they are incomprehensible. However, if we now try to adopt an artistic perspective, we may experience the emotion of wonderment, making these examples awe-inspiring. In fact, the emotion of awe in this case has the same origin as the previously mentioned quality of incomprehensibility. It is exactly due to our lack of ability to comprehend temporal disorientation, that we feel awe as an emotional consequence. Such an awe-inspiring device is commonly found in the arts as well as religions. For example, rising from the dead (i.e.~resurrection) is one of the cornerstones of Christianity, directly resulting in Easter being the most important day in the liturgical calendar. The same concept finds its importance in music as we take into consideration religious and religion-inspired repertoire. In section \ref{section:resurrection as recurrence in action}, we will examine Bach's St Matthew passion and the Christmas oratorio as a musical case study of the concept of resurrection, which is itself a powerful musical tool that takes full advantage of temporal disorientation.

\section{Further decomposition}
\label{section:further_decomposition}

We need to take a step back, and realize that when we state that ``time flows from the past to the future'', the statement can be further decomposed into two substatements. The first states that, when viewed from the present, the future looks vastly different from the past. The sheer distinguishability between the past and the future gives rise to the notion of temporal asymmetry. Metaphorically, we may imagine an old style weighing balance, with the past and the future respectively placed at both ends. Then temporal asymmetry amounts to assigning the past and the future different weights. The second substatement further adds a condition that assigns a specific arrow of time relating temporal events in a particular order, that time flows definitively from the past to the future, never the other way around. Extending the weighing balance metaphor, the second substatement specifies that the future weighs more than the past, such that if we were to place a drop of water on top of the balance, it would naturally flow toward the future end of the balance.

In the following section, we will use the two substatements previously mentioned as a point of departure. The goal is to discuss a few general remarks relating to the notion of temporal asymmetry from a philosophical and scientific perspective, thereby providing necessary prerequisites and insights for a musical discourse that follows.

\part{A discursive quest}

\chapter{Necessity of extramusical discussion}

Before discussing and exploring how temporal asymmetry is at work in music, it is deemed necessary to get a sense of how commonly we encounter the topic of temporal asymmetry across various fields. By expanding the universe of discourse so that the scope of discussion is no longer limited to music, we may better understand how we are situated in discussing the topic of temporal asymmetry. Poetically speaking, we may view the collection of all fields of study as comprising an atlas. Then our objective is to have a birdview of the entire landscape of such an atlas. Consequently, when the discussion eventually returns to music, we are equipped with a holistic understanding of the topic.

We would like to inquire how various disciplines observe and explain the phenomenon of temporal asymmetry. In particular, the focus is on the key concepts that frequently emerge during the inquiry. By focusing on these key concepts, we discover essential topics that a fruitful discussion must presuppose. For example, one important aspect of temporal asymmetry involves the idea of trace: records of the past that make our inference of the past different from prediction of the future. Once we know the role of trace in a scientific context, the same topic can be used to guide our musical inquiry: how the role of memory shapes the way we listen to music and anticipate upcoming musical events. For example, we may ask about how musical repetitions take advantage of our memory capacity. That is, after knowing the key concepts relevant to temporal asymmetry in non-music subjects, we obtain the knowledge of what to look for in music. One may challenge such indirect strategy by claiming that it is entirely feasible to conduct an inquiry into the issue of temporal asymmetry in music without consulting other fields of study, treating it as a purely musical quest. While it is true that confining the topic exclusively to music is feasible, it may be nevertheless limiting. The limiting factor can be stated in two ways: the topic of time is often ignored in music, and the topic of time is actively discussed in other fields.

\section{Temporality as an ignored topic in music}

We start with a counterintuitive observation: we often overlook the role of temporal structure in music. It might sound unexpected at first. After all, in instrumental lessons, one of the most frequently discussed topics is timing. Additionally, meter, rhythm and formal analysis, being temporal structures, are indeed essential components of music theory and analysis. Finally, music is, by definition, dependent on time as its essential medium of expression. However, there are two justifications backing up the claim that temporal structure is an overlooked topic, corresponding to two perspectives:

\begin{itemize}
\item
  Temporality in comparison to other musical topics
\item
  Temporality in music compared to other disciplines
\end{itemize}

\subsection{Temporality in comparison to other musical topics}

Within the realm of musical discourse, if we consider conventional music theory and analysis, an important rule of thumb states that pitch is more important than rhythm. In other words, the majority of theories, as far as Western art music is concerned, focus on pitch contents and pitch relations more than their rhythmic counterparts. For example, theoretical constructs central to Schenkerian analysis often omit rhythmic structures in the sense that a typical Schenkerian graph reveals more insights about pitch relations (i.e.~Ursatz and prolongational spans) while discarding most of the durational information. Because the durational information is filtered out, we are typically unable to recover the temporal proportion of the original piece it represents. In other words, Schenkerian graph as a graphical representation of analytical reduction, filters out durational information because the theory considers rhythmic structures as constructs subordinate to pitch contents and pitch relations, hence are subject to omission in analysis. In fact, we may even find abuse of notation in a typical Schenkerian graph: note values become time-irrelevant. For example, Salzer puts a footnote in his Schenkerian treatise ``Structural hearing'':

\begin{quote}
The use of half-notes for chords of the structural progression in the graph is not intended to indicate time-values, but to differentiate structural points from chords having a different function. \autocite[12]{salzer_1952}
\end{quote}

Therefore, rhythmic structure may be described as, borrowing ideas from Schenkerian analysis itself, foreground elaborations that can be omitted due to its lesser importance.

We obtain a similar conclusion if we consider serialism as a case study. One of its central constructs is the tone row along with rules for manipulating the tone row. However, the tone row, evident from its name, is an exclusively pitch-based construct. It is possible that the development of total serialism is a response to the limitations of the pitch-based serialism, by extending the same technique to musical parameters other than pitch, in particular, rhythm. However, simply seen from the historical development itself, we can observe the implied subordination: serialism was developed with pitch manipulation in mind, and it was later on extended and ported to cover rhythm and other parameters such as timbre and dynamic. Therefore, rhythm is a subordinate to (or alternatively, a derivative of) pitch. More explicitly, when we apply serialist techniques to rhythmic parameters, we necessarily ignore temporal structures because serial rules were not originally developed with the concept of temporal structures in mind.

Additionally, if we inspect the overall historical development of music theory and analysis, then we also discover that most of the elaborate theories focus on pitch relations. To name a few examples, species counterpoint is appropriate for pre-tonal music; tonal harmony is, by definition, suited to tonal music; serialism and pitch-class set theory are designed for post-tonal music. Moreover, as illustrated in the case of total serialism, theories of temporality are often extensions to pitch-centric theories, i.e.~as a by-product. In particular, the procedure of theory-building can be described as follows: we start with formulating symbolic representation of pitch entities (e.g.~tone row as representation of the twelve pitch classes) and rules for its symbolic manipulation (e.g.~inversion, retrograde, and retrograde-inversion). Then we try to apply the same theory, as the collection of symbolic representation and rules for symbolic manipulation, to non-pitch parameters. For example, we try to see what happens if the tone row is populated by parameters other than pitch classes. We may establish an arbitrary mapping in which pitch class 0 is substituted by durational value of eigth-note, pitch class 1 by quarter-note, and so on. At this point, we may already see a potential issue: because the theory is not developed originally with rhythmic parameters in mind, many constructs in the theory are questionable. For example, for a rhythmic ``tone'' row, does it make sense to employ a collection of twelve elements? For pitches the collection of twelve is justified because of tuning constraints in equal-temperament. However, the choice of twelve becomes arbitrary once rhythmic parameters are in question.

We may argue that, it is exactly for the reason of subordination that David Lewin, a proponent of neo-Riemannian theory, writes in his treatise ``Generalized Musical Intervals and Transformations'' the following statement:

\begin{quote}
This chapter takes as its point of departure a figure showing two points s and t in a symbolic musical space. The arrow marked i symbolizes a characteristic directed measurement, distance or motion from s to t. It intuits such situations in many musical spaces, and i is called ``the interval from s to t'' when the symbolic points are pitches or pitch classes. \autocite{lewin_2007}
\end{quote}

In order to present the central theme of the book, i.e.~the mathematical model named ``Generalized Interval System'' (GIS), the author clearly understands that it is the most natural to base the model on pitch relations first, hence providing an intuitive motivation for the GIS model. Evident in its naming, the GIS model is built upon the notion of interval, which has very specific connotations: in music, the term interval unequivocally refers to intervallic distance between pitches. If, instead, we would like to refer to temporal intervals, we need to further specify the term by prefixing it with additional qualifiers, e.g.~time-span interval. After introducing GIS using its originating idea of the interval, the model naturally extends, or using its own terminology, generalizes to rhythmic parameters.

Many authors also acknowledge the general neglect of temporal structure in the scholarly field. As the Grove article on ``Theory, theorists'' points out, the 1985 (which is considered a recent publication in the field of music academia) issue of the journal ``Music Theory Spectrum'' dedicates its entirety under the title ``Time and Rhythm in Music''. \autocite{grove_theory_theorists} The editor of the issue, in the opening words, states that the temporal dimension of music remains ``the less explored of the music's two major domains''.\autocite{music_theory_spectrum_1} Within the same issue, Kramer suggested in his article ``Studies of Time and Music: A Bibliography'', that musical ``time has not been widely recognized as an independent field of study''.\autocite{music_theory_spectrum_2} Then, he enumerates specific observations as evidence:

\begin{quote}
    The New Grove has no article on time; RILM has no separate category for time; The Music Index has only recently begun to list articles under the heading ``Time.''\autocite{music_theory_spectrum_2}\label{quote:grove_no_time}
\end{quote}

Additionally, in the preface of the 1960 book ``The Rhythmic Structure of Music'', the authors outrightly point out that the study of the temporal aspect of music ``has been almost totally neglected in the formal training of musicians since the Renaissance. There are many textbooks on harmony and counterpoint but none on rhythm''. \autocite[v]{cooper_meyer_1960} It is true that their statements might turn out to be outdated by the current century as the development of music scholarly research has been constantly ongoing. Furthermore, ethnomusicologists are justified in arguing that the neglect of rhythm might be peculiar to Western art music only. In other words, the neglect is not universal, but applicable to only a particular subset of the entire human society. However, the fact that temporality (i.e.~rhythmic structures) had been a relatively overlooked topic within the realm of mainstream Western music theory and analysis for several centuries is itself noteworthy.

\subsection{Temporality in music compared to other disciplines}

\label{music_as_subjective_experience}

If we compare music externally to other fields, i.e.~when comparing it to science and philosophy, we observe that musical discourse often avoids explicitly discussing time in itself. To make the statement clear, consider the very term of temporal asymmetry. In the context of music, many readers would consider it a borrowed term in the sense that it appears foreign to us. Upon encountering the term, we find it more natural to interpret it as a term borrowed from physics: in particular, the thermodynamic temporal asymmetry. In other words, time is a built-in topic that is native to physics, whereas it is foreign to music. Upon contemplation, it is worth emphasizing again how ironic and peculiar the case is. On the one hand, time as a topic is overlooked both within the field of music and in comparison to other fields of study, to the extent that discussing time in itself would seem out of place in a musical discourse. On the other hand, the role of time is in fact, more crucial in music than in physics. Firstly, the physical reality of music, when interpreted as a physical process, is temporal in a strictly physical sense. Therefore, since musical events are themselves physical phenomena, what applies to general physical phenomena must equally apply to musical events, if not more. Secondly, in addition to being physical, music also attains its psychological reality, such that we must take into consideration our subjective experience of time. It is exactly for this reason that music theory is considered a completely different discipline from acoustics, even though both are ultimately studies of sound.

Therefore, time plays another role in music: it governs the mental representations of music, in addition to the physical reality of music. One may argue that the same statement is true in physics where mental representations are necessary if we were to understand anything at all: we need mental representations of the universe in order to conduct any study of it. However, the crux of the statement resides in that physical theories often intentionally omit subjective experience of time altogether. Suppose that for reasons analogous to optical illusions, we happen to believe that time flows more slowly today. Meanwhile, we have the same degree of belief that a time-measuring device is credible. However, we find ourselves in a conflicting situation since our belief of slower time contradicts the evidence given by the time-measuring device. In such a case, omitting subjective experience of time is equivalent to saying that we necessarily reject the belief of slower time in favor of the evidence given by the time-measuring device.

The above thought experiment may sound unsurprising in a physical context. However, if the same situation happens in music, the outcome may turn out to be vastly different. Suppose that we encounter a similar situation, with the time-measuring device being a metronome. The situation is given that when listening to a performance of a piece, most of the listeners in the room believe that the performer is unable to maintain a steady tempo. The performer responds by showing evidence from an accurate metronome that the tempo is steady in the metronomic sense. However, in this case, musicians would still keep the conclusion that the tempo is not steady with the following justification: it is true that the tempo is steady in the physical sense, yet the performer still fails to maintain a steady tempo musically as the harmonic content, dynamic, register and texture may require some nuanced timing for the listeners to experience a musically steady tempo. Notice that we are not discussing timing nuances such as rubatos that listeners can detect. Instead, we are talking about the counterintuitive performance technique that, sometimes in order to achieve the perception of a steady pulse, performers need to do the exact opposite by avoiding a metronomically accurate playing. 

\begin{figure}[!h]
   \centering
   \includegraphics[width=.8\textwidth]{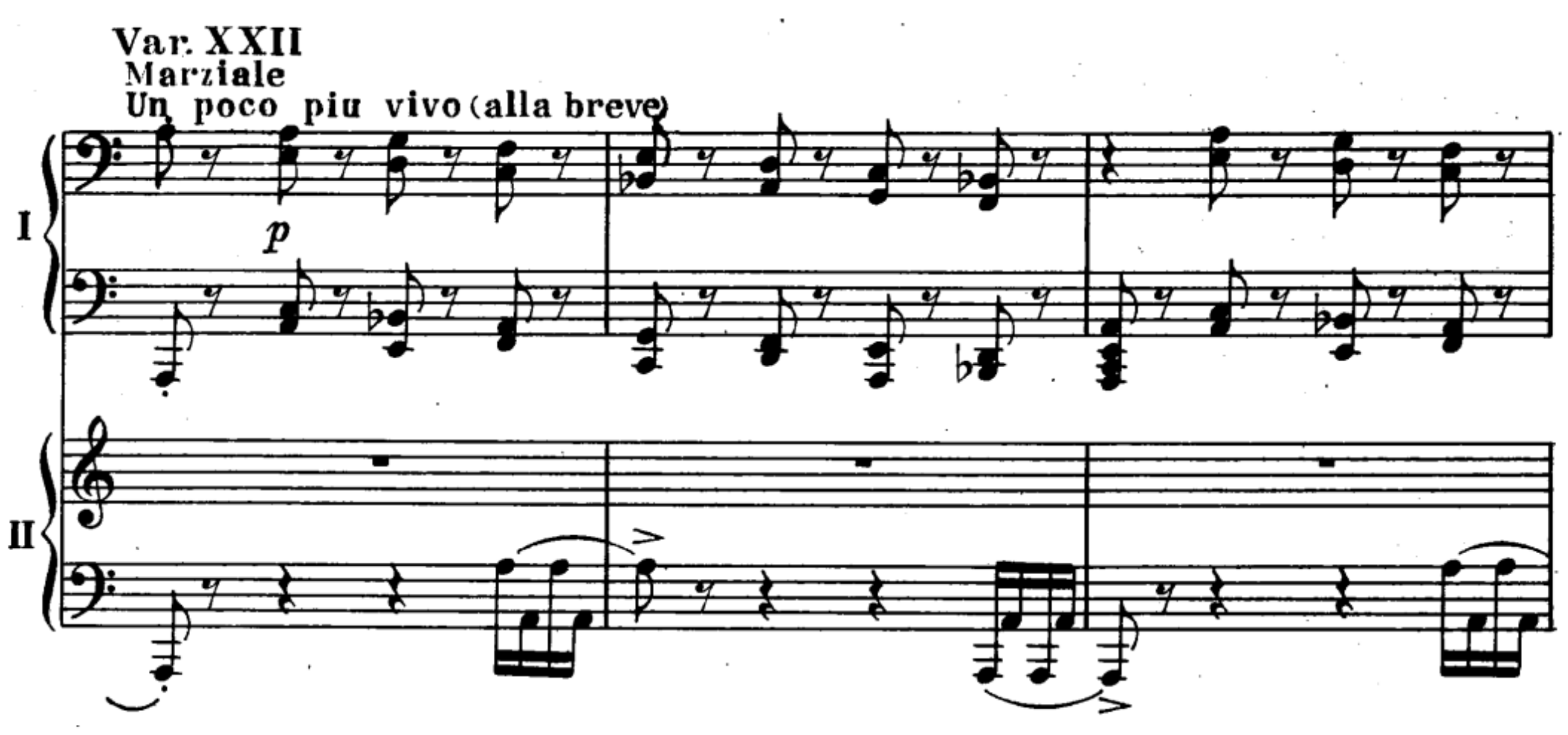}
   \caption{Rachmaninoff, Rhapsody on a Theme of Paganini, Op. 43, variation 22, beginning (two-piano reduction arranged by the composer)}
   \label{fig:rachmaninoff op43 var22 beginning}
\end{figure}

\begin{figure}[!h]
   \centering
   \includegraphics[width=.8\textwidth]{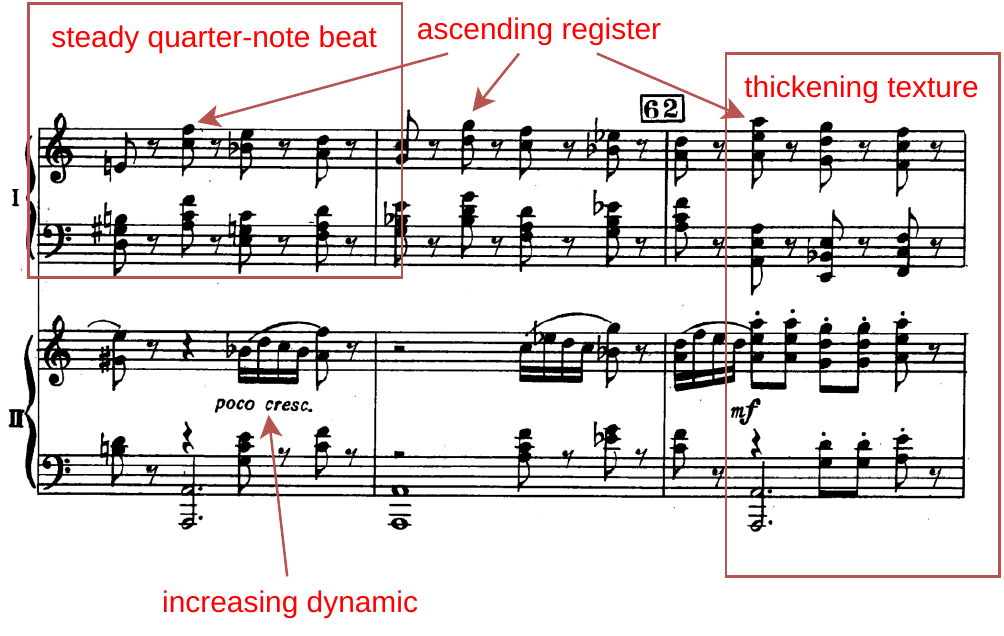}
   \caption{Rachmaninoff, Rhapsody on a Theme of Paganini, Op. 43, variation 22, excerpt (two-piano reduction arranged by the composer), annotated to enumerate musical features of interest}
   \label{fig:rachmaninoff op43 var22 annotated}
\end{figure}

For example, consider variation 22 from Rachmaninoff's Paganini rhapsody (see figure \ref{fig:rachmaninoff op43 var22 beginning}). The entire variation can be interpreted as a huge crescendo in many senses of the word: the dynamic is increasing, the texture (i.e.~number of simultaneous notes played) is thickening, and the register is ascending. Meanwhile, the steady quarter-note beat is the rhythmic pulse throughout the entire passage (see figure \ref{fig:rachmaninoff op43 var22 annotated}).

Because so many musical factors contribute to the feeling of musical accumulation, if we keep the metronomic steady tempo throughout, we may have the auditory illusion (analogous to optical illusion) that the performance is speeding up, as the crescendo in dynamics, thickening in texture and ascending pattern suggest accelerando without actually playing one. As a result, with this expansive musical passage, the performer may have to judiciously make the interpretive decision of stretching the tempo if the goal is to convey a sense of a steady march-like progression, in order to cancel out the effect of speeding up implied by the non-temporal parameters mentioned above.

The rhythmic nuances which performers frequently exercise show that listener's perception is prioritized over scientific measurement. For the exact reason, we have the controversy of the metronome\label{metronome_controversy}: a metronome that measures physical time gives equality of durations, yet it is not strictly equivalent to the notion of steady pulse musicians have in mind. This is why, as commonly observed in instrumental lessons, one of the main challenges in rhythmic training involves reconciling the musical notion and the metronomic notion of rhythmic pulse. On the one hand, instructors emphasize the importance of metronomes as an indispensable aid to understanding rhythmic pulse. However, on the other hand, the emphasis on metronomes almost always accompanies an equally urgent reminder concerning the potential misuse that, by excessive use of the metronome, students may lose the organic quality of rhythm that is vital to true musicianship. Therefore, students are repeatedly told that a musically steady pulse does not equal mechanically metronomic playing (the same way the slogan ``correlation does not imply causation'' is repeatedly spelled out in science classrooms).

In his book ``Sound and symbol'', Zuckerkandl makes a clear distinction between physical time and musical time. According to his schematic comparison, the primary distinction between the two is that physical time is the ``form of experience'' whereas musical time is the ``content of experience'', \autocite[202]{zuckerkandl_1973} invoking the dichotomy between form and content. As such, physical time serves as mere measurement of events while musical time acts as ``an active force''\autocite[247]{zuckerkandl_1973} that produces musical events. One of the specific consequences of his claim is that musical time resists measurement attempts. In his own words, time ``knows no equality of parts''.\autocite[208]{zuckerkandl_1973} In other words, the statement claims that talking of equal beats is ill-defined for musical time. Speaking of ``an equality of times, or of parts of time, has no reasonable meaning in the realm of meter''. \autocite[210]{zuckerkandl_1973} By contrast, in the physical world, equality of time depends on its measurement, so that the ``equality of hours is the equality of the distances traveled by clock hands''. \autocite[209]{zuckerkandl_1973} In fact, measurement of moving body can be taken as the very definition of time, as Zuckerkandl points out:

\begin{quote}
The motion of one body, if it is taken as the measurement of the motion of another body, is called time.\autocite[209]{zuckerkandl_1973}
\end{quote}

\begin{figure}[!h]
   \centering
   \includegraphics[width=.3\textwidth]{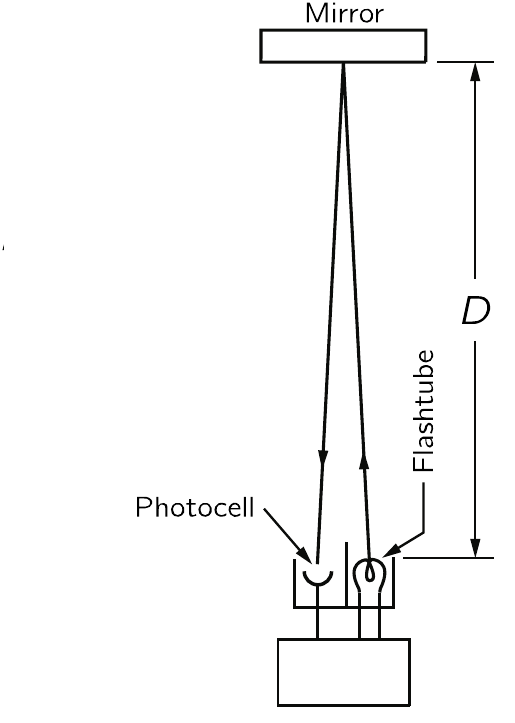}
   \caption{A hypothetical clock given by Feynman that measures time using the roundtrip of lightbeam}
   \label{fig:relativity clock}
\end{figure}

We can further justify taking measurement of moving body as the definition of time if we consider a common thought experiment in understanding special relativity. The thought experiment is set up in the way given by Feynman in his lectures about ``Transformation of time''.\autocite{feynman_leighton_sands_2013} We may consider a railway train. Inside the train, we have a laser pointer placed on the floor of the train, pointing at the ceiling that has a mirror (see figure \ref{fig:relativity clock}).

\begin{figure}[!h]
   \centering
   \includegraphics[width=.6\textwidth]{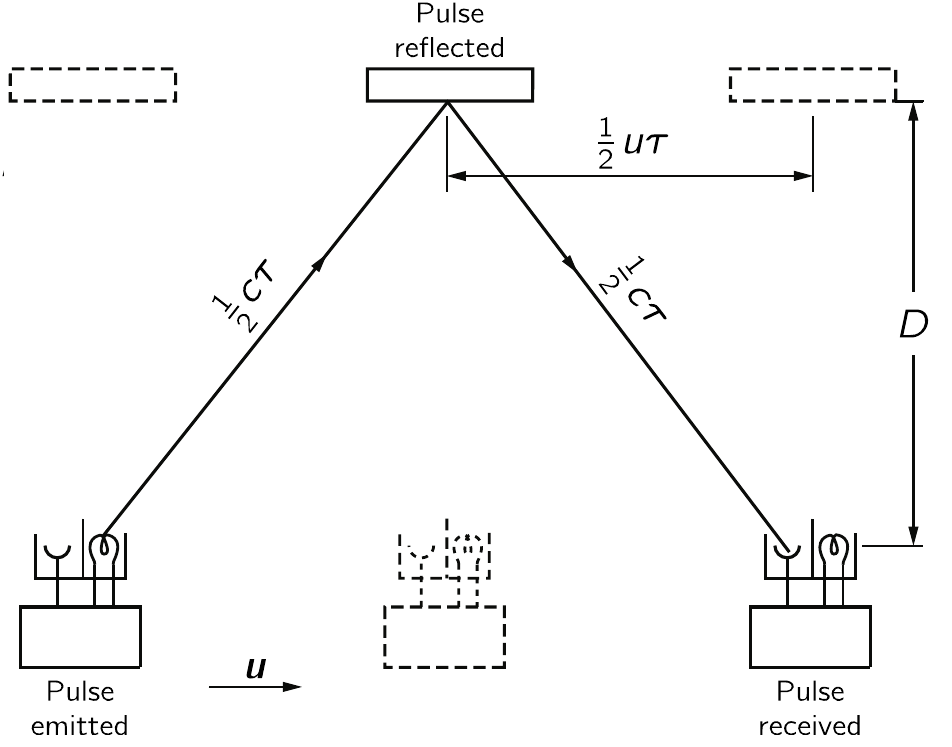}
   \caption{From the perspective of observers on the platform, quoting Feynman's words: ``when the external observer looks at the clock going by, he sees that the light {[}\ldots{]} is `really' taking a zigzag path, since the rod {[}i.e.~the train{]} is moving sidewise all the while''.}
   \label{fig:relativity clock moving}
\end{figure}

The laser pointer emits light that returns to its original place after reaching the mirror. Then, we define time by the motion of the laser beam as follows: we record how long it takes to make a roundtrip, and call it one unit of time. Meanwhile, we call the setup of the laser pointer and mirror the clock. Now, as physicists tend to do in their creative thought experiments, the train moves outrageously fast, in fact, at the speed of light. Special relativity tells us that the speed of light is the same for all observers. Additionally, we defined one unit of time using the roundtrip of the laser beam. Consequently, an observer on the ground will find that the clock stops on the train, precisely because the laser beam on the train travels a diagonal path (see figure \ref{fig:relativity clock moving}). In order to for the laser beam to reach the mirror in the ceiling, it has to travel a distance greater than the horizontal displacement of the train (because the diagonal path corresponds to the hypotenuse of a right triangle, which is the longest side of a triangle). However, because the train moves horizontally at the speed of light, the laser beam can never reach the mirror in the ceiling by completing the diagonal path, let alone making a roundtrip.

This thought experiment shows the slogan that ``moving clocks run slower''.\autocite{feynman_leighton_sands_2013} Meanwhile, it also shows that physically, time is defined by its measurement. In other words, it is not that physical time favors measurement, but rather, measurement is the very definition of time. In Feynman's words:

\begin{quote}
Now if all moving clocks run slower, if no way of measuring time gives anything but a slower rate, we shall just have to say, in a certain sense, that time itself appears to be slower in a space ship. All the phenomena there---the man's pulse rate, his thought processes, the time he takes to light a cigar, how long it takes to grow up and get old---all these things must be slowed down in the same proportion, because he cannot tell he is moving. The biologists and medical men sometimes say it is not quite certain that the time it takes for a cancer to develop will be longer in a space ship, but from the viewpoint of a modern physicist it is nearly certain; otherwise one could use the rate of cancer development to determine the speed of the ship! \autocite{feynman_leighton_sands_2013}
\end{quote}

On the other hand, music rejects measurement. The mechanical pendulum-based metronome is an example of devices that achieve durational equality by using the motion of the pendulum. In fact, Zuckerkandl might have exactly the example of the metronome in mind when writing, so in later passages he invokes arguments similar to the example of steady tempo given above, in order to demonstrate his schematic comparison:

\begin{quote}
What do we mean, then, when we demand that musicians play in time; demand, that is, that they preserve equality of measures and beats? The poor performer who takes all sorts of liberties with time is censured for the capricious inequality of his measures and beats. By what concept of equality do we measure this inequality? Certainly not by the concept of measurably equal lengths. {[}\ldots{]} There is no such thing as a musician whose performance does not depart from mathematical equality within certain limits; accurate experiments have given amazing proof of how great such departures can be without even being noticed by the listener. \autocite[210]{zuckerkandl_1973}
\end{quote}

In comparison, the discussion of time gains its ontological status in philosophy and natural science in the way time is often explicitly recognized and spelled out. Without loss of generality, we consider a few selected examples in the field of philosophy and physics to illustrate how time is explicitly spelled out.

\subsubsection{Philosophy: transcendental time determination}

As a philosophical figure, Kant is accredited for his formulation of transcendental idealism, which is considered by many to be an important milestone in the development of epistemology. One of the central concerns in transcendental idealism is about construing the adjective ``transcendental'' that prefixes a considerable portion of his terminology. For example, by observing the chapter outline of his magnum opus ``Critique of pure reason'', we readily find that the work is divided into two halves bearing the titles ``Transcendental Doctrine of Elements'' and ``Transcendental Doctrine of Method'', respectively. The first half is subsequently divided into two parts: ``Transcendental Aesthetic'' and ``Transcendental Logic''. Finally, the part of ``Transcendental Logic'' is further divided into ``Transcendental Analytic'' and ``Transcendental Dialectic''. Without regarding the specific meaning of each term, the sheer prevalence of the adjective prefix ``transcendental'' indicates its importance to understanding the theory as a whole. Therefore, it is necessary to discuss the term ``transcendental'' in the context of a Kantian framework, if one is to fully understand and appreciate its philosophical endeavor.

Whenever attaching the prefix ``transcendental'' in its Kantian sense to some terms, we invariably deal with the necessary and universal conditions through which the very existence of our cognitive experience is possible. In the introduction to the ``Critique of Pure Reason'', Kitcher interprets two terms that are transcendental-prefixed. Firstly, for transcendental philosophy, its goal is to

\begin{quote}
investigate the necessary conditions for knowledge with a view to showing that some of those necessary conditions are a priori, universal and necessary features of our knowledge, that derive from the mind's own ways of dealing with the data of the senses. \autocite[xxxi]{critique_of_pure_reason}
\end{quote}

Similarly, for the specific chapter on transcendental deduction, its goal is to

\begin{quote}
show that certain concepts that are a priori, in the sense that they cannot be derived from sensory data, are necessary for all cognition, and so are a priori in the sense that they describe universal and necessary features of all the objects of which we can ever have any knowledge. \autocite[xliii]{critique_of_pure_reason}
\end{quote}

Finally, in Kant's own words, he begins the introduction to first edition with a section on ``The Idea of Transcendental Philosophy'':

\begin{quote}
even among our experiences there is an admixture of cognitions that must originate a priori, and that serve perhaps only to give coherence to our presentations of the senses. For even if we remove from our experiences everything belonging to the senses, there still remain certain original concepts, and judgments generated from these, that must have arisen entirely a priori, independently of experience. These concepts and judgments must have arisen in this way because through them we can {[}\ldots{]} say more about the objects that appear to the senses than mere experience would teach us; and through them do assertions involve true universality and strict necessity, such as merely empirical cognition cannot supply. \autocite[44]{critique_of_pure_reason}
\end{quote}

By comparing the three excerpts above, one can summarize a few common qualities that a topic must enjoy in order to be considered transcendental. Firstly, it must deal with a priori concepts, i.e.~concepts that are independent from empirical observations. In order to make his claims specific, he stresses that they are ``not those that occur independently of this or that experience, but those that occur absolutely independently of all experience''. \autocite[45]{critique_of_pure_reason} Secondly, those concepts must be necessary (as opposed to contingent) and universal. Finally, they must be concepts through which our cognitive experience is made possible, i.e. being necessary conditions for the very existence of cognitive experience.

Therefore, Kant's philosophical ambition is to lay the epistemological foundation upon which we can have any cognitive experience at all. How transcendental idealism achieves such foundational work is to examine the mechanism through which our cognitive faculties enable and shape our experience. Consequently, all experiences (whether internal or external) as sensory data must necessarily and universally conform to the structure and form imposed by our cognitive faculties. In fact, the shift in focus is precisely what justifies Kant to compare ``his revolution in epistemology to the Copernican revo­lution in astronomy''. \autocite[xxxi]{critique_of_pure_reason} It follows that the epistemological foundation laid by transcendental idealism is also the foundation of essentially all sciences: it is through the structure and form of our internal cognition that we can possibly make claims about the external world.

Now, given the importance and ambition of transcendental idealism, it is noteworthy to observe the role of time within the framework of the theory. One may be surprised at how the issue of time has been elevated to a status of unrivaled prominence. To see how time is a topic in focus, we consider two aspects offered by the Critique.

The first aspect identifies time as ``the formal condition of inner sense''. \autocite[153]{critique_of_pure_reason} The very first part of the Critique, titled ``Transcendental Aesthetic'', consists solely of explicit discussion of space and time. In other words, Kant immediately brings into discussion the topic of space and time, thus marking them for consciousness. We may safely speculate that dedicating the entire opening portion of the book proves how space and time are highly esteemed topics. In fact, he makes the claim that ``transcendental aesthetic cannot contain more than these two elements, i.e., space and time''. \autocite[93]{critique_of_pure_reason} Being fundamental topics in his philosophical construction, he presents space and time as the two ``pure forms of sensible intuition''. \autocite[75]{critique_of_pure_reason} He claims that space and time are transcendental in the sense that they are non-empirical: they are forms without contents, and they necessarily and universally give rise to the possibility of empirical contents. For a better understanding, we can refer to the container metaphor mentioned on page \pageref{container_metaphor_of_time} once more: space and time are containers in which everything is arranged, related, organized and cognized. However, space and time themselves do not contribute to the content of experience. Therefore, the metaphor enables us to cognize the world in a bottom-up manner: the process starts with space and time without content that are analogous to an empty container. Then we provide space and time with empirical content to make it perceptible to us. What Kant does involves a similar thought experiment, except that it proceeds in a top-down manner. Our task is to ask: what are we left with, if we try to remove all empirical content from our experience? He continues that

\begin{quote}
if from the presentation of a body I separate what the understanding thinks in it, such as substance, force, divisibility, etc., and if I similarly separate from it what belongs to sensation in it, such as impenetrability, hardness, color, etc., I am still left with something from this empirical intuition, namely, extension and shape. These belong to pure intuition, which, even if there is no actual object of the senses or of sensation, has its place in the mind a priori, as a mere form of sensibility. \autocite[73]{critique_of_pure_reason}
\end{quote}

By comparing the way Kant argues for space and time, we may arrive at the conclusion that time is a pure form of sensible intuition that is more fundamental than space. At first glance, space and time appear complementary: space is ``nothing but the mere form of all appearances of outer senses'', \autocite[81]{critique_of_pure_reason} whereas time is ``nothing but the form of inner sense''. \autocite[88]{critique_of_pure_reason} Note that the parallelism in wording suggests that space and time are responsible for outer senses and inner sense, respectively. However, he then immediately clarifies his point by asserting that time is (indirectly) responsible for all outer senses as well:

\begin{quote}
Time is the formal a priori condition of all appearances generally. Space is the pure form of all outer appearances; as such it is limited, as a priori condition, to just outer appearances. But all presentations, whether or not they have outer things as their objects, do yet in themselves, as determinations of the mind, belong to our inner state; and this inner state is subject to the formal condition of inner intuition, and hence to the condition of time. Therefore time is an a priori condition of all appearance generally: it is the direct condition of inner appearances (of our souls), and precisely thereby also, indirectly, a condition of outer appearances. \autocite[88]{critique_of_pure_reason}
\end{quote}

This passage vividly suggests that time is more fundamental than space, as it is the formal condition of both inner and outer appearances, whereas space is limited to ``just outer appearances''. \autocite[88]{critique_of_pure_reason}

The second aspect identifies time as ``the transcendental time determination''. Later in the Critique, Kant embarks on an investigation of a puzzling concept: schematism. Schematism involves the study of the ``transcendental schema'', \autocite[211]{critique_of_pure_reason} which mediates between the ``category'' and the ``appearance''. \autocite[210]{critique_of_pure_reason} In particular, he examines how ``an object is subsumed under a concept''. \autocite[209]{critique_of_pure_reason} For example, in music analysis, we often subsume musical objects (which are technically sensory data, i.e.~appearance) that bear rondo-like properties under the concept (or category) of rondo. Kant sees a potential issue here: while subsuming object under concept produces no problem in most cases, it may produce problems in cases involving a special kind of concept, that is, pure concept of understanding. According to him, in subsuming an object ``under a concept, the presentation of the object must be homogeneous with the concept''. \autocite[209]{critique_of_pure_reason} However, pure concepts of understanding are special because they ``are quite heterogeneous from empirical intuitions''. \autocite[210]{critique_of_pure_reason} Therefore, the problem reads:

\begin{quote}
How, then, can an intuition be subsumed under a category, and hence how can a category be applied to appearances{[}\ldots{]}? \autocite[210]{critique_of_pure_reason}
\end{quote}

In order to solve the problem, the idea of a transcendental schema is devised. Metaphorically, it serves as a third-party broker, such that it is ``homogeneous with the category, on the one hand, and with the appearance, on the other hand''. \autocite[210]{critique_of_pure_reason} Without going into technical details of terminology, we readily see that transcendental schema is devised as a solution to an urgent philosophical problem: reconciling the experiential content of the world and our cognitive faculties. It is then natural to ask what the candidates of a transcendental schema are.

Now comes the surprising part: transcendental schema is nothing but transcendental time determination. It is surprising because the statement explicitly identifies time as the basis of all cognition and understanding, including those that are considered non-temporal. He argues:

\begin{quote}
Time, as the formal condition for the manifold of inner sense and hence for the connection of all presentations, contains an a priori manifold in pure intuition. Now, a transcendental time determination is homogeneous with the category (in which its unity consists) insofar as the time determination is universal and rests on an a priori rule. But it is homogeneous with appearance, on the other hand, insofar as every empirical presentation of the manifold contains time. Hence it will be possible for the category to be applied to appearances by means of the transcendental time determination, which, as the schema of the concepts of understanding, mediates the subsumption of appearances under the category. \autocite[211]{critique_of_pure_reason}
\end{quote}

Note that this is a strong claim: transcendental time determination is not \emph{a} candidate (among others) for transcendental schema, but \emph{the} ``schema of concepts of understanding''. \autocite[211]{critique_of_pure_reason} In short, the term transcendental schema is nothing but an alias for transcendental time determination. It follows that in his subsequent enumeration of transcendental schemata (i.e.~transcendental time determinations), time cries out for attention. For example, the schema of actuality ``is existence within a determinate time'', and the schema of necessity ``is the existence of an object at all time''. \autocite[217]{critique_of_pure_reason} Moreover,

\begin{quote}
the schema of magnitude, the production (synthesis) of time itself in the successive apprehension of an object; the schema of quality, the synthesis of sensation (perception) with the presentation of time--or, i.e., the filling of time; the schema of relation, the relation of perceptions among one another at all time (i.e., according to a rule of time determination); finally, the schema of modality and of its categories, time itself as the correlate of the determination of an object as to whether and how it belongs to time. Hence the schemata are nothing but a priori time determinations according to rules; and these rules, according to the order of the categories, deal with the time series, the time content, the time order, and finally the time sum total in regard to all possible objects.\autocite[217]{critique_of_pure_reason}
\end{quote}

\subsubsection{Ubiquitous time parameter: a brief survey of physics with respect to the role of time}

Among disciplines of natural science, physics stands as one of the major fields of study. In fact, one might go as far as to claim that physics is the most fundamental discipline of natural science. For example, physicist Feynman once made a remark that might cause elevated debate due to its alleged implication of condescension:

\begin{quote}
Physics is the most fundamental and all-inclusive of the sciences, and has had a profound effect on all scientific development. In fact, physics is the present-day equivalent of what used to be called natural philosophy, from which most of our modern sciences arose. Students of many fields find themselves studying physics because of the basic role it plays in all phenomena.\autocite{feynman_leighton_sands_2013}
\end{quote}

Within the taxonomy of physics, one major branch is mechanics. In the narrow and etymological sense of the word, mechanics concerns the ``motion of material bodies'',\autocite[1]{classical_mechanics2002} i.e.~change in spatial position of material bodies in time. However, in its general sense, it concerns time evolution: change in states of physical systems in time. One should note that time evolution is a generalization of motion (or inversely, motion is a special case of time evolution), as motion of material bodies is precisely change in positional states in time. However, we may be interested in issues other than positional state. For example, in quantum mechanics, we are interested in the time evolution of wavefunctions. Therefore, taking time evolution as the general definition of mechanics enables us to subsume under the term a variety of theories, as what constitute ``states'' and ``physical systems'' is open to interpretation.

Now comes the crucial observation: the role of time remains prominent throughout the development of mechanics. The claim runs parallel to the previous section, that time is explicitly spelled out in the field of physics. In particular, it has very specific meaning in physics: the time parameter is, in the most literal sense, spelled out in writing down equations of motion.

Poetically speaking, the concept of time remains ubiquitous in mechanics. Ubiquity of time implies two statements: 1) time remains to be a fundamental concept in all formalisms, surviving through theoretical developments; 2) within debates on interpreting physical formalisms, time is a topic that often gains consensus among different parties. We will show what the two statements mean in light of the three examples of formalisms: Newtonian mechanics, Lagrangian formulation, and quantum mechanics. \footnote{See appendix \ref{appendix:evolution of equations of motion} for supplementary discussion of the three examples.}

Throughout the history of mechanics, people argue and debate about representations of the world, and formulate various formalisms that try to capture different aspects of it. In particular, they often argue about ways of encoding states of physical systems: how should a theory represent and specify the instantaneous state of a physical system? Newtonian mechanics and Lagrangian formulation (i.e.~classical mechanics) use position (hence velocity as the time derivative of position) as the conceptual basis, while quantum mechanics uses wavefunctions. Now, one must realize that no matter how the concepts of position and wavefunctions differ, they are nothing but mathematical functions parametrized by time. Therefore, people seem to reach consensus on the interpretation of time. It is true that one might argue about the rate at which time flows, but everyone must agree that time evolution of physical state depends explicitly on time. The dependency on time is best illustrated in notation, where the parameter \(t\) representing time is visible in all mathematical representations of physical state. Similarly, the equations in their general form governing the time evolution (e.g.~equations of motion) of physical state are all differential equations containing time-derivatives. Now, even without knowing the semantics of the following equations or the meaning of the term ``time-derivative'', one can readily see the presence of letter \(t\):

\begin{center}
    \begin{tabular}{|c|c|c|}
        \hline
        Newton's second law & Euler-Lagrange equation & Schrodinger's equation \\
        \(F = m\frac{dv}{dt}\) & \(\frac{\partial L}{\partial x} = \frac{\mathrm{d}}{\mathrm{d} t}\left(\frac{\partial L}{\partial \dot{x}}\right)\) & \(i\hbar \frac{\partial \Psi}{\partial t} = H\Psi\)\\
        \hline
    \end{tabular}
    \label{equations of motion}
\end{center}

Simply from the syntactical appearance of these equations, one can observe that symbols differ as fundamental concepts of different theories vary. However, the symbol of letter \(t\) that is reserved for time is persistent across all equations. Time stands the test of time (no pun intended).

One can be provocative and claim that time is a completely imaginary concept. Nevertheless, in the end of the day, one has to deal with the letter \(t\) in practice. The ubiquity of time manifests itself in the painstaking process of solving the differential equations. In contrast, we should note how easily (or even necessary) musicians can bypass time altogether. Instead of thinking about issues of timing directly, performers are encouraged to tackle them using techniques of breathing, bodily gestures and pictorial imaginations. In instrumental lessons, we are unlikely to hear instructors say: ``don't rush at the rest! In fact, the timing of the rest should be linearly proportional to the magnitude of your musical intensity just now. Also note that the musical intensity is also a function depending implicitly on time''. It is more common to hear the following hypnotic-style argument: ``don't rush at the rest! Now, close your eyes and take a deep breath. You should imagine that you are on a field trip, and you see the expansive landscape. You can even hear the birds sing! Do you want this quiet moment to last longer, or you just want to go home and sleep? Of course you want to take time here!''

\section{A justification for ignoring temporality: musical time}

One potential explanation for avoiding explicit discussion of time in music is that, instead of discussing time as a concept in itself, we often focus on its derivatives: constructs built on top of time that describe experienced time, as opposed to objective time. Music, as essentially a study of human expressions which just happens to use the medium of time, prioritizes time in its wrapped form: experienced time. The statement has two layers of meaning.

The first layer pertains to the idea that musicians differentiate between physical time and musical time. For example, rhythm and meter are constructs built on top of the metronomic notion of physical time. However, they are not interchangeable, as the metronome controversy discussed on page \pageref{metronome_controversy} indicates. To this end, Zuckerkandl dedicates a new term ``metric wave'' to describe musical rhythm. He then used this newly coined term to explain what it means to have musical equality of time, given that durational equality of time is an ill-defined idea in music:

\begin{quote}
To play in time musically does not mean to play tones that fill equal lengths of time, but tones that give rise to the metric wave. \autocite[210]{zuckerkandl_1973}
\end{quote}

The exact meaning of the metric wave is open to interpretation. It is possible that the metric wave is more of a poetic metaphor than a term to be taken literally. However, the sheer fact that the author designates a unique term shows the necessity of describing rhythm in a way that is exclusive and specialized to music.

The second layer pertains to the idea that musicians differentiate between temporality and atemporality in music, which eventually abstracts away the topic of psychological time altogether. More specifically, given the musical notions of rhythm and meter, we can construct derivative concepts: harmonic rhythm and hypermeter (more generally, large-scale rhythmic organization). At this point, we observe an interesting shift of focus. Harmonic rhythm primarily focuses on harmonic organization. However, harmony (in particular, tonality) itself is atemporal because it prioritizes pitch relations. By the same token, hypermeter focuses on grouping structures. The question remains to discover what the grouping criteria are. In particular, the question addresses ways in which we determine hypermetrical boundaries. As we move from meter to hypermeter, i.e.~from microscopic metrical structures such as a phrase, to macroscopic hypermetrical structures, the discussion inevitably incorporates additional considerations such as harmonic progression and melodic contour due to the increased complexity. For example, establishing a metrical pattern may simply require durational information (i.e.~knowing the onset and duration of each note-producing event) as well as description of dynamics using strong and weak beats, while disregarding other information such as pitch content, let alone harmonic progression and melodic contour. However, determining a hypermetrical pattern (or more generally, large-scale musical pattern spanning an extended period of time) requires much more information. For example, how do we decide if a moment is a so-called ``structural downbeat''? What does it mean to have a hypermetrical strong beat? Durational and dynamic information alone will not suffice to explain. To arrive at a decision, one must necessarily address additional concerns such as the tonal and motivic scheme of the piece. It is for this cause, the very term of hypermeter stirs up debates and confusions in the theoretical literature. Yust, in his book ``Organized time", points out that ``the concept of hypermeter seems to change its colors depending on the analytical situation, coming to mean different things to different people''. \autocite[123]{yust_2018} For example, many authors treat hypermeter and phrase structure interchangeably. As indicated by Krebs, Lester ``appears to equate hypermeasures and phrases''. \autocite[84]{equate_hypermeasures_phrases} More interestingly, as observed by Smyth, some authors arrive at self-contradictory conclusions about hypermeter:

\begin{quote}
Schachter blurs the distinction between a ``group of measures'' (grouped by virtue of similar surface rhythms and accentual patterning) and a ``phrase'' (a musical segment ending with a cadence). Failing to retain the crucial distinction he drew in his first article between durational and tonal rhythm, he (like Berry) effectively turns the phrase into a hypermeasure. \autocite[82]{hypermeter_self_contradiction}
\end{quote}

One should not dismiss the blurring boundary between phrase and hypermeasure as failures authors commit. Instead, to think positively, the blurring boundary reveals a praiseworthy quality of musical organization: variety. As we move toward large-scale musical structures, i.e.~structures spanning an extended period of time, we necessarily have to consider more musical factors contributing to their analysis and interpretation. The increased complexity of music's internal organization corresponds exactly to the notion of musical variety that is praised and valued throughout music history. For example, Zarlino brought up the idea of variety \label{musical_variety} as early as 1558, noting the importance of ``variety in the movement of the parts and in the harmony; for harmony is nothing other than diversity of moving parts and consonances, brought together with variety''. \autocite[52]{zarlino_1976} Later on, Niedt spelled out the importance of musical variety by dedicating an entire chapter titled ``On the Necessity and Grace of Variation in General'', echoing how variety was cherished during the Baroque period:

\begin{quote}
nothing in human life can be more pleasant and necessary than variety, in artistic as well as in natural things. Were it not for summer and winter, sowing and harvesting, frost and heat, day and night, and so forth, what creature would be able to endure this mortal life? {[}\ldots{]} Indeed, the ear knows of no greater pleasure than in the variety of many tones, songs, and melodies. {[}\ldots{]} the greatest charm rests in Variation, whether it be performed by the human voice or by various instruments. \autocite[73]{niedt1989}
\end{quote}

In order to see how the importance of variety is style-agnostic, consider a more recent figure. Schoenberg treated the dichotomy between variation and repetition as the determinant for effective composition, thereby highlighting the role of variety in music:

\begin{quote}
A motive appears constantly throughout a piece: it is repeated. Repetition alone often gives rise to monotony. Monotony can only be overcome by variation. {[}\ldots{]} Variation means change. But changing every feature produces something foreign, incoherent, illogical. It destroys the basic shape of the motive. {[}\ldots{]} Accordingly, variation requires changing some of the less-important features and preserving some of the more-important ones. \autocite[8]{schoenberg1970}
\end{quote}

The term ``increased complexity'' above refers to the increasing number of possible interactions between musical events. Simply put, large-scale musical structures provide more ways in which musical events can be defined. Increasing number of well-defined musical events then enables analysts to examine a richer collection of musical interactions. For example, in discussion of form, we inevitably examine large-scale musical events bearing different names. Sometimes formal sections bear generic letter names such as ``AAB'' for bar form, while on other occasions formal sections acquire dedicated names indicating their functions such as recapitulation in sonata form. Then, we are entitled to speak about the ``recapitulating event'', referring to either the moment where recapitulation section begins (i.e.~timepoint) or the recapitulation section itself (i.e.~timespan). With either interpretation, we may now relate the recapitulating event to other musical events in the piece. For example, we can ask conventional questions about the relationship between recapitulation and development: how does the retransition at the end of the development lead the music into the recapitulation, creating a strong sense of return? Furthermore, we can be less conventional by asking questions that are more creative. It is a perfectly valid question, for example, to ask: how does the development section of the piece as a whole relate to the opening three-note motive? It is a creative question precisely because we are now inspecting relationship between musical events across different organizational levels. Using the language of Schenkerian graphs, we are connecting elements between a foreground graph and a background graph.

\subsection{A musical case study: Variations and Fugue on a Theme by Handel, Op. 24}
\label{section:musical case study op24}

\begin{figure}[!h]
   \centering
   \includegraphics[width=.5\textwidth]{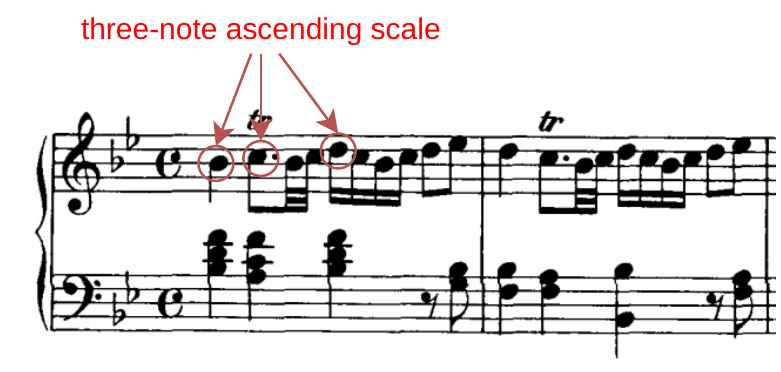}
   \caption{Brahms, Variations and Fugue on a Theme by Handel, Op. 24, beginning, annotated to indicate the three-note ascending scale}
   \label{fig:brahms op24 beginning annotated three-note}
\end{figure}

To show how the creative question above is at work in music, consider the following example. The opening theme of Brahm's ``Variations and Fugue on a Theme by Handel, Op. 24'' can be characterized locally (i.e.~focusing solely on the melodic segment) by its three-note ascending scale figure: B-flat, C, and D (see figure \ref{fig:brahms op24 beginning annotated three-note}).

\begin{figure}[!h]
   \centering
   \includegraphics[width=.8\textwidth]{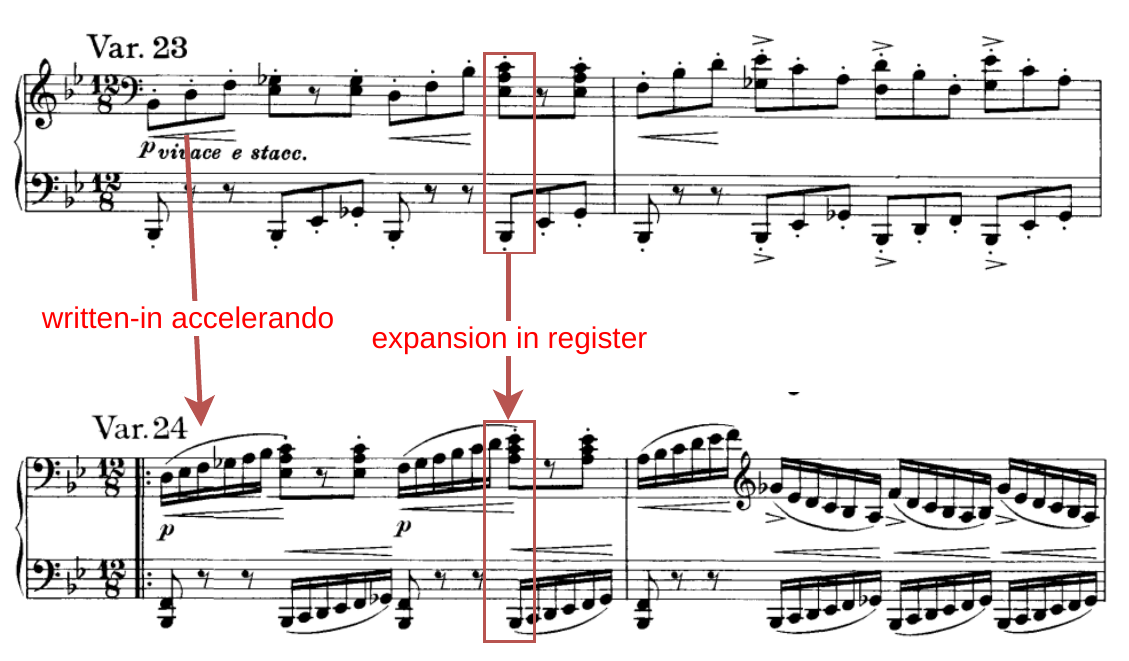}
   \caption{Brahms, Variations and Fugue on a Theme by Handel, Op. 24, variation 23 and 24, side by side comparison of the beginning}
   \label{fig:brahms op24 var23 24}
\end{figure}

Our task is to see what this local structure of the three-note ascending scale reveals about musical events occurring at a larger level. Therefore, we look at the distant future: the last variation before the final fugue (i.e.~variation 25). The reason for choosing this specific excerpt is not arbitrary. We make the choice due to its musical significance. In fact, one may even claim that variation 25, in the context of its large-scale formal organization, is more significant than the prolonged fugue that follows. The justification resides in the preceding two variations (variation 23 and 24). From a performer's perspective, variation 25 sounds like a major arrival point, precisely because it is well prepared and anticipated by the preceding two variations. Variation 23 and 24 are unique among other variations, in that they can be treated as a single variation, forming a two-variation long build-up leading to variation 25. In fact, for performers, the two variations as a whole make up a giant creascendo: variation 24 is the intensified version of variation 23 in the sense that it is equipped with a written-in accelerando (i.e.~replacing eighth notes by sixteenth notes) as well as an expansion in register (see figure \ref{fig:brahms op24 var23 24}). It is for this reason, it would be absolutely inappropriate to take time in between variation 23 and 24, as one normally would not interrupt a crescendo.

Given the importance of variation 25, it follows that one should examine its internal structure. For performers, the task is vital, since one must find justified ways to bring out the climax and convince the audience musically (as opposed to analysts who can rely on verbal means). The variation, consisting of eight measures, has three structural moments. One might ask for the precise definition of structural moments. For now, we may safely take a performer's perspective and define it operationally: moments are structural if they are less fault-tolerant. Such an operational definition of structural moments is self-evident if we apply it to daily life. Moments in life are important (i.e.~structural) if they are events that we, using everyday language, cannot afford to screw up. As a result, a DMA entrance exam at Juilliard is a structural moment when compared to, say, practicing the 86th measure of some Beethoven sonata in room 481f.

\begin{figure}[!h]
   \centering
   \includegraphics[width=.8\textwidth]{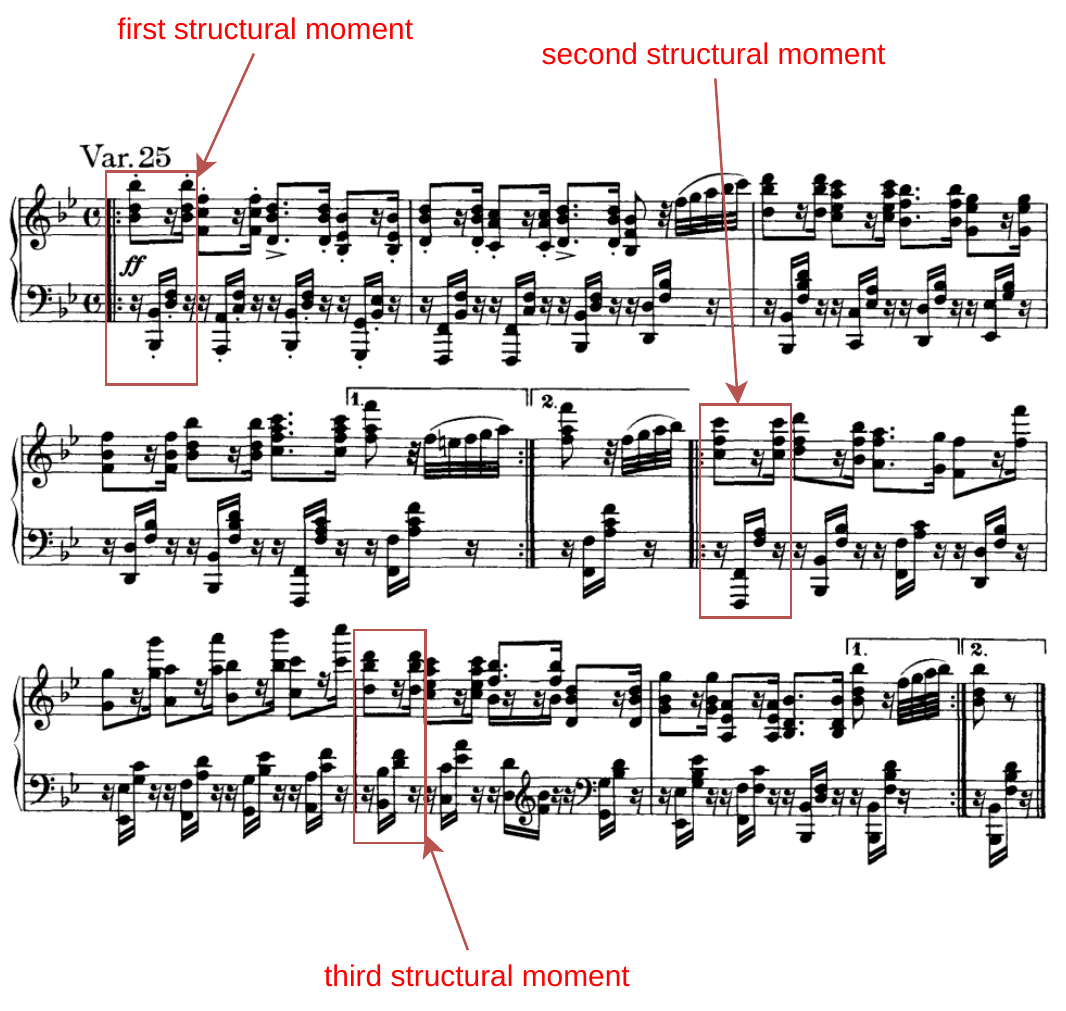}
   \caption{Brahms, Variations and Fugue on a Theme by Handel, Op. 24, variation 25, annotated to indicate the three structural moments}
   \label{fig:brahms op24 var25 annotated}
\end{figure}

Musically, performers can then evaluate the following question: in variation 25, what are the moments in which we are least willing to make a mistake? Interpreting the question positively, what are the moments we must bring out perfectly in order to be satisfied with the performance? The three candidates for such moments are: the beginning of the first four-measure phrase, the beginning of the second four-measure phrase, and the downbeat of the penultimate measure (see figure \ref{fig:brahms op24 var25 annotated}). The first two choices are straightforward because they correspond to phrase boundaries, performers need to clearly present them in order to make the audience aware of groupings that are vital to musical understanding. Now, the third is a structural moment because, just like variation 25 itself, it is well-prepared by the measure before: a virtuoso measure that is characterized by ascending scales and outrageous octave leaps, forming a bursting crescendo that finally arrives at the downbeat of the penultimate measure.

Note now that the three structural moments of variation melodically delineate a three-note ascending scale figure: B-flat, C and D. The figure is identical to the opening three-note figure aforementioned. Poetically speaking, the opening figure organically grows through time, and eventually becomes the Leviathan who is capable of causing the climax of the piece, the same way it stirs up gigantic waves in the sea (taking its biblical definition) or in society (in the Hobbesian sense).

\begin{figure}[!h]
   \centering
   \includegraphics[width=.5\textwidth]{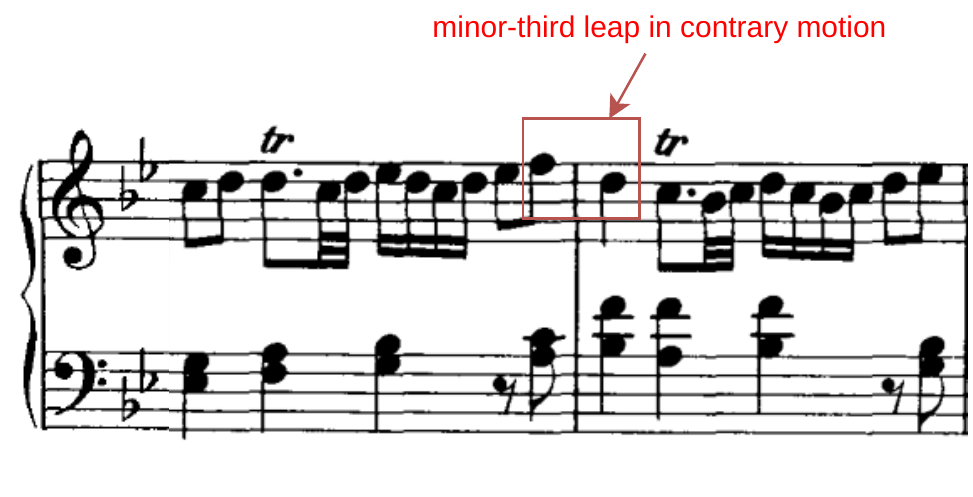}
   \caption{Brahms, Variations and Fugue on a Theme by Handel, Op. 24, theme, excerpt, annotated to indicate the minor-third leap in contrary motion at the penultimate measure}
   \label{fig:brahms op24 theme minor-third leap}
\end{figure}

The projection of the opening three-note figure into variation 25 is reinforced if we incorporate a larger context by considering preceding variations. A common argument is that one does not have to find variation that far from the opening theme to establish parallelism. For example, the theme (the first eight measures of the piece) itself also possesses three structural moments identical to variation 25. However, notice that the statement above is not true: for the theme, we are not justified to assign the same structural moments as those of variation 25, namely, downbeats of measure 1, 4 and 7. In particular, the downbeat to the penultimate measure in the theme is not a structural moment. In fact, it is intentionally masked if we examine the preceding measure: its melody ascends linearly to F. Then the voice is abruptly and haphazardly cut off by the minor-third leap, in contrary motion, downward to D (see figure \ref{fig:brahms op24 theme minor-third leap}).

\begin{figure}[!h]
   \centering
   \includegraphics[width=.8\textwidth]{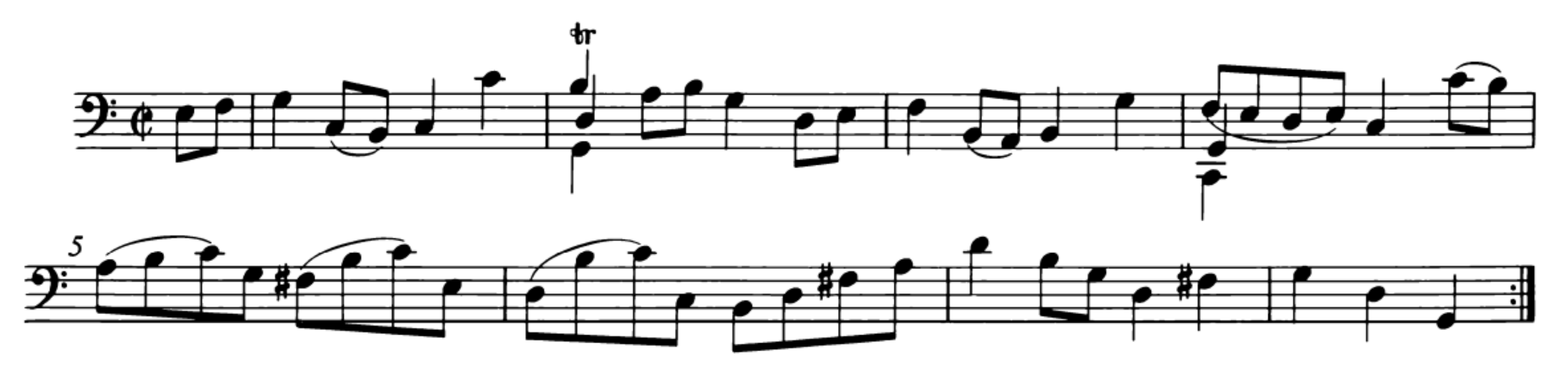}
   \caption{Bach, Cello Suite No.3 in C major, BWV 1009, Bourree I, opening}
   \label{fig:bach cello suite 3}
\end{figure}

Because of the leap, we get a sense of interruption and discontinuity. Lasser's theory of contrapuntal voices may help explain why the leap conveys a sense of interruption. In his ``The spiraling tapestry'', Lasser proposes a musical structure termed ``contrapuntal voice''. A seemingly monophonic voice of melody may be decomposed into possibly many ``contrapuntal voices'',\autocite[8]{lasser2008} thereby creating ``\,`single-line' counterpoint''\autocite[8]{lasser2008} within a monophonic texture. In Lasser's words, a melody ``which we normally consider to be a single voice, is in fact made up of a multiplicity of Contrapuntal Voices engaged in counterpoint with each other within the melody''.\autocite[8]{lasser2008} By proposing the idea of contrapuntal voices, we are equipped with better analytical tools in analysis of melodic contours. Specifically, we are able to understand monophonic melody from a more contrapuntal and polyphonic perspective. In practice, some instruments have limited capability in performing polyphonic texture. For example, it is physically impossible for a string instrument to play five moving voices simultaneously (whereas such task is simple for keyboard instruments such as piano and organ). However, it would be musically untrue to claim that compositions for string instruments are incapable of carrying out counterpoint. From a listener's perspective, we vividly recognize, through our musical instinct, that in a piece for solo string instrument, there are multiple events happening concurrently. Figuratively, we may imagine that a monophonic voice played by the string instrument is capable of conveying a theatrical sense of conversation between voices, personified as characters. For example, we may consider Bach's third cello suite (see figure \ref{fig:bach cello suite 3}).

For the example, we are justified to assume that Bach invariably carries over his polyphonic compositional style and method into this work for solo cello. However, due to the physical constraints of the cello as an instrument, he has to flatten the polyphonic texture into a monophonic one. Therefore, our task is to restore polyphony (because the piece is by a polyphonic composer) from monophony (because cello is inherently limited). For such a task, Lasser's formulation of contrapuntal voices provides insights into identifying and interpreting contrapuntal voices necessary for restoring polyphonic understanding of a monophonic voice.

For identification of contrapuntal voices, Lasser points out that the crux lies in the distinction between leap and step in a monophonic texture:

\begin{quote}
When discussing melodic contour, we habitually treat conjunct {[}i.e.~stepwise{]} and disjunct {[}i.e.~leap{]} motion as two distinct but nonetheless equal ways in which adjacent pitches can relate in a melody.{[}\ldots{]} Though visually, we see notes move by step or by leap, to the ear, notes connect \emph{only} by step, not by leap.\autocite[7]{lasser2008}
\end{quote}

Thus, the life cycle of contrapuntal voices is controlled by the melodic motion classified into step and leap motions. In particular, leap motion creates a new contrapuntal voice:

\begin{quote}
When a ``leap'' appears on the musical surface, it is aurally understood as the cessation of one Contrapuntal Voice and the beginning of another Contrapuntal Voice. Defined in this way, leaps or disjunct motion, disappear from the actual experience of a melody {[}\ldots{]}\autocite[8]{lasser2008}
\end{quote}

Applying the idea of contrapuntal voices to our analysis of figure \ref{fig:brahms op24 theme minor-third leap}, by moving the melody in linear motion up to F, listeners are following closely on the contrapuntal voice. However, the leap effectively breaks the contrapuntal voice and creates a new contrapuntal voice. Therefore, the D on the downbeat, being the beginning of the new contrapuntal voice, cannot serve the role of arrival point since no event comes before it. The leap causes the D on the downbeat to behave drastically differently from that in variation 25. In variation 25, the D is prepared using a leap-free linear motion, which gives it the role of inevitable arrival. The minor-third leap in this case, thus breaks the three-note figure that is the basis  for establishing parallelism.

\begin{figure}[!h]
   \centering
   \includegraphics[width=.8\textwidth]{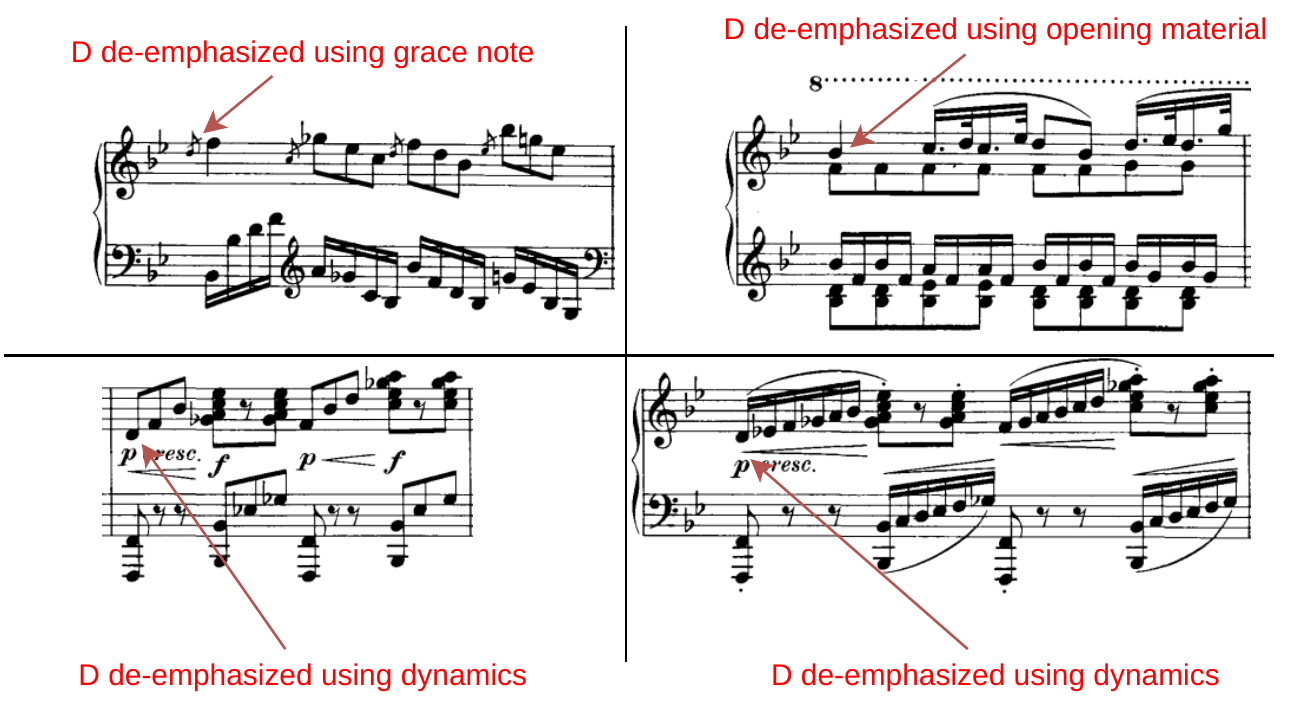}
   \caption{Brahms, Variations and Fugue on a Theme by Handel, Op. 24, variation 21 (top-left), 22 (top-right), 23 (bottom-left), 24 (bottom-right), side by side comparison of respective penultimate measures, annotated to illustrate methods of de-emphasizing the downbeat of the penultimate measure}
   \label{fig:brahms op24 var21 22 23 24}
\end{figure}

By the same token, we find that in variations leading to variation 25, the downbeat on the penultimate measure is often evaded. It is then possible to speculate that the evasion is intentional, with purposes similar to evaded cadence: by leaving the tension hanging and unresolved, the music can then delay its resolution, which effectively strengthens the resolution when it finally arrives. To see how evasion is achieved, we start our analysis from variation 21. In variation 21, the downbeat D in the penultimate measure is evaded by the trick of grace notes: D is technically the beginning of the measure, yet turning it into grace note successfully shifts the real downbeat to F. Similar technique happens with variation 22. On the penultimate measure, the music experiences a sudden pullback in the sense that the music is being reset to the beginning of the variation in the literal sense: the first half of the penultimate measure is identical to that of the first measure of variation 22. Now for variations 23 and 24, D finally becomes the downbeat of the penultimate measure. However, it is evaded again due to figuration: the scale with crescendo on the first and third beat causes the motion to be directed to the second and fourth beat of each measure (see figure \ref{fig:brahms op24 var21 22 23 24}).

The musical case study above illustrates how musical structures across different organizational levels may interact, contributing to the increased complexity as musical time climbs the ladder of abstraction. In fact, it also answers a potential objection. One may object by claiming that concepts of musical time at a higher organizational level does not lead to increased complexity. On the contrary, it leads to simplification. The claim is that large-scale musical structures serve as reductions that in fact simplify analysis. Consider the example of Schenkerian graphs: as we traverse from foreground graphs to background graphs, we are essentially in a process of moving from a low organizational level (i.e.~surface level that includes all musical events) to a high organizational level (i.e.~level that includes only major events). In other words, we are climbing the ``ladder of abstraction'' precisely in its originating and etymological sense: the linguistic concept, introduced by Hayakawa, that words subsume ideas of ``greater generality and applicability'' \autocite[159]{hayakawa1990language} as we metaphorically climb the abstraction ladder. One may then argue that such process is reductional, as it terminates with a background graph consisting of relatively few major events (i.e.~the Ursatz) because many of the details (i.e.~elaborations) are being omitted during the process. What we eventually obtain from the process is a simplified representation consisting of number of notes (for example, six for a typical Ursatz) that is orders of magnitudes lower than its foreground representation (by literally counting number of notes that appear on the printed score). Similarly, from a formal perspective, we observe that in analysis, a piece is often reduced to its formal components. For example, analysts subsume a variety of pieces into the catalog of rounded-binary form, ignoring pitch contents specific to each piece. How can a simplified representation reconcile with the idea that large-scale structures introduce additional complexity?

To counter such potential objection, one must recognize how multiple organizational levels introduce more possibilities for cross-level interactions. It is true that the background level of a Schenkerian graph is simple, which mostly contains just the Ursatz itself. However, the essential part of the analysis is to see how different levels interact. In particular, in Schenkerian analysis, we are entitled to say that its value resides in comparing graphs side by side, showing how events across different organizational levels inform each other. It is for this reason that by convention, Schenkerian graphs are often notationally presented in a way that they are aligned vertically, in order to highlight the reductional process, rather than the individual graphs (e.g.~musical illustrations found in the second volume of Salzer's ``Structural Hearing''\autocite{salzer_1952_2}, see figure \ref{fig:salzer schenkerian graphs} on page \pageref{fig:salzer schenkerian graphs}). In other words, it is through the relationship between Schenkerian graphs representing different organizational levels, that the theory attains its explanatory power. Similarly, knowing the three formal divisions of the sonata-allegro form alone is pointless, because the real musical value concerns how these formal divisions are brought to life through composers' crafts, which requires us to closely examine how musical details relate to the formal divisions.

\begin{figure}[!h]
   \centering
   \includegraphics[width=.8\textwidth]{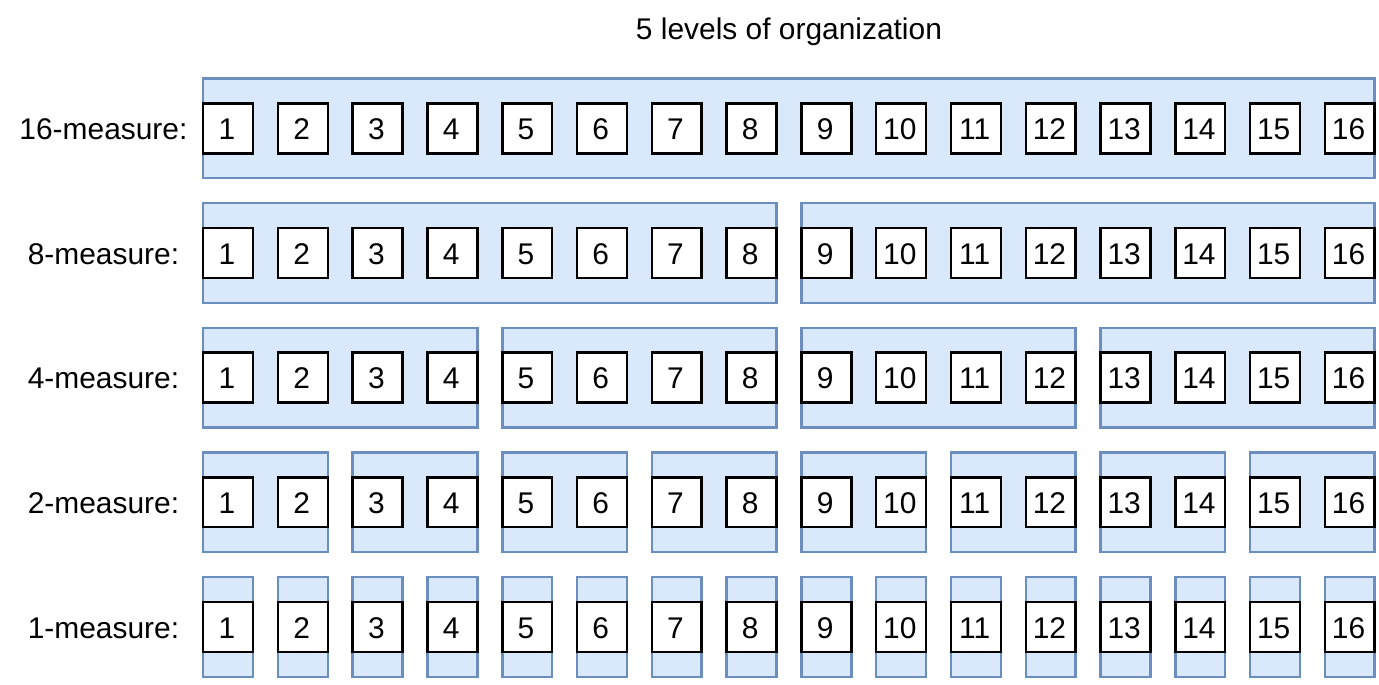}
   \caption{Diagram showing 5 levels of organization as a result of recursively dividing a 16-measure passage by a factor of 2}
   \label{fig:16-measure grouping}
\end{figure}

In order to illustrate how interactions between multiple organizational levels introduce more musical possibilities, we present a simple combinatorial argument here. Consider a 16-measure metrical model that is recursively divided into groups of two, giving us a collection of the following musical groupings (see figure \ref{fig:16-measure grouping}): a 16-measure phrase group, two eight-measure phrases, four four-measure subphrases, eight two-measure phrase segments, and 16 one-measure measures. The collection consists of five levels of organization: 16-measure level, eight-measure level, etc. We can do analysis with this metrical model in three different ways. For this toy example, what constitutes analysis here is simply counting pairwise relations within the collection of musical groupings enumerated above. For this example, we focus on unordered pairs, in the sense that the relationship between phrase A and phrase B is the same as that between phrase B and phrase A. Combinatorics gives us useful mathematical tools for counting, making this example a combinatorial argument.

The first approach is what can be called a forgetful simplification. With this approach, we are forgetful such that when moving to a higher level of organization, we forget about lower levels of organization. Applying to Schenkerian analysis, this is to say that we keep only the background Schenkerian graph, and feed all other graphs (i.e.~foreground and middle ground graphs) into the paper shredder. Therefore, we start with the 16-measure metrical model, and through the analysis, are left with only one thing, namely, the 16-measure phrase group. Everything at a lower level of organization (e.g.~eight-measure phrase) is lost. With this approach, the analysis is not really informative because the number of possible pairwise relations is zero (assuming that we disregard self-relating pairs).

The second approach is more telling. The analysis is no longer forgetful, because we realize that every organizational level is born equal: a four-measure subphrase is not inferior to a 16-measure phrase group. Therefore, by principle of indifference, we should not forget about four-measure subphrases in favor of the 16-measure phrase group. However, at this stage, we are still reluctant to consider pairwise relations across levels of organization because we think such relations cause categorical error: we are comparing apples and oranges. In this case, how many pairwise relations can we produce? Introducing tools from combinatorics, we have a function \(\binom{n}{r}\) that gives us the number of possible ways to select \(r\) items from a collection of \(n\) items where the order of selection does not matter. With the introduced tool, we can count now:

\[
\binom{16}{2} + \binom{8}{2} + \binom{4}{2} + \binom{2}{2} = 155
\]

The third approach takes into consideration pairwise relations across levels of organization. Suppose now, after reading the musical case study presented in section \ref{section:musical case study op24}, we realize that comparing a four-measure subphrase to an 16-measure phrase group is not comparing apples and oranges. Instead, such comparison might be musically insightful. In such a case, how many pairwise relations can we produce? The answer is given by:

\[
\binom{16 + 8 + 4 + 2 + 1}{2} = 465
\]

By comparing the numbers, we can readily see what it means to have more possibilities. It literally refers to more number of pairwise relations through counting.

Therefore, we arrive at a seemingly counterintuitive conclusion: reduction of musical time is not about simplifying analysis. Rather, it makes us better appreciate the complexity of music as an inherently ``hierarchical and temporal" \autocite[6]{yust_2018} construct. In fact, Salzer arrived at the same conclusion by stressing the interdependence and symbiosis between structure and prolongation. On the one hand, he points out that the central concern of Schenkerian analysis is the ``distinction between structure and prolongation''. \autocite[13]{salzer_1952} Meanwhile, he warned the reader with a beautifully written passage:

\begin{quote}
It is wrong to assume, however, that finding the structural framework constitutes the sole purpose of this [Schenkerian] approach. On the contrary, structural hearing implies much more. It enables us to listen to a work musically, because by grasping the structural outline of a piece we feel the full impact of its prolongation which are the flesh and blood of a composition. Thus the structural outline or framework represents the fundamental motion to the goal; it shows the direct, the shortest way to this goal. The whole interest and tension of a piece consists in the expansions, modifications, detours and elaborations of this basic direction, and these we call the prolongations. \autocite[14]{salzer_1952}
\end{quote}

Performers are in a position to resonate with the above statement. As one studies a piece of music in order to present it for on-stage performance, the symbiosis is particularly evident. While the knowledge of formal structure is necessary for a better musical understanding, it is through elaborations in the Schenkerian sense that music manifests its manifold of expressive variety aforementioned (see page \pageref{musical_variety}), bringing into life its ``flesh and blood''. \autocite[14]{salzer_1952} Knowing that a piece conforms to the formal schema of exposition-development-recapitulation is informative, yet the true musical interest rests in how it carries out the schema, making it stand as a unique work of art among the oeuvre of all works. In short, realization and (more importantly) deviation of formal schema in practice (in terms of both composition and performance) are the force behind the expressive power listeners readily feel. Poetically speaking, if one is to say that the value of music analysis by reduction is not the end result, but the very process itself, then for performers, the process of preparing a piece for performance becomes a narrative itself. In fact, it is a metanarrative whereby performers produce a series of analyses (e.g.~Schenkerian graphs) in order to become intimately familiar with the piece. Most importantly, the significance of such metanarrative is precisely the very personal experience of living with the music, along with all the joy and frustration associated that take place off-stage. When compared to the totality of this metanarrative, its end result, i.e.~a successful performance of the piece, becomes less significant.

\section{Summary}

Throughout this chapter, we focused on a particular statement: time is often overlooked in music. The way we elucidated the statement was to interpret it using multiples perspectives. With the first perspective, we explored how within the field of music, time receives relatively little attention. With the second perspective, we explored how explicit discussion of time is often deliberately avoided in music. To highlight the deliberate avoidance, we contrast the treatment of time in music to that in other disciplines where time as an essential concept is often spelled out. Finally, with the third perspective, we attempt to justify the observation that time is overlooked in music by proposing that the explicit notion of time has been abstracted away as we focus on increasingly abstract conceptions of music that are based on notions such as musical phrase and musical variety. Therefore, the focus of musical discourse has shifted away from time.

With the statement that time is overlooked in music, the natural necessity of looking elsewhere arises. After all, we may argue that if a topic in a field is overlooked, then it is also likely that the field would lack terminologies and methodologies required to address the topic. On the other hand, disciplines with explicit focus on the topic of time are likely to inform and inspire us on the topic. One natural place to continue the current discussion is the disciplines where the notion of temporal asymmetry acquires its etymological origin, to which we should refer even for the sake of faithfully introducing the concept. Therefore, we examined the topic of temporal asymmetry as explicated in the fields of science and philosophy, which will be our focus of chapter \ref{chap:time_arrow_science}.

\chapter{The arrow of time in science and philosophy}
\label{chap:time_arrow_science}

Recall that in section \ref{section:further_decomposition}, the notion that time flows implies two separate statements. The first states that the past and the future are distinguishable. The second states that they are distinguishable in a unique way such that we are entitled to say that time necessarily flows from the past to the future. We should note that the first statement is a necessary condition for the second statement, but the converse is not true: it is possible to have distinguishability of the past and the future, yet time flows in an opposite direction (if the term ``opposite'' means anything at all).

This chapter is dedicated to the discussion of these two statements from the perspective of philosophy and science. In particular, in section \ref{sec:record}, we focus on the philosophical perspective of the distinguishability (i.e.~the first statement) between the past and the future, with emphasis on epistemology. In section \ref{section:arrow of time}, we focus on the scientific perspective of the arrow of time (i.e.~the second statement), with emphasis on thermodynamics.

\section{Record and trace}
\label{sec:record}

We begin by considering a simple musical question: how is the beginning distinguished from the ending? The question may appear silly at first because everyone knows its answer, the same way everyone knows what time is according to Augustine (see page \pageref{augustine_time}). But if we continue along the way Augustine argues, we then ask: do we \emph{really} know their distinction? We should acknowledge that this simple musical question is real as it is what musicians must seriously consider in practice. The beginning and the ending of a piece of music must be constructed differently and under different set of premises. For example, more specifically, how is an introduction different from a coda? Knowing the difference between the two directly affects one's interpretive decisions. Musicians may give various creative answers. For example, one may argue that the introduction is more anticipatory in function, whereas coda is more conclusive. Meanwhile, others may argue that a coda can also be anticipatory if we take into consideration that it may serve as the transition to the next movement.

\subsection{A motivational musical case study: introduction and coda in the first movement of Beethoven's piano sonata, Op. 111}

\label{sec:op111}

\begin{figure}[!h]
   \centering
   \includegraphics[width=.8\textwidth]{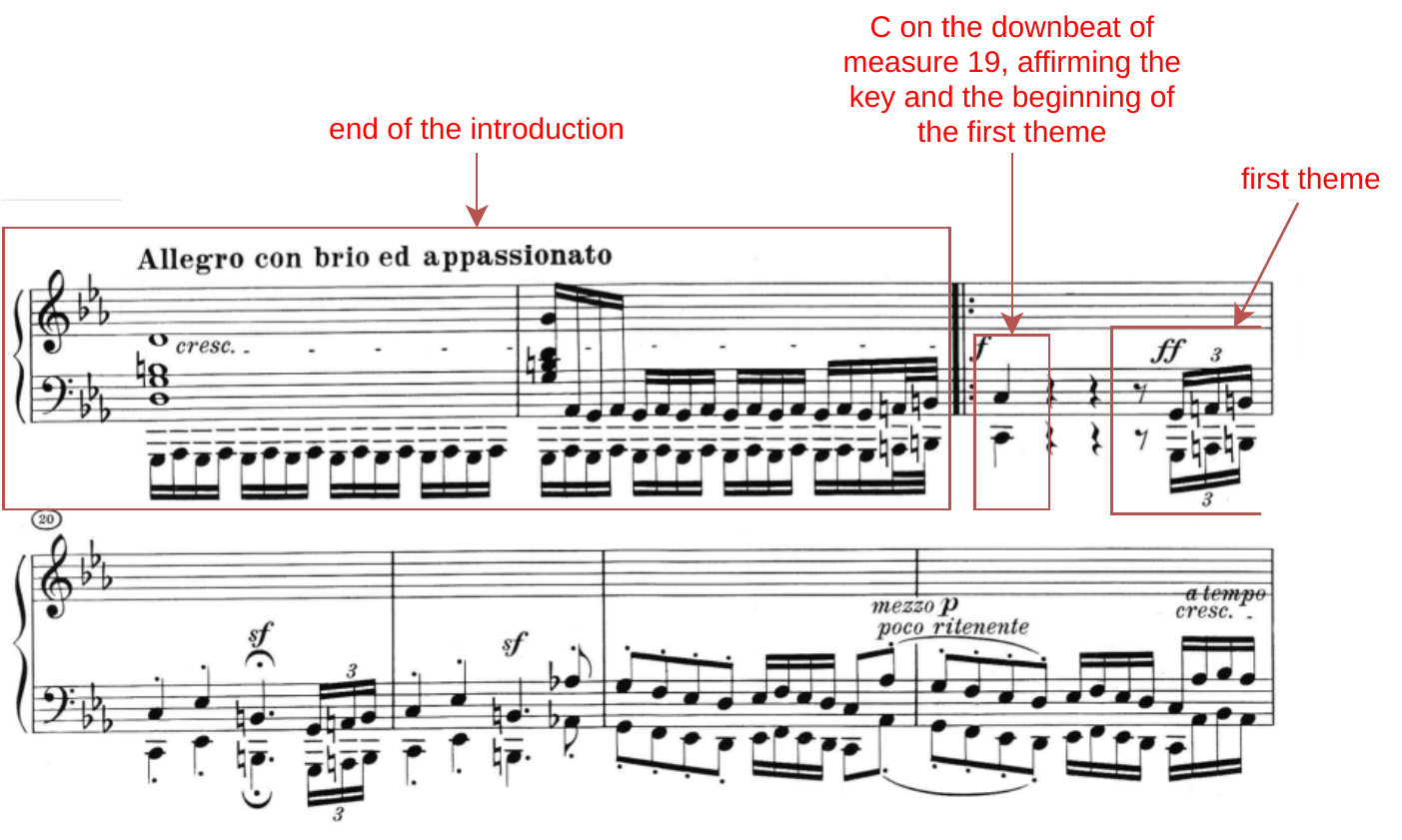}
   \caption{Beethoven, Piano Sonata No.~32, Op. 111, first movement, excerpt, annotated to identify the transition between introduction and the first theme}
   \label{fig:beethoven op111 mov1 19}
\end{figure}

The first movement of Beethoven's piano sonata Op. 111 serves the purpose of illustration. Conveniently, the movement simultaneously contains both an introduction and a coda, enabling us to compare them for differences. The role of the introduction in this example is non-arbitrary: it introduces, prepares and anticipates the theme on measure 19 (see figure \ref{fig:beethoven op111 mov1 19}). Knowing its anticipatory function, we can then claim a provocative statement: the theme is motivically captured and represented by a single note, the note C representing its downbeat as well as the tonic of the key of the whole movement. One can further notice that for this sonata, the single pitch C plays a more crucial role than C minor as a key by observing its two movements: the first movement (C minor) and the second movement (C major) are tonally related by parallel relationship. For parallel keys, the commonality primarily rests in the shared tonic \emph{note} instead of \emph{scale}. Motivically speaking, the introduction then represents a process of affirming the note of C, thereby achieving its anticipatory function, i.e.~it helps listeners to anticipate the arrival of C as the downbeat of the exposition.

\begin{figure}[!h]
   \centering
   \includegraphics[width=.8\textwidth]{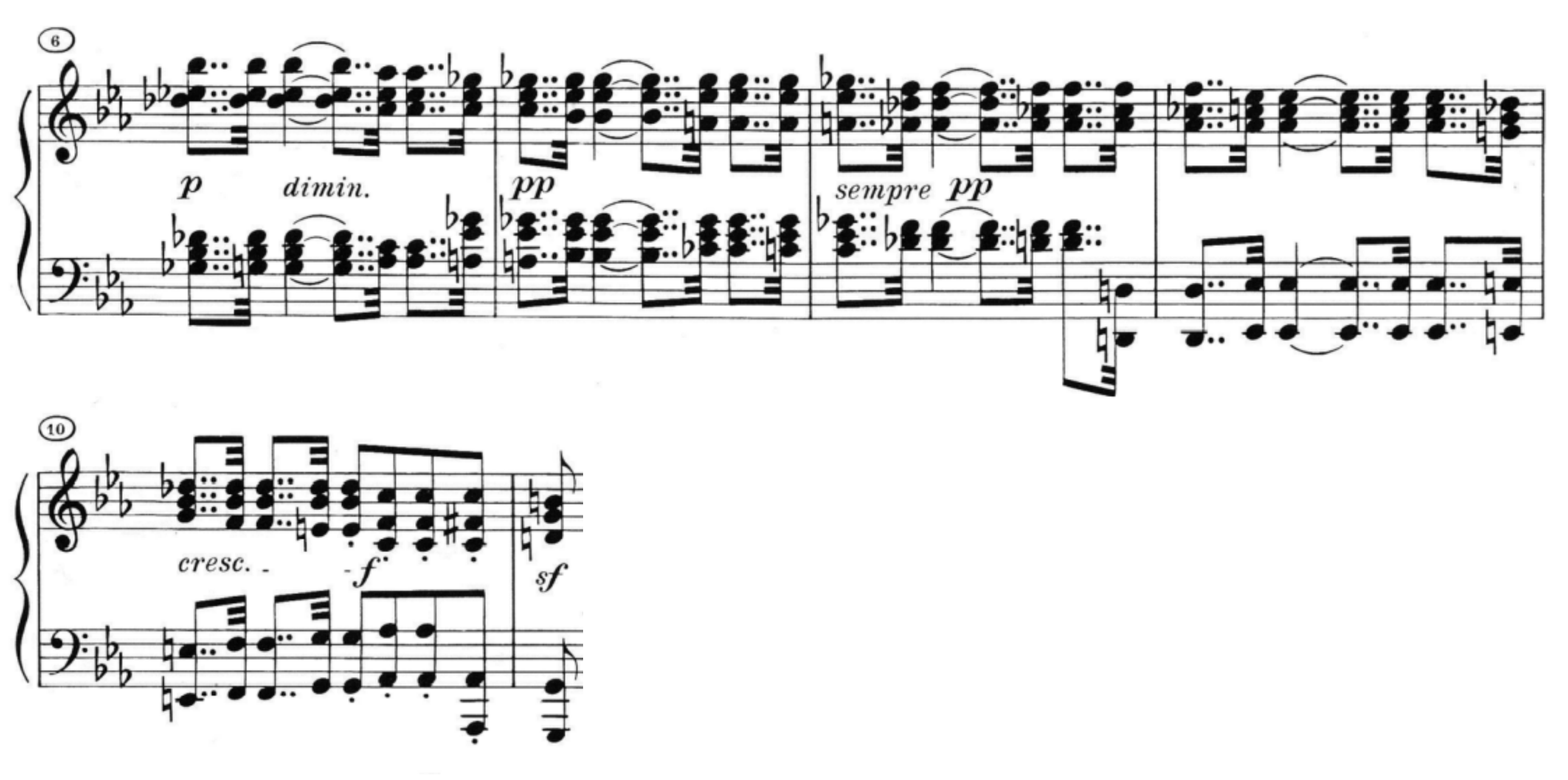}
   \caption{Beethoven, Piano Sonata No.~32, Op. 111, first movement, measures 6 - 10}
   \label{fig:beethoven op111 mov1 6-10}
\end{figure}

The musical strategies to achieve the anticipatory function are the following. Our goal is to prepare the theme represented by the note C. Firstly, we adopt a performer's perspective and examine a crucial section within the introduction: measures 6-10 (see figure \ref{fig:beethoven op111 mov1 6-10}). The section is crucial because for performers, it is tricky both technically and musically. Technically, the section is challenging as it requires precise sound control (voicing of six simultaneous notes with pianissimo) and rhythmic control (i.e.~the double dotted rhythm, again with pianissimo). More important, the section is musically challenging because one easily produces a monotonic performance primarily due to the double dotted rhythmic pattern throughout. In response, listeners might be disoriented and wonder what this section is musically about. Ironically, a convincing performance of this section represents a musical sense of disorientation: a process of searching, as if the music is poetically personified, and wanders aimlessly in the dark. The distinction between a musical sense of disorientation and disorientation as expressing a listener's confusion is in the same manner we have good and bad surprises: musical surprises and disturbing surprises due to suboptimal performance. Through depiction of musical disorientation, the anticipatory function is achieved in this section by a successful representation of searching: we anticipate the theme because the theme is previously unclear, hidden in the musical mist. As a result, the anticipatory function of the introduction is nothing but the urge to clear the musical mist as one musically searches for C: a clearly presentation of the note C at the downbeat by itself may sound abrupt and haphazard. However, if the note is prepared by the introduction such that it results naturally as the inevitable consequence of the introduction, listeners are satisfied in the same way one finds satisfaction in solving puzzles.

\begin{figure}[!h]
   \centering
   \includegraphics[width=.8\textwidth]{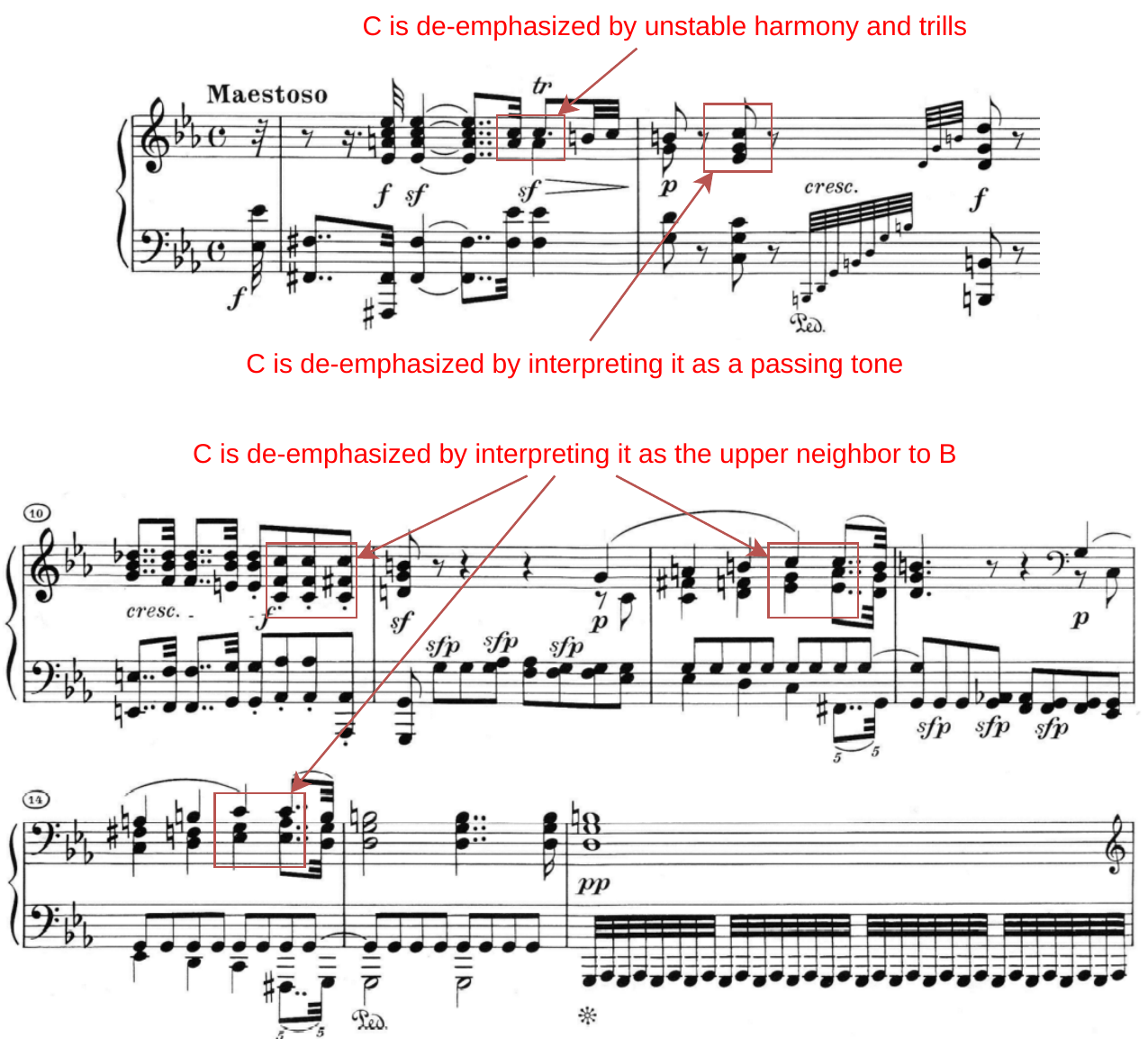}
   \caption{Beethoven, Piano Sonata No.~32, Op. 111, first movement, enumerating occurrences of note C, annotated to illustrate methods of de-emphasizing the note}
   \label{fig:beethoven op111 mov1 enumerate}
\end{figure}

Secondly, in addition to the section highlighted by measures 6-10, the entire introduction prepares the arrival of note C by deliberately avoiding it until the onset of measure 19. We observe the avoidance by enumerating all musical events involving note C in the introduction (see figure \ref{fig:beethoven op111 mov1 enumerate}). The trill in measure 1 is the first melodic appearance of C, however, it is de-emphasized and destabilized by two musical devices: the trill representing melodic instability and unstable harmonic support (i.e.~diminished seventh). Then in measure 2, C occurs in the second quarter beat. However, notice how it is musically passing: melodically it acts as a passing tone between B on the downbeat and D on the strong third beat (not only in terms of metrical accents if we notice the dynamic marking); dynamically it is explicitly marked by p; metrically it occurs on the weakest quarter beat (note that the fourth beat is empty, making the second beat the weakest beat that listeners can perceive). After measure 2, surprisingly (unsurprising to our analysis), note C is completely absent until measure 10. However, C is again de-emphasized in measure 10, albeit having a dynamic mark of f: the sole purpose of the three last eighth notes of the measure is to lead into the downbeat of measure 11. Between measure 11 and measure 18, the interpretation of C is more evident by noticing its role as the upper neighbor of B. Despite the harmonic evidence that B is supported by a dominant harmony throughout the introduction (in which case B is the leading tone to C), locally (between measure 11 and measure 18) B is harmonically stabilized by repeating localized harmonic resolutions: B on the downbeat of measure 11 is the resolution of an augmented sixth, and the two occurrences of B on the downbeat of measure 13 and measure 15 are the resolutions of diminished fifth between the outer voices. As a result of B appearing as a stable pitch locally in the section between measure 11 and measure 18, C can be locally interpreted as the upper neighbor of B.

The section consisting of measure 11-18 is one of the best examples illustrating how musical interpretation is entirely contextual: the local musical evidence in the section indicates that C is the upper neighbor of B. Only through retrospection after hearing the downbeat of measure 19, one can realize the reversal of role: B is in fact, the leading tone to C, restoring our understanding that B is supported by a dominant pedal throughout the introduction.

The coda that begins on the third beat of measure 146, on the other hand, is more controversial concerning its function within the piece. For this example, the coda is less definitive in function because we can argue that its function may be interpreted in two distinctive ways: conclusive or anticipatory.

\begin{figure}[!h]
    \centering
    \includegraphics[width=.8\textwidth]{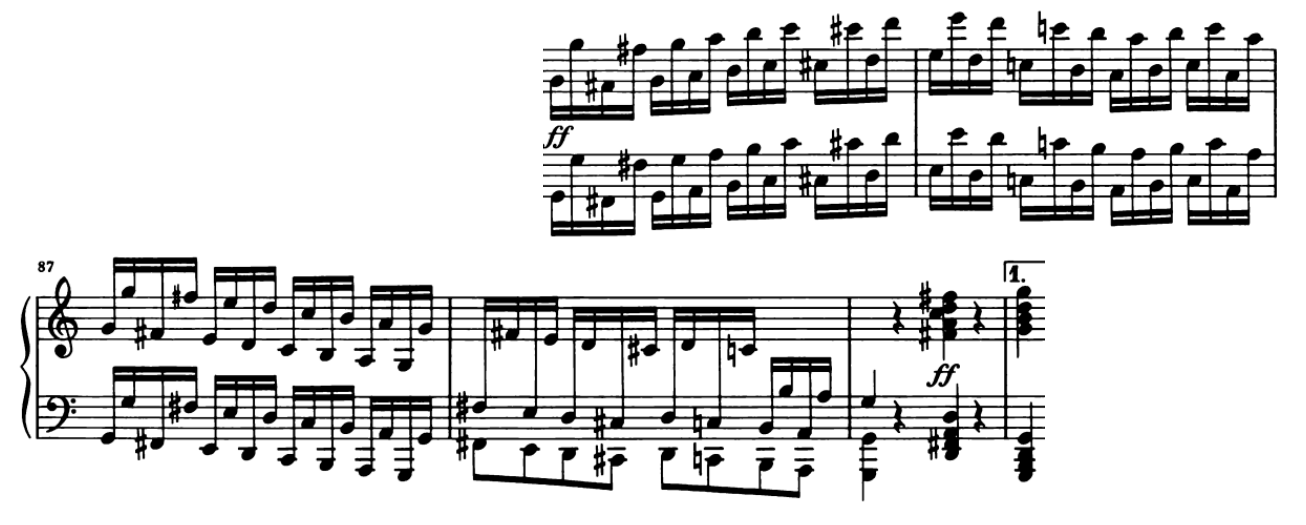}
    \caption{Beethoven, Piano Sonata No. 3, Op. 2, No. 3, first movement, ending of the exposition}
    \label{fig:op2no3expo}
\end{figure}

\begin{figure}[!h]
    \centering
    \includegraphics[width=.8\textwidth]{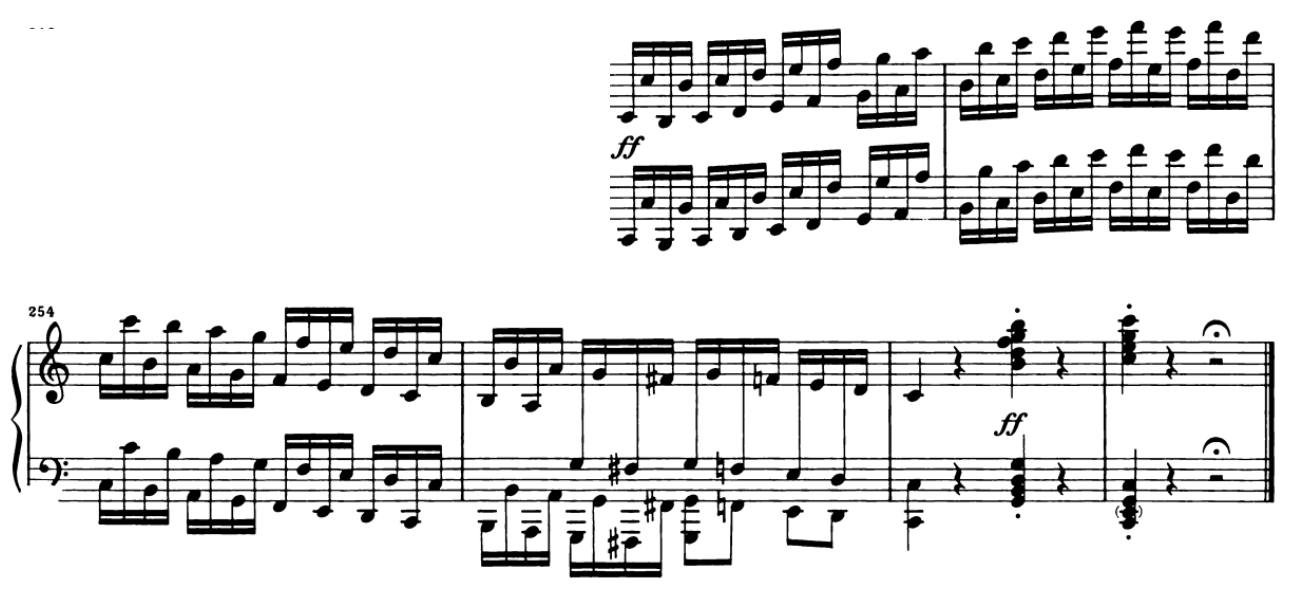}
    \caption{Beethoven, Piano Sonata No. 3, Op. 2, No. 3, first movement, ending of the entire movement}
    \label{fig:op2no3recap}
\end{figure}

A word is needed to justify the identification of the coda. We can identify coda using musical parallelism (or more generally, pattern matching) between the end of the exposition section and the end of the recapitulation section. In sonatas without a coda, by definition, the end of the recapitulation is also the end of the entire work. Conversely, if there is musical material after the end of the recapitulation section, then we identify it as the coda. Now the question is reduced to the identification of the end of the recapitulation section. Empirically, the match between the end of the exposition and that of the recapitulation is exact, in the sense that the last phrase of the exposition is the same to that of the recapitulation, up to a difference in keys (i.e.~the exposition often ends with a ``wrong'' key that is later corrected by the recapitulation). For example, consider the first movement of Beethoven's piano sonata Op. 2, No.~3. The final phrase of the movement (see figure \ref{fig:op2no3recap}) is the transposed version (disregarding minor difference in musical details) of the final phrase of the exposition (see figure \ref{fig:op2no3expo}).

\begin{figure}[!h]
    \centering
    \includegraphics[width=.8\textwidth]{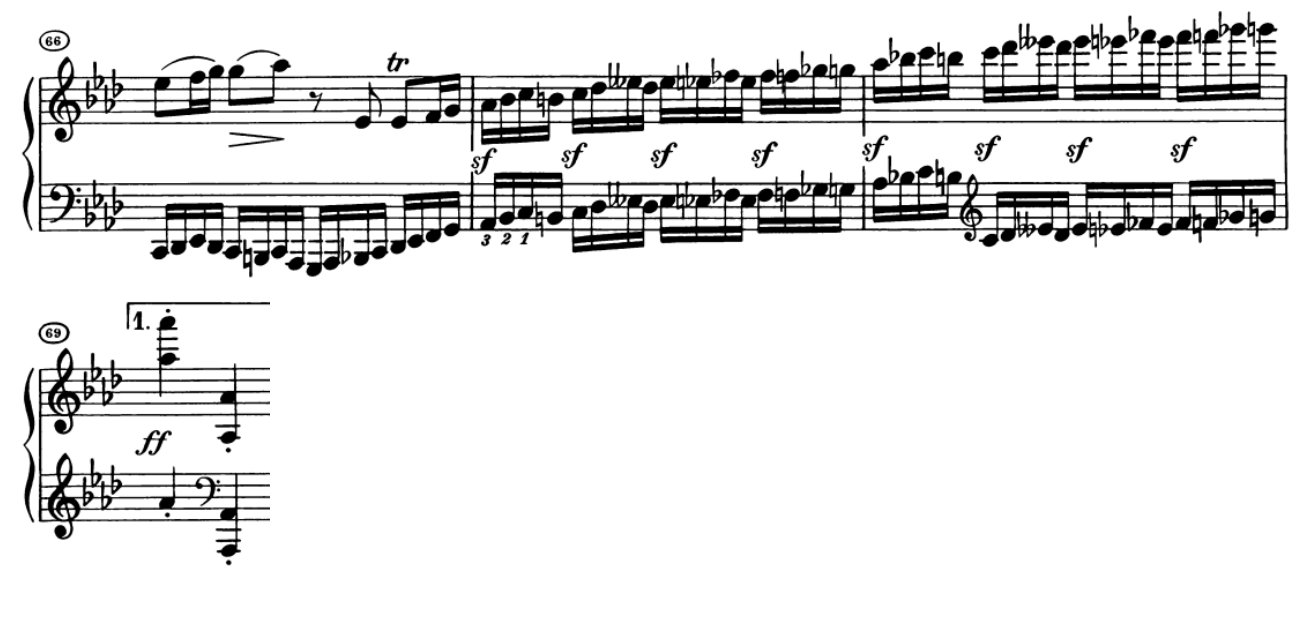}
    \caption{Beethoven, Piano Sonata No. 32, Op. 111, first movement, ending of the exposition}
    \label{fig:op111expo}
\end{figure}

\begin{figure}[!h]
    \centering
    \includegraphics[width=.8\textwidth]{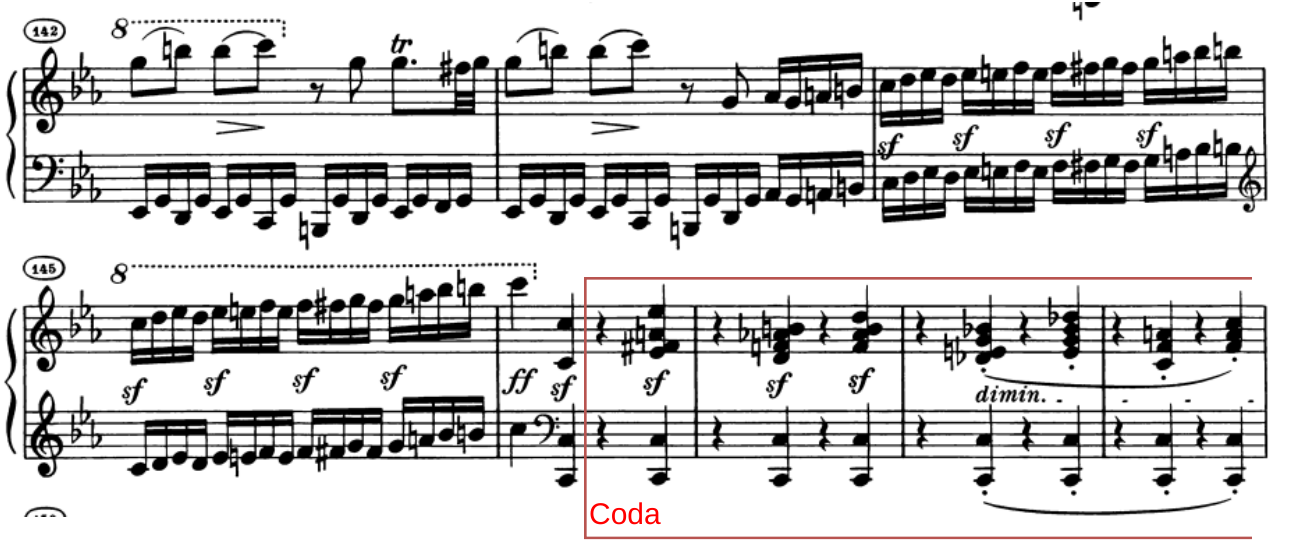}
    \caption{Beethoven, Piano Sonata No. 32, Op. 111, first movement, ending of the recapitulation, annotated to identify the coda}
    \label{fig:op111recap}
\end{figure}

In both places, the ending is characterized by four measures of stormy sixteenth notes consisting of broken octaves, followed by two measures of cadential closure. Such parallelism between the ending of the exposition (see figure \ref{fig:op111expo}) and that of the recapitulation (see figure \ref{fig:op111recap}) is also found in the first movement of Op. 111.

In fact, we may even notice how the musical characterization is similar for the two sonatas: a few measures of stormy running notes, followed by a decisive cadential closure. Now, we see how the parallelism is broken: if the movement contains no coda, then it should end in the same manner as the first movement of Op. 2, No.~3. In other words, the final phrase of the exposition indicates that the movement should end on the second beat of measure 146 if it contains no coda. It follows that all musical material after that moment constitutes the coda.

\begin{figure}[!h]
   \centering
   \includegraphics[width=.8\textwidth]{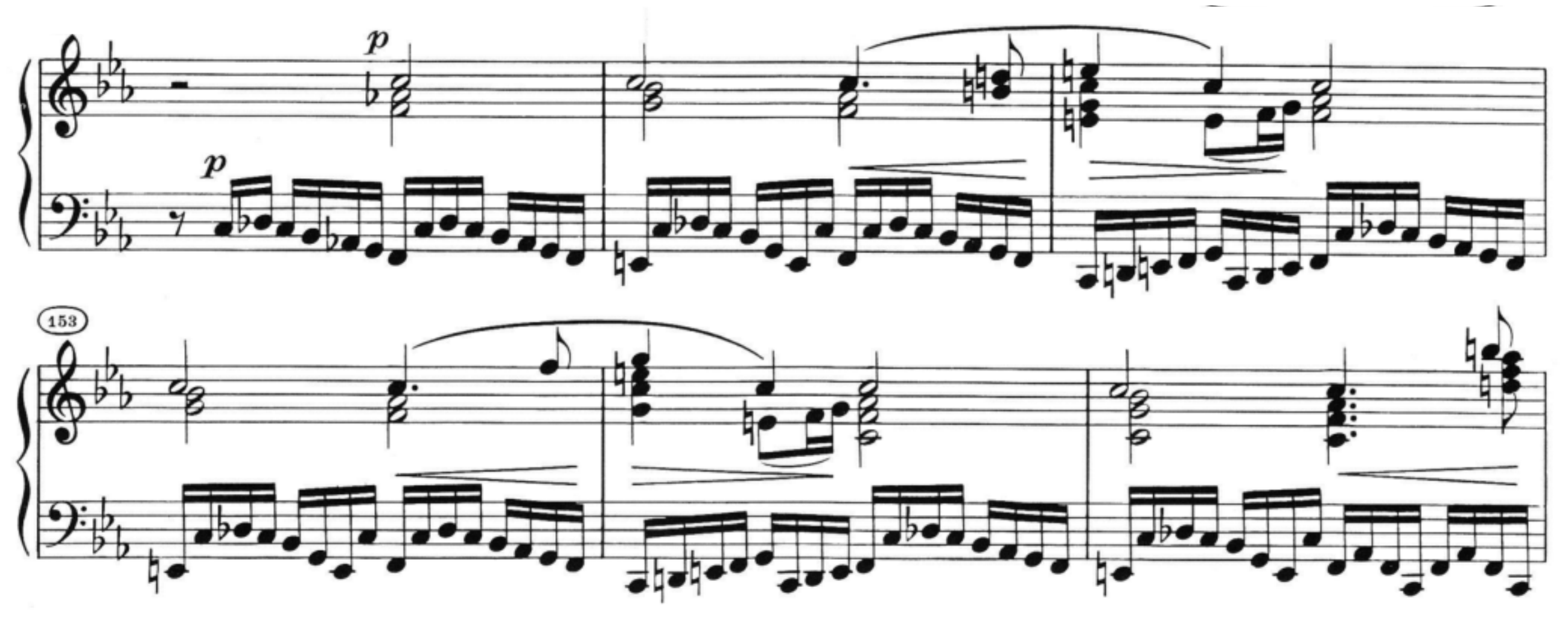}
   \caption{Beethoven, Piano Sonata No.~32, Op. 111, first movement, coda (excerpt)}
   \label{fig:op111 mov1 coda running figure}
\end{figure}

\begin{figure}[!h]
   \centering
   \includegraphics[width=.8\textwidth]{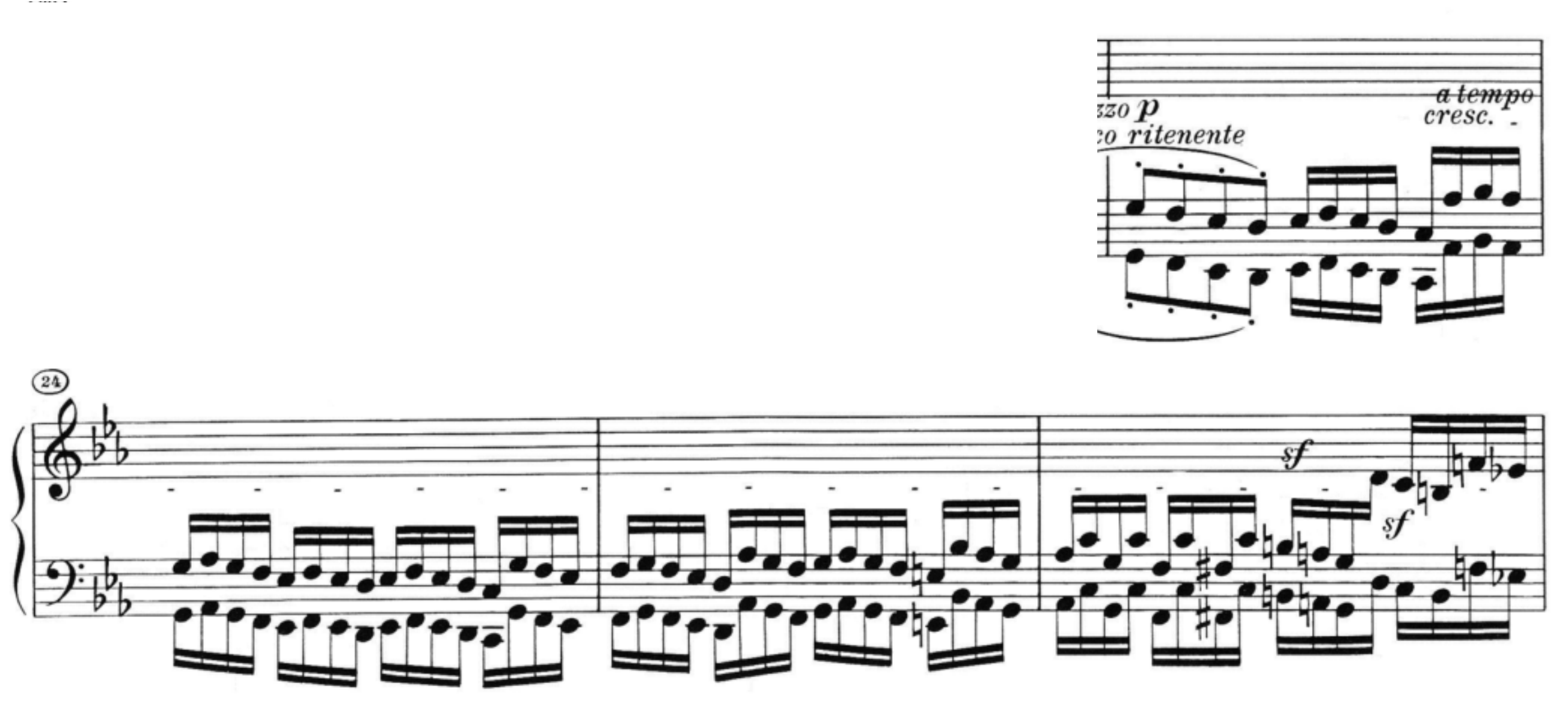}
   \caption{Beethoven, Piano Sonata No.~32, Op. 111, first movement, measure 23-26}
   \label{fig:op111 mov1 theme running figure}
\end{figure}

By the terminological definition of the word, we naturally interpret coda as a conclusive passage as the term literally means ``tail''. In other words, it musically summarizes a movement. Summary, which often occurs at the end of a work (note that it does not make sense to talk about summary of the future, i.e.~events which have not yet taken place), looks behind into the past and refers to previous materials. One common challenge in a coda is to find the balance: on the one hand, it must be sufficiently unrelated to the rest of the work in order to be considered a coda as an appended section. On the other hand, it must at the same time refer to previous materials in order to be considered a conclusion that summarizes the work as a whole. In this example, Beethoven achieves the balance through recombination of musical elements from previous passages. The recombination makes the coda reminiscent of previously heard passages, suggesting parallelism. For example, the running sixteenth-note figure in the left hand (see figure \ref{fig:op111 mov1 coda running figure}) reminds the listeners of the running sixteenth notes between measure 23 and 26 (see figure \ref{fig:op111 mov1 theme running figure}).

\begin{figure}[!h]
   \centering
   \includegraphics[width=.8\textwidth]{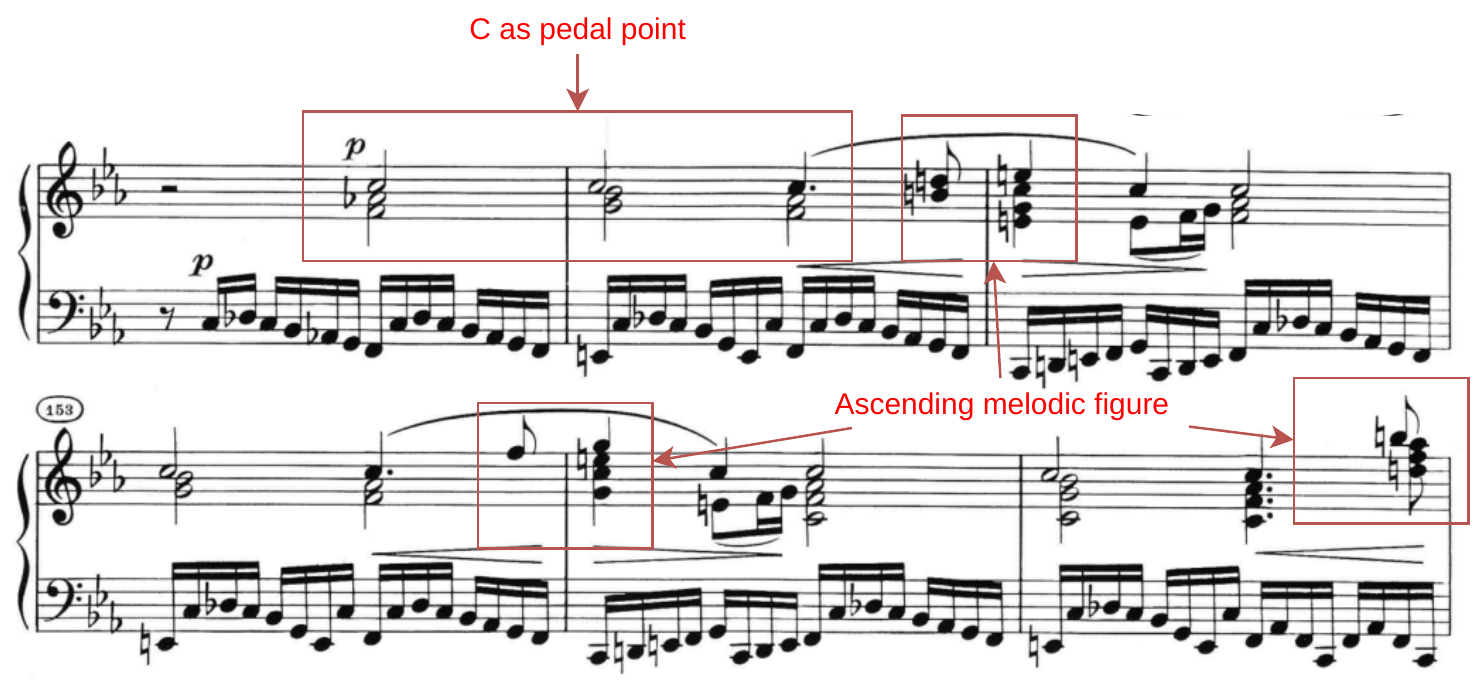}
   \caption{Beethoven, Piano Sonata No.~32, Op. 111, first movement, coda (excerpt), annotated to identify ascending melodic figure (interrupted by pedal point C) delineated by the following pitches: D - E - F - G - B - (C, not included in figure)}
   \label{fig:op111 mov1 coda ascending figure}
\end{figure}

\begin{figure}[!h]
   \centering
   \includegraphics[width=.8\textwidth]{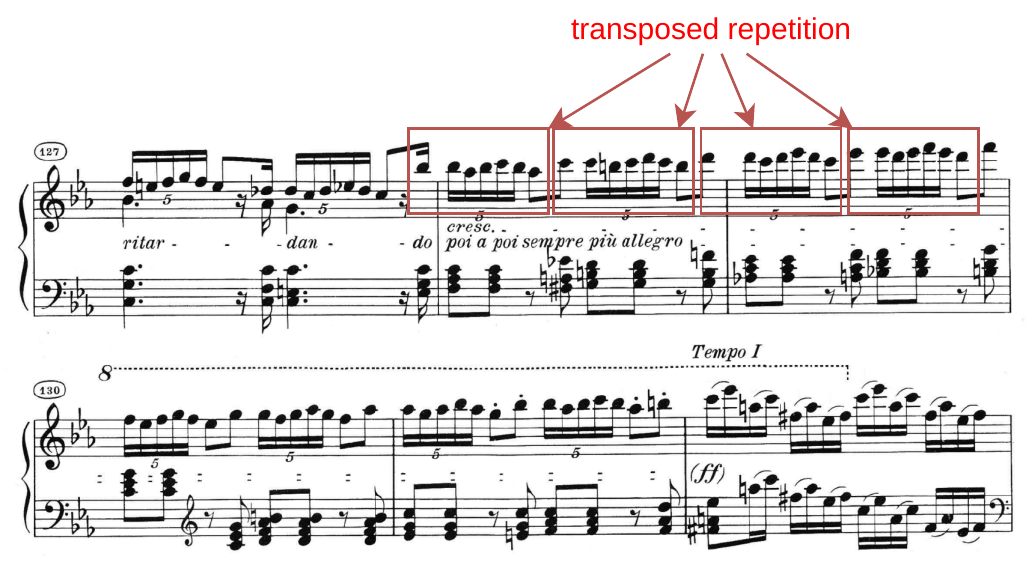}
   \caption{Beethoven, Piano Sonata No.~32, Op. 111, first movement, excerpt, annotated to identify transposed repetition}
   \label{fig:op111 mov1 transposed repetition}
\end{figure}

Additionally, the ascending melodic figure (see figure \ref{fig:op111 mov1 coda ascending figure}) in the right hand (ignoring the haunting pedal point on note C) is reminiscent of multiple places in the movement characterized by sequentially elevating a melodic segment in register. For example, the section between measure 128 and 131 shows such pattern where the same melodic segment (i.e.~a quintuplets followed by two eighth notes) is repeated with transposition (see figure \ref{fig:op111 mov1 transposed repetition}).

\begin{figure}[!h]
   \centering
   \includegraphics[width=.5\textwidth]{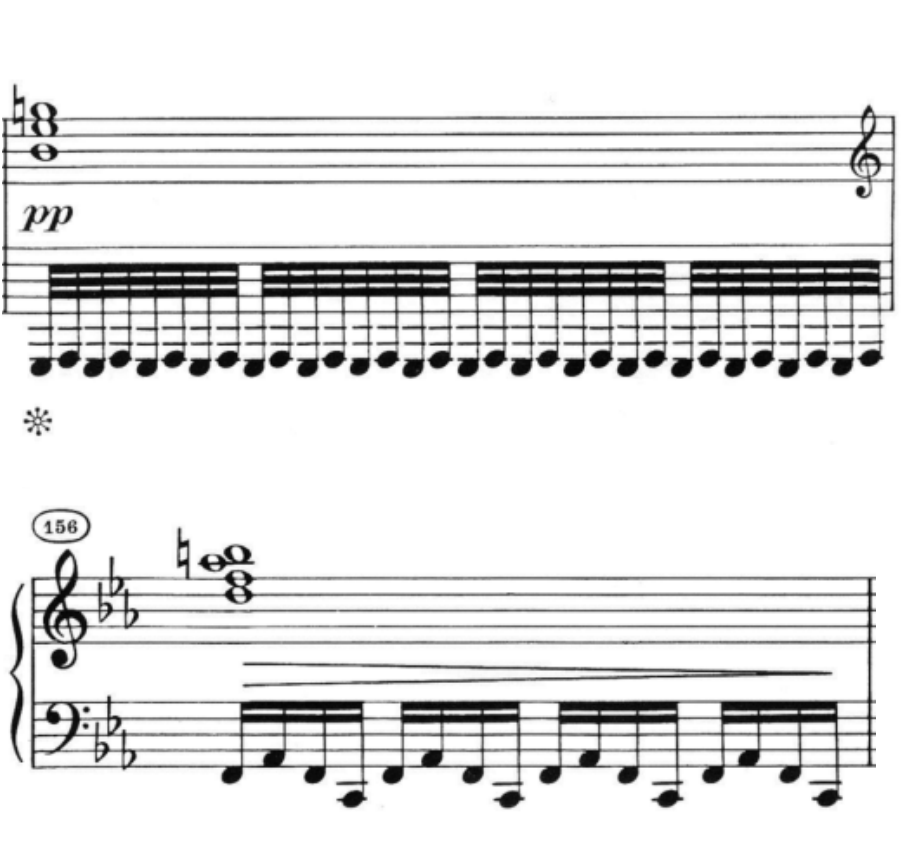}
   \caption{Beethoven, Piano Sonata No.~32, Op. 111, first movement, top: measure 16, bottom: measure 156}
   \label{fig:op111 mov1 long note}
\end{figure}

Finally, the whole note in measure 156 makes direct reference to the whole note in measure 16 (see figure \ref{fig:op111 mov1 long note}). In fact, those are the only two occurrences of the leading tone B in the form of long note values throughout the movement. One should note from the examples of recombination above that the effect of recombination is exactly the following. Musical elements are firstly broken into fragmentary pieces, then they are recombined in the coda, such that each fragmentary piece refers to some earlier details yet there is no obvious parallelism in large-scale (e.g.~phrase level) musical patterns.

However, the coda in this example can also serve an anticipatory function. It is anticipatory if we consider it as a transition between the first and second movement. That is, if we consider inter-movement narrative, which is common in late-Beethoven sonatas (e.g.~the last movement of sonata Op. 101 begins in direct reference to the opening of the first movement), then the coda in this example functions similarly to the opening introduction: it introduces the arrival of the second movement.

The musical discussion above, through the specific concepts of introduction and coda, introduces the distinction between the past and the future. It shows that in music we invariably treat the past and the future differently, and such difference is hard-coded in our musical thinking. For now, however, we examine how the difference between the past and the future is handled in philosophy. In particular, we focus on the epistemological aspect of the difference, as music concerns the subjective experience of time more than the physical reality of time (see section \ref{music_as_subjective_experience}). Things related to our subjective experience, as knowledge of the world, are categorized as epistemological.

\subsection{Epistemic difference between the past and the future}
\label{section:epistemic difference between the past and the future}

The previous musical discussion in secion \ref{sec:op111} can act as a springboard for motivating a more general question: how is the past distinguished from the future? Notice that this question must be discussed before we can define the arrow of time at all, as the tail and head of the arrow correspond exactly to the past and the future.

In his book ``Time and chance'', Albert points out that the ``sort of epistemic access we have to the past is different from the sort of epistemic access we have to the future''.\autocite[113]{time_and_chance} Furthermore, the difference is not about the content of our knowledge. For example, we have the knowledge whose content is simply the sentence ``sun rises in the morning'', and the same content applies equally to our knowledge of yesterday (in the past) and of tomorrow (in the future). Additionally, we have exactly the same degree of belief that the sun rose yesterday and that the sun will rise tomorrow. Similarly, it is not so much about the difference in the amount of knowledge we have for the past and the future either. Therefore, the author continues:

\begin{quote}
{[}If{]} it's said that we know more of the past than we do of the future, this seems (according to the usual way of talking) true enough, but (as it stands) not particularly informative.\autocite[113]{time_and_chance}
\end{quote}

In fact, we can easily construct a thought experiment that presents a counterexample where one somehow knows more about the future. Consider a degenerate scenario in a mythological style: a fortune teller has the incredible ability to make predictions. Like a muse, she sings about the future events naturally as water flows out of the fountain. However, the gods are jealous of her gift and make her virtually memoryless as trade-off to her predictive gift. Therefore, the fortune teller can predict the future accurately, yet at the same time, cannot even remember what was just said seconds ago. In this case, we are quite confident that the fortune teller knows more about the future than the past. But does this difference in amount of knowledge confuse her perception of time? Moreover, does it reverse the direction of her perceived time? The answer is no. In particular, causal relations are preserved such that cause must happen before effect in her predictions. For example, she may predict poetically that the end of the world's history is an inferno-like scene filled with sorrow and destruction, \emph{as a result} of the gods' wrath and jealousy.

Thus, according to Albert, the difference lies in ``the methods by which we come to know things about the past and the future''.\autocite[113]{time_and_chance} In other words, the issue is not about the content of knowledge, but the process of acquiring that content. The methods (or processes) of acquiring the knowledge of the past and of the future somehow involve inferences (or calculations) that are fundamentally different in mechanism.

We remember the past. However, the peculiar implication is that we do not remember the future. In other words, we have the memory of the past, which has no counterpart in the future. One may claim that this is merely a play of words, since the term ``remember'', by definition, refers to the past. However, one immediately realizes that it is not a play of words if we consider the idea of retrodiction as the time-reversal counterpart to prediction. According to Albert, the epistemic difference between the past and the future is illustrated by considering the unequal treatment of prediction and retrodiction in reality:

\begin{quote}
{[}Whatever{]} we take ourselves to know of the future, or (more generally) whatever we take to be knowable of the future, is in principle ascertainable by means of prediction. Some of what we take ourselves to know about the past (the past positions of the planets, for example) is no doubt similarly ascertainable by means of retrodiction---but far from all of it; rather little of it, in fact. Most of it we know by means of records.\autocite[116]{time_and_chance}
\end{quote}

The past is special because it gives rise to things called record and trace, through which we can know the past through a special method called recalling. It is special because, as Smart puts it, ``\,`prediction' is not the future analogue of `memory'\,''.\autocite[80]{smart_temporal_asymmetry_of_the_world} Therefore, what we are encountering is a peculiar case that we have three ways of knowing the past and the future: record-reading (e.g.~memory), retrodiction, and prediction. The asymmetry corresponds entirely to the uneven assignment of methods: while there are two methods to know the past, there is only one method to know the future. Prediction and retrodiction are conceptually dual, which can be symbolically cancelled out as they are paired the same way the past and the future are paired, leaving record-reading unpaired. The core issue here is summarized concisely as follows. Firstly, why is there no future-analogue of records? One should note that this question is exactly what we mean by we not remembering the future: we have no way to acquire knowledge of the future through future-analogue of records. Secondly, one may ask alternatively that, why in reality, we almost never use retrodiction (as the past-analogue of prediction) which is a conceptual counterpart more natural record-reading to prediction, to know the past?

Let's see what that means in music. As classically trained musicians, upon hearing a dissonance, we immediately predict and anticipate that it will be eventually resolved. We might not know when exactly the dissonance will be resolved. For example, late-Romantic works might not resolve the dissonance until the very end, leaving the dissonance unresolved for hours. However, we know with a high degree of confidence that it will be resolved eventually and we have the patience to wait. Moreover, we know that the resolution will be unique: it will resolve to some form of the tonic chord. How would we know that? Because it is the definition of harmonic resolution. If it were to resolve to some other sonority such as a vi in deceptive cadence, then it would not actually be considered a resolution. Therefore, anticipating a harmonic resolution is an instance of musical prediction, representing our epistemic method of knowing the future. The specific rules for prediction may be complicated, but the nature of prediction is straightforward: we are given the present state of the world (e.g.~a heard dominant seventh chord), as well as a set of inference rules (e.g.~voice-leading rules) that allow us to speculate about the future state of the world (e.g.~a tonic chord that follows). Albert describes the nature of prediction similarly, albeit with a different set of terminology:

\begin{quote}
    Everything we can know about the future {[}\ldots{]} can in principle be deduced from nothing over and above the dynamical equations of motion and the probability-distribution which is uniform, on the standard measure, over the world's present directly surveyable condition. \autocite[114]{time_and_chance}
\end{quote}

Now consider the method of acquiring knowledge of the past. In practice, we know the heard musical events simply by recalling: we know that we are in the recapitulation section of a piece because we remember the opening of the piece. Therefore, by comparing what we currently listen to and our memory of the opening, we note the similarity that enables us to claim that we are at the recapitulation section in the piece. But how about retrodiction? Suppose that we hear the final tonic chord of the piece, can we retrodict, without relying on the convenience of memory, and somehow expect the consonance to be musically un-resolved to a dissonance in the past? Essentially, we are asking the question: can we analyze voice-leading backwards in time? To be more specific, we are not talking about the idea that dissonance comes after consonance (as in the case of back-relating dominants). All musical events and score are given as they appear now: the content of musical events is presented in the original temporal sequence such that the resolution of a dissonance comes after it. Instead, we are interested in a time-reversal \emph{description} (i.e.~music analysis) of the same \emph{content} (i.e.~music performance). In particular, we are doing analysis in a right-to-left fashion.

Before moving on, we need to further clarify the distinction between retrodiction and recalling from memory, as both refer to the past. One may argue that in the example of Beethoven Op. 111 given above, we encounter instances of retrodictive analysis (see section \ref{sec:op111}). In particular, one may claim that it is due to retrodiction that listeners may retrospectively reinterpret the whole note B in measure 16 as the leading tone to C on the downbeat of measure 19. Therefore, one may start the analysis with measure 19 and proceed to the left of the score, and ``retrodict'' that the identity of B is the leading tone of C. However, it is not correct to call this process of reinterpreting note B a retrodiction. By definition of retrodiction, it must be identical with prediction in mechanism. In particular, we must predict and retrodict with the sole help of inference rules. However, in reinterpreting the note B which lies in the past, we have the additional help of memory. We are inferring the identity of B retrospectively using additional information, i.e.~the memory of listening to B. Therefore, the basic analytical procedure of reinterpreting B can be briefly summarized as follows: the analysis begins from the first measure and proceeds forward in time. Upon encountering the note B in measure 16, its identity is inferred by the musical context to be the central note to which note C is an upper neighbor. Then, when we encounter the note C in measure 19, we necessarily recall our experience in measure 16 from memory, in order to reinterpret the identity of B. The essential step of recalling from memory is what makes the analysis no longer retrodictive. Memory gives us a shortcut by granting us immediate access to the past: in reinterpreting B, we are given the present moment (i.e.~measure 19) as well as the memory of the past (i.e.~measure 16) as known conditions. In particular, we must retain the memory of note B until at least measure 19 (although we might not be aware of its presence), keeping it known to us. Only through retaining the memory, we are then able to confirm and reinterpret the identity of a previously heard chord. On the contrary, if the analysis is to be considered a retrodiction, we must be given the present moment as well as (potentially) future moments (i.e.~music after measure 19) as known conditions, while keeping the experience of the past largely unknown. In short, to qualify for retrodiction, one must have more epistemic access to the future than to the past. However, in reality, the opposite is often true: we more often encounter cases where our epistemic access of the past is (believed to be) more than that of the future. In fact, it is precisely due to the difference in epistemic access, that we are entitled to interpret memory of the past as a \emph{known} condition, while the future remains largely \emph{unknown}. Therefore, the retrospective (not retrodictive) analysis of reinterpreting the note B is of very different nature than a true retrodictive analysis: it does not need to infer the unknown (e.g.~the candidates of a harmonic resolution). Instead, it serves to further clarify what we already know (i.e.~the experience of listening to measure 16).

Now, memory is considered additional help to reinterpretation of the note B because it is not domain-specific to music analysis, i.e.~recalling a heard musical event does not constitute a rule in the realm of music analysis. As such, it is additional to the rules of music analysis. For example, given the dominant chord, we predict that it is followed by a tonic chord based on voice-leading rules. However, given a tonic chord, remembering that it follows a dominant chord does not involve any voice-leading rules. Instead, memory involves a general-purpose and non-domain-specific method of inquiry. We need less knowledge of music theory to realize that we just heard a dissonance, than to predict possible resolutions upon hearing a dissonance.

Retrodictive voice-leading is technically possible, yet seldom encountered in practice. The possibility is warranted by the fact that, when the focus of analysis lies largely on musical notations, the analysis of temporal relations is reduced to that of spatial relations between notated symbols. As mentioned in section \ref{space_vs_time}, spatial relations attain no inherent orientation due to the principle of indifference. Consequently, a right-to-left analysis should be equally justified to a left-to-right analysis (whether they make equal sense is another matter), resulting in equal popularity of prediction and retrodiction concerning methods of music analysis. However, we almost never discuss backward harmonic progressions in terms of backward voice-leading rules in practice (except for a few special settings such as the crab canon), let alone right-to-left analysis in general. As a result, we lack terminologies to discuss voice-leading rules (if they make any sense at all) that describe right-to-left musical relations. For example, what is the past-analogue of resolving the leading tone and chordal seventh? Notationally, we are justified to use left-to-right arrows to denote the resolution process. In a typical resolution scheme of dominant seventh chord, we have three such arrows: scale degree \(\hat5\) (i.e.~the root) points to scale degree \(\hat1\), scale degree \(\hat7\) (i.e.~the leading tone) points to \(\hat1\), and scale degree \(\hat4\) (i.e.~chordal seventh) points to \(\hat3\). On the contrary, concerning the same resolution process, it makes less musical sense to construct the reciprocal description using three right-to-left arrows where, given the tonic chord, arrows point to the preceding dominant chord on the left to retrodict the voice-leading rules. In fact, retrodictive music analysis is generally uncommon due to two potential reasons. The first is that it is not musically natural: why would analysis begin with the last measure and proceed backward when the music is to be performed from the first measure onward? Such a question is particularly appealing to performers who insist that analysis must be performance-centric, in the way that analysis of notated symbols must explain (as opposed to challenge) intuitions listeners have in response to performances. In other words, after reading an analysis of a piece of music, the readers should experience an ``aha'' moment, which enables us to say: ``this is why I felt that way when I heard the piece!''. The second reason is that it is not analytically effective because one has a better method to know the past: through memory rather than retrodiction. Since recalling from memory is far more effective than retrodicting and speculating about the past, one simply avoids analytical methods using retrodiction altogether for practical purposes. In other words, we do not use right-to-left voice-leading rules to calculate and speculate the past, because the past is simply in our memory as a given.

In short, memory (or more generally, record) is an exceptionally powerful and efficient method by which we obtain knowledge of the past. It is due to its exceptional power and efficiency of inference that we are accustomed to saying we remember the past instead of retrodicting it. Nothing is technically unorthodox about retrodicting the past in ways identical to predicting the future to give us knowledge of other points in time. However, memory is practically more useful so that we simply do not use retrodiction in practice when memory is available. As a result, we have distinct epistemic methods through which we acquire knowledge of the past and the future, thus making them distinguishable.

The power of memory is vividly captured in the words of Smart's thought experiment, as he further examines the difference between the past and the future in his article ``The temporal asymmetry of the world'':

\begin{quote}
Suppose that like Laplace thought of the universe as such that an infinite intelligence who knew the total state of it at time \(t_0\) could deduce its at any other time before or after \(t_0\). Then a sufficiently powerful intelligence who made a sufficiently thorough examination of the state of a person's brain would be able to deduce (admittedly not with certainty, but very nearly with certainty) that, say, he shot a rabbit ten miles away three weeks ago. However much it examined a man's brain, however it could not deduce that in three weeks time he would shoot a rabbit which is now ten miles away. In order to do so it would have to consider a far wider region of space than the person's brain: it would have to consider the present state of the rabbit and the intervening country, and after superhuman calculations predict that the rabbit will travel ten miles, enter the gentleman's paddock just as the gentleman was out shooting, that the shot would occur at just such a time that the bullet would enter the rabbit's head, and so on. \autocite[81]{smart_temporal_asymmetry_of_the_world}
\end{quote}

Albert argues similarly using the thought experiment of billiard balls instead. He argues that reading records of the past through measuring record-bearing devices

\begin{quote}
    can be immensely more powerful, that they can be immensely more informative, than inferences of the predictive/retrodictive variety. Think (for example) of an isolated collection of billiard balls moving around on a frictionless table. And suppose that billiard ball number 5 (say) is currently at rest; and consider the question of whether or not, over the past ten seconds, billiard ball number 5 happens to have collided with any of the other billiard balls. The business of answering that question by means of retrodiction will of course require as input a great deal more information about the present --- it will require (in particular) a complete catalogue of the present positions and velocities of all the other billiard balls in the collection. But note that the question can also be settled, definitively, in the affirmative, merely by means of a single binary bit of information about the past; a bit of information to the effect that billiard ball number 5 was moving ten seconds ago.\autocite[117]{time_and_chance}
\end{quote}

The author gives an information-theoretic perspective which can shed light on the power of memory. When we say that memory is more powerful than retrodiction, the implication is that it greatly simplifies problems concerning the past, making their solutions easier. The information-theoretic perspective quantifies the notion of difficulty: memory makes acquiring knowledge of the past less difficult because it, in the above example the billiard board, requires only one binary bit (0 or 1) to represent the state of the particular billiard board in question. On the contrary, retrodiction requires many more binary bits to be considered in order to compute the same state that is represented using a single bit. To show how information can quantify difficulty, consider the simple example of species counterpoint. In a classroom setting, instructors and students alike agree that counterpoint exercises involving more voices are more difficult than those involving fewer voices. For example, a two-part third-species counterpoint exercise is considered easier than a four-part third-species counterpoint given the same cantus firmus. It is for this reason that the pedagogical order of introducing the subject is to begin with two-part writing, and gradually move to writings involving more voices. As the number of voices increase, students necessarily have to consider more information (i.e.~relationship between voices) in order to complete tasks such as avoiding parallel motion.

After recognizing the effectiveness of memory and records in general, a question naturally follows: what makes records so uniquely powerful in probing the past? Why can memory greatly simplify questions concerning the past? In other words, we need to closely examine the mechanism through which memory is used and justified as a valid method of knowing the past. It appears to us that we can magically read the past from either memory or record-bearing devices (such as a video footage) and trust the reading in the following sense: our memory and record-bearing devices may be instrumentally unreliable as to make technical glitches. However, once we are certain that there is no technical error with the records, we unconditionally trust its reading as the record of the past. In essence, we firmly believe that objects are capable to deterministically carry records of the past.

To continue the tradition in philosophical arguments that one often constructs absurd scenarios (for a famous example: brain in a vat) to illustrate a case, we consider a thought experiment concerning criminal investigations. In a criminal scene, the trace of fingerprint is a definitive proof of one's presence in the scene. We might question the instrumental reliability of reading the fingerprint trace. For example, the equipment might lack the precision required to give an accurate reading, or the fingerprint trace might be sufficiently deteriorated over time such that it is no longer intelligible. However, once we overcome the technical difficulties, obtain a reading of the fingerprint, and find a match to some suspect, then we can definitively assert his/her presence. In other words, we firmly believe that the fingerprint trace is, somehow, fundamentally capable of documenting a person's presence as a record of the past. The likelihood, that this fundamental (i.e.~ignoring the issue of instrumental reliability) correlation between the record (i.e.~the fingerprint trace) and the represented past (i.e.~the event that the suspect was present in the scene) is broken, is abysmally small to the extent that it is almost absurd to say that the fingerprint trace is formed in any ways other than the act of physical touching the place with the finger carrying that exact fingerprint. Similarly, Smart gives a more poetically tranquil metaphor:

\begin{quote}
Consider a stretch of sand on which there are footprints. We can deduce that a man walked that way because there is practically only one way in which an indentation like a footprint can come about. It is by the actual pressing of a foot into the sand. But if we see a virgin stretch of sand can we deduce that there will be a footprint? Obviously not. Virgin sand is compatible with there later on being footprints, claw prints, heel marks, wave marks, or anything else, including nothing at all. \autocite[80]{smart_temporal_asymmetry_of_the_world}
\end{quote}

Note that in both examples presented above, we have embedded a type of a priori assumption, that records faithfully give knowledge of the past. In particular, the assumption further implies that records are brought into reality with the sole cause of its representing events in the past. Therefore, the fingerprint trace cannot be due to \emph{any} means (e.g.~forming the fingerprint trace as the result of pure chance) other than the event of physical touching with a finger. Similarly, the footprints are record-bearing devices that come into existence with the sole cause of the actual foot-pressing event. It is so unlikely that we may say it is practically impossible to have the footprints that are due to spontaneous formation under the influence of wind or, say, butterflies. Such assumption is what justifies the use of records as a valid method of knowing the past. It assumes a causal relationship that necessitates events of the past to be the cause of their result records (i.e.~the effect). Notice how we do not hold the same assumption for the future: nothing in the present can be called a record of the future because, as the metaphor given by Smart illustrates, any present state can be associated with infinitely many future states. It is indeed possible (in fact, very likely) to have footprints now and they spontaneously deteriorate into virgin sand in the future. In mathematics, the simplification of a problem is often achieved using a stronger premise by introducing additional assumptions, making the problem enjoy more well-behaved properties. Now by the same token, an assumption we hold uniquely for the past but not the future, makes the premise stronger in inferring the past. In particular, this additional assumption translates to additional known conditions. As Albert points out, inference of the past using record-bearing devices in fact uses two known conditions:

\begin{quote}
    The sort of inference one makes from a recording is not from one time to a second in its future or past (as in prediction/retrodiction), but rather from two times to a third which lies in between them. \autocite[117]{time_and_chance}
\end{quote}

In prediction/retrodiction, one takes for granted the present condition of the world, whereas in reading record-bearing devices, an additional condition is given: the past moment in which the recording-bearing device interacted with its environment and became the bearer of records.

Upon contemplation, we should find the assumption peculiar. Firstly, it is universal and necessary (hence described as a priori). The assumption is naturally used in inferences of the past, regardless of the specific content of inquiry. In other words, we can have records to anything in the past, and they are not limited a certain set of objects. Therefore, for any inquiry about the past, we do not have to ask about the actual statement of the inquiry. As long as the inquiry concerns events in the past, we are justified to use the assumption and trust record-bearing devices to give us the answer. We may even claim that it is precisely due to our trust in record-bearing devices, that we have the very perception of time. For example, if we are asked the question ``which side is the left side'', we then actively feel our body and say ``it is the side where my hand plays the accompaniment of many pieces''. Analogously, if we are asked the question ``which side is the past'', we then look at whatever object is in front of us, and try to see if it can be the record-bearing device of any event. Then our answer would be that ``the past is the temporal side of the event to which the object in front of me is its record-bearing device''.

Secondly, we seem to have an exceptionally strong degree of belief in the assumption yet are neither aware of its justifications nor its origins. One may consequently ask due to curiosity: where does this assumption that records faithfully give knowledge of the past originate from? How can a mere assumption attract such a high degree of belief? On the one hand, this assumption that record-bearing devices are credible methods of knowing the past, lies entirely in our empirical world: it is a statement about the empirical universe. However, on the other hand, the assumption is exceptionally strong in the sense that it is also the justification for causality, which is fundamental to all sciences including non-empirical science such as logic.

Philosophers and scientists conventionally call this assumption the past-hypothesis. It can be stated in various forms under different contexts. In addition to the form stated above (that records faithfully give knowledge of the past), one often invokes a statement using thermodynamical terms: the initial macroscopic condition of the universe is low in entropy. \autocite[96]{time_and_chance} Albert further points out the link between the two ways of stating the past-hypothesis as two sides of the same coin:

\begin{quote}
{[}It{]} turns out that precisely the thing that makes it the case that the second law of thermodynamics is (statistically) true throughout the entire history of the world is also the thing that makes it the case that we can have epistemic access to the past which is not of a predictive/retrodictive sort; the reason there can be records of the past and not of the future is nothing other than that it seems to us that our experience is confirmatory of a past-hypothesis but not of any future one. \autocite[118]{time_and_chance}
\end{quote}

As the name suggests, it is a hypothesis we assume uniquely about the past, but not about the future. It is solely due to the past-hypothesis that we are justified to use records as a primary way of knowing the past, differentiating the past from the future. Additionally, the past-hypothesis gives rise to the temporal ordering of the world's narrative (i.e.~history), which scientists often encode using the second law of thermodynamics. This very temporal ordering, bearing close relationship to thermodynamics, is referred to as the arrow of time in its most etymological and original form.

\section{Arrow of time and its forms}
\label{section:arrow of time}

In this section, we consider the term ``arrow of time'' originally spelled out in science and examine a few candidates for its explanation. The temporal asymmetry represented by the difference between the past and the future is, as discussed previously, closely related to the unique and powerful role of record and trace.

\subsection{Thermodynamic arrow of time}
\label{section:thermodynamic arrow of time}

Now, we shift our attention and focus on the other aspect of temporal asymmetry: the temporal asymmetry of the world's narrative that is often described in terms of the second law of thermodynamics and the concept of entropy. The thermodynamic arrow of time is often considered the canonic form of temporal asymmetry in science. The term ``canonic'' in this context specifically means the following: 1) etymologically, thermodynamic is one of the first scientific studies of temporal asymmetry. Therefore, for historical reasons, the arrow of time refers to the thermodynamic arrow. Consequently, we are justified to classify various arrows of time into the thermodynamic arrow and other arrows. 2) the thermodynamic arrow of time along with its core concept (i.e.~entropy) can be interpreted as \emph{the} arrow through which all other arrows of time can be explained. As Carroll points out:

\begin{quote}
    \label{quote:carroll_arrows}
    {[}There{]} are many different ``arrows of time,'' {[}\ldots{]} We've been dealing with the thermodynamic arrow of time, the one defined by entropy and the Second Law. There is also the cosmological arrow of time (the universe is expanding), the psychological arrow of time (we remember the past and not the future), the radiation arrow of time (electromagnetic waves flow away from moving charges, not toward them), and so on. These different arrows fall into different categories. Some, like the cosmological arrow, reflect facts about the evolution of the universe but are nevertheless completely reversible. It might end up being true that the ultimate explanation for the thermodynamic arrow also explains the cosmological arrow (in fact it seems quite plausible), but the expansion of the universe doesn't present any puzzle with respect to the microscopic laws of physics in the same way the increase of entropy does. Meanwhile, the arrows that reflect true irreversibilities---the psychological arrow, radiation arrow, and even the arrow defined by quantum mechanics we will investigate later---all seem to be reflections of the same underlying state of affairs, characterized by the evolution of entropy. \autocite{from_eternity_to_here}
\end{quote}

Recall from the footprints-in-sand metaphor above, the crux of the argument lies in the observation that the present state of having footprints in sand corresponds to uniquely one possibility in the past, while it corresponds to many possibilities in the future. It is for such reason that the future appears unknown and to-be-determined to us: given the present state, anything (among many possibilities) in the future can happen. The observation suggests that, given the world now as a collection of record-bearing devices, the past is somehow more orderly because it corresponds to few possibilities, whereas the future is more chaotic because it corresponds to many possibilities. This characterization is considerably informal because it contains vague ideas such as ``orderly'' and ``chaotic'', and they are haphazardly associated with the number of possibilities. Nevertheless, it intuitively agrees with our everyday experience that the future appears indeterminate (i.e.~with many possibilities) and tends to get worse (i.e.~become chaotic).

Meanwhile, the ideas of orderliness and possibilities are useful tools in explaining the second law of thermodynamics, which is often accredited as the scientific root of temporal asymmetry we experience daily. In particular, the idea of orderliness corresponds to the concept of entropy; possibilities correspond to the concept of microstates. Among various forms of its statement, the second law is conventionally stated using the concept of entropy:

\begin{quote}
    The entropy of an isolated system either remains constant or increases with time. \autocite[32]{from_eternity_to_here}
\end{quote}

In layman's term, entropy can be interpreted as a measure of orderliness of arrangement: the higher the entropy, the more disorderly the arrangement is. Therefore, the second law is translated to the statement that the world tends to become less orderly over time. In fact, the second law is often used to define time. Physicist Eddington is often accredited for coining the very term ``arrow of time'', giving rise to its etymological origin. In his book ``The nature of the physical world'', he introduces the metaphor of the arrow using the idea of orderliness:

\begin{quote}
    Let us draw an arrow arbitrarily. If as we follow the arrow we find more and more of the random element in the state of the world, then the arrow is pointed towards the future; if the random element decreases, the arrow points towards the past. \autocite[79]{nature_of_the_physical_world}
\end{quote}

It is true that entropy and orderliness are not technically identical concepts, but it suffices to illustrate how the second law is ubiquitous in everyday experience. For example, consider a jigsaw puzzle where one's task is to recover the original picture by re-arranging scrambled pieces. In this case, we have a clear sense of what an orderly arrangement means: it means that the pieces are arranged exactly in the way that depicts the original picture. Notice how the second law comes into play in this case: during factory production, the original picture is printed on a single paperboard and the paperboard is then cut into pieces. If we are lucky, by the time the package is prepared for delivery, the pieces are still arranged in their solved state. However, during transit, the package will inevitably experience countless collisions at random, and these collisions effectively scramble the pieces into less orderly arrangements, i.e.~an unsolved state. The process in which the arrangement of pieces becomes increasingly disorderly is essentially an effect of the second law in reality. Similar processes are commonly found in everyday experience, so that people come up with pessimistic life lessons in the form of Murphy's law and its variants with the following assumption: the universe tends to evolve against our favor. Therefore, we are entitled to say that it is harder to make things right than to make them wrong.

One should be careful not to haphazardly personify nature. It is not that a state of affairs tends to evolve to a chaotic one, as if the universe is somehow evil and purposefully favors chaos. Rather, nature evolves indifferently, treating all possibilities equally. However, we, as humans, appreciate only a tiny portion of states and call them ``orderly'' in contrast to disorder. For example, in his book ``From eternity to here'', the author gives a motivating scenario:

\begin{quote}
You're out one Friday night playing pool with your friends. We're talking about real-world pool now, not ``physicist pool'' where we can ignore friction and noise. One of your pals has just made an impressive break, and the balls have scattered thoroughly across the table. As they come to a stop and you're contemplating your next shot, a stranger walks by and exclaims, ``Wow! That's incredible!''

Somewhat confused, you ask what is so incredible about it. ``Look at these balls at those exact positions on the table! What are the chances that you'd be able to put all the balls in precisely those spots? You'd never be able to repeat that in a million years!''

The mysterious stranger is a bit crazy---probably driven slightly mad by reading too many philosophical tracts on the foundations of statistical mechanics. But she does have a point. With several balls on the table, any particular configuration of them is extremely unlikely. Think of it this way: If you hit the cue ball into a bunch of randomly placed balls, which rattled around before coming to rest in a perfect arrangement as if they had just been racked, you'd be astonished. But that particular arrangement (all balls perfectly arrayed in the starting position) is no more or less unusual than any other precise arrangement of the balls. What right do we have to single out certain configurations of the billiard balls as ``astonishing'' or ``unlikely,'' while others seem ``unremarkable'' or ``random''? \autocite[157]{from_eternity_to_here}
\end{quote}

In the jigsaw puzzle example, out of all possible arrangements, only one arrangement counts as the orderly one, namely, the one that recovers the original picture.

Here is where the association between number of possibilities and orderliness enters. In everyday experience, one might wonder: what does it mean to say ``harder to make things right''? How should one go about quantifying the notion of difficulty? Number of possibilities gives a potential solution: we experience the effect of the second law by counting. The finding from the jigsaw puzzle can be equally applied to other processes in life: we call only a small number of possibilities ``orderly'', while all other possibilities are considered disorderly. In the essay ``Harmony explained'', the author shares an ``engineering adage''\autocite{harmony_explained} bearing salient resemblance to the finding from the jigsaw puzzle above:

\begin{quote}
There is one way for things to go right, but many ways for things to go wrong. \autocite{harmony_explained}
\end{quote}

He accompanies the adage with a convention in the field of computer science:

\begin{quote}
For example, in the Unix operating system each program upon exit returns to the system a number. In theory it could be used to mean anything, but all of the tools of Unix are set up to enforce the convention that the returned number be interpreted as an error code, as follows: zero is the code meaning ``ok'' and any other number means some kind of error occurred (where the map from the number code to the exact meaning of the error depends on the program returning the number). This convention works because of the above observation: the one uniquely distinct number, namely zero, is the code meaning that things went the one way for things to go right, and the rest of the numbers encode the many ways that things can go wrong. \autocite{harmony_explained} 
\end{quote}

It would be a mistake to consider the adage as a convention unique to the field of engineering. It is essentially a statement of life in disguise. In fact, similar observations are widespread throughout history and across fields of study. For example, in his ``Nicomachean ethics'', Aristotle observes that:

\begin{quote}
{[}It{]} is possible to fail in many ways (for evil belongs to the class of the unlimited, as the Pythagoreans conjectured, and good to that of the limited), while to succeed is possible only in one way (for which reason also one is easy and the other difficult - to miss the mark easy, to hit it difficult) \ldots{} For men are good in but one way, but bad in many. \autocite[38]{nicomachean}
\end{quote}

Similarly, Tolstoy begins his novel ``Anna Karenina'' with the following line:

\begin{quote}
Happy families are all alike; every unhappy family is unhappy in its own way. \autocite{anna_karenina}
\end{quote}

Now, the same argument (that the past is more orderly, corresponding to few possibilities; the future is more chaotic, corresponding to many possibilities) can be applied to physics with some modifications in wording. Firstly, the term ``possibilities'' is replaced by the concept of microstates. Without loss of generality, we can consider the example of the jigsaw puzzle again. A microstate in this example is simply a possible arrangement of pieces: the position of pieces in relation to one another. In general, a microstate is a detailed description that completely specifies the state of the physical system. The description must be as detailed as possible, insofar as the question at hand is concerned. For example, for the jigsaw puzzle, specifying the positional arrangement of each piece gives the most detailed description with respect to this puzzle: on the one hand, knowing physically more detailed description (e.g.~the position and velocity of individual molecules) than the positional arrangement of pieces does not further specify the state of the puzzle; on the other hand, knowing less detailed description (e.g.~the total weight of all the pieces) results in incomplete specification of the state, insofar as solving the puzzle is concerned. Given the concept of microstates, we consider the collection of all microstates. With the jigsaw puzzle example, we consider all possible positional arrangements of pieces. We observe two facts: 1) solving the puzzle amounts to a series of transitions between microstates. If we imagine each microstate as a point, then solving the puzzle is essentially about finding a path that connects multiple microstates such that it starts with the initial state of the puzzle and ends with the solved state. 2) many microstates are considered equivalent (or indistinguishable) to one another. For example, there are many possible positional arrangements (i.e.~microstates) corresponding to the description ``50\% completed'': a half-solved puzzle with the left half solved, right half solved, etc. Now, descriptions such as ``50\% completed'' are called macrostates: each macrostate corresponds to a collection of microstates. Again, if we visualize each microstate as a point, then macrostates specify regions containing collections of points. One crucial role of macrostate is that it contains microstates which are somehow considered equivalent and indistinguishable (mathematically, macrostates are equivalent classes that partition the set of microstates). At this point, we should note how macrostates can greatly simplify our jigsaw puzzle: all possible positional arrangements can be classified into macrostates corresponding to ``0\% completed'', ``1\% completed'', \ldots, ``100\% completed''.

The next crucial step involves considering the sizes of macrostates (i.e.~number of microstates belonging to individual macrostates), again by counting. In fact, this step is where the insight of physicist Boltzmann lies. Firstly, he refers to the concept of entropy:

\begin{quote}
Entropy is a measure of the number of particular microscopic arrangements of atoms that appear indistinguishable from a macroscopic perspective. \autocite[37]{from_eternity_to_here}
\end{quote}

According to his statement, entropy as a quantity is proportional to (the logarithm of) the size of macrostate: higher entropy means larger macrostate, and vice versa. Boltzmann's claim is then reduced to the statement that ``there are more ways to be high-entropy than to be low-entropy, just by counting microstates''. \autocite[166]{from_eternity_to_here} Going back to the jigsaw puzzle example, we notice how sizes of different macrostates differ: the macrostate corresponding to 100\% completed contains exactly one microstate, namely, the arrangement of pieces such that all of them are in the correct position. On the other hand, the macrostate corresponding to 0\% completed contains vastly many possible arrangements: an empty board, a 1\% filled board with all pieces in the wrong position, a fully filled board unfortunately with all pieces in the wrong position, and so on. By counting the size of macrostates, we know that the macrostate corresponding to 100\% completed has the least amount of entropy because it contains the least number of microstates. The counting argument then explains our intuition that it is harder to make things right (i.e.~100\% completed jigsaw puzzle) than making them wrong. Even if we pick microstates at random with equal probability, landing in a small macrostate is unlikely. It is through the different sizes of macrostates that nature appears to prefer increase in entropy over time.

We now consider two examples in music which illustrate the intuitive idea of orderly arrangement in relation to entropy discussed above.

\subsubsection{Entropy as seen from a performer's perspective}

Consider a thought experiment with the following setting: we are pianists who are currently practicing. Suppose additionally that we are technical perfectionists. As such, we play through a piece and record places where we miss a note. Piano is suitable for the purpose of illustration because we know for sure that a note is either played accurately, or missed (whereas string instruments have a more subtle notion of intonation, so that one's playing can somehow be in tune and out of tune at the same time). For now we only focus on missing notes because it can be easily quantified. One can surely invoke the argument that playing all notes right does not make you a true musician, and vice versa, but that is not the issue here because we can say that fixing missing notes (i.e.~the focus of this thought experiment) comprises only a tiny portion of our musical training. Insofar as the task of fixing missing notes is concerned, we can realize that, there is only one way to have a good performance: a performance where we miss zero notes. However, there are many ways to miss a note (or more): missing a note on the first beat of the 295th bar is in the same macrostate of ``performances with missed notes'' as missing a note on the last beat of the 162nd bar.

Moreover, we can have a more continuous description: the worse the situation gets, the more ways to achieve the situation. Playing two notes wrong is considered worse than playing one note wrong. Meanwhile, there are more ways to play two notes wrong than playing one note wrong. Using the terminology of macrostate, we say that the macrostate of playing two notes wrong is larger (equivalently, having higher entropy) than that of playing one note wrong. Again, one can invoke the classic argument saying that, ``maybe the music with more wrong notes sounds more appealing because it's more approachable and intimate since, in the same way tragedy is appealing, it reveals human vulnerability and the inevitable devastating force of fate!''. It's certainly an artistically valid argument, the same way one may argue that professional training is not necessary anymore because being amateur reveals a considerable amount of human vulnerability. Now, one may argue that, ``wait a second, how about the worst case scenario where you literally miss every single note in the performance? Is not there only one way to achieve this also?''. The answer is no. We can accomplish the tragedy of missing every note by playing the adjacent key to the correct one on the first beat, missing the note entirely by producing awkward silence on the second beat, etc. We can see the underlying principle: various places in a piece can go wrong, and for each place, we also have various ways to go wrong.

The second law of thermodynamics then translates to a common mindset performers have: while on-stage performance is exciting and enjoyable, it is inevitably fear-inducing. Individuals may differ in terms of what is feared. Some performers are fearful of memory slips, while others may be fearful of sound control. However, all types of fear share a common assumption, that the stage is to be feared as if a little demon is living there. The little demon constantly tries to interfere with our performance whenever it has a chance, in moments performers are underprepared or distracted. Additionally, the little demon finds its way to daily practice sessions. Consequently, without deliberate and diligent practice, performance tends to increase its entropy (characterized by increasing number of wrong notes for example discussed previously) over time. It is precisely due to the effect of the second law (or its personified form, the little demon) that performers over generations tirelessly emphasize the necessity of practicing.

\subsubsection{Entropy as seen in the dissonance/consonance dichotomy}

Dissonant and consonant harmonies are one of the most fundamental constructs in music. By applying the concept of entropy, we can observe the following: there are more ways to obtain dissonant harmonies (i.e.~high-entropy) than consonant harmonies (i.e.~low-entropy). In other words, the macrostate corresponding to dissonance is larger in size than the macrostate corresponding to consonance.

We build harmonies in a bottom-up fashion, starting with the simplest case consisting of two notes, namely, intervals. By convention, out of the 12 intervals, we have 7 consonant intervals (unison, minor third, major third, perfect fourth, perfect fifth, minor sixth, major sixth) and 5 dissonant intervals (minor second, major second, tritone, minor seventh, major seventh). So far, consonance outnumbers dissonance, which seems to contradict our claim that there are more ways to obtain dissonance.

From intervals, we are then able to build three-note structure, namely, trichords. Trichords are simply stacking of two intervals. However, we notice one crucial rule: trichords consisting of two consonant intervals are not necessarily consonant because the interval between outer voices might be dissonant. One simple example is the trichord made of two perfect fourths, where the interval between outer voices is a minor seventh. Meanwhile, any appearance of dissonant interval will make the entire trichord dissonant. Therefore, dissonant trichords can be divided into three classes: stacking of two dissonant intervals, stacking of one dissonant interval and one consonant interval, and stacking of two consonant intervals which happens to form dissonant interval between outer voices.

The question now is a combinatorial one: how many dissonant and consonant trichords are there, respectively? Because a trichord is uniquely determined by two intervals, we have \(12^2 = 144\) trichords in total. Out of the 144 trichords, we count the number of dissonant trichords (note that a trichord is either dissonant or consonant). As discussed above, dissonant trichords are divided into three classes. The class of dissonant trichords characterized by stacking of two dissonant intervals contains \(5^2 = 25\) members. The class of dissonant trichords characterized by stacking of one dissonant interval and one consonant interval contains \(5 \times 7 \times 2 = 70\) members. The third class of dissonant trichords is more complicated to calculate because we need to find out those trichords that are at the same time 1) stacking of two consonant intervals and 2) forming dissonant interval between the outer voices. With the aid of brute-force enumeration using a computer program, we learn that 16 out of 49 trichords as stacking of two consonant intervals are dissonant. After collecting the results, we observe that out of 144 trichords, we have 111 dissonant ones and 33 consonant ones. As we consider more number of voices, the dominance of dissonant sonority is increasingly evident: dissonant sonority (i.e.~the size of the macrostate corresponding to dissonance) outnumbers consonant sonority with increasingly higher percentage as number of voices increases.

Without going into further technical details, we should notice the crux of the matter: the reason that our example starts with consonance outnumbering dissonance, but as number of voices increase, dissonance eventually becomes dominant, is nothing other than the fact that we have stricter criteria to call something a consonant. In the example of trichord, a consonance requires that \emph{all} of the intervals are consonant, while a dissonant only requires that \emph{any} of them are dissonant.

\subsection{Other arrows of time}

\begin{figure}[!h]
\centering
\includegraphics[scale=1.5]{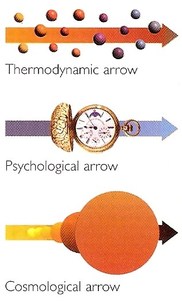}
\caption{Three arrows of time as discussed by Hawking}
\label{fig:hawking_arrows}
\end{figure}

The preceding section discusses the thermodynamic arrow of time as the canonic form. However, as the quote by Carroll points out on page \pageref{quote:carroll_arrows}, arrow of time has various forms such as cosmological arrow, psychological arrow, and causal arrow. For another example, in his book ``A brief history of time'', Hawking enumerates three arrows of time (see figure \ref{fig:hawking_arrows}):

\begin{quote}
    There are at least three different arrows of time. First, there is the thermodynamic arrow of time, the direction of time in which disorder or entropy increases. Then, there is the psychological arrow of time. This is the direction in which we feel time passes, the direction in which we remember the past but not the future. Finally, there is the cosmological arrow of time. This is the direction of time in which the universe is expanding rather than contracting. \autocite[184]{brief_history_of_time}
\end{quote}

It is true that all other arrows might be explained through the thermodynamic arrow using the concept of entropy, making them different sides of the same coin. Nevertheless, we should appreciate the value of other arrows of time as they offer unique perspectives and insights into the problem of time. In other words, all arrows of time are, to some extent, equally valuable. In fact, depending on the specific question being asked, some arrows may appear more canonic than others. For example, in discussing music, one often finds the psychological arrow more useful to describe the experience of listening to music, whereas thermodynamic arrow or cosmological arrow are less relevant in such context. It is true that the evolution of entropy as a physical quantity (i.e.~thermodynamic arrow) or the expansion of the universe (i.e.~cosmological arrow) might play a crucial role in the perception of music, but the link is not as natural as the link between the psychological arrow of time and our immediate appreciation of music.

In fact, according to Hawking, one of his claims is exactly that the psychological arrow of time is nothing but the effect of the thermodynamic arrow of time ``within our brain''.\autocite[189]{brief_history_of_time} Therefore, the thermodynamic arrow is proposed to be the explanation of the psychological arrow of time, as the latter is a special case of the former. If the psychological arrow can be reduced to its general form, i.e.~the thermodynamic arrow, what then is the point of recognizing the psychological arrow at all?

The answer is that the psychological arrow of time provides unique insights that are not obvious in discussion of thermodynamics. For example, section \ref{sec:record} is a discussion focusing on the psychological arrow of time: how the arrow is made by the difference in epistemic methods of acquiring knowledge about other times. In particular, the section focuses on the exceptionally powerful role of memory and record-bearing devices. It is true that the same result can be deduced if the discussion starts with the thermodynamic arrow of time instead, where record-bearing devices are made possible as we consider the role of entropy. However, insights about the epistemic aspect of time's arrow are lost in the midst of entropy-centric arguments.

By considering the psychological arrow of time, we can further make two observations that are not obvious from a thermodynamic point of view. The first is the idea of causality that cause always precedes effect in time. Furthermore, the psychological arrow of time involves a special type of causality: causality achieved through free will. In essence, we believe that there is something more to the temporal ordering of cause and effect in causal relations. In addition to the definitive temporal ordering, we believe that cause somehow has the active and organic force to actuate effect. In particular, we have the power, through will and action, to deliberately make things happen. According to Albert, this is the second fundamental epistemic conviction (the first being record and trace discussed in section \ref{sec:record}) ``we have about the difference between the past and the future, which is that the future depends on what happens now---that the future depends on what we do now---in a way that the past does not''. \autocite[125]{time_and_chance} In other words, the psychological arrow of time combines physical causality and free will.

One may argue that both causality and free will are mere illusions, thus participating in another round of heated philosophical debate on these controversial topics. However, we should notice that, regardless of the deep nature (if it means anything at all) of causality and free will, they give rise to the psychological arrow of time. More importantly, in the context of discussing the psychological arrow, the idea that we can cause things to happen through free will (i.e.~intervene the future) shapes our attitude about the past and the future. In particular, we have the attitude that we can somehow cause things to change in the future according to free will, whereas the past is outside of our causal influence. As such, the future is mutable and uncertain, hence our attitude is either excited or fearful. Meanwhile, the past is immutable and certain, hence our attitude is either satisfied or regretful. As a result, we ``dread future but not past headaches and prison sentences''.\autocite{sep-time-thermo} The difference between the past and the future is then translated to the difference in sentimental attitudes. For example, it is simply absurd to say that ``I look forward to having done that yesterday'', or ``I'm regretful about the decision I will make tomorrow''. Our psychological arrow of time, then, functions as a set of criteria to specify what the allowed emotional attitudes are. On the one hand, some emotional attitudes are time-neutral, such as anger. On the other hand, some emotional attitudes are time-sensitive, such as anticipation and regret.

In summary, the psychological arrow of time consists of two important components: the past and the future give rise to 1) vastly different epistemic methods of acquiring knowledge and 2) vastly different attitudes. Note that the two components are complementary in that they correspond, respectively, to the rational aspect and sentimental aspect of the human ego.

\chapter{Asserting the arrow of time: a comparative study}

One of the main objectives in chapter \ref{chap:time_arrow_science} is to assert the significance of temporal asymmetry by showing how it is acknowledged in the field of science and philosophy as one of the indispensable foundations of the world. In particular, we show that temporal asymmetry is not a mere social construct that is imaginary and ad-hoc, whose illusory appearance may motivate potential nihilists to dismiss it as irrelevant to the discourse. Instead, temporal asymmetry is as real and significant as to give rise to the very existence of our observed universe. To dismiss its importance is to deny virtually all imaginable aspects of our cognitive experience (in particular, artistic and musical experience).

\section{Social science}

In order to see how we must make the conscious effort to assert the significance of temporal asymmetry, we consider a case in the social science as a counterexample. It demonstrates how a concept, once it is interpreted as a social construct, becomes subject to potential deconstruction and challenge of its significance. We begin by considering the concept of nation. Among various interpretations, one particular interpretation defines nations as ``imagined communities'',\autocite{anderson2006imagined} a term originally proposed by Anderson in his book bearing the same name. According to Anderson, a nation is ``an imagined political community - and imagined as both inherently limited and sovereign''.\autocite[6]{anderson2006imagined} We should note that the author is by no means to deny the political importance of nation by calling it imagined. In fact, he introduces the idea of nations as imagined communities with the intention of acknowledging its virtue, by refuting the condescending attitudes his contemporaries hold against nationalism. In particular, his definition of nation is an attempt to fill the scholarly ``vacuum'' as he observes that

\begin{quote}
nationalism has never produced its own grand thinkers: no Hobbeses, Tocquevilles, Marxes, or Webers. This `emptiness' easily gives rise, among cosmopolitan and polylingual intellectuals, to a certain condescension. {[}\ldots{]} It is characteristic that even so sympathetic a student of nationalism as Tom Nairn can nonetheless write that: ``Nationalism'' is the pathology of modern developmental history, as inescapable as ``neurosis'' in the individual {[}\ldots{]}\autocite[5]{anderson2006imagined} 
\end{quote}

However, rather ironically, the term itself provides basis for its own deconstruction. The subtlety of the term ``imagined communities'' lies in that the adjective ``imagined'' suggests that it is a social construct, which can be arbitrarily fabricated. Anderson recognizes the potentially negative connotation associated with the term and tries to make a distinction between ``fabrication'' and ``imagination'':

\begin{quote}
Gellner makes a comparable point when he rules that `Nationalism is not the awakening of nations to self-consciousness: it invents nations where they do not exist.' The drawback to this formulation, however, is that Gellner is so anxious to show that nationalism masquerades under false pretences that he assimilates `invention' to `fabrication' and `falsity', rather than to `imagining' and `creation'.\autocite[6]{anderson2006imagined}
\end{quote}

Anderson tries to clarify that in his definition of imagined communities, the adjective ``imagined'' is to be interpreted in a constructive way, such that the legitimacy of a community is strengthened, rather than challenged, when it is an imagined one. But to no avail, his explanation does not alter the implications of the adjective ``imagined''. The reason for his ineffective explanation is evident if we recognize two layers of meaning pertaining to ``imagined''. The first layer is the one defined by the author, that nation as a community is the product of the imagination of its members, since ``the members of even the smallest nation will never know most of their fellow-members, meet them, or even hear of them, yet in the minds of each lives the image of their communion''. \autocite[6]{anderson2006imagined} The second layer, on the other hand, is the idea that nation as a community is the product of the imagination of us, as readers and spectators, who try to understand and discuss the concept, as a social construct. It is the second layer of meaning that makes the term subject to deconstruction and challenge. If an object is imagined, then it can be imagined in a multitude of equally justified ways, including the special case: its absence. In other words, once we accept the multitude of imaginations, we are equally justified to challenge the very act of imagining. In particular, as all social constructs can be deconstructed, the act of imagining can be challenged through deconstruction. Arguably, this underlying implication is what justifies activism on, for example, the topic of LGBT: one begins by realizing that various types of identities are socially constructed, so that the process of identification is an act of imagining. As such, we may challenge the norm by deconstructing identities such as gender, socio-economic status, and so on. Consequently, for the topic of nationalism, the concept of nation as imagined community would often fall victim to an ``identity crisis'', where one challenges, deconstructs, and casts doubt on the significance and legitimacy of a community that is \emph{imagined} rather than \emph{given}. On the one hand, one may acknowledge the instrumental importance of nation, as it may be necessary for utilitarian reasons. Namely, a nation organizes people socially to achieve (ideally) the greatest happiness for majority of its members. On the other hand, because a nation is imagined, it is no longer indispensable in the fundamentalist sense, that nations are themselves irreducible individuals which, analogous to biological organisms, attain their own consciousness and purpose. The term ``irreducible individuals'' here specifically refers to the observation that individuality as a social concept cannot be deconstructed (hence the name ``irreducible'') as a social construct. In particular, few in the social science is entitled to challenge the value and significance of individual human right. For example, we would not interpret the pursuit of happiness as a social construct whose value is to be challenged and imagined in a multitude of ways. Therefore, from a fundamentalist perspective, once nations are interpreted as irreducible individuals, they become the \emph{teleo} in themselves, to whose well-being the happiness of the majority is a by-product. Consequently, citizens become a derived concept, to the extent that, instead of interpreting nations as imagined communities, one now interprets citizens, reciprocating Anderson's terminology, as ``imagined individuals''.

The case in social science illustrated above serves as a counterexample. In comparison, the arrow of time is significant precisely in the sense that it cannot be interpreted as a social construct, hence deconstructed. It is true that music, when interpreted as an inherently social activity, is a social construct whose deconstruction is under active research in the field of ethnomusicology. However, the arrow of time, as an \emph{aspect} (whose meaning is to be taken from section \ref{section:aop}) of music (as well as other parts of the world, since it is a cross-cutting concern), is not subject to potential deconstruction and challenge of its significance. One of the primary findings in chapter \ref{chap:time_arrow_science} is that the arrow of time is not a concept that can be imagined arbitrarily in a multitude of ways: in philosophy and science, as chapter \ref{chap:time_arrow_science} illustrates, we have a rather clear (albeit incomplete) picture of what the arrow of time specifically refers to.

\section{Visual art}

Supposing that we would like to challenge the significance of the arrow of time in music, how would one go about deconstructing and casting doubt on its significance? One potential strategy is to invoke an argument similar to that with the imagined communities, by recognizing its instrumental importance yet deny that it is somehow a fundamentally indispensable part of music. Specifically, one can invoke the following argument: the role of time in music, along with all its properties such as temporal asymmetry, is essentially the same as in the visual arts. We look at works of visual art, and realize the also instrumentally uses the medium of time. In particular, as our gaze visits different parts of a painting in temporal sequence, we actively participate in the act of creating a narrative. Because of this implied narrative, the arrow of time, such as our epistemic asymmetry between the past and the future, remains critical in forming the narrative. Painters often use visual elements to serve as cues that direct our attention, thus shaping the viewer's temporal experience. For example, in a beautiful rendition of the rural scenery, is it not due to the horizon afar along with its vanishing point that our attention is directed and translated to a yearning for a temporal journey inside the depicted landscape? Metaphorically, we may think of the interplay between the spatial features (such as scenery, altitude, difficulty of the trail) of a hiking trail and the temporal journey of the hiker shaped by those spatial features: spatial features, analogous to visual elements in a work of visual art, have the ability to form (or more precisely, suggest) temporal narrative. Through careful design and placement of visual elements, one can thus create narrative and temporality in visual art (which is ironic in the context of our current discussion because, in music, we try to deny temporality). In his essay ``The American action painters'', Rosenberg highlights the temporal aspect of visual arts:

\begin{quote}
At a certain moment the canvas began to appear {[}\ldots{]} as an arena in which to act - rather than as a space in which to reproduce, re-design, analyze or ``express'' an object, actual or imagined. What was to go on the canvas was not a picture but an event. \autocite{rosenberg1991american}
\end{quote}

Even though Rosenberg writes the paragraph with a very specific artistic movement and time period in mind, we can nevertheless apply this view to a broad spectrum of art history.

Specifically, in analyzing a work of visual art, one interprets it as part of a larger context, i.e.~as a moment in a narrative (which is necessarily temporal by definition). To give a motivational real-life example, consider the act of taking pictures in everyday life. The purpose is to capture and remember cherishable moments in life. However, the picture by itself is meaningless, unless one takes into consideration the surrounding life stories. A compelling work of visual art should then have the power to inform the viewer, from the work alone, about the implied narrative. Note that the very process of informing about the implied narrative constitutes the act of constructing temporality. Of course, there is always the question that is not limited to visual arts: to whose narrative do we refer? It could be the viewer's (meta)narrative as he/she walks inside the gallery space, relating his/her gallery experience with his/her life experience. This is exactly the narrative highlighted by Mussorgsky's ``Pictures at an Exhibition'' as he interweaves promenade movements with the rest of the work. The narrative could also be the autobiographical narrative of the author, as Rosenberg points out that a painting ``that is an act is inseparable from the biography of the artist''.\autocite{rosenberg1991american} Finally, treating an artwork as an autonomous object in itself (in a similar manner Hanslick advocates for absolute music) disregarding the previous two types of narratives, we obtain the actual story represented by the artwork. This is the type of narrative to which we commonly refer in visual arts. Because visual art, unlike music, is mostly representational, the majority of this type of narrative is analogous to what musicians call a program (as in program music). However, regardless of how we answer the ``whose narrative'' question, the outline of argument remains the same: given a work of visual art, we notice two premises.

The first premise states that the temporal order in which viewers inspect the work is arbitrary in the sense that artists are unable to force any temporal ordering. The viewer is free to inspect details in any sequence at will: one may start from the top-right corner and proceed to the center, and the opposite is equally justified. In contrast, temporal ordering is ingrained in musical works, as it is physically impossible to experience the ending before the performance undergoes, in sequential order, all moments before the ending. In other words, the first premise is reduced to the statement that temporal ordering in visual arts can only be achieved suggestively, i.e.~it can only be \emph{suggested} rather than \emph{imposed}.

The second premise subsequently concerns the methods through which temporal ordering can be suggested. It reads that visual artists have nothing but spatial arrangements of strokes and colors on canvas at their disposal. Therefore, if artists are to suggest temporal ordering (i.e.~a narrative), they can achieve the goal solely with the help of strokes and colors. With the two premises above, the claim is that, even though the trajectory of the viewer's gaze, i.e.~temporal ordering of looking into details, is arbitrary, it can nevertheless be guided through spatial arrangements of strokes and colors.

\begin{figure}[!h]
   \centering
   \includegraphics[width=.8\textwidth]{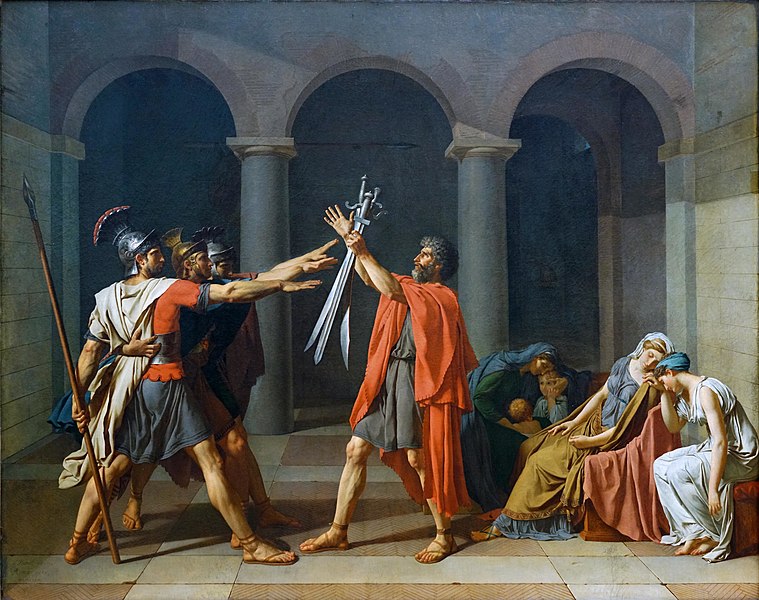}
   \caption{David, Oath of the Horatii (1784)}
   \label{fig:david_oath}
\end{figure}

Consider the painting ``Oath of the Horatii'' by David (see figure \ref{fig:david_oath}). It is programmatic in the sense that it intends to depict a moment of a story. As the title suggests, it involves a theatrical setting: we have an action related to taking oath, as well as the characters relevant to this action, namely, the Horatii warriors taking the oath. Upon encountering such painting, our first reaction is likely a temporal one: what happens before and after this moment? Our curiosity for stories naturally motivates us to complete the story based on this picture as a hint. Furthermore, our curiosity is strengthened by the unrestfulness conveyed. Firstly, the unrestfulness is physical: the four men in the picture are in an unstable posture, suggesting that they are in the midst of active and ongoing movements. Therefore, there are events before (e.g.~raising the swords) and after (e.g.~announcing the oath) this instantaneous moment. Secondly, the unrestfulness is psychologically portrayed by unsolved puzzles: the depiction of the three women is unsettling because they introduce more curiosity. Besides the obvious questions such as ``who are they'', one may ask: is their sorrowful presence an indicator of the upcoming danger and difficulty implied in the oath? If so, what is to be expected in the story after this depicted oath-taking moment? The viewer is more eager to know the contextual story in order to explain the presence of the three women. Now, the author carefully designs and places visual elements in order to guide our attention, suggesting a temporal order in which we inspect the details. In particular, our gaze may initially land at any place in the painting arbitrarily. However, the visual elements direct our gaze such that we eventually arrive at the ``point of convergence'' of the artwork. As a result, the temporal ordering means precisely the fact that no matter where one starts the gaze, the attention is eventually directed to the ``point of convergence''. Where is this point? Consider how eyes are portrayed in this painting. If we begin our inspection of details with any of the three brothers on the left, then their eyes direct our attention to where they stare at: the swords in the center. Alternatively, if we begin with the right-hand side of the painting, we are lost as the three women are too sorrowful to make eye contacts. However, we may interpret it as a deliberate choice made by the author: it motivates us to actively look for characters with eyes open. Unsurprisingly, we find such character at last: a child in the shadow, unaware of the world in which grown-ups struggle, stares at the man in the middle, who in turn transitively directs our attention to the swords at his hand. Therefore, by observing the visual elements of characters' eyes, we are able to find that they are visually constructed in a way that uniformly direct our attention to the swords in the center.

The discussion above indicates that temporal narrative can be a crucial component of visual arts, such that we recognize the instrumental importance of time in construing and appreciating visual artworks. However, by convention, we still refuse to classify visual arts as temporal arts. When compared to temporal arts such as music, we feel strongly that something is fundamentally missing in visual arts, and this missing ``something'' makes up the difference between painting and music, thus enabling us to subsume music (but not painting) under the category of temporal arts. The difference lies precisely in the discussion above, that temporality can only be suggested rather than imposed in visual arts, whereas in music artists are additionally equipped with the ability to impose temporal ordering. Moreover, the difference between suggesting versus imposing temporality is not a mere play of words.

To show how it is not a mere play of words, we start the argument from a rather vague intuition: when we construe a work in visual arts in terms of its content, presentation and delivery, temporality in the form of suggested narrative seems to be an attribute of us, instead of the artwork. In other words, visual arts induce narrative but are not themselves narrative-bearing objects. Note that this is essentially a restatement of the idea that temporal narrative is suggested. As visual elements suggest temporal narrative, they (by definition of the term ``suggest'') express mere opinions. Opinions may be strong in degree (i.e.~the visual elements are constructed in a strongly opinionated way such that there is essentially one way to interpret the work), yet it is up to the viewer's own discretion to evaluate (or more accurately, decipher) these opinions and construct his/her temporal narrative in response. During this process, the viewer takes control of all the decision-making, thus qualifying our intuition that temporality is ``an attribute of us'': the viewer is the sole party constructing temporality, using the mental faculty of imagination. In the example of ``Oath of the Horatii'' discussed above, the narrative in terms of our speculations of the happenings before and after the depicted moment of oath, as well as the narrative in terms of the trajectory of our gaze directed by visual elements, are both the sole product of the viewer's imagination. Poetically, they are like a piece of cloud in the sky, undergoing constant transformations in shape. At one moment, it is morphed to take a shape that saliently resembles a giant donut, i.e.~it \emph{suggests} a donut the same way visual elements suggest narrative. However, to our dismay, the donut is the sole product of our imagination, such that if we take a picture of the cloud and ask ``where is the donut found \emph{inside} this picture?'', it is nowhere to be found. It is true that interpretation of narrative in music is also heavily based on the listener's imagination (after all, music without listener's active imagination is reduced to mere sound). However, it is plainly false to claim that musical narrative is the \emph{sole} product of the listener's imagination: our anticipation of the future unheard musical events is not a product of imagination, but the lack of epistemic access to the future.

The term ``imagination'' serves a pivotal role here, as it allows us to draw parallels between imagined communities in social science and suggested narrative in visual arts. In particular, recall the criterion that if an object is imagined, then it can be imagined in a multitude of equally justified ways. Now, temporal narrative in visual arts fits this criterion: as the temporal ordering is suggested rather than imposed, it can be imagined arbitrarily in a multitude of ways, including its absence. As discussed above, the viewer is free to inspect details of a painting in any order, while disregarding what the visual elements suggest. In comparison, note how this is not applicable to music. If somehow we manage to transform music such that we are allowed to begin inspecting a musical work at arbitrary places and to inspect details in any order (other than that imposed by the performance), then what does this kind of ``music'' look like? In fact, this is not a hypothetical question, as the answer is readily found in reality. Musicians are all familiar with this kind of non-temporal music: notated music scores. As we perceive the notated score as a purely visual construct, all imposed temporality is lost: we may start our gaze from anywhere on the score, and moreover, we can construe the entire score in simultaneity (by printing all pages and laying them side by side). However, the crux of the matter lies in the fact that notated music score cannot live independently from its performance, and its performance necessarily imposes temporal ordering. Therefore, we need to further specify the question as follows: what kind of music would it be, if it is to be made completely free from any kind of imposed temporal order whatsoever? The answer, in fact, is quite straightforward: such task is certainly possible, simply because we would no longer call the end result ``music''.

The attempt to nullify the imposed temporal ordering associated with performance is found in music history, and unsurprisingly, the strategy is to make music visual-art-like. In music history, such a trend has a rather self-evident name: eye music (alternatively, Augenmusik). The Grove dictionary dedicates an article to the term ``eye music'' (which is ironic, as it dedicates articles to such specialized term, yet fails to do so for the term ``time'' (see page \pageref{quote:grove_no_time})), defining the term as follows:

\begin{quote}
Musical notation with a symbolic meaning that is apparent to the eye but not to the ear (e.g.~black notes for words such as `darkness' and `death'). Since its effects are derived from notation it is the concern of composers and performers rather than listeners. \autocite{eye_music}
\end{quote}

\begin{figure}[!h]
   \centering
   \includegraphics[width=.6\textwidth]{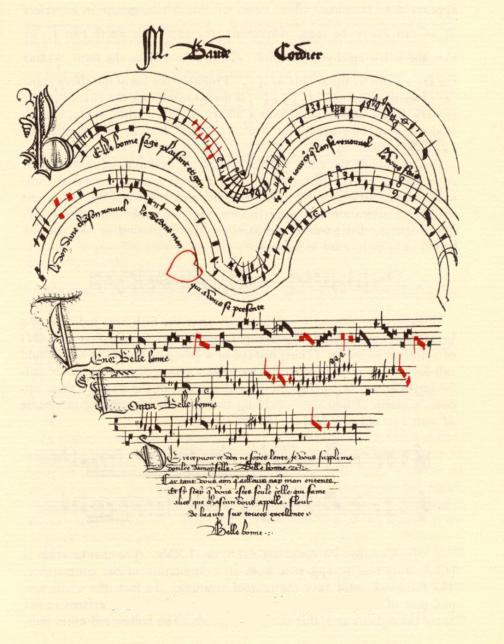}
   \caption{Cordier, ``Belle, bonne, sage" (15th century)}
   \label{fig:eye_music_1}
\end{figure}

For example, one ``type of eye music involves writing out the music of a love-song on a staff bent into a heart shape'' (see figure \ref{fig:eye_music_1}).\autocite{eye_music}

\begin{figure}[!h]
   \centering
   \includegraphics[width=.7\textwidth]{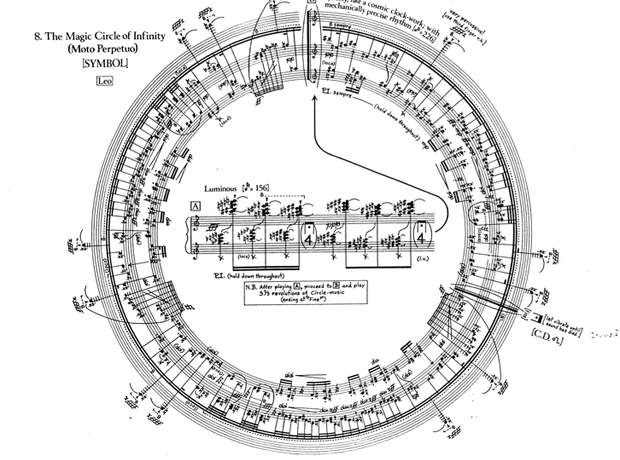}
   \caption{Crumb, The Magic Circle of Infinity (Moto Perpetuo), from ``Makrokosmos", volume 2 (1972)}
   \label{fig:eye_music_2}
\end{figure}

One must recognize that the term eye music refers loosely to a broad range of musical works across different time periods and styles. Therefore, the term potentially introduces controversy as we try to come up with a precise definition and criteria, according to which a work may be regarded as eye music. For example, as Taruskin observes in Crumb's music (see figure \ref{fig:eye_music_2}):

\begin{quote}
Some works of his are notated on circular staves, others on spirals, still others in the shape of crosses. But the intention, as Crumb has explained it, is not only symbolic (as in the Augenmusik or ``eye music'' of old). Circular staves can represent all kinds of mystical notions, it is true, but they are also the most efficient way of notating a moto perpetuo (and were used for that purpose since the early fifteenth century). \autocite{taruskin2006music_late_20th_century}
\end{quote}

Therefore, according to the way Taruskin interprets Crumb's intention, circular staves that Crumb employs are technically not eye music, because they are the more effective ways to encode a musical message (i.e.~moto perpetuo) which is indeed audible. In other words, unconventional notations are not indicators of eye music, but the exact opposite of what we think of eye music: they are visual notations that, ironically, serve a better musical purpose. Therefore, it is possible that according to Crumb, the repeat sign we encounter in conventional staff notation is what should be called eye music: it notates circularity of time in an exceptionally unnatural way, so that the musically connected measures are haphazardly notated with discontinuity (simply imagine the frustration when performers have to turn several pages at once at the end of the exposition). However, due to notational conventions, we are nurtured in a way that we become desensitized of how the repeat sign is unnatural. Instead, we call notations with which we are unfamiliar a disdaining name: eye music.

After recognizing the complexity and ambiguity entailed by the term eye music, the scope of the discussion here is limited to ideas associated with eye music, rather than trying to participating in the debate of how eye music is specifically defined. In particular, we focus on the idea that one can nullify the imposed temporal ordering associated with performance by highlighting visual elements on score that are often irrelevant to performance (i.e.~cannot be heard). As a result, these visual elements do not translate into audible ones as part of the listening experience. As such, they are to be construed more appropriately as visual elements in the sense of visual arts: they are strokes and colors whose arrangements direct and guide our attention. Therefore, as far as the visual elements in works dubbed ``eye music'' are concerned, they transform the music and push it into the realm of visual arts. In the hypothetical extreme case, the notated score of eye music is to be appreciated independently from its performance, whose sole value is the visual representation of the score. An artwork in this case becomes a work of visual art: it is to be analyzed using methods and terminology of visual art. In particular, temporal ordering is no longer imposed, but merely suggested. In short, ideas associated with eye music showcase the consequence of trying to achieve the ``emancipation of the temporal ordering'' (borrowing the term from Schoenberg's ``emancipation of the dissonance''): as we decouple music from its performance, the temporal ordering imposed by performing is no longer ingrained in music. Consequently, temporal ordering from the listener's perspective (i.e.~the ordering of perceived musical events in time) becomes suggested rather than imposed. However, at the same time, ``music'' that is free (i.e.~emancipated) from performance and imposed temporal ordering is simply no longer music. It can be interpreted as anything (such as visual art as in the example of eye music, or other undiscovered fields) but music, since it loses the very defining feature of music we are so compelled to believe: performance and the arrow of time brought forth by performance.

\section{Theology}
\label{section:theology}

Instead of being imagined arbitrarily in a multitude of ways, the arrow of time should be acknowledged, almost religiously, as the axiomatic, unconditional, and intrinsic aspect of our experience (in particular, musical experience). In fact, one may arguably take the term ``religiously'' literally if we compare the arrow of time with the theological idea of eternal election and observe how both reveal an uncompromising attitude. Calvin, one of the leader figures in the Reformation movement, proposes the idea of eternal election, claiming that one's salvation is predetermined through the eternal election, which is completely independent of work. His argument proceeds by firstly recognizing how God is omnipotent as he knows and predetermines all of eternity in one single timeless moment:

\begin{quote}
When we attribute foreknowledge to God, we mean that all things always were, and perpetually remain, under His eyes, so that to His knowledge there is nothing future or past, but all things are present. And they are present in such a way that He not only conceives them through ideas, as we have before us those things which our minds remember, but He truly looks upon them and discerns them as things placed before Him. And this foreknowledge is extended throughout the universe to every creature. \autocite[184]{hillerbrand1968protestant}
\end{quote}

As a side note, his argument about God's time completely undermines the very thesis of this project. However, it does no harm to our inquiry because theological arguments maintain a very different set of premises to begin with. For example, arrow of time is not applicable to God as he does not view the past and the future differently due to the fact, according to Calvin, that his epistemic access is omnipotent in all of space and time.

Now, eternal election means that one's life path, in particular, the decision on whether one is saved (i.e.~elected) or damned is predetermined for all eternity:

\begin{quote}
He determined with Himself what He willed to become of each man. For all are not created in equal condition; rather, eternal life is foreordained for some, eternal damnation for others. Therefore, as any man has been created to one or the other of these ends, we speak of him as predestined to life or to death. \autocite[184]{hillerbrand1968protestant}
\end{quote}

Meanwhile, God's decision is completely free in the sense that it does not take into account any human concerns such as justice, fairness and comprehensibility:

\begin{quote}
We assert that, with respect to the elect, this plan was founded upon His freely given mercy, without regard to human worth; but by His just and irreprehensible but incomprehensible judgment He has barred the door of life to those whom He has given over to damnation. \autocite[189]{hillerbrand1968protestant}
\end{quote}

In order to make his theological interpretation clear, Calvin highlights that one's work (i.e.~what we do in life) bears absolutely no effect to the election:

\begin{quote}
Surely the grace of God deserves alone to be proclaimed in our election only if it is freely given. Now it will not be freely given if God, in choosing His own, considers what the works of each shall be. \autocite[192]{hillerbrand1968protestant}
\end{quote}

Now, eternal election would appear as completely incomprehensible from a human perspective that is inherently limited. However, this is precisely the point Calvin tries to clarify: the incomprehensibility shows our weakness, not God's. For example, one might say that eternal election is unequal as some people are predestined to be denied salvation. However, according to Calvin, the ``very inequality of His grace proves that it is free''. \autocite[187]{hillerbrand1968protestant}

From Calvin's interpretation, we can vividly feel the uncompromising attitude entailed by the term eternal election. One must unconditionally accept eternal election as the unquestionable premise, from which one then discusses, under the premise of eternal election, common topics such as the issue of theodicy. For example, given the eternal election unconditionally, how would one proceed to explain the existence of evil in the world? Now, we cannot alter the premise of eternal election but to treat it unconditionally true. Meanwhile, we cannot ignore the observation that righteous people are sometimes mistreated (which, unsympathetically, is the inspirational source of many great works of tragedy). Therefore, the only way to reconcile the uncompromising premise and the factual observation stated above, is to conclude that human perspective is too limited to understand the omnipotence of God. This is in fact, one of the main doctrines of Calvinist theology, that God's omnipotence surpasses the human understanding of justice and fairness. The incomprehensibility of his decisions precisely highlights our limit and humbleness. As such, theodicy is no longer an issue that is frustrating, but rather, it humbles us to recognize the glory of God.

Now we can observe how the eternal election and the arrow of time are similar with regard to their uncompromising attitude. The general procedure runs as follows: given a premise (i.e.~arrow of time and eternal election), we treat it axiomatically by unconditionally assuming its truth. The question then becomes the following: how can we make sense of the world, given the premise \emph{as is}? In the case of the arrow of time, we are to make sense of the happenings we observe and experience in terms of the evolution of entropy as well as the epistemic interplay between memory and expectation, whereas in the case of eternal election, we are to make sense of the issue of theodicy. In both cases, the premise itself remains insusceptible to challenges. In particular, it is unchallenged in the sense that we are willing to insist on keeping the premise at the expense of adjusting the conclusion (even when the conclusion is unfavorable, such as admitting our mortal limit in theology, as well as our ignorance of the future).

Now, a mathematical method comes in handy: in mathematics, arguments are in the form of deductions, where one assumes a set of premises. Through inference, one can then deduce conclusions based on the given premises. One special quality of mathematics as a formal science (rather than natural science) is that no proposition (i.e.~either the premise or the conclusion) is intrinsically true: premises are assumed to be true such that all mathematical statements are in the form ``\emph{if} A, then it follows that B'' (notice the term ``if''). Metaphorically, premise and conclusion form the two ends of a rope. Depending on our goal, we can fix either end of this rope: we can keep the conclusion fixed if we highly value the conclusion, and try to adjust the premise such that the valued conclusion follows more naturally. By the same token, we are justified to do the opposite. Clearly, as discussed above, the arrow of time and eternal election exhibit a strong insistence on the premise.

In comparison, the unchallenged status of the premise does not hold for social constructs such as gender. In fact, the basis for, for example, the LGBT movement is the very idea that we have the right to challenge and deconstruct the normative premises of gender and other identities, and see how the world potentially becomes a better place as a result. In short, the contrast between the arrow of time and eternal election on the one hand, and social constructs such as gender on the other hand, is reduced to difference in preference: if the premise is preferred over conclusion, then one obtains the examples of the arrow of time and eternal election; if the conclusion is preferred instead, one obtains the examples of social constructs.

\chapter{Temporal arts as narrative}
\label{chapter:temporal arts as narrative}

The arrow of time (i.e.~temporal asymmetry), in particular, the psychological arrow of time, gives rise to the expressive power of all temporal arts. Now, as the arrow of time is ubiquitous throughout all aspects of our cognitive experience, the expressive power resulting from the ubiquitous arrow should also relate to statements universal to all temporal arts. Here we propose a universal statement as an attempt to address all forms of temporal art. In order to propose a universal statement, one natural strategy is to take a bottom-up approach. Being musicians, the most natural point of departure is the first-hand experience and observations in the field of music. By observing the expressive power commonly attributed to music as an instance of temporal art, our goal is to generalize our observation and propose a statement that is applicable to music as well as other temporal arts. In particular, we take a frequentist approach: by observing and collecting recurring themes that are frequently brought up in musical discourses (in particular, instrumental lessons), the task is to see if these recurring themes can potentially shed light on temporal arts in general, thus providing insights into attributes universal to all temporal arts.

In music lessons, disregarding idioms specific to instruments (such as fingering for keyboard instruments and bowing for strings), musicians often describe performance as an analogue to story-telling. Consequently, to be a good performer is equivalent to being a good story-teller. The metaphor continues: a good story-teller is a spiritual guide, who invites the audience to the spiritual world of music, in which the audience is guided and actively participating in an adventurous spiritual journey. In such a journey, every listener becomes Odysseus, who

\begin{quote}
\centering
    wandered and was lost

    when he had wrecked the holy town of Troy,

    and where he went, and who he met, the pain

    he suffered in the storms at sea, and how

    we worked to save his life and bring his men

    back home \autocite{odyssey}
\end{quote}

A successful performance is then a breathtaking journey such that when listeners are back in reality, they invariably cry out ``what an experience!''. We should note that interpreting music performance metaphorically as an act of story-telling and spiritual journey is a universal statement in the sense that it does not take into account what the musical content is. Rather, the focus is solely on the very act of performance. Therefore, the division between program music and absolute music is irrelevant to the story and spiritual journey discussed here. Music performance is considered an act of story-telling regardless of whether the piece being performed has a program. Performers are story-tellers even when the performed piece is classified as absolute music without apparent program. In other words, the concept of ``story'' here makes no reference to the program of the music. From here we can see the subtlety: the musical narrative, i.e.~object of story-telling, is self-contained and autonomous precisely in the sense Hanslick coins absolute music:

\begin{quote}
It is a specifically musical kind of beauty. By this we understand a beauty that is self-contained and in no need of content from outside itself, that consists simply and solely of tones and their artistic combination. \autocite[28]{hanslick1986musically}\label{quote:hanslick}
\end{quote}

Therefore, we are arriving at a seemingly paradoxical conclusion: by suggesting that music is analogous to narrative (as a term borrowed from literary studies), we are in fact asserting the autonomy of music, such that it has its own narrative whose ``plot'' cannot be translated to extramusical terms.

By interpreting music performance metaphorically as an act of story-telling, we gain several advantages as a result.

\section{Narrative as pedagogical tool}

The first advantage is instrumental, in the sense that such metaphor is an effective pedagogical tool. To interpret performance as story-telling, is pedagogically similar to using visual aids to inspire imagination.

\begin{figure}[!h]
\centering
\includegraphics[width=.8\textwidth]{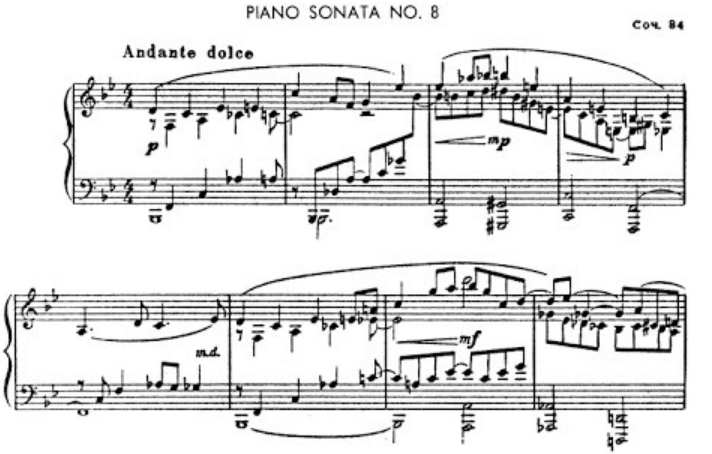}
\caption{Prokofiev, Piano Sonata No.~8, Op. 84, opening of the first movement}
\label{fig:prokofiev_8_opening}
\end{figure}

For example, consider the opening of Prokofiev's 8th piano sonata (see figure \ref{fig:prokofiev_8_opening}). One may interpret the general mood of the opening as a mixture of gloominess and expansiveness. The gloominess is expressed through the counterpoint resulting from multiple voices crawling, where the movement in semitones is particularly alarming. Meanwhile, due to the spacing between left-hand and right-hand voices, the music is expansive in that it creates a sense of spatiality. The expansive spatiality is maximal in the third measure, where the crescendo is accompanied by contrary motion between the two hands. Contrary motion creates expansiveness in two regards. The first regard reads: contrapuntally, outward contrary motion has a typical example, i.e.~the resolution of the augmented sixth interval to an octave, which creates a sense of expansion because it is intervallically expanding in the literal sense, as well as easing the harmonic tension. The same is true in the case of our example: the downbeat of the third measure makes up a tritone between the outer voices, which expands in register through contrary motion, magically arriving at a minor third on the third beat of the same measure. The second regard is more performance-oriented: it concerns the bodily gesture which is an inseparable part of performer's understanding of music. As both hands expand in register, we create two gestures that are both suggesting expansiveness. One gesture is technical, that it is physiologically more effective to transfer body weight (which is necessary in the modern theory of piano performance, whose emphasis lies in injury prevention) to the keyboard, when hands are placed symmetrically around the performer's center of mass (i.e.~the key that we can play most conveniently with our nose, which approximately corresponds to the middle C) and are reasonably far apart. While it is tedious to explain technical gestures in words, one can readily see how the statement above is true by experimenting push-up exercise with varying distance between hands. Another gesture is, on the other hand, symbolic: hands placed far apart on the keyboard are symbolic of the concept embrace and hug, which is often associated with the feeling of expansiveness. The association is rather more literal than symbolic: when we embrace, arms naturally expand, creating a feeling of coziness. This is exactly the reason that in his neo-Hellenistic endeavor, Holderlin conveys the poetic sense of warmth and coziness with the simplest choice of wording:

\begin{quote}
\centering
    The euphony of the rustling
    
    meadow was my education;
    
    among flowers I learned to love.
    
    I grew up
    
    in the arms of the gods.\autocite{holderlin2004poems}
\end{quote}

We may claim that the poetic climax of the entire quote is the word ``arms'', as it poetically gives life to all other elements in the poem: through the arms of the gods, we are embraced by the gods, meadow, flower, and more importantly, the infinitely expansive universe.

The mixture of gloominess and expansiveness is indeed a strange emotional mixture, as if we are juxtaposing two opposing magnets. On the one hand, the gloominess suggests introspection concerning contemplation of one's inner state. On the other hand, the expansiveness suggests something outgoing concerning the external environment. It is for this reason that we have different dispositional interpretations of the following two statements: ``the weather is gloomy today'', and ``the mountain scenery is expansive''. Note that they are both factual statements. However, gloomy weather is equipped with a certain degree of sentimental implication, to the extent that gloomy weather \emph{reflects} our mood. Now, this strange emotional mixture translates to challenges in musical understanding. From a pedagogical perspective, how would instructor inspire the student to grasp the musical representation of this emotional mixture? The primary challenge concerns the process of synthesis. The student may be familiar with musical depiction of gloominess as well as that of expansiveness separately. However, how should one go about synthesizing the two moods into a unified musical presentation, especially when the two moods seem to be in conflict with one another? This challenge about synthesis has its analogue in the technical aspects of the piano: one may know how each hand should play its part, but putting together (i.e.~synthesizing) both hands requires extra practice because it poses a challenge of body coordination which is found nowhere in single-hand practice. Therefore for students, such task of synthesizing two (conflicting) musical moods may be as contradictory as playing fortissimo using the dynamic of pianissimo (note that this is a restatement of challenges pianists encounter on everyday basis: playing powerfully without hammering the keyboard).

In such a pedagogical scenario, visual aids come to the rescue. The instructor inspires the student to think associatively about a scene in which one can see the gloomy and cloudy weather. At the same time, one stands at the center of an open field. From the standpoint, one can see a mountain near the horizon. However, the image of the mountain is blurred by the cloud so that one is unable to see its peak. Now, such scene is spatially expansive and sentimentally gloomy at the same time. As the student's imagination fills in the details, he/she might associate additional senses and feelings such as melancholy or disquiet. Then a common pedagogical strategy is to let student start playing on the piano, while retaining the imaginary scene mentally. The goal is to let the playing somehow musically approximate the imaginary scene. In this example, we are able to see how visual aids are helpful. First and foremost, they provide an ``anchor'' against which students are able to make conscious adjustments to the playing. Secondly, visual aids are more approachable in a twofold sense: 1) we understand visual aids more easily because they frequently appear in everyday life, whereas musical thinking and performance are much more specialized and domain-specific skills; 2) one's mental faculty of imagination is somehow more capable in visual processing. The specific cause for our exceptional capability in visual imagination is up to debate. For example, one might argue that it is due to an evolutionary cause (i.e.~survival in the wild) that we must prioritize the development of visual processing and imagination. In particular, one must be able to predict how wild animals move by imagining their position in the next moment in order to know how one should fight (in case of hunting) or flight (in case of being chased). Without offering further speculations about the cause, we simply acknowledge an observation from the Prokofiev example discussed: while it is challenging to musically synthesize gloominess and expansiveness, it is relatively straightforward to synthesize gloomy weather and expansive scenery simply by juxtaposing them in our imagination. Furthermore, we are able to juxtapose them side by side, and construe them as a whole. This holistic picture is what helps the student in understanding musically the required mood.

By the same mechanism as visual inspiration, story-telling is an equally effective pedagogical tool. Instead of imagining visual scenery, students now participate in the activity of creative writing. In particular, one constructs plots that are reflective of the mood required for the playing. Note that the plots can be completely arbitrary and ad-hoc, as long as they fulfill the purpose of stimulating the musical mood students are expected to grasp. For the example above, one may brainstorm a story into which gloominess and expansiveness are incorporated. For example, we may simply turn the visual scenery discussed above into a narrative:

\begin{quote}
The weather is gloomy today. As I walk in the wild, clouds along with mountains afar speak of my hollow heart. A running brook asks about my sorrow and gloominess. I look around and see no answer. The unsettling clouds are not to be blamed. They smile at the world, in the shape of a giant donut. Thus I look inside and find the answer: my gloominess finds its cause in the distant past, as I tried to grasp the meaning of gloominess and expansiveness at the piano. Suddenly an angelic fox with nine tails visits me to give me heavenly revelation: ``by the grace of god, I'm here to give you the doctrine of perpetual motion. It reads: The weather is gloomy today. As I walk \ldots{}''.
\end{quote}

In the spirit of creative writing, the story is to be taken with a virtual repeat-sign at the end so that it becomes cyclical and continues indefinitely. Disregarding details (such as the random appearance of the word ``giant donut'') that are specific to creative writing, the mechanism by which story-telling is an effective pedagogical strategy is the following: on the one hand, expressing musical affects requires musical understanding of the corresponding emotional states (e.g.~the mixture of gloominess and expansiveness). On the other hand, stories are effective means to induce these emotional states. When we say ``imagine happiness'', the term ``happiness'' itself is meaningless in the sense that we cannot directly imagine happiness by itself.\label{idea:emotions cannot stand on their own} Instead, happiness is \emph{represented} by happiness-bearing experiences. In other words, happiness is a property of happy experience as a process (which gives us the saying that ``process is more important than result''). Now, experience is nothing but story, and to recall an experience is to tell a story. Therefore, it is through stories that we can effectively feel the emotions therein contained.

We should note that story-telling as a pedagogical strategy is based on an assumption. In fact, this assumption concerns a widespread belief in the performing arts that we may claim that it is universally assumed for all performing arts. It reads: in order to communicate a certain mood, the performer has to possess it first. For example, in order to musically express sorrow, the performer has to understand and possesses the sorrowful emotion. This assumption essentially underlies one of the most valued qualities of artistic expression: sincerity. One's performance is sincere, if it communicates a musical message that is faithful to the performer's disposition. The opposite to sincerity is hypocrisy, where one's performance may be considered excellent to the extent there is nothing wrong in the performance, both technically and musically. Therefore, the performance appears convincing as it is flawless. However, listeners still feel that something is fundamentally missing, as if the performer delivers a tragedy with all the splendid staging and spectacle, but without a soul. What is it that is missing? It is nothing but the very possession of emotions deemed necessary for the work. By faithfully feeling and resonating with emotions contained in the work, the performer is then able to deliver a sincere performance. As a result, listeners are able to perceive the performance as a complete work of art: in addition to containing nothing wrong (musically and technically), the performance contains extra qualities related to sincerity, such as compassion, sympathy and honesty. Note that these qualities are uniquely human: they indicate the organic quality of artists, as humans, behind the work. In other words, they indicate trace of life behind the work. When we say that something is sincere, we necessarily interpret it as related to a human act. For example, it does not make sense to say ``the rain yesterday was sincere''. Meanwhile, many qualities attributed to good performance are not necessarily uniquely human. For example, we often use the term ``breathtaking'' to describe compelling performance. However, we are also justified to say ``the thunderstorm yesterday was breathtaking''.

The very assumption that, given an emotional state, the performers must possess it in order to present it, is not singular to music performance. As mentioned, it is acknowledged in virtual all performing arts. For example, it lies at the heart of Stanislavsky's system of acting. For him, the role of emotion is vital in carrying out expressive performance. In his writing ``An actor prepares'', Stanislavsky makes the following remark:

\begin{quote}
To me, as a spectator, what was going on inside of you was of much greater interest. Those feelings, drawn from our actual experience, and transferred to our part, are what give life to the play. {[}\ldots{]} All external production is formal, cold, and pointless if it is not motivated from within. \autocite[164]{stanislavski1989actor}
\end{quote}

Note how this statement can be applied to music performance without modifying a single word. Namely, we can imagine that the quote appears, word for word, in a piano masterclass without realizing that it is a comment originally made about acting. Therefore, a true performing artist whose ambition is to bring about try artistry, must somehow generate the required emotion before he/she can act at all. A considerable portion of Stanislavsky's system is then dedicated to how one can generate appropriate emotions necessary for the stage. To this end, he proposes two concepts: emotional memory and the method of physical actions.

According to Stanislavsky, emotional memory is memory consisting of emotions that can be recalled, retrieved and recreated on-demand during on-stage acting. Coincidentally, we can use an observation we make previously (see page \pageref{idea:emotions cannot stand on their own}), that emotions cannot stand on their own, but must be accompanied (or more technically, encoded) by emotion-bearing experiences. Emotional memory is then a collection of useful memories from personal life, useful in the sense that they effectively bring up the mood required for on-stage acting. According to Moore's interpretation of Stanislavsky, emotional memory concerns the ``actor's own experience, transformed into that of the character''.\autocite[42]{moore1984stanislavski} Therefore:

\begin{quote}
To enrich the emotional memory, the actor must observe what is happening around him; he must read, listen to music, go to museums, watch people. Well developed emotional memory is the most important requirement for the actor's work in the theater of living experience. It is the storage of past experiences and the only source for emotions on stage.\autocite[44]{moore1984stanislavski}
\end{quote}

To cultivate emotional memory, one consciously collects and catalogues individual pieces of memories from his/her own life experience, in order to build his/her personal collection of valuable memories. For example, consider the following fictional scenario:

\begin{quote}
You are an actor but at the same time you are enthusiastic about classical music. Recently you read about Messiaen and became fascinated by his idea of modes of limited transposition. However, unlike him, you are not particularly fond of birds. In fact, you hate birds. One day, you woke up and found that there was a giant pigeon in the room. It was the biggest pigeon you had ever seen in your whole life so far. You panicked and ran outside.
\end{quote}

In this scenario, a true actor who has the goal of cultivating emotional memory in mind would take one extra step: reflecting and contemplating on his panicking experience. Therefore, the story continues:

\begin{quote}
You ran outside, yet the image of the giant pigeon is still lingering in your mind like a spectre. Suddenly you recall Stanislavsky's idea of emotional memory and find that this panicking experience is worth remembering. Therefore you tell yourself: ``indeed I should panic, but meanwhile as an aspired actor, I have to remember my current mood as vividly as possible. I should remember as detailed as possible the pigeon, how it causes my anxiety, and how my subjective experience evolves after encountering the pigeon''. You then decide to write down your instantaneous feeling and take a selfie as records. After this experience, you have one addition to your personal collection of useful memories, under the label ``panic-related emotions''.
\end{quote}

The actor then participates in a drama workshop on Hamlet, where he/she is assigned the role of Horatio. Now the rehearsal focuses on the opening scene, in particular, the moment of apparition:

\begin{quote}
    
\begin{dialogue}
    \speak{BERNARDO} Last night of all,
    
    When yond same star that's westward from the pole
    
    Had made his course to illume that part of heaven
    
    Where now it burns, Marcellus and myself,
    
    The bell then beating one,--

    \direct{Enter Ghost}

    \speak{MARCELLUS} Peace, break thee off; look, where it comes again!

    \speak{BERNARDO} In the same figure, like the king that's dead.

    \speak{MARCELLUS} Thou art a scholar; speak to it, Horatio.
    
    \speak{BERNARDO} Looks it not like the king? mark it, Horatio.

    \speak{HORATIO} Most like: it harrows me with fear and wonder.
    
    \speak{BERNARDO} It would be spoke to.
    
    \speak{MARCELLUS} Question it, Horatio.
    
    \speak{HORATIO} What art thou that usurp'st this time of night,
    
    Together with that fair and warlike form
    
    In which the majesty of buried Denmark
    
    Did sometimes march? by heaven I charge thee, speak!
    
    \speak{MARCELLUS} It is offended.
    
    \speak{BERNARDO} See, it stalks away!
    
    \speak{HORATIO} Stay! speak, speak! I charge thee, speak!
    
    \direct{Exit Ghost}
    
    \speak{MARCELLUS} 'Tis gone, and will not answer.
    
    \speak{BERNARDO} How now, Horatio! you tremble and look pale:
    
    Is not this something more than fantasy?
    
    What think you on't?
    
    \speak{HORATIO} Before my God, I might not this believe
    
    Without the sensible and true avouch
    
    Of mine own eyes.\autocite[293]{shakespeare2014arden}
    
\end{dialogue}

\end{quote}

How would the actor understand the character of Horatio and, specifically, his reaction to the ghost? Based solely on the textual evidence in the excerpt above, we understand his mood as a mixture of ``fear and wonder''. However, meanwhile, he has to pretend he is brave, by trying to overcome his fear with loud voice: ``Stay! speak, speak! I charge thee, speak!''. We know that he is pretending to be brave because his comrade Bernardo notices his fear: ``How now, Horatio! you tremble and look pale''. This discrepancy between his mood and his action indicates that he is panicking. The actor recalls his horrific experience of encountering the giant pigeon. Consequently, when he/she acts out the line ``Stay! speak, speak! I charge thee, speak!'', the actor thinks of the image of the giant pigeon, such that the ghost of the king is morphed into the shape of a giant pigeon (the same way that performers, in overcoming stage fright, think of the audience members as potatoes). The actor starts to panic; his/her voice starts trembling; his/her face turns pale. The timing is so right, that when Bernardo speaks ``How now, Horatio! you tremble and look pale'', the actor appears exactly in the state described by Bernardo: trembling and pale. As a result, the audience is astounded and praises the actor's performance as sincere and organic: the character is understood from within, such that the acting faithfully portrays the character in every possible aspect. The example above involving pigeon is arbitrarily constructed in order to show a point. Even though it sounds absurd (and even ridiculous) to associate the appearance of a ghost that has so much dramatic significance (i.e.~from which the rest of Hamlet's plot is made possible) with the everyday encounter of a pigeon, nevertheless the example shows that the focus is on utility. Both the actor's encounter of pigeon and Horatio's encounter of ghost stimulate the feeling of panic, despite the drastic difference in content.

Emotional memory involves strategies similar to the pedagogical strategy in music lessons explored above. In music lessons, instructors use stories to motivate students in finding the right musical mood. A particular characteristic of this strategy is that we introduce external help: the stories are often based on materials from our personal experience, and they are considered external to music as they are not about relationships between tones insofar as music is considered an autonomous object (see page \pageref{quote:hanslick} for Hanslick's remark). By the same token, emotional memory introduces external help. According to Stanislavsky's system, every actor lives two lives: stage life and personal life. In particular:

\begin{quote}
An experience of the actor on stage is different from an experience in life. The difference lies in the fact that the actor lives on stage as the character and also as the actor who creates the character. \autocite[42]{moore1984stanislavski}
\end{quote}

In one's stage life, the actor lives poetically and dramatically. It is a life where, as in Aeschylus' dramatic depiction ``Oresteia'', Orestes is chased by wrathful Furies. Or, as in Wagner's ``Götterdämmerung'', the dramatic force of vengeance, wrath, and greed brings doom to the gods in the final scene. Alternatively, if characters were to break the third wall, then it is the sheer musical intensity that consumes Valhalla and the gods into dust. On the other hand, one's personal life is less eventful. Instead of being chased by methodological figures such as the Furies or Rhinemaidens, one is more likely chased by debt collectors. Instead of being consumed by wrath of the gods, one is more likely consumed by the realization that, like Beethoven, he/she is left with ``insufficient money for his living expenses''.\autocite{grove_beethoven}

How are the two lives with contrasting styles reconciled? Stanislavsky suggests that personal life is the source of emotional memory for stage life. In other words, personal life as external help (i.e.~experience in personal life is not part of the dramatic plot) is introduced to shape one's stage life. Additionally, Stanislavsky's claim is stronger: experience from personal life is not only useful, it is indispensable, to the extent that this external help is ``the only source for emotions on stage''.\autocite[44]{moore1984stanislavski} Specifically:

\begin{quote}
The actor must be capable of bringing out the imprint of a past experience and of making it respond to the conditioned stimulus on stage at the moment he needs it. Through rehearsals, the actor develops a conditioned reflex in which his emotion is stirred through the stage stimulus. \autocite[43]{moore1984stanislavski}
\end{quote}

Later in the development of Stanislavsky's system, the focus shifts from emotional memory to the method of physical actions. For method of physical actions:

\begin{quote}
Instead of forcing an emotion before going on stage {[}as in emotional memory{]}, the actor fulfills a simple, concrete, purposeful physical action which stirs the psychological side of the psychophysical act, thus achieving psycho-physical involvement. \autocite[19]{moore1984stanislavski}
\end{quote}

In other words, the method relies on the assumption that bodily gestures, i.e.~behavioral patterns, have psychological impacts to the extent they can shape our emotional state. Music performers should resonate with this assumption: to overcome stage fright, we have two viable strategies (other than the brute force strategy: practice more). The first is psychological, that before going to stage, we can imagine listeners as potatoes sitting in the audience. The second is behavioral, or in Stanislavsky's term, related to physical actions, that we take a deep breath, and walk to the stage in a posture that is relaxed and confident. Essentially, we are completing a set of pre-performance routines that may be different for each person. On the one hand, these routines are ritual-like. By a sequence of bodily gestures, performers hope to conjure the spirit called ``optimal condition for performance''. On the other hand, they are akin to the field of therapy that has been increasingly popular among musicians in recent years. For example, many types of therapies put emphasis on the conscious act of breathing and bodily relaxation (as in, for example, meditation), and believe that they can have definitive impact on mental well-being.

We should acknowledge that emotional memory and the method of physical actions are vastly different approaches. In fact, one may argue that he formulates the method of physical actions as a replacement of his theory of emotional memory, overthrowing his previous findings as he became ``disappointed in the results of his earlier developments''.\autocite[17]{moore1984stanislavski} Nevertheless, it is noteworthy that emotional memory and the method of physical actions are different approaches to achieving the same goal: generating appropriate inner emotional states from which acting is made possible at all. Articulated plainly, as Moore observes, emotions ``are stirred in an actor in order to stir the spectator's emotions in turn''.\autocite[14]{moore1984stanislavski} Therefore, emotional memory and the method of physical actions can be interpreted as complementary, as both are ``conscious means which would stir the actor's emotions''.\autocite[17]{moore1984stanislavski} More importantly, both methods are heavily based on the notion of narrative: for emotional memory, one makes use of the narrative of past experience; whereas for the method of physical actions, one treats the sequence of bodily movements as a narrative without words, since it tells a meaningful story which has the power to shape one's disposition.

To sum up, we have discussed how narrative (i.e.~story-telling) is an important pedagogical tool in the performing arts (note that performing arts are a subset of temporal arts). In particular, narrative can facilitate a performer's (e.g.~actors and musicians) preparation process by inspiring his/her emotions required for an expressive performance. Moreover, because emotions are indispensable in the sense that performers must possess the emotion before they can communicate it with the audience, narrative as an (or in Stanislavsky's view, the only) emotion-generating tool is also indispensable to the performing arts.

\section{Narrative as foundation}
\label{sec:narrative as foundation}

However, we may not be fully satisfied in acknowledging the pedagogical utility of narrative. Instead, when we say that music is about story-telling, we feel that narrative somehow describes music (and all temporal arts) at a more fundamental level. In other word, the discussion of narrative in a musical context should not be classified as pedagogy of music, but ontology of music. Narrative is not something one relates to music, but something that is part of music.

Therefore, when interpreting music performance metaphorically as an act of story-telling, we are not only suggesting that performing a piece is \emph{like} telling a story, but also suggesting that it \emph{is} telling a story. Treating narrative as an inherent part of music allows us to examine the concept using a perspective that is not confined to pedagogical usefulness. In particular, one looks at features of story-telling in general and realizes that they shed light on our understanding of music. By interpreting music as narrative and performing music as story-telling, we potentially gain new insights into aspects of music that are otherwise undiscovered.

As discussed previously, story-telling enables us to extend the metaphor to the idea of a spiritual journey. In short, narrative mediates between music and spiritual journey. One may argue that metaphorically interpreting music as narrative and spiritual journey is merely a play of words, as they do not inform us of anything beyond what we already know about music. Therefore, one might say ``OK I see that listening to music is like embarking on a journey, so what?''.

The answer is that features of narrative and journey reveal important properties of music itself. Moreover, these properties are often overlooked in cases where the metaphor of narrative and journey is absent. For example, consider how the idea of a journey gives music a proactive force. Specifically, proactive force refers to the presence of some conscious agent by whose will and deliberate choice musical events are \emph{proactively produced} instead of being \emph{mere happenings}. In other words, proactive force is the quality in music that enables us, as listeners, to attribute it uniquely to human endeavors. Note that proactive force concerns a statement that is not scientific in that it cannot be falsified. For example, one may ask: ``we now know that the proactive force essential to music is a trace of human endeavors, but \emph{exactly} where can we find it? Moreover, what \emph{exactly} do you mean by human endeavors?''. A series of questions would then lead us again to the unfruitful discussion on the physical basis of free will. As our discussion enters the realm of performing arts, we necessarily encounter propositions that are of a strange mix: on the one hand, we strongly feel and believe in the living quality (i.e.~proactive force) of music; on the other hand, one cannot find it anywhere in music. In particular, one cannot physically tell (i.e.~using measuring equipment) if musical events are proactively produced or are mere happenings. According to the infinite monkey theorem:

\begin{quote}
If an army of monkeys were strumming on typewriters they might write all the books in the British Museum. \autocite[82]{nature_of_the_physical_world}
\end{quote}

Consequently, the notion of proactive force should be understood as strictly limited to an artistic discourse. In particular, one should approach it with the mindset of a performer, where the lively quality characterizing the proactive force is one of the (if not the most) essential elements of arts, with respect to which arts are considered worthy of anything at all.

In a journey, there is a constant presence of the protagonist (i.e.~the identification of the protagonist who embarks on the journey is spelled out), by whose actions and conscious decisions the journey is made possible. Moreover, as readers, we are so naturally put in the role of the protagonist instead of a spectator who merely witnesses the journey. Therefore, the protagonist fulfills the function of proactive force, and it is vividly recognized by readers. On the contrary, in a landscape painting or a piece of music, the presence of the protagonist is less obvious. In a landscape painting, we do not find anyone inside the picture because the protagonist lies outside the picture: the spectator. Similarly, in a piece of music, we tend to be confused as to who the protagonist is.

Therefore, because the presence of a protagonist is not immediately obvious in music, one must actively remind himself/herself of the proactive force vital to music by construing music as journey. Without construing music as journey, we are in danger of neglecting the proactive force vital to the expressiveness of music. Consequently, we are in danger of reducing musical events (as proactively produced by conscious actions) to mere happenings. An example of mere happenings is chemical reactions, in which we have events but not decisions, causal relations but not conscious actions. For example, it is poetic yet absurd to imagine oneself as electric current in the chemical process of electrolysis. The reason for the absurdity is simply that we are unable to find the protagonist to whom we can substitute in our identity. As a result, chemical reactions lack the proactive force brought by a protagonist. When lacking in proactive force, a piece of music unfolding in time is not substantially different from observing a chemical reaction, such that we are spectators detached from the event instead of being part of the event.

We are surely aware that music is different from chemical reactions, in the same way a journey as a process is different from chemical reactions. They are different, to the extent that we call music but not chemical reactions an art. However, without the metaphor of journey that constantly reminds us of the proactive force so essential to music, even performers are sometimes at a loss. For example, in discussing effective practice, performers commonly refer to a type of ineffective practice, where one's attention starts to wander around, entering the so-called ``autopilot mode''. In this case, performers are physically moving their hands and bringing out all articulations notated on the score. However, their attention is detached from their playing. At first glance, the issue seems to be about attention in its medical sense (such as in the context of ADHD). However, artistically, the issue is exactly that performers lose sight of the proactive force that gives performance its truly human quality.

It is very different to perform with the idea of spelling out notes on the score, than with the idea of proactively exploring some musical space. Therefore, the metaphor of a journey fundamentally changes (or reminds us of) the way music is construed, bringing a completely altered mindset. It is not just a pedagogical tool that serves as tricks (i.e.~emotional memory) to bring about effective performance. Instead, it is a statement concerning what music \emph{is} and is \emph{about}. With the metaphor of a journey, we thus construe music performance as an active exploration of an open world. This is a world in which both listeners and artists are invited and welcomed (analogous to a safe space) to participate and collaborate in an active exploration. In this world, we no longer say that ``tension is followed by release''. The statement is too cold to be considered artistic, as it is not substantially different from the statement ``observation is followed by the collapse of wavefunction''. It is ``cold'' because we do not find ourselves in the statement . Therefore, we insert and assert a declarative subject ``I'', saying that ``I encounter a harmonically unstable passage of music. As a result, I become emotionally challenged as it creates a sense of tension. In response, I, through my sole will and power, decide to release this tension by ending this harmonically unstable passage with an assertive tonic chord''. We should notice the crux of the matter, that we are turning passive voice into an active one. Therefore, instead of saying that ``dissonance is resolved'' as if the resolution is spontaneous, we say that ``I resolve the dissonance''.

One might argue that the difference between passive and active voice is merely in wording. However, we should see how the shift in mindset particularly benefits performer's musicianship. Recall that the metaphor of journey reveals the proactive force inherent in music, giving it a lively quality that is highly valued by many to be the essence of musical beauty. Now, this proactive force puts performers in the mindset that every musical event in performance is caused by one's active will and action. Therefore, performance is not about musical happenings, but about musical purposes and their fulfillment. With this mindset, performers are in the position to ask, with an active voice: what can I do to make this moment a better musical instant? For a socially responsible performer, the question naturally extends: what can I do to make this world a better place? The crux is in the observation that active voice naturally makes us to think about what to do, instead of what happens. Therefore, the mindset shifts from a descriptive perspective (i.e.~musical happenings) to a prescriptive one (i.e.~musical actions). Consequently, in practice, performers are constantly aware of how every musical moment is the result of deliberate interpretive choice in fulfillment of certain purposes. Performers then become, borrowing Stanislavsky's words, ``motivated from within''.\autocite[164]{stanislavski1989actor} As every interpretive decision is the result of some inner motivation, there must be no ``dangling'' musical moments that are unmotivated. Note that as consequence, we naturally arrive at familiar concepts that musicians often discuss. Specifically, as performance is motivated from within, it is naturally \emph{sincere}. Moreover, as performers must constantly think about every moment of the performance in relation to their deliberate and purposeful interpretive decisions, the performance is more likely \emph{justified} as one actively tries to make sense of each moment. Finally, the music is \emph{convincing}: if we treat interpretive decisions as musical statements, then a convincing performance is nothing other than the idea that it has the ability to make the audience believe in the statement. Because of the active voice, we know that the performer firmly believes in his/her musical statements. As the performance is sincere, the performer can then pass the belief in the musical statements to the audience.

In short, the proactive force of music implied by the metaphor of journey makes the performance sincere, justified, and convincing. It is this lively quality of music that gives rise to musical knowledge whose term means something very specific here. The term ``knowledge'' is to be interpreted as a music analogue to the knowledge defined by the epistemological theory of JTB. In epistemology, one interpretation of what constitutes knowledge states that knowledge is ``justified true belief'' (abbreviated as JTB).\autocite{sep-knowledge-analysis} In particular, given any person \emph{S} and proposition \emph{p}:

\begin{quote}
\emph{S} knows that \emph{p} iff

\begin{enumerate}
\item
  \emph{p} is true;
\item
  \emph{S} believes that \emph{p};
\item
  \emph{S} is justified in believing that \emph{p}.\autocite{sep-knowledge-analysis}
\end{enumerate}
\end{quote}

Disregarding technical details, we can see how the criteria applies to music with little modification. Sincerity corresponds to truth (notice how musical truth is special in that it is not experimental), as to be sincere is to be true to oneself. Meanwhile, being convincing corresponds to belief, as the performer firmly believes in his/her musical statements in a convincing performance. Finally, justified performance corresponds to the justification condition. One may notice the subtlety here: in JTB, the justification condition refers to the state of being justified in believing (i.e.~referring to the belief condition instead of the truth condition). Meanwhile, a justified performance makes no reference to the belief in musical statements. However, this subtlety does not harm our discussion here. To see why, we notice that even though musical justification makes no reference to belief, the belief is in fact implied. In order to justify a particular interpretive decision (such as the placement of a crescendo at a particular position), the performer shows that it is musically effective to the extent the listener is emotionally satisfied. Now, because of the assumption that the performer must possess the emotion he/she wishes to communicate, the performer must be emotionally satisfied by the particular interpretive decision first, which necessarily implies the firm belief in the decision.

\section{What should narrative of music be about?}
\label{section:what should narrative of music be about}

The previous discussion explains how narrative (i.e.~story-telling and metaphor of journey) in music can serve practically as a pedagogical tool, and ontologically as the source of the lively quality (i.e.~the proactive force) essential to music.

Despite the crucial role of narrative in interpreting music, we must nevertheless recognize the potential pitfall of interpreting music too literally as narrative. Specifically, one must be careful not to interpret too rigidly the term ``narrative'' in its literary sense. As the Grove article on ``Narratology'' warns us of the potential pitfall once we wish to apply narratology as the study of narrative to music:

\begin{quote}
One recurring issue in such discussions of narrative and music concerns the identification of agents or actors. Ordinary story-telling normally concerns characters, and musicologists who explore analogies to narrative often identify fictional agents such as themes or instruments.\autocite{grove_narratology}
\end{quote}

At this point, we can already recognize the potential issue: music narratology is in danger of an identity crisis. The issue is that, narrative is, like temporal asymmetry, not a concept native to the field of music. Instead, it owes its etymological origin to the literary tradition. Consequently, it extensively borrows terminology and concepts from its literary counterpart. Therefore, a potential risk is that one borrows from other fields without constructing a truly original theory that is native to music. The issue is made clearer as the article continues:

\begin{quote}
Descriptions offered in support of a narrative analogy may remain close to ordinary technical analysis, but often they become anthropomorphic and sometimes, as in Newcomb's account of Mahler's Symphony no.9, musical events may be translated into a detailed, almost novelistic story about an individual protagonist.\autocite{grove_narratology}
\end{quote}

The main issue that emerges is not about the use of analogy or anthropomorphic descriptions, but about the act of translating music analysis into literary terms. In potentially offensive terms, narratology in music is in danger of being annexed to literary narratology, thus losing its autonomy (which is, ironically, what musicians take pride in, suggesting that music as non-representational art, makes no reference to external objects). Specifically, consider the following commentary:

\begin{quote}
The taxonomic and rule-orientated qualities of formalism and structuralism already seemed dated to many literary and cultural scholars by the 1980s, and the subsequent promotion of narratology by musicologists may have been untimely.\autocite{grove_narratology}
\end{quote}

Readers should raise eyebrows due to the implication made by the term ``untimely''. The implication states that musicologists are somehow still making promotion of what others have considered dated (notice that both ``promotion'' and ``dated'' appear in the original commentary). Therefore, the absurd usage of the word ``untimely'' suggests that the study of narratology in music somehow falls behind with its upstream, i.e.~literary and cultural studies from which narratology in music borrows concepts and terminology. Hence the very word ``untimely'' reflects and defines a hierarchical relationship, in which musicologists are followers who have to keep up with the research progress in other fields.

As a result, we often encounter unnatural analysis when the borrowed literary terms about narrative make little sense in music. In other words, a music analysis becomes contrived and ad-hoc if we borrow literary narratology for the sake of borrowing it. For example, in literary narratology, the role of narrator is crucial in presenting the narrative. A fictional narrative in first person is very different from that in third person. Meanwhile, the role of narrator is less important in music. It is true that we can always forge, through analogy, a narrator in music analysis and claim it is crucial. Nevertheless, such forgery is contrived and ad-hoc. In literature, the choice of first person versus third person is crucial because it is what every writer must consciously deliberate before even starting the writing. On the other hand, we can indeed find some musical features and claim that they show a music analogue of first person narrative. For example, we might say that an off-tonic beginning of a piece signifies a third person narrative, as the tonal remoteness from the tonic corresponds to the feeling of remoteness often encountered in a third person narrative. However, we can always respond to such interpretation concisely: ``are you \emph{really} sure about that?''. In precise terms, the issue of first person versus third person in literature is not susceptible to intentional fallacy (i.e.~the work is written in first person because the author deliberately means it), whereas in music the issue is more complicated. Therefore, just as we should ignore idioms specific to each instrument when discussing musical expressiveness in general, we should be careful not to bring in any concepts deemed unhelpful to musical inquiry.

Then, what should narratology in music be about? More specifically, what should a narratology that is developed specifically with music in mind look like? In other words, we are interested in a musical narratology, instead of the application of literary narratology in music. One potential strategy is to shift (or more precisely, reverse) the mindset in ways analogous to the Copernican revolution. All else being equal, the Copernican revolution states that, maybe instead of treating earth as the center around which the sun orbits, we can treat the sun as the center around which the earth orbits. By the same token, maybe instead of putting music at fault to which literary narratology does not apply, we can put literary narratology at fault for which music does not qualify. Therefore, instead of identifying the issue as interpreting music too literally as (literary) narrative, it might be more appropriate to identify the issue as interpreting narrative too narrowly whose scope is confined to literary studies. Consequently, our goal is to find alternative formulation of narrative which can capture essential aspects of music as well as other temporal arts.

To sum up, we notice the interesting trajectory of our inquiry: we start with analyzing music with a tool called ``narratology'' by treating music metaphorically as a narrative. However, we find potential pitfalls in such analysis. Instead of maintaining the status quo, we choose to challenge and reform the very notion of narrative itself. We ask the following question: what should a narratology look like if it is to be uniquely musical, to the extent that when we talk about the term ``narrative'', it does not sound like a concept borrowed from the literary tradition? In order to answer such question, we need to find essential ingredients of a purely musical narrative, and they would become the core concepts of a purely musical narratology. In order to achieve this goal, we have to return again to music where, as musicians believe, one can find statements universal to all temporal arts. In particular, owing to the fact that the author of this project identifies himself as a performer, our discussion will be invariably biased toward a performer's perspective. Now, from a performer's perspective, the pattern of tension and release is of utmost importance to music, which is the focus of the upcoming discussion.

\section{Tension-release as a universal theme for all narratives}
\label{section:tension-release}

In music lessons, one of the (if not the most) frequently discussed topics is the pattern of tension and release. The terms ``tension'' and ``release'' are to be bundled as a single concept, hence ``tension and release'' will be discussed as a singular noun hereafter, denoted by the hyphenated term ``tension-release''. Few would challenge the vital role of tension-release, particularly among performers. Moreover, upon hearing the term tension-release, we immediately perceive it as a concept uniquely native to music. In other words, it is not borrowed from any other fields of study. One of our objectives is to show that tension-release, though particularly favored by performers, is in no way limited to discussions of music performance. In fact, it is universal to all temporal arts.

Being a concept unique to music, tension-release derives its power from its omnipresence. In particular, we should note that musicians discuss tension-release under various contexts. In harmony and counterpoint, tension-release is expressed through controlled dissonances. Resolution in tonal harmony, suspension in fourth species counterpoint, passing and neighbor tones are all examples where dissonances are treated and controlled with preparation and resolution. As a result, the treatment of dissonance constitutes what listeners perceive as tension-release.

Additionally, the pattern of tension-release is observed across all hierarchical levels.\label{idea:tension-release is observed across all hierarchical levels} We enumerate a few examples for illustrative purposes. At the level of measures, we recover the prototypical examples of harmonic resolution and contrapuntal suspension. For example, illustrations of dominant-tonic resolution often use a simplified model consisting of two chords: a dominant chord, followed by its resolution within a single measure. Additionally, articulations such as slurs contribute to the measure-level tension-release. For performers, these are the details that determine how polished the performance will be.

For tension-release on the level of phrases, musicians are interested in the topic of phrase model which is closely related to the idea of tension-release. For example, performers focus on the frequently discussed topic of phrasing. The idiom of ``shaping'' the phrase is simply a restatement of tension-release on the phrase level. In particular, when we shape a phrase, we identify a goal, namely, the moment in a phrase with the highest concentration of tension, to which performers must direct the performance, giving it a sense of direction.

On a more global level, tension-release within a piece gives rise to the topic of formal structure. For example, in standard sonata form, we often identify the end of the development section as the point of maximal tension. The tension originates precisely from the dominant-tonic resolution seen in tonal harmony, as ``the standard development culminates on an active dominant''. \autocite[19]{hepokoski2006elements} Additionally, for performers, tension-release operating on the entire span of a piece means something more than music analysis. It gives the piece the structural coherence which enables us to identify it as a complete work instead of a loose collection of excerpts. In performer-centric terms, tension-release operating at a large scale tests the performer's ability to handle large-scale musical works holistically. In particular, one of the criteria for performer's artistic competence is the general (i.e.~repertoire-agnostic) ability to maintain musical tension for an extended period of time, thus holding an audience's attention for the entire duration of the concert. To this end, performers often invoke the metaphor of breathing and phrasing to describe such ability: one's performance is breathtaking if he/she is able to present the entire piece in one breath, as a single phrase. Therefore, for performers, the ability to control tension-release is essential yet challenging, since it requires a rich set of skills: careful planning that shows musical understanding through analysis, mental concentration that can relate musical events far from one another in time, and physical stamina.

Pattern of tension-release continues climbing the ladder of hierarchical levels. One can interpret multiple pieces as a single entity. For example, we may consider inter-movement connections in a multi-movement work. Alternatively, we may group multiple works as a cycle which defines a single narrative. Therefore, tension-release operates across musical works. From a music history perspective, we may argue that one of the historical tendencies is to expand formal structures. Therefore, the rounded binary form that is short in length and simple in tonal scheme grows into the sonata-allegro form that is characterized by expansive length and tonal complexity. Owing to the tendency of expanding formal structures, composers experiment with removing boundaries between movements in an attempt to give formal unity to a multi-movement work. Poetically speaking, composers assign a single grand narrative to a collection of movements, thus turning the work into an organic whole.

\subsection{A musical case study: Liszt's B-minor sonata}
\label{sec:liszt sonata}

For example, one of the defining characteristics of Liszt's B-minor sonata is its thematic unity across movements, to the extent that it might be more appropriate to describe it as a single-movement sonata-like work with movement-like sections, than a multi-movement sonata with thematic unity. The possibility of admitting two contrasting descriptions is, according to the Grove article on Liszt, what gives the work a ``double-function'', such that the ``material is constantly making contributions to two sonata forms simultaneously''.\autocite{grove_liszt} In ``A history of Western music'', the authors make similar remarks about the piece:

\begin{quote}
{[}Liszt's{]} Sonata in B Minor (1853), modeled on Schubert's Wanderer Fantasy, is a masterpiece of formal innovation, using four main themes in one extended sonata-form movement subdivided into three large sections analogous to the movements of a Classic-era sonata. \autocite{norton_history}
\end{quote}

\begin{figure}[!h]
   \centering
   \includegraphics[width=.8\textwidth]{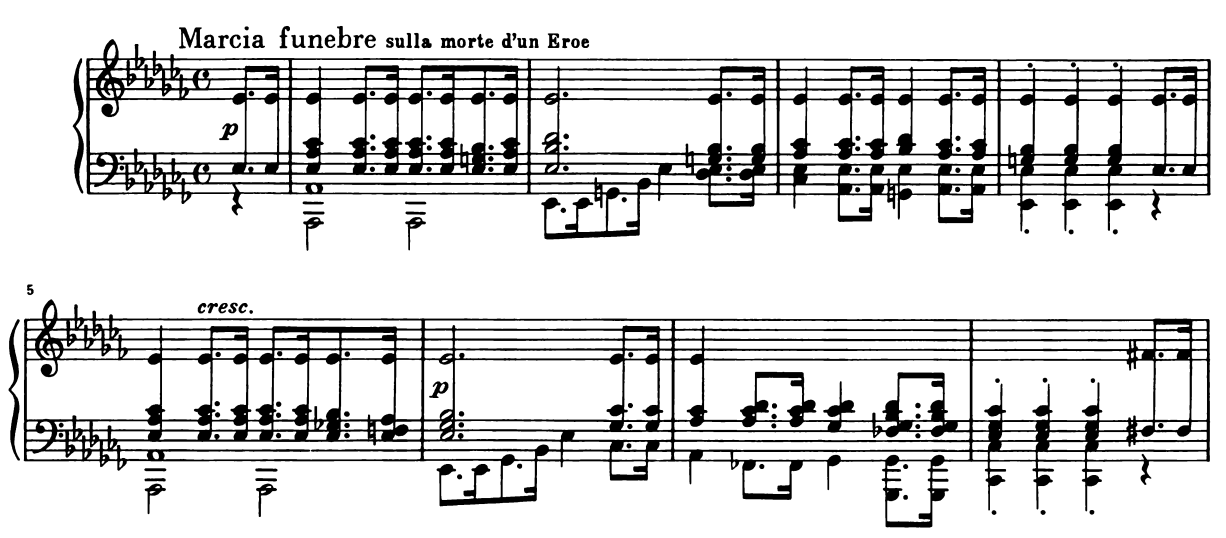}
   \caption{Beethoven, Piano Sonata No.~12, Op. 26, third movement, beginning}
   \label{fig:beethoven_op26_3mov}
\end{figure}

\begin{figure}[!h]
    \centering
    \includegraphics[width=.8\textwidth]{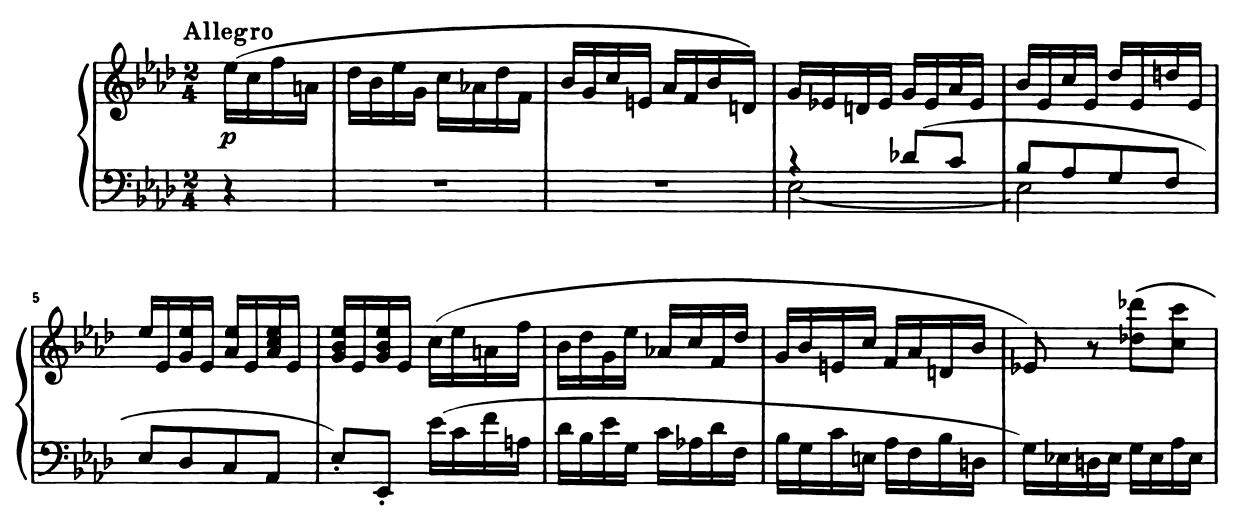}
    \caption{Beethoven, Piano Sonata No.~12, Op. 26, fourth movement, beginning}
    \label{fig:beethoven_op26_4mov}
 \end{figure}

Tension-release operating across movements poses greater challenges to performers. It requires performers to actively retain memory of the distant past (e.g.~the beginning of the piece) in order to maintain the musical tension. In a multi-movement work where one finds little inter-movement connections, each movement can be considered as an isolated work of music, regulated by its own pattern of tension-release. As a result, forgetting the previous movements as one begins playing the new movement is not only forgiven, but also encouraged. One does not have to recall the beginning of the first movement in order to begin the finale. For example, consider Beethoven's piano sonata No.~12. The third movement (see figure \ref{fig:beethoven_op26_3mov}) is a funeral march, followed by a lighthearted rondo movement as finale (see figure \ref{fig:beethoven_op26_4mov}).

This example is typical to the Classical period in that the slow movement and the lighthearted finale are often juxtaposed without any transition. One of the potential reasons for such arrangement is extramusical: a joyful finale concludes the performance, as a social event, with joy and exuberance. Therefore, the discontinuity between the finale and its preceding movement is arguably intentional: it expresses the good will that listeners should go home with a positive mood. It is for the same reason that popular repertoire of encores consists largely of virtuosic, lighthearted and joyful pieces.

\begin{figure}[!h]
    \centering
    \includegraphics[width=.8\textwidth]{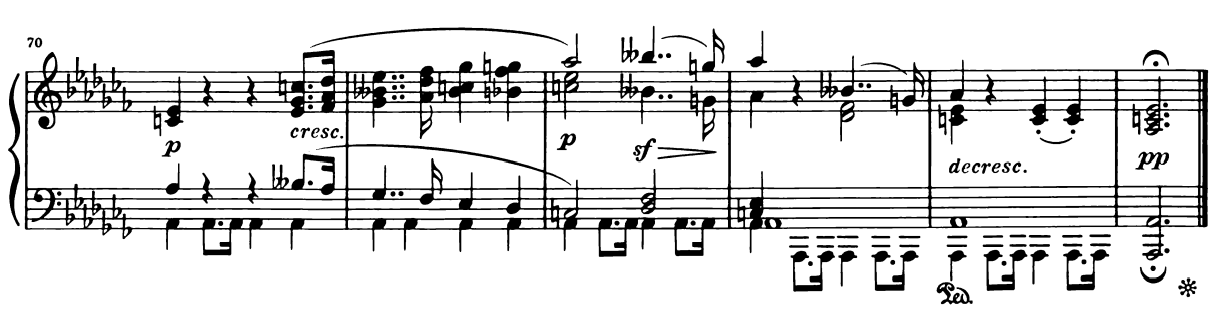}
    \caption{Beethoven, Piano Sonata No.~12, Op. 26, third movement, ending}
    \label{fig:beethoven_op26_3mov_ending}
 \end{figure}

Now, as performers, it would be inappropriate to either show any lighthearted mood of the finale in any part of the slow movement (i.e.~spoilers), or carry the solemn mood over to the finale. Therefore, performers are encouraged to play the third movement in isolation, as if it is the entire program. The entire movement is then performed with the deliberate choice of avoiding reference to other movements. By the end of the third movement, the entire concert hall is filled with the contemplative A-flat major sonority (see figure \ref{fig:beethoven_op26_3mov_ending}), causing the listeners to revisit the movement using their imagination. Metaphorically, the listeners are invited into a imaginary space where the movement is replayed, echoing the title ``funeral march'' as a replay of the hero's life. As everyone (including the performer, because as discussed, he/she must possess the emotion he/she wishes to communicate) is consumed by grief and sorrow, the A-flat major sonority gradually decays into silence. The performer at this point must avoid any physical movement because the listeners are still awestruck by what they just heard. Therefore, the performer freezes his/her movement as if time on stage has stopped, leaving the audience in total silence. After the momentary silence, the performer should then completely remove hands from the keyboard and prepare to perform the finale anew. The gesture of removing hands from the keyboard is a signal of psychological reset, announcing that the upcoming movement is in no way related to the previously heard movement. Therefore, everyone (the performer in particular) must erase from memory the sorrowful journey of the third movement. Upon receiving the gestural signal, listeners who were too haunted and startled by the silence to physically move, are now able to stretch their body and cough. It is a challenge for the performer to be emotionally sincere, but at the same time transition from deep grief to lighthearted joy within blink of an eye.

 \begin{figure}[!h]
    \centering
    \includegraphics[width=.8\textwidth]{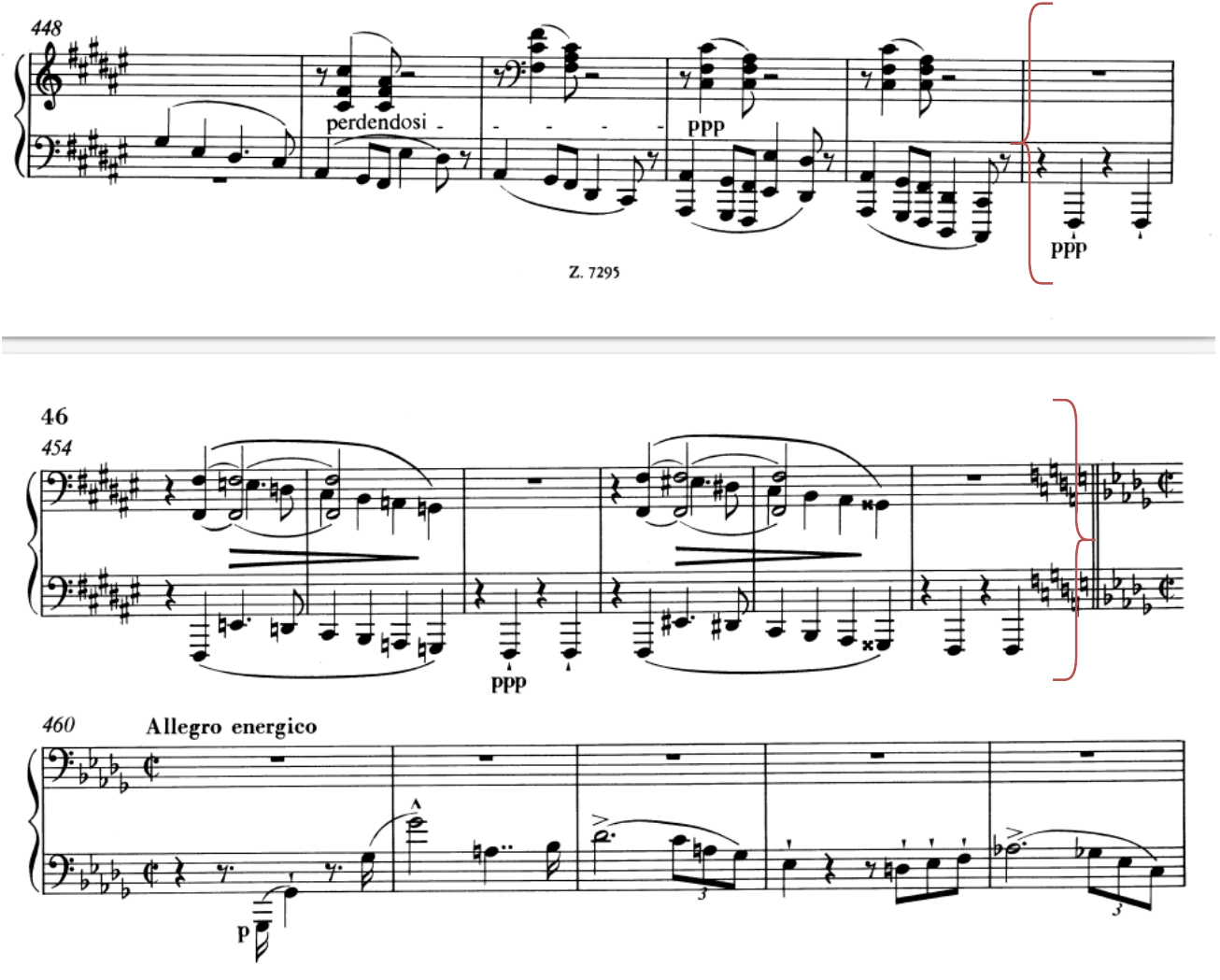}
    \caption{Liszt, Piano Sonata in B minor, S. 178, excerpt, curly brackets in red indicate parallelism to figure \ref{fig:liszt_sonata_opening}}
    \label{fig:liszt_sonata_before_finale}
 \end{figure}

 \begin{figure}[!h]
    \centering
    \includegraphics[width=.8\textwidth]{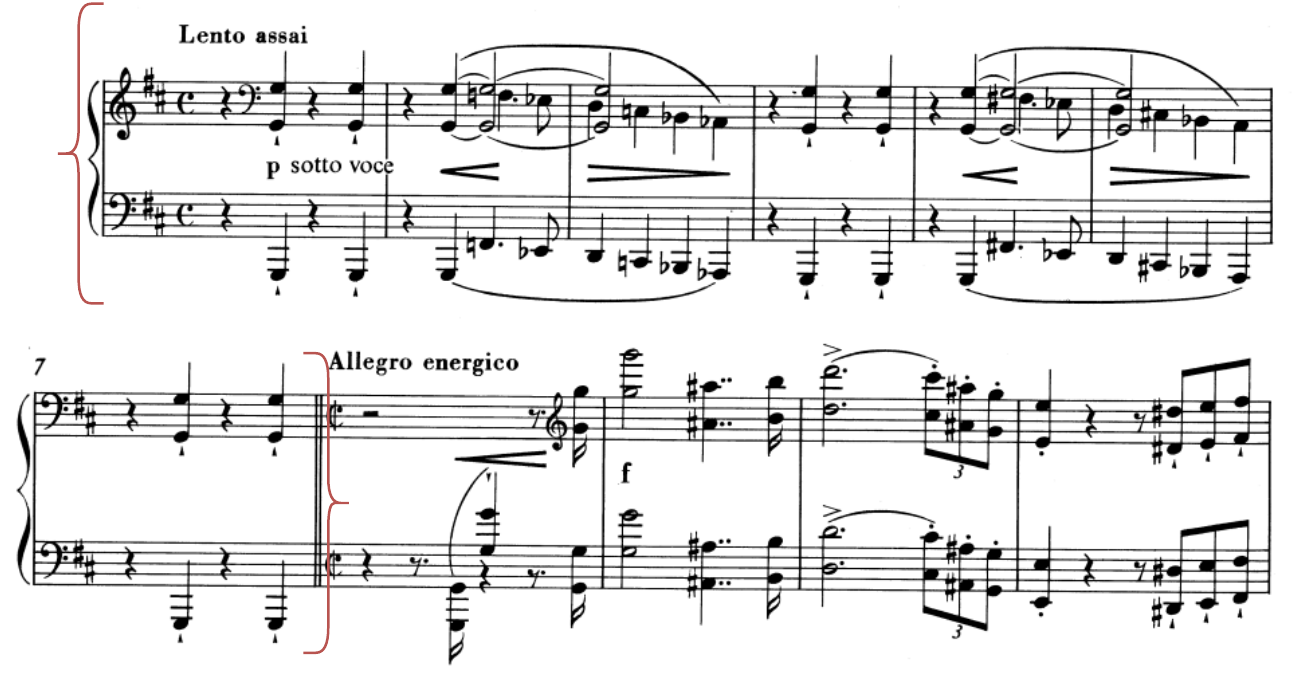}
    \caption{Liszt, Piano Sonata in B minor, S. 178, beginning, curly brackets in red indicate parallelism to figure \ref{fig:liszt_sonata_before_finale}}
    \label{fig:liszt_sonata_opening}
 \end{figure}

\begin{figure}[!h]
   \centering
   \includegraphics[width=.8\textwidth]{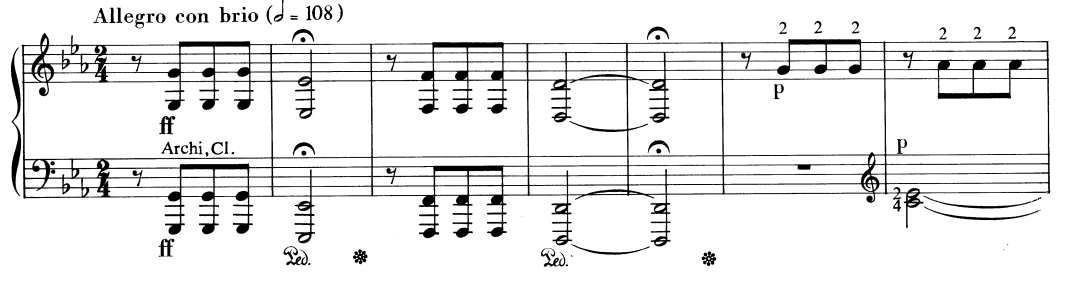}
   \caption{Beethoven, Symphony No.~5, Op. 67, beginning (piano reduction by Liszt)}
   \label{fig:beethoven op67 beginning}
\end{figure}

\begin{figure}[!h]
   \centering
   \includegraphics[width=.8\textwidth]{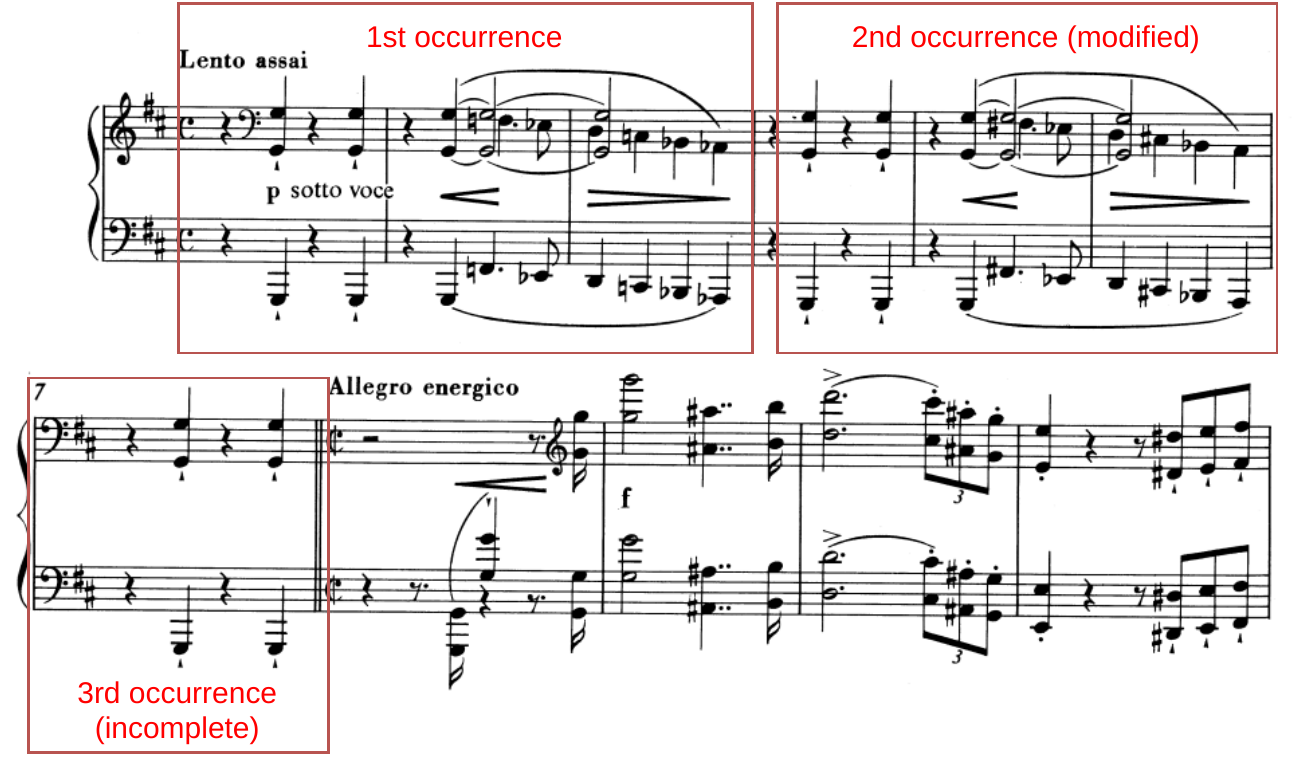}
   \caption{Liszt, Piano Sonata in B minor, S. 178, beginning, annotated to show three occurrences of the musical gesture}
   \label{fig:liszt sonata beginning annotated}
\end{figure}

\begin{figure}[!h]
   \centering
   \includegraphics[width=.8\textwidth]{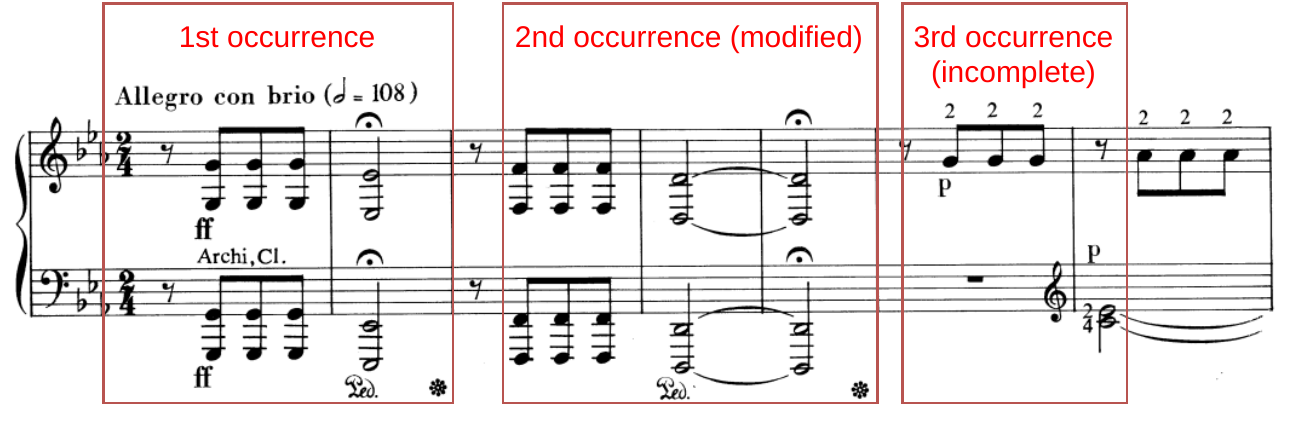}
   \caption{Beethoven, Symphony No.~5, Op. 67, beginning (piano reduction by Liszt), annotated to show three occurrences of the musical gesture}
   \label{fig:beethoven op67 beginning annotated}
\end{figure}

By contrast, consider the additional mental concentration required to perform Liszt's B-minor sonata. Instead of being forgetful of previous movements, one must make conscious effort to address previous materials. In particular, we consider a similar location: the moment leading into the finale (see figure \ref{fig:liszt_sonata_before_finale}). The onset of the note F-sharp in measure 453 immediately reminds us of the very opening of the sonata (see figure \ref{fig:liszt_sonata_opening}). In fact, the opening pattern (characterized by the repetition of note G in unison, followed by a descending scale) is so characteristic that its occurrence anywhere in the piece would infallibly reminds us of the opening. One potential reason for the memorability of the opening gesture is its simplicity. In other words, it is more appropriate to interpret it as representing the act of knocking on the door or heartbeat, than as a musically sensible melody. In fact, we may draw a parallel between the opening of Liszt's B-minor sonata and the opening of Beethoven's fifth symphony (see figure \ref{fig:beethoven op67 beginning}). Firstly, both are openings that express a gesture rather than complete melodic or harmonic statements. As a result, attempts to analyze using melodic or harmonic terms would be unfruitful by the sheer fact that both examples intentionally resist analysis in melodic or harmonic terms by the use of unisons. Secondly, we may observe additional similarities by considering phrase structure in details. As illustrated in figure \ref{fig:liszt sonata beginning annotated} and \ref{fig:beethoven op67 beginning annotated}, both openings feature three occurrences of the musical gesture in succession with the following schema:

\begin{itemize}
   \item
     The first occurrence presents the musical gesture (unmodified by definition)
   \item
     The second occurrence presents a modified version of the gesture (in Beethoven's example, the modification is implemented using transposition; in Liszt's example, the inner voice is modified)
   \item
     The third occurrence presents an incomplete version of the gesture, leading the music into new material (in Beethoven's eample, the final note of the gesture is interrupted by running notes, undermining the completeness of its note value; in Liszt's example, the gesture is truncated at the double bar)
   \end{itemize}

\begin{figure}[!h]
   \centering
   \includegraphics[width=.8\textwidth]{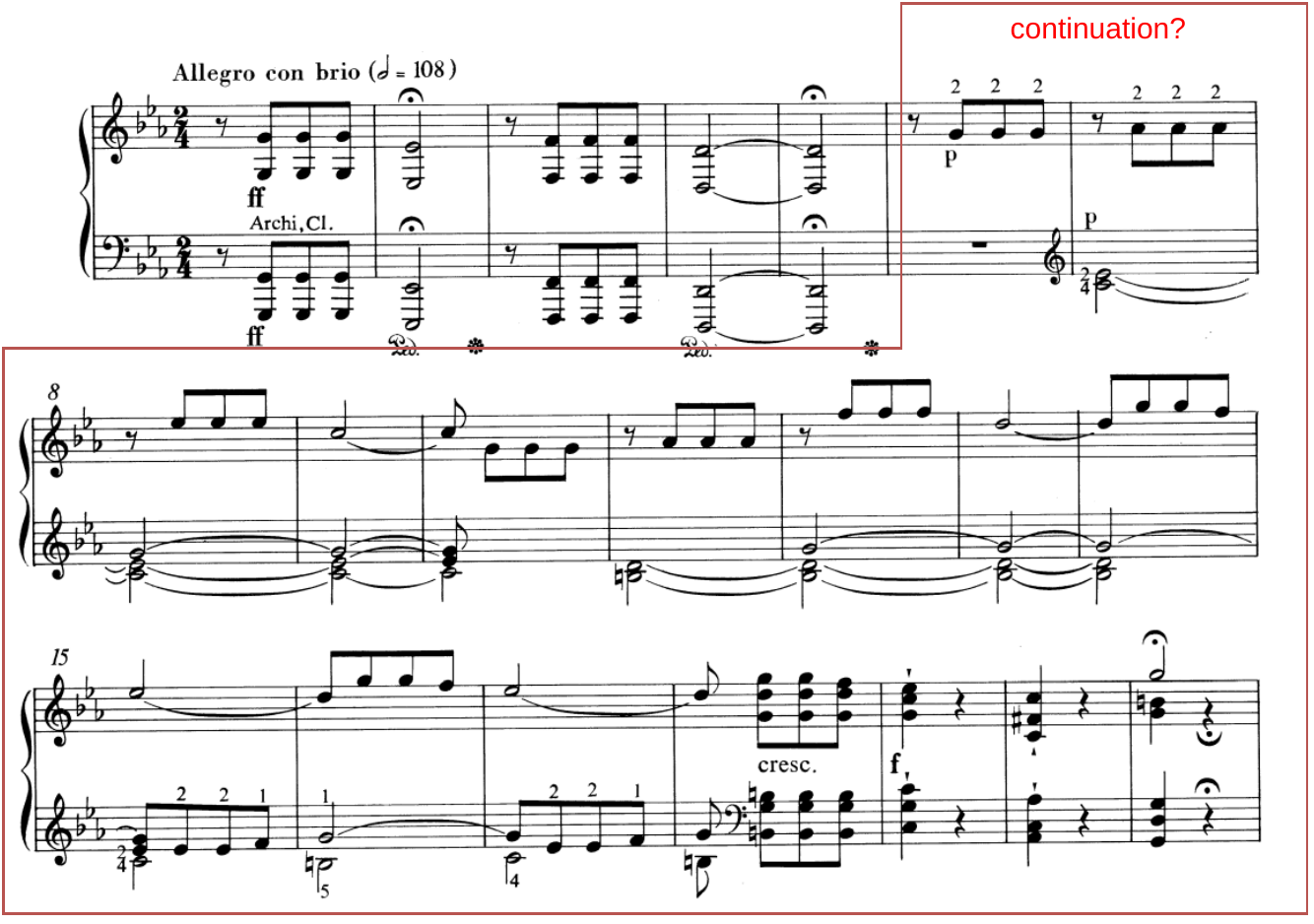}
   \caption{Beethoven, Symphony No.~5, Op. 67, beginning (piano reduction by Liszt), annotated to show that sentence structure is unsuitable for this example}
   \label{fig:beethoven op67 beginning annotated sentence structure}
\end{figure}

One might interpret both openings as instances of a sentence structure where the music follows the pattern of presentation followed by continuation. However, using sentence structure or any conventional phrase structure is inappropriate in this case because one cannot explain the irregular phrase length in both examples. For example, in the Beethoven example, a phrase structure analysis would give a 16-measure continuation, which is unbalanced when compared to the 5-measure presentation (see figure \ref{fig:beethoven op67 beginning annotated sentence structure}). In the Liszt example, the analysis using conventional phrase structure is even more nonsensical because we are unable to identify a cadential moment in the opening section. Therefore, it is more natural to interpret both examples using the idea of three occurrences of the musical gesture illustrated above. Additionally, the three occurrences serve as an introduction, i.e.~a prepending structure that is not part of the main musical body, but an musical anacrusis. The interpretation that the three occurrences of the opening gesture is anacrusis is backed by the observations in both examples. In the Liszt example, the double bar line at measure 8 indicates that it is the beginning of a new phrase; in the Beethoven example, the 16-measure unit between measure 6 and 22 indicates that it is to be considered as a complete phrase. Therefore, in both examples, the three occurrences of the opening gesture are singled out. They are immediately memorable in multiple ways: they are gestures readily recognized; they are placed in the beginning; they are presented three times in succession to reinforce listener's memory.

The discussion of two musical examples above serves as a proof that the opening gesture in Liszt's sonata is readily recognizable and memorable. Upon hearing the F-sharp in the bass, we (performer and audience alike) are naturally reminded of the beginning. We then actively search our memory and recall the beginning. Moreover, we actively compare and contrast the beginning with the current moment of measure 453. In other words, we are participating in an active interplay between listening and recalling, by comparing side by side, on the one hand, what we are listening to and, on the other hand, what we remember.

\begin{figure}[!h]
   \centering
   \includegraphics[width=.8\textwidth]{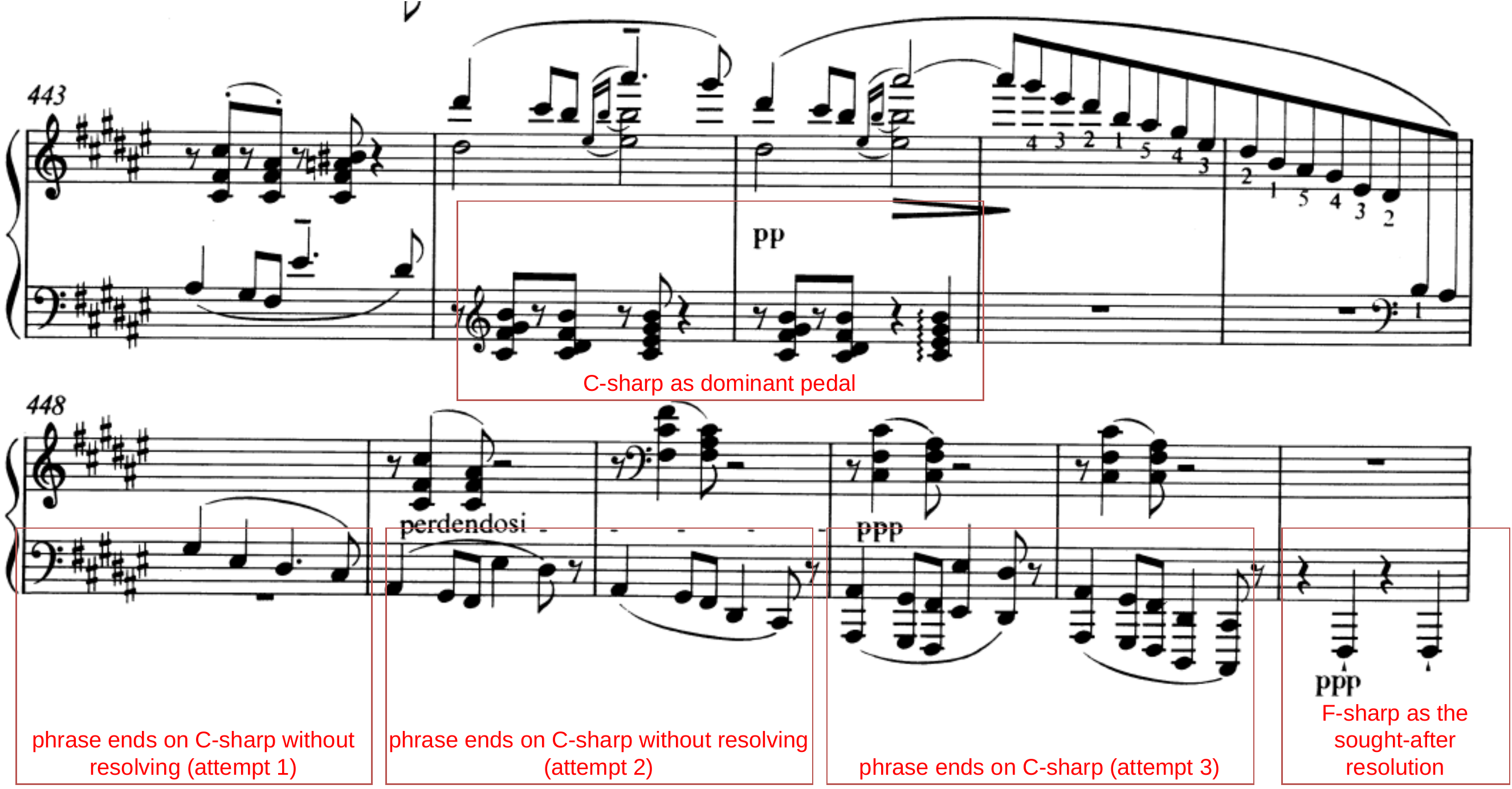}
   \caption{Liszt, Piano Sonata in B minor, S. 178, passage before the final movement, annotated to show how F-sharp is prepared by repeated visits of C-sharp}
   \label{fig:liszt sonata f-sharp prepared}
\end{figure}

As we actively compare, we make several observations. The first observation is about the identity of the F-sharp in measure 453. In its immediate context, it is to be interpreted as the tonic, i.e.~a tonally stable state that is considered resolved. In fact, F-sharp as the tonic is extensively prepared by the dominant note C-sharp (see figure \ref{fig:liszt sonata f-sharp prepared} for an annotated illustration).

Specifically, the note C-sharp is visited multiple times. However, each time as we feel that the note is going to be resolved, the resolution is evaded by pulling the music back to previous material. What we get as a result is analogous to a dominant pedal: repeated visit of unresolved dominant in order to highlight the final moment of resolution. In order to highlight the arrival of F-sharp, we should also note how in addition to the repeated visit of C-sharp, F-sharp is deliberately avoided before measure 453. All of the aforementioned compositional strategies are serving the purpose of portraying the F-sharp as a point of arrival such that it is well prepared and anticipated to the extent that upon hearing the note, listeners are psychologically relieved.

However, the interpretation of F-sharp as resolution is not musically intuitive. Upon hearing the F-sharp in the bass without any harmonic support, we do not feel home-like warmth (which is often associated with resolution, hence the term ``home'' key). Instead, one naturally perceive a sense of unrest. Metaphorically speaking, the F-sharp renders a scene similar to that in a horror movie:

\begin{quote}
    It is a lovely summer day and everyone enjoys the field trip (metaphorical of the musical warmth found in passage before the F-sharp in measure 453). However, suddenly, a wave of exceptionally chilly wind (metaphorical of the F-sharp in measure 453) causes everyone to shiver. It is unexplainable why there would be such chilly wind during summer time. Following the trace of the wind, the protagonists discover a deserted house. ``It's weird, I come here often but have never seen this house before'', says one protagonist. Driven by curiosity, the protagonists decide to knock (metaphorical of musical gesture) on the door. Everyone is excited but at the same time unsettled: what is on the other side of the door?
\end{quote}

\begin{figure}[!h]
   \centering
   \includegraphics[width=.8\textwidth]{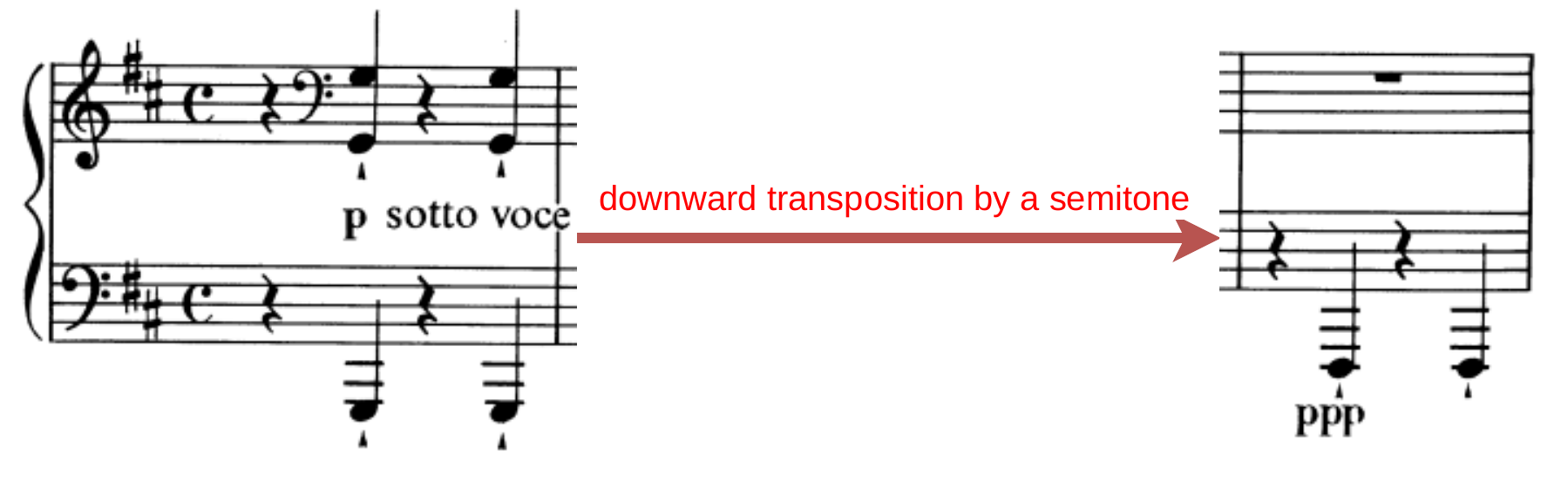}
   \caption{Liszt, Piano Sonata in B minor, S. 178, comparison between the beginning (on the left) and measure 453 (on the right), annotated to show the semitone relation}
   \label{fig:liszt sonata comparison semitone relation}
\end{figure}

Therefore, there must be additional destabilizing factors not yet considered. We notice that F-sharp as a point of arrival is a valid assertion only when we are considering its immediate surroundings. As discussed previously, the magic of F-sharp in measure 453 is that it reminds us of the beginning of the piece. When comparing the F-sharp to the note G in the beginning, its identity is no longer a point of arrival, but a semitone descent (see figure \ref{fig:liszt sonata comparison semitone relation}).

Even though listeners are unlikely to notice the semitone descent consisting of notes which are more than 400 measures apart, performers can be consciously aware of it in order to present a musically expressive performance. The semitone descent signifies that the recapitulation begins with a transposed (instead of identical) restatement of the opening. In order to bring out his/her awareness of the semitone descent and the transposed restatement of the opening theme, the performer actively devises interpretive decisions which listeners are able to notice. For example, the performer can choose to play the phrase in measure 453 with a darker and more gloomy sound by bringing out the bass voice in the left hand. Regardless of interpretive decisions the performer makes, he/she must actively retain the memory of his/her performance of the opening, in order to make an interpretive decision on the transposed restatement.

\begin{figure}[!h]
   \centering
   \includegraphics[width=.8\textwidth]{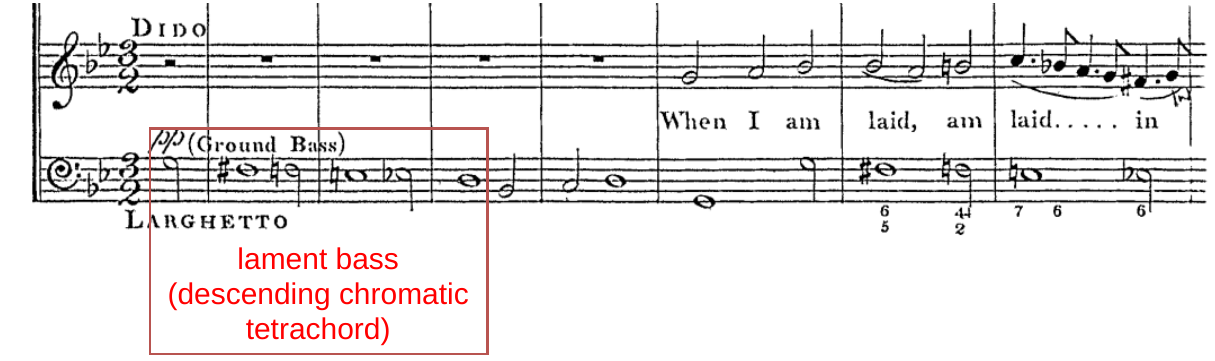}
   \caption{Purcell, aria ``When I am laid in earth'', from opera ``Dido and Aeneas'', beginning, annotated to indicate the descending chromatic tetrachord}
   \label{fig:purcell lament bass}
\end{figure}

The unsettled feeling is reinforced by a distinct feature of the semitone descent: semitone descents can represent emotional unrest associated with negative connotations such as grief, death, and destruction. Such association possibly originates historically from the word-painting (or eye-music) tradition in which the visual elements in notation are translated to musical meanings. In particular, we refer to the practice of lament bass, consisting of ``descending (chromatic) tetrachord as an `emblem of lament'\,''.\autocite{grove_word_painting} For example, consider Purcell's aria ``When I'm laid in earth'' (see figure \ref{fig:purcell lament bass}), in which the use of the lament bass as a compositional device expresses Dido's grief.

Why is a descending figure capable of expressing grief? The explanation lies in the mechanism of word-painting. On the one hand, word-painting associates certain compositional devices to some visual elements. For example, Bach uses ``notes marked with a sharp sign (Kreuz) in works whose text refers to the Cross''.\autocite{grove_word_painting} On the other hand, the visual elements are symbolic of some sentiments. For example, the Cross may be symbolic of redemption. Therefore, word-painting works transitively: a compositional device is symbolic of some sentiments through some visual elements as mediator. For example, Bach's use of sharp sign is symbolic of redemption, through a mediator: the visual element of the Cross. Concerning our current discussion, what is the mediator between descending figure and grief? It is precisely the physical act of falling. To proceed, we start with an observation: the legacy of word-painting which associates visual (or more precisely, spatial) concepts with compositional choice is hard-coded into the language of music theory itself. For example, the word ``descent'' (as in descending scales) is not just a metaphor referring to the spatial notion of downness. Instead, it is a purely technical term in music theory: there is no term in music theory that can describe the concept of descending scales more precisely than actually using the very word ``descending''. As a result, we naturally associate downward intervallic motion to its metaphorical origin: downward physical motion. The crux of the matter is then in extending the metaphor further: what is the implied meaning of downward physical motion? Notice that such question is not entirely a musical one but a cultural-historical one. In many cultures, downness (as a spatial concept) is often associated with destructive forces that are threatening to life. The reason for such association might be surprisingly simple: prehistoric people observe that animals as well as fruits fall to the ground and rot in earth. For example, many cultures consider hell to be positioned underground, such that entering hell is equivalent to falling (whose meaning here is to be interpreted, in addition to its physical meaning, also in its biblical sense, as in Luke 10:18 (KJV) ``I beheld Satan as lightning fall from heaven'') or descent. In Dante's ``Inferno'', the connection between hell and descent is spelled out:

\begin{quote}
    \centering
    ``Now let us descend into the blind world below,''

    the poet said, appearing pale and drawn.

    ``I will be first, you second, as we go.''\autocite[37]{inferno}
\end{quote}

\begin{figure}[!h]
   \centering
   \includegraphics[width=.5\textwidth]{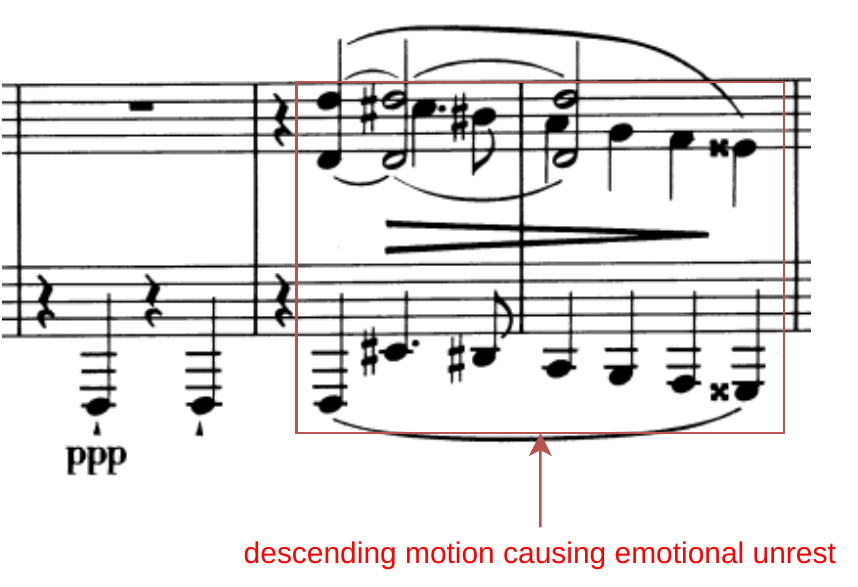}
   \caption{Liszt, Piano Sonata in B minor, S. 178, excerpt, annotated to indicate the descending scale}
   \label{fig:liszt sonata descending scale}
\end{figure}

Therefore, to fall or to descent, is symbolic of entering the underworld. Now, the underworld is characterized by death and unknown, which cause moods underlying the unsettled feeling listeners perceive upon hearing the enigmatic F-sharp as part of the descending motion. Besides the semitone descent indicated in figure \ref{fig:liszt sonata comparison semitone relation}, the F-sharp is part of another descending motion: its containing phrase itself (see figure \ref{fig:liszt sonata descending scale}).

The two descending motions associated with F-sharp collectively contribute to listeners' emotional unrest. As a side note, the previous discussion of the biblical meaning of falling as well as Dante's literary work is intimately relevant to Liszt's compositions. In particular, Liszt's composition ``Dante sonata'' shows his acquaintance with Dante's literary works; additionally, his religious affiliation in his late years is evidence of his involvement in theology, and hence topics associated with the biblical idea of falling: death, sin, and demons. Moreover, rather ironically, Liszt is often portrayed as, albeit his religious devotion, a demonic figure, as a result of the Romantic era. During the Romantic era, musicians often invoke the imagery of the supernatural and the demonic to represent the Romantic spirit (it suffices to consider operas at the time by, for example, Weber and Meyerbeer to see the point). The rise of Romantic virtuoso intensifies such demonic obsession, to the extent that one often praise virtuosos as ``demonic possessor of all who heard him''.\autocite[268]{taruskin2009oxford_19century} For example, Taruskin makes the following remark about Paganini:

\begin{quote}
With his gaunt and gangling appearance and his demoniac temperament, Paganini almost single-handedly forged the romantic mystique of virtuosity as a superhuman, even diabolical endowment. He was Faust come to life---a role model for countless geniuses, charlatans, entertainers, and adolescents ever since his first appearances abroad.\autocite[254]{taruskin2009oxford_19century}
\end{quote}

As the leading virtuoso alongside Paganini at the time, Liszt received similar compliments:

\begin{quote}
As Liszt sat before the piano, the first impression of his personality was derived from the appearance of strong passions in his wan face, so that he seemed to me a demon nailed fast to the instrument whence the tones streamed forth -- they came from his blood, from his thoughts; he was a demon who would liberate his soul from thraldom\autocite{grove_liszt}
\end{quote}

Therefore, with the demonic obsession lurking around the Romantic era, as well as Liszt's personal acquaintance with theology and the Romantic virtuosity, it would be no exaggeration to say that the opening gesture of the B-minor sonata portrays devil's knocking on the door, and the descending motions represent the descent to hell. In fact, in a historically-informed perspective, such interpretation using the imagery of the supernatural and the demonic might be considered the preferred answer. As such, our emotional unrest is fully justified with little danger of intentional fallacy. Meanwhile, we should recognize that musically, the emotional unrest is partly due to the active process of comparison between the moment before the final movement and the very beginning, delineating a semitone descent illustrated in figure \ref{fig:liszt sonata comparison semitone relation}.

\begin{figure}[!h]
   \centering
   \includegraphics[width=.8\textwidth]{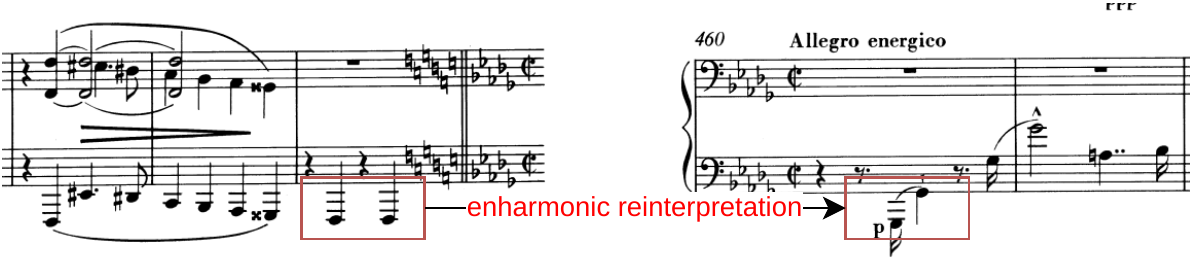}
   \caption{Liszt, Piano Sonata in B minor, S. 178, the enharmonic reinterpretation at the beginning of the final movement}
   \label{fig:liszt sonata enharmonic}
\end{figure}

\begin{figure}[!h]
   \centering
   \includegraphics[width=.8\textwidth]{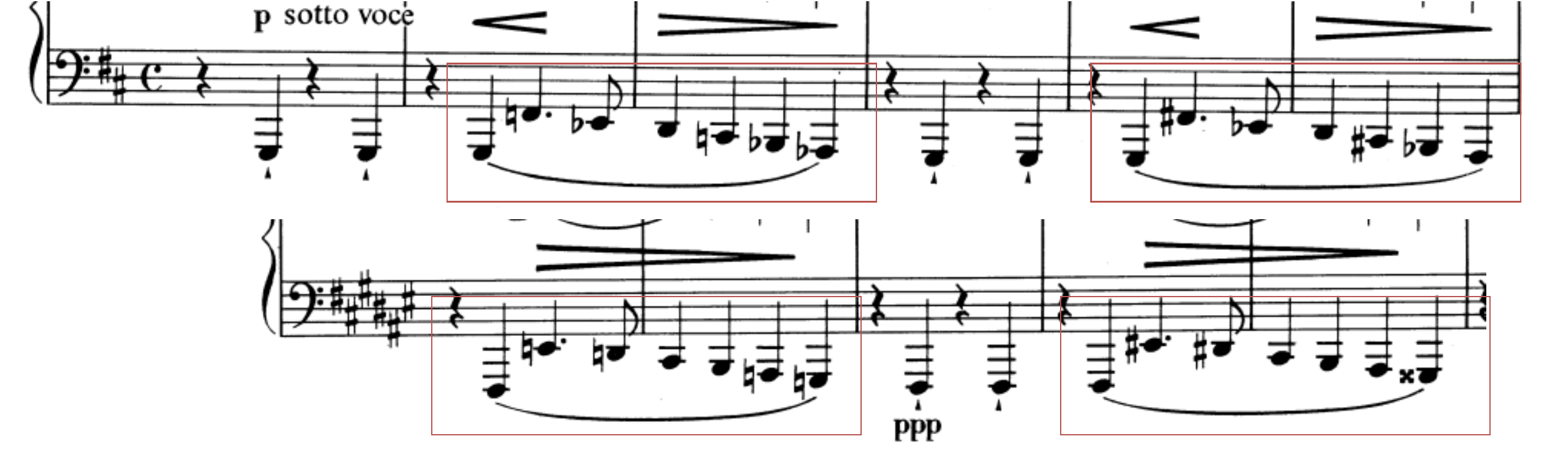}
   \caption{Liszt, Piano Sonata in B minor, S. 178, side by side comparison of four occurrences of the descending scale, note the difference in intervallic contents, first occurrence (top-left): 221222, second occurrence (top-right): 131131, third occurrence (bottom-left): 221222, fourth occurrence (bottom-right): 122211}
   \label{fig:liszt sonata four scales}
\end{figure}

On a positive note, the emotional unrest due to the F-sharp prepares the entrance of the final movement, which is characterized by an enharmonic reinterpretation of F-sharp (see figure \ref{fig:liszt sonata enharmonic}). Unlike the Beethoven example in which one is encouraged to use moments of silence to erase memories of previous movements, the moment before the enharmonic reinterpretation requires one to use all his/her mental power to synthesize heard materials in order to present the enharmonic reinterpretation. In particular, one asks the following series of questions:

\begin{itemize}
\item
  How is the enharmonic reinterpretation a consequence of the semitone descent? In particular, how is the enharmonic reinterpretation related to the beginning of the piece?
\item
  How is the enharmonic reinterpretation a result of four transformations (performers are consciously aware of them, since they are sources of confusion during memorization) of the descending scale (see figure \ref{fig:liszt sonata four scales})?
\item
  How should we treat our emotional unrest due to descending motions and (metaphorically) demonic imagery at the moment of the enharmonic reinterpretation? Is it to be an act of casting out the demon or compromise?
\end{itemize}

Therefore, moods, especially the emotional unrest from the previous movement, are necessarily carried over to the finale. Such emotional residue is precisely what makes Liszt's sonata a continuous and organic whole. It is true that in analysis we often use observations on the score (e.g.~thematic parallelism) as a basis of discussion. However, those observations alone do not entitle us to say that the piece is a continuous whole. Instead, the essence lies in their psychological effects: we say that the piece is a continuous whole because, through the use of compositional devices such as thematic parallelism, we are able to retain a single continuous flow of emotional states. In the Beethoven example above, as the finale starts, we can be assured that the sorrows found in the funeral march would not come back. However, in Liszt's sonata, we are unable to claim the same. Instead, materials from previous movements are, again, like a demon who lurks around haunted places, whose potential recurrence throughout the piece causes the interplay between our memory and anticipation.

\section{Tension-release as psychological narrative}
\label{section:tension-release as psychological narrative}

One outstanding feature of the analysis in section \ref{sec:liszt sonata} is that, as we discuss tension-release, we invariably focus on our emotional response to musical events. For example, a considerable portion of the analysis is to find musical explanation of what causes our emotional unrest upon hearing the recurrence of the opening gesture in measure 453. As a result, the analysis of the narrative using the concept of tension-release becomes an analysis of our own psyche. Instead of focusing on fictional characters and events that are core concepts in conventional narratology, our interest lies in the analysis of our emotional states in response to musical events.

Tension-release is not a property of music, but a property of our psyche in response to music. Therefore, if we are to find tension-release, it is nowhere to be found \emph{in} music but in our response to music. In other words, it does not make sense to talk about tension-bearing events or release-bearing events since tension-release is about our emotional reactions. For instance, in the dominant-tonic harmonic resolution as one of the textbook examples of tension-release, we often tend to label the dominant harmony using the term ``tension'' while labeling the tonic harmony using the term ``release''. However, such act of labeling shares the same potential pitfall as music notation itself: we tend to misunderstand the act of labeling. In particular, labeling a dominant harmony using the term ``tension'' serves as a convenient notational shorthand. However, the shorthand also suggests that tension is a property of the dominant harmony (simply by juxtaposing the spelled-out term ``tension'' next to the dominant harmony), whereas the full statement should be that tension is a property of our emotional state in response to the dominant harmony. Similar misunderstanding is found in natural languages: as a grammatical shorthand, we often say that ``this paper intends to discuss {[}\ldots{]}'', whereas the full statement we mean is expanded as ``the author intends to discuss {[}\ldots{]} through this paper''. In short, we are in danger of interpreting notational shorthand too literally, thus failing to see the full statement. For example, music in notated scores can mislead us into equating notation with music, which can be greatly limiting, as many of the lively features of music are suppressed by notation: where can we find the subtle rhythmic nuances when all we see are written notes enclosed by rigid bar lines?

Thus it is worth reiterating the psychological aspect of narrative. As most of analytical traditions (of almost all disciplines) invariably rely on notation-based systems, we are in danger of misinterpreting notations. In terms of narrative, the notations (in music, notated scores; in literature, sequence of written symbols) often draw our attention to notated symbols, instead of the lively quality behind notations (e.g.~rhythmic nuances behind notated bar lines) that is essential to music. In short, we are at the risk of focusing solely on notated symbols and symbolic manipulations as the object of study. Consequently, plot becomes sequence of events describing mere happenings; music analysis becomes a series of annotations. In both cases, we lose qualities that are quintessentially human: sentiments such as tenderness, struggle, and striving are nowhere to be found in plot-as-happening or music-analysis-as-annotation. In such context, tension-release comes to the rescue as it is an attempt to reiterate the psychological aspect of narrative. A narratology based on tension-release is then functioning as a reminder that the dynamics of narrative primarily lie in our psychological self. It is a reminder that narrative as story-telling is not about stories per se, but about how our emotions fluctuate in response to stories. In short, taking tension-release as the foundation of narratology highlights an important shift in mindset: the focus of the narrative is on the journey of the inner self.

To this end, physics offers, ironically, a more poetic interpretation of the world that we can borrow: the narrative of the world as time evolution of states. As discussed above, we are interested in the psychological aspect of narrative. Therefore, we can define narrative with the goal of capturing its psychological aspect as follows:

\begin{quote}
Narrative is time evolution of psychological states regulated by patterns of tension-release.
\end{quote}

One might be puzzled by the trajectory of our inquiry: how is it possible that we start with tension-release which is a concept mostly limited to music, and somehow take it as a universal (i.e.~applicable to all temporal arts) definition of narrative? In other words, it appears that defining narrative universally using the idea of tension-release is either contrived or a result of sheer luck. On the one hand, it may be contrived that we, as musicians, are biased toward tension-release as it is a familiar concept to musicians. As such, we insist on a biased view and assume that tension-release is equally important in other forms of narrative (such as literary narrative), even though the assumption might be false. On the other hand, suppose that tension-release is indeed acknowledged universally in all forms of narrative, then our discussion is too lucky to be reasonable: as musicians, we pick tension-release as the basis of discussing narrative solely for the sake of convenience, as it is a frequently discussed concept we are most familiar with. Somehow, it luckily has the explanatory power to be considered the universal definition of narrative, generalized to all temporal arts. In short, we are puzzled by the unusual explanatory power of tension-release found in music.

However, the unusual explanatory power of tension-release becomes understandable if we observe a special quality of music. When compared to other arts, music takes pride in its non-representational quality. Music, as a non-representational art, invariably directs our attention to the psychological side of the narrative. The reason is that non-representational art acts transparently on its artistic medium: it is transparent because there is no represented object standing in between the artwork and our psyche (metaphorically as opaque object which blocks light), so that the artwork can directly influence us, i.e.~can shed light on us.

In representational arts such as visual art and literature, we are overwhelmingly preoccupied (and to some extent, distracted) by the represented objects. For example, when we inspect a painting, our focus is on the objects being portrayed in the painting, such that the painting is representational of possible objects in reality. Consequently, we might overlook the psychological aspect of our artistic experience as an audience because our attention is directed to the world of the represented. Metaphorically, representational art is not transparent because the represented object stands in between the artwork (e.g.~the painting as a collection of photons that reaches our retina) and our psyche (i.e.~our emotional state): the artwork does not move us directly, as it must firstly make us cognize the objects being painted, for example, a scenery. Then, as we focus on the scenery, we are less aware of our psychological state. Therefore, when we are asked to describe the painting, we naturally describe what is painted, instead of what we feel.

On the contrary, upon encountering non-representational art such as music, our attention is directed to the \emph{representing} (i.e.~us who experience the work) instead of the \emph{represented}. The reason is simply that we are unable to find what is the represented object. We try to say a piece of music is about the represented objects, but what exactly is represented? The answer is much less obvious than with works of visual art. As music is non-representational, we are left with no choice but to focus on our inner world. We try to find what music depicts, yet invariably come to the realization that it reflects our inner world. Similarly for the topic of narrative in music, we ask: ``what is the story about''? For literature, we have a definitive answer: ``of course the story is about the interactions between characters, the journey of the protagonist, and the struggle of mankind''. However, for music, we are left speechless, and our despair is turned into the awareness and conscious exploration of our psychological world.

In short, the advantage of taking a musician's perspective is that the non-representational quality of music translates to our preference for the psychological aspect of narrative in music. Given the emphasis on our inner experience, we can make claims about narrative that are more universal. To understand this, we consider how statements concerning the psychological aspect of our experience can attain universality and generality, contrary to the intuitive view that subjective experience is mostly biased and unreliable. A statement concerning our subjective experience, such as describing music using tension-release, is in little danger of being subject to cultural relativism. In particular, few would challenge the fact that general emotions involved in our emotional responses to music are culture-independent (although the specific mechanisms underlying our emotional responses to music can be culture-dependent: listeners from different backgrounds might emotionally react to the same piece of music differently). Emotions such as anxiety, excitement and nostalgia are universal and well-understood across cultures and time periods. It is true that listeners are educated and nurtured to be accustomed to specific idioms of music which are culture-dependent (for example, one becomes sensitive to tonal resolution as a result of repeated exposure to tonal repertoire). Moreover, different musical idioms may employ distinct mechanisms to influence listener's psyche (for example, tonal music relies on tonality, while other traditions may focus on the rhythmic aspect of music). However, we note that despite differences in mechanisms, it is universally acknowledged that music has the power to emotionally move the listeners. For example, tension-release, in addition to being observed across all hierarchical levels (see page \pageref{idea:tension-release is observed across all hierarchical levels}), is also style-agnostic to the extent that it is observed across a majority of time periods (even though musicians in different time periods might have different interpretations and implementations of how tension-release is specifically achieved). We can use tension-release to describe medieval motets and post-tonal repertoire alike, in the same way we describe tonal music in terms of tension-release with the particular use of tonal resolutions. To this end, we may say that tension-release is a theme that stands the test of time. By contrast, if we use tonality as basis for discussing narrative in music, then it is necessarily style-sensitive because once we move to non-tonal music (either pre-tonal or post-tonal), the discussion falls victim to anachronism.

To sum up, our discussion conveniently starts with a musician's term: tension-release. However, the goal is ambitious as it tries to show that tension-release underlies all narratives universally. The universality of tension-release has its root in its emphasis on the psychological aspect of musical experience: music as narrative of our inner self (i.e.~time evolution of psychological states). Statements about the psychological aspect of our experience are universal because emotions are universal: as long as humans, as biological species, are concerned, emotions as a result of evolution (such as fear of the uncertain future and nostalgia of the lost past) are felt and understood universally. Finally, music is particularly interested in the psychological aspect of experience due to its special quality: music as an inherently non-representational art. That concludes a musician's derivation of what is essential in a narrative. One should notice that such derivation can be entirely independent without referencing any other disciplines. In short, the conclusion that tension-release is essential in a narrative can be a result of a purely musical discourse.

Rather surprisingly, musicians are not alone. The conclusion we have discussed is not singular to the field of music. The previous discussion is a unilateral assertion that tension-release is universally applicable to narratives in other art forms. If we now inspect other disciplines, in particular, literary studies as the de facto origin of narratology, we find similar trends: the shift in mindset that puts emphasis on the internal narrative of the reader. One may argue that such an observation is due to confirmation bias: we come to the belief that tension-release is essential in a narrative first, after which we selectively and actively search for evidence to our favor while disregarding everything else that does not confirm our bias. However, we should nevertheless acknowledge another possibility, that researchers in different fields independently arrive at the same statement, precisely because the statement is universal.

For example, in discussing literary narrative, Bridgeman points out that one should always keep in mind the crucial role of the reader's active world-building in response to the unfolding story:

\begin{quote}
Narratives unfold in time, and the past, present, and future of a given event or action affect our interpretation of that action, while the characters who populate narrative texts move around, inhabit and experience different spaces and locations, allowing readers to construct complex worlds in their minds. \autocite[52]{bridgeman_2007}
\end{quote}

\begin{figure}[!h]
   \centering
   \includegraphics[width=.5\textwidth]{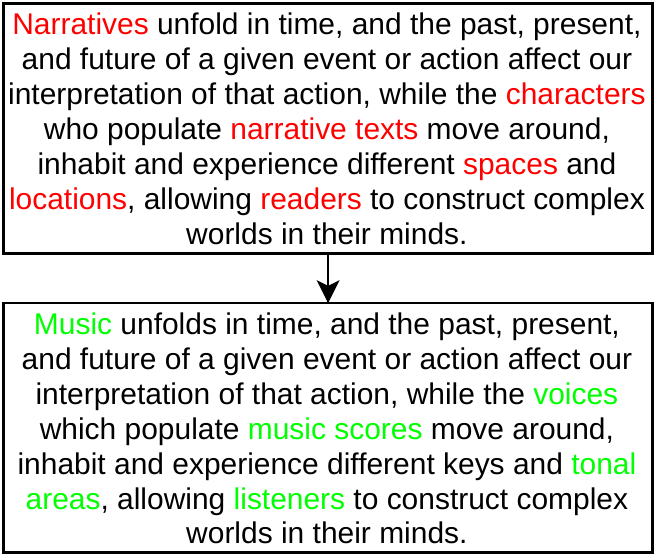}
   \caption{A colored differentiation illustrating modification required to transform Bridgeman's literary statement into a musical one}
   \label{fig:literary statement to musical one}
\end{figure}

As a side note, apart from the emphasis on the psychological aspect of narrative which this section extensively discusses, the word ``unfold'' should be particularly outstanding to readers familiar with Schenkerian analysis since a considerable portion of the theory focuses on the idea of unfolding. On the one hand, the term can be technical and mean specific compositional technique in Schenkerian analysis. In the Grove article on the very term ``unfolding'', the author defines it to be ``a method of Prolongation whereby the separate voices of a contrapuntal idea are amalgamated as a single line''.\autocite{grove_unfolding} On the other hand, we should note that it is also a beautiful metaphor full of poeticness: it represents the organic growth of prototypical music structures, through whose unfolding in time, fulfills the \emph{teleo} of music. It is more likely due to this poetic reason (instead of technical reason) that Schachter gives his Schenkerian treatise the title ``Unfoldings''\autocite{schachter1999unfoldings} (the same way this project is titled: arrow of time is a technical term, yet its poetic quality is what justifies its role as the main title). In fact, the parallelism inspired by the very term ``unfold'' extends further, such that with minor modifications we are able to transform Bridgeman's comment into a purely musical statement (see figure \ref{fig:literary statement to musical one} for a colored differentiation):

\begin{quote}
Music unfolds in time, and the past, present, and future of a given event or action affect our interpretation of that action, while the voices which populate music scores move around, inhabit and experience different keys and tonal areas, allowing listeners to construct complex worlds in their minds.
\end{quote}

It is worth reiterating that the transformed statement is not ad-hoc. Instead, it is a purely musical statement that can be naturally placed in the midst of any musical discourse. In fact, musicians frequently spell out the importance of temporal unfolding in relation to the interplay between the past and the future. For example, Lasser remarks that:

\begin{quote}
Music is a temporal art. It unfolds in time; its medium or canvas, as it were, is time. {[}\ldots{]} In music, the present is represented by the moment of listening, the past by memory, and the future by a set of constantly shifting expectations and predictions for the events to come.\autocite[3]{lasser2008}
\end{quote}

In more details, Bridgeman spells out the unifying theme underlying narratives bearing different styles and from different time periods:

\begin{quote}
Whatever the temporal patterns set out within fictional worlds - whether they are those of a nineteenth-century novel that moves toward a defined and anticipated ending, or whether they are those of a postmodern narrative, operating by disjunctions, loops, and effacements - it is inescapable that these patterns will be set against the reader's temporal experience of the text, founded on memory and anticipation. And the reader's attempt to relate these two kinds of temporality will be an important part of the effect of the text. \autocite[54]{bridgeman_2007}
\end{quote}

In a similar manner, we are able to replace a few words in order to transform this literary statement to a musical one. The reason we are able to transform with such little effort is that both statements share the same set of core concepts: audience's temporal experience of the artwork, and the interplay between memory and anticipation. Therefore, these core concepts (in addition to tension-release) are universal in that they are not only able to address narrative of different styles and time periods, but also different art forms such that music and literature are treated alike.

\section{Temporal asymmetry in narrative}
\label{section:temporal asymmetry in narrative}

At this point, we should be aware of the similarity between the two concepts discussed above: tension-release (introduced in section \ref{section:tension-release}) pertaining to the study of narrative, and arrow of time (introduced in section \ref{section:arrow of time}) pertaining to the history of the universe. They are two concepts whose similarity is warranted by the attribute of omnipresence. In other words, they are universally observed under different contexts: in an artistic discourse, tension-release is observed universally in narratives; whereas in a scientific discourse, arrow of time is observed universally in macroscopic physical phenomena.

It turns out that their universality may share the same root cause. We may speculate that tension-release and the arrow of time are two sides of the same coin. Specifically, they may concern the same thing which is described in different wordings as a result of distinct conventions employed by science and art. For example, the idea of transposition in music theory is essentially the same with the idea of translation in geometry (transposition literally becomes translation in geometric representations of pitch such as the Tonnetz): they both represent transformations that preserve the notion of ``distance'' (musically, intervallic relations). Therefore, we may say that the difference between transposition and translation is merely an issue of wording as a result of different historical developments.

To see how tension-release and the arrow of time may be ascribed to the same cause, we invoke a simple poetic metaphor:

\begin{quote}
\label{prop:history as narrative}
The history of the universe is nothing but another form of story-telling.
\end{quote}

This metaphor is an artist-centric one. Analogous to the Copernican revolution, it reverses the commonly perceived relation between the subjective (narrative as in art) and the objective (events as in science). Instead of describing story as a sequence of events on a physical basis (whose truth artists cannot deny), we treat the history of the universe artistically just as we treat music: a narrative whose dynamics are governed by the psychological construct of tension-release. Instead of entering another round of unfruitful philosophical debate on the topic of materialism versus idealism, it suffices to notice that the crux of the matter is discussed previously: both narrative and the history of the universe are expressed by the time evolution of something. Hence, the difference primarily lies in what this ``something'' refers to: it refers to the psychological state in expressing narrative, and to the physical state (in particular, the quantity called entropy) in expressing the history of the universe.

One particular concept that lies at the heart of both tension-release and the arrow of time is the interplay between memory and anticipation. We have discussed the concept in section \ref{section:epistemic difference between the past and the future} using the ideas that our epistemic access to the past and to the future consists of fundamentally different mechanisms. On the other hand, tension-release in narrative operates on the interplay between memory and anticipation. Besides the discussion of tension-release in section \ref{section:tension-release}, we further notice that it also of interest to literary studies:

\begin{quote}
   It is important to consider the effects on the reader of temporal patterns. Sternberg is particularly interested in these, suggesting that we should consider the story - discourse relationship in terms of the universals of suspense, curiosity, and surprise, which are generated by the gaps between story time and discourse time (or communicative time). Suspense arises from the gap between what we have been told so far and what we anticipate lies ahead. Curiosity arises from the gap between what we have been told of the past and what else we imagine might have happened. Surprise arises when a twist in the order of narrative conceals from us an event which is subsequently revealed. For Sternberg, ``the play of suspense/curiosity/surprise between represented and communicative time'' defines narrativity.\autocite[54]{bridgeman_2007}
\end{quote}

According to Bridgeman's interpretation of Sternberg's ideas, the frequently used words for describing narratives, namely, suspense/curiosity/surprise (note how they all focus on the psychological aspect of narrative) correspond to different ways in which the past and the future interact, giving rise to the interplay between memory and anticipation.

Therefore, if we were to trace the origin of temporal asymmetry in narrative, it is to be found precisely in the psychological arrow of time. In particular, the epistemic access to the past through memory (as record-bearing devices) and to the future through anticipation (as predictions) differs fundamentally, so that narrative is shaped asymmetrically in time.

The meaning of temporal asymmetry is again twofold. We first recall the twofold statements in section \ref{section:further_decomposition}:

\begin{itemize}
\item
   \textbf{Condition 1 (distinguishability)}: the future appears to be fundamentally different from the past.
\item
   \textbf{Condition 2 (unique orientability)}: time invariably flows in the direction pointing to the future.
\end{itemize}

Now, the two conditions can be translated to the corresponding statements about narrative. In particular, we focus on two specific statements:

\begin{itemize}
\item
   \textbf{Instantiation of condition 1 with respect to narrative}: components of narrative, depending on their temporal context within the narrative, exhibit different behaviors and perceptual effects.
\item
   \textbf{Instantiation of condition 2 with respect to narrative}: placement of the climax is uniquely determined by the arrow of time.
\end{itemize}

The upcoming discussion is split into two sections in order to address these two statements individually.

\subsection{Distinguishability condition: perceptual difference due to temporal context}
\label{sec:distinguishability}

\begin{figure}[!h]
   \centering
   \includegraphics[width=.8\textwidth]{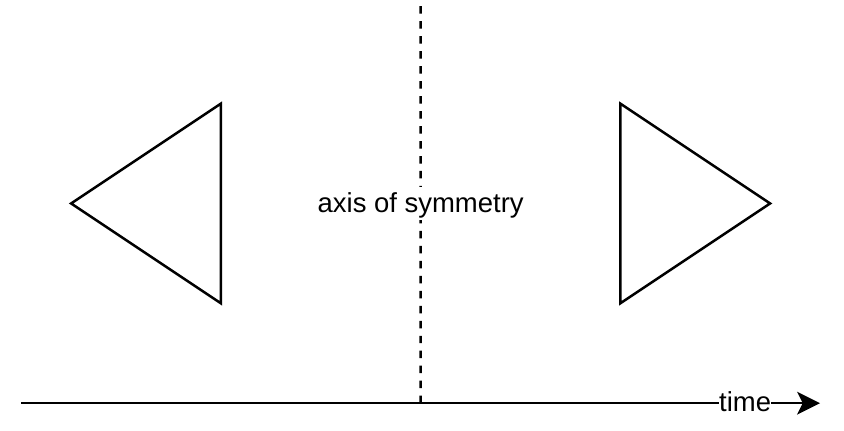}
   \caption{A visual representation of the temporal relations between paired constructs. The two triangles represent paired constructs in narrative (e.g.~rising action and falling action). They are placed around the axis of symmetry (i.e.~the climactic moment of the story). Additionally, they form mirror images (one leads to another through flipping around the axis of symmetry).}
   \label{fig:narrative structure symmetric}
\end{figure}

In a narrative, we observe that many of its parts are paired. Structurally, the beginning is paired with ending; poetically, rising is paired with falling; and musically, introduction is paired with coda (see section \ref{sec:op111}). As such, we are able to use these pairings to address musical topics in terms of dichotomy and duality in general. The above examples of pairing share a common feature that they relate similar constructs with opposite temporal features in terms of placement and direction (see figure \ref{fig:narrative structure symmetric}). In other words, we are interested in a special subset of paired constructs: parts of a narrative that are related by similar contents whose temporal placement and direction in the text are reversed with respect to an axis of symmetry. For example, rising and falling are placed at the opposite end of the narrative's timeline. Additionally, their directions are opposite in the sense that falling can be interpreted as the time-reversal of rising, which essentially undoes what is achieved by rising. Metaphorically, if rising is like playing a movie forward in time, then falling is rewinding the tape such that it is played backward in time.

As illustrated in figure \ref{fig:narrative structure symmetric}, without considering the specific content of narrative, schematically they are often placed symmetrically around the axis of symmetry: the climactic moment. Visually, they are images mirrored about the axis of symmetry. Consequently, one may wonder: where does temporal asymmetry arise in this visually symmetric representation?

Asymmetry arises once we take into consideration contents of narrative. Paired constructs are different in nature, to the extent that we instinctively feel that the difference is not just in the mere rearrangement of the same thing (e.g.~take the plot of the rising action, reverse its course and copy it to the later part of the story). But rather, it should additionally be attributed to something else. Because of the rearrangement, paired constructs become entirely different things, to which we must employ completely different sets of premises, analytical methods, and perceptual attitudes as a consequence of the psychological arrow of time (i.e.~epistemic difference between the past and the future). Therefore, the factor contributing to our instinctive impulse, that the difference between paired constructs in narrative must be due to something else, is nothing other than the psychological arrow of time. Specifically, as the effect of the arrow of time, temporal rearrangement of events has the power to shape the very identity and definition of those events. In short, paired constructs may be similar in content, yet the difference in temporal context render them entirely different in identity. To clarify the point, we consider two examples.

Consider an apple placed on a table. If we move it to the floor and rotate it such that it is placed up-side-down, it is still an apple, in the sense that nothing significant about its identity has changed. Therefore, we call the act of moving and rotating the apple \emph{mere} rearrangement because the identity of the apple remains unaffected by such spatial rearrangement. However, if we take into consideration temporal context by putting the apple into a narrative unfolding in time (for example, as in the snow white fairy tale), the very occurrence of an apple at different places in the plot makes it no longer the same apple, because the psychological arrow of time alters its significance so strongly such that its identity changes. The physical object called apple appearing in a conspiring scene is, with respect to the narrative structure, considered a completely different symbolic object from the exact same physical object appearing in the happy ending. Therefore, we recognize the difference between apple-in-conspiring-scene and apple-in-happy-ending to the extent that the physical object of apple becomes a dramatic placeholder: whether it is an apple or donut is irrelevant to the narrative. The crux of the matter is that, the psychological aspect of narrative (e.g.~memory and anticipation, tension-release) is so crucial that the arrangement of events has the power to define these events. For the apple, when appearing at two places in the plot, we acknowledge that it is the same (physical) object at different moments. Nevertheless, because narrative depends so heavily on the interplay between our memory and anticipation, our psychological perception takes precedence in defining the object (i.e.~the apple). As such, the apple appearing at different places in the plot is interpreted differently, whose psychological interpretation \emph{is} its dramatic definition.

Therefore, the same plot in terms of the content of events, can be interpreted and defined in drastically different terms, as an effect of temporal asymmetry, under different placements. For another example, consider the following hypothetical story:

\begin{quote}
The protagonist is practicing one of Chopin's nocturnes at home. He/she then stops practicing and enters the kitchen. In the kitchen, he/she finds that the oven has caught fire, as a result of his/her experimenting with a new recipe: a giant donut. Luckily the fire is not severe, so that the protagonist puts out the fire using a fire extinguisher. He/she then leaves the kitchen and starts practicing. The protagonist is practicing one of Chopin's nocturnes.
\end{quote}

This story is simplified and reduced to absurdity in order to highlight the palindrome-like construction (or in musical terms, following an ABA form). In particular, we should note that the first and last sentences are identical, therefore consisting of identical events: practicing one of Chopin's nocturnes. If one is to adapt the story to a film, the footage for the beginning and ending can be conveniently reused. Therefore, when the nocturne practicing scene is inspected in isolation (i.e.~without considering the narrative context in terms of what comes before and after), the beginning and the ending are identical in content. However, when we interpret the story as a whole, i.e.~when every event occurs in relation to the rest of the story, the scene of nocturne practicing differs as its placement in the story varies. When the nocturne is heard again after the event that the kitchen is caught on fire, our perceptual attitude has changed drastically. We see that the protagonist is practicing a nocturne, but given what has happened previously, we have additional sentiments mixed in. In addition to enjoying the tranquility of the nocturne, we are at the same time concerned about the donut part: is the fire completely extinguished? Does the overbaked donut smell bad such that it makes the nocturne sound like a donut nocturne (analogous to the coffee cantata by Bach)? Therefore, juxtaposing the nocturne scene with our concerns about the kitchen makes the moment particularly absurd (even comical). The simple tranquility of the nocturne found in the opening scene is completely lost . Instead, one interprets the musical beauty of nocturne as an irony. It is for the same reason that in constructing horror movies, one typical technique is to create a surreally quiet scene after the terrifyingly turbulent scene one commonly associate with the category of horror movies. However, such quietness in the context of the story is exceptionally disquieting because it highlights the terrifying aspect of quietness: the fear of uncertainty.

Therefore, in analyzing narrative structures, we must be careful with identifying and labeling structural components. One potential pitfall is to neglect temporal context that is, as discussed above, the defining feature of structural components. In fact, we observe similar ideas in music analysis: students are repeatedly reminded that harmonic analysis is not about mere labeling, in Roman numerals, of vertical chords in isolation. Instead, one must actively interpret each chord in the context of its surrounding passage in order to define its identity at all. On the other hand, if we are given an analysis, then we must interpret the written symbols (e.g.~the written symbols ``ABA'' in rounded binary form or sonata form represent an analysis of formal structures) with care. In particular, we must be careful not to interpret the equality of written symbols too literally. For example, in using the written symbols ``ABA'' to describe a piece's formal structure, we are in danger of claiming that A section (before B) is somehow the same as A section (after B) because they are represented by the same written symbols, namely, A. Granted that symbol-wise equality does not imply literal equality, we are able to observe temporal asymmetry that is often disguised, borrowing Riemann's wording, ``under the cloak of''\autocite[22]{riemann1895harmony} apparent symmetry.

Literary analysis of narrative, surprisingly (or unsurprisingly, since all temporal arts are instances of narrative), addresses ideas similar to the ABA form in music. In his ``Structural analysis of narrative'', Todorov proposes a schematic model that can be characterized by a threefold division: equilibrium-disequilibrium-equilibrium. Specifically:

\begin{quote}
The minimal complete plot can be seen as the shift from one equilibrium to another. This term ``equilibrium,'' which I am borrowing from genetic psychology, means the existence of a stable but not static relation between the members of a society; it is a social law, a rule of the game, a particular system of exchange. The two moments of equilibrium, similar and different, are separated by a period of imbalance, which is composed of a process of degeneration and a process of improvement.\autocite[75]{structural_analysis_of_narrative}
\end{quote}

Musicians should immediately recognize that this is a statement also suitable for describing the ABA ternary form in music. In particular, when interpreting equilibrium in its music-theoretic term as harmonic stability, disequilibrium as harmonic instability, one recovers the basic design principle of sonata-allegro form: the exposition (expressing harmonic stability), the development (expressing harmonic instability), and the recapitulation (expressing harmonic stability) form the outline of sonata-allegro form commonly labeled as ABA. 

Now, both the equilibrium-disequilibrium-equilibrium model in literary analysis as well as the ABA form in music analysis fall victim to the rigidity of notation. Firstly, written symbols are necessarily reductions, and in the process of reduction we may lose crucial information so that the description fails to capture essential aspects of narrative (i.e.~temporal context). In particular, as discussed above, we tend to interpret symbolic equality too literally: when we label two events in a sequence using the same written symbol (e.g.~equilibrium and symbol ``A''), we acknowledge that there must be substantial parallelism between the events that justify our use of the same written symbol to denote the events. However, it can lead us into overlooking contextual information not encoded by the symbols alone (since we tend to think that symbolic representations \emph{are} the universe of discourse). Therefore, when examining the two occurrences of equilibrium in a narrative, noticing the recurrence of the equilibrium state is certainly interesting, yet it is more crucial to notice the difference between the two occurrences:

\begin{quote}
Sequence is perceived by the reader as a finished story; it is the minimal narrative in a completed form. This impression of completion is caused by a modified repetition of the initial clause; the first and the last clause will be identical but they will either have a different mood or status, for instance, or they will be seen from different points of view.\autocite[74]{structural_analysis_of_narrative}
\end{quote}

Two keywords here are, borrowing Cooper and Meyer's wording, ``marked for consciousness''.\autocite[8]{cooper_meyer_1960} The first is the idea of a modified repetition. Temporal asymmetry in narrative is partly a result from modified repetition, such that in ABA form, the two occurrences of A section constitute a variation in the literal sense: modified content. For example, in sonata-allegro form, the recapitulation is a restatement of the exposition. However, the restatement is never exact in content. One of the construction scheme for the exposition section is that the music has to purposefully go to the ``wrong'' state (e.g.~modulating to the dominant) such that the ``mistake'' is to be ``corrected'' (e.g.~cadence in the tonic) in the recapitulation. Therefore, the recapitulation, in terms of the musical content (i.e tonal trajectory), is a modified repetition of the exposition (or inversely, the exposition is the modified version of the recapitulation, since the exposition is more ``abnormal'').

The second keyword is the idea of ``different points of view''. As illustrated above, even when recurrence is literal, i.e.~containing literally the same content, the identity of the same event varies as its temporal placement (i.e.~temporal context) in the plot differs. In Todorov's view, the difference is attributed to different points of view. One may interpret the meaning of ``points of view'' literally, such that different points of view amount to different in-story characters' perspectives: characters Alice and Eve witness the same event to produce two points of view called ``Alice's recounting'' and ``Eve's recounting''. However, a more universal approach is to take the result of our previous discussion, that narrative is shaped by the experience of its audience, such that the psychological metanarrative of the reader is the defining feature of narrative. In this case, the source of different points of view is nothing other than the very psychological arrow of time.

The threefold division of equilibrium-disequilibrium-equilibrium echoes the literary tradition of trilogy, in that the division into three is a prominent theme found in many literary practices. One may argue that the particular interest in threefold division is attributable to the theological (more precisely, Christian) notion of trinity. However, such claim cannot hold once we realize that such tradition is also observed in pre-Christian time periods. To trace its root, one may observe Aristotle's ``Poetics'', in which he characterizes the narrative structure of a tragedy:

\begin{quote}
Tragedy is an imitation of an action that is complete, and whole, and of a certain magnitude; for there may be a whole that is wanting in magnitude. A whole is that which has a beginning, a middle, and an end.\autocite[31]{aristotle_poetics}
\end{quote}

One should be perplexed: why would Aristotle spell out such a self-evident statement? We all know that a story, by definition, has a beginning, a middle, and an end. To spell out the threefold division is analogous to spelling out the statement ``something called the Manhattan school of music is a school of music located in Manhattan''. However, in the absurdly self-evident statement addressing the threefold division of a tragedy, Aristotle has a point. He then explicates each of the dramatic components:

\begin{quote}
A beginning is that which does not itself follow anything by causal necessity, but after which something naturally is or comes to be. An end, on the contrary, is that which itself naturally follows some other thing, either by necessity, or as a rule, but has nothing following it. A middle is that which follows something as some other thing follows it. A well constructed plot, therefore, must neither begin nor end at haphazard, but conform to these principles.\autocite[31]{aristotle_poetics}
\end{quote}

After the explication, the self-evident statement of stories having threefold division is no longer trivial. In particular, Aristotle hints upon the temporal asymmetry in the construction scheme of the tragedy (or of all narratives unfolding in time). In particular, his explanation points out the difference between the three parts as the sole effect of the arrow of time. It is the sole effect of the arrow of time because the statement does not mention any content but focuses on the pure form of dramatic structure. We translate the statement as follows:

\begin{itemize}
\item
  ``A beginning is that which does not itself follow anything by causal necessity, but after which something naturally is or comes to be'': nothing appears prior to the beginning because of our lack of the epistemic access to the past (i.e.~no record-bearing devices available), therefore at this moment, the narrative is solely governed by our anticipation.
\item
  ``A middle is that which follows something as some other thing follows it'': at this moment, we have epistemic access to both the past and the future through memory (as record-bearing devices) and anticipation (as predictions).
\item
  ``An end, on the contrary, is that which itself naturally follows some other thing, either by necessity, or as a rule, but has nothing following it'': as a narrative unfolds, our memory of the past (of the story) accumulates, resulting in a reversal between memory and anticipation in terms of their roles. In contrast to the beginning, now our epistemic access to the past outweighs that to the future.
\end{itemize}

\begin{figure}[!h]
   \centering
   \includegraphics[width=.5\textwidth]{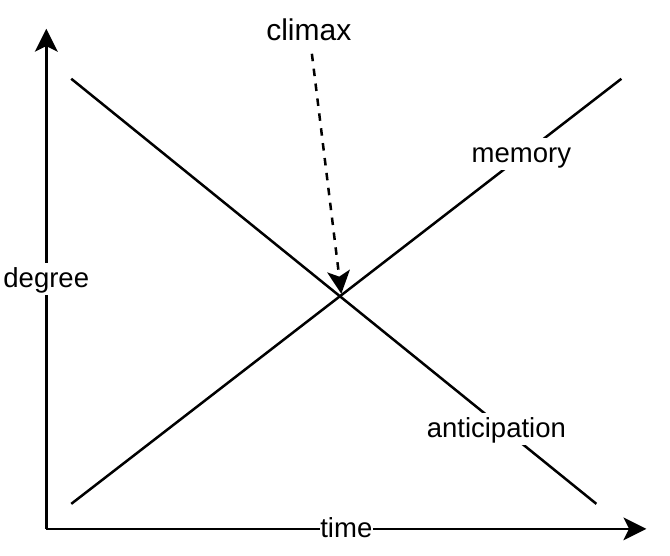}
   \caption{A generalized model of narrative in terms of the interaction between memory and anticipation}
   \label{fig:narrative memory anticipation}
\end{figure}

In summary, beginning-middle-ending as the interplay between memory and anticipation can be visualized using a schematic graph (see figure \ref{fig:narrative memory anticipation}).

Therefore, Aristotle's statement can be transformed in a way to indicate non-trivial conclusions: constructing a narrative which is complete with respect to beginning-middle-ending, is equivalent to the construction of the narrative as time evolution of psychological states, whose dramatic ``equations of motion'' (as transition between states) are subject to the laws specified by the arrow of time.

\subsection{Unique orientability condition: placement of the climax}
\label{section:unique orientability}

Whenever we invoke the metaphor of the arrow, we often perceive a sense of force that pushes whatever object of interest along the direction of the arrow. On a less metaphorical level, we may interpret ``force'' and ``object'' literally, such that we recover the visual notion for the concept of force field in physics: at each point in space, we assign an arrow to visually represent the underlying force at that point in space. As a result, if we put a hypothetical particle in the visualization, it would be literally pushed by the arrow. The use of arrow becomes valuable pedagogically because it captures the essential qualities of force: arrow, like the force it represents, has direction and magnitude.

\begin{figure}[!h]
   \centering
   \includegraphics[width=.7\textwidth]{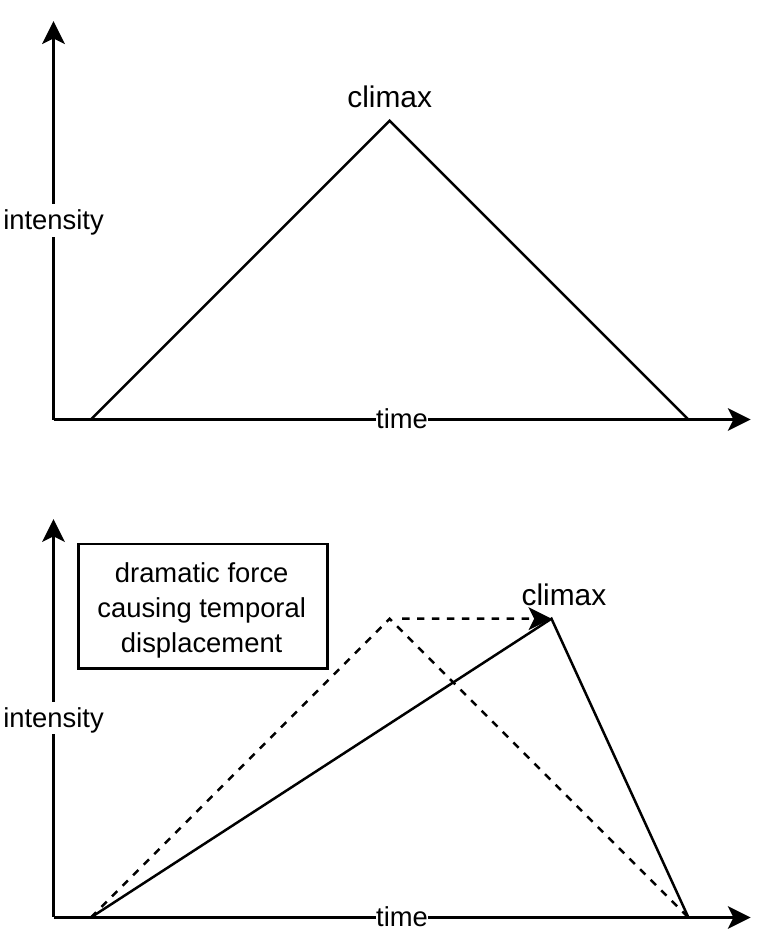}
   \caption{Visual illustration of temporal asymmetry represented by temporal displacement of the climax, producing a skewed shape}
   \label{fig:temporal asymmetry as displacement}
\end{figure}

Therefore, when discussing arrow of time as a metaphor, we also imagine a sort of force (call it musical force for music, and dramatic force for theatrical art) that pushes the object called ``climax'' along the direction of time's arrow, producing a skewed shape (see figure \ref{fig:temporal asymmetry as displacement}). Therefore, in musician's terms, the climax is delayed.

In fact, delaying the climactic point is one of the most canonical forms of temporal asymmetry. By the term canonical here, we mean intuitive. The visualization illustrated in figure \ref{fig:temporal asymmetry as displacement} captures the everyday meaning of asymmetry in the following sense. If we are to teach the term asymmetry to a five-year-old (in the spirit of ``explain like I'm five'' internet trend, where ``an online community provides answers to questions which are comprehensible by five year olds''\autocite{eli5}), the most intuitive method would be a pictorial illustration similar to figure \ref{fig:temporal asymmetry as displacement}: a skewed shape such that after flipping it, the resulted shape does not overlap the original one. In other words, the visually skewed shape is the most intuitive representation of asymmetry in everyday life.

Before discussing temporal asymmetry as delaying the climax, we should observe that the visualization in figure \ref{fig:temporal asymmetry as displacement} has another metaphorical implication: the shape resembles a pyramid. Now, the use of pyramidal structure as a visual representation is spelled out in dramatic studies due to its descriptive utility. For example, in Freytag's ``Technique of the drama'', the author spells out the use of a pyramidal structure:

\begin{quote}
{[}The{]} drama possesses --- if one may symbolize its arrangement by lines a pyramidal structure.\autocite[114]{freytag1895technique}
\end{quote}

With the visual representation of a pyramidal structure, the author models the narrative scheme of a five-act drama as follows:

\begin{quote}
{[}The drama{]} rises from the introduction with the entrance of the exciting forces to the climax, and falls from here to the catastrophe. Between these three parts lie (the parts of) the rise and the fall.\autocite[114]{freytag1895technique}
\end{quote}

\begin{figure}[!h]
   \centering
   \includegraphics[width=.3\textwidth]{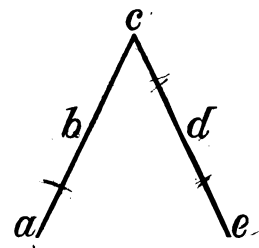}
   \caption{The pyramidal structure of a five-act drama (figure in original text of ``Technique of the drama''), quoting Freytag's words: ``These parts of the drama, (a) introduction, (b) rise, (c) climax, (d) return or fall, (e) catastrophe, have each what is peculiar in purpose and in construction.''}
   \label{fig:freytag pyramid}
\end{figure}

In the illustration (see figure \ref{fig:freytag pyramid}), apart from the five parts labeled by letters, he further introduces three structural moments:

\begin{quote}
Between them stand three important scenic effects, through which the parts are separated as well as bound together. Of these three dramatic moments, or crises, one which indicates the beginning of the stirring action, stands between the introduction and the rise; the second, the beginning of the counteraction, between the climax and the return; the third, which must rise once more before the catastrophe, between the return and the catastrophe. They are called here the exciting moment or force, the tragic moment or force, and the moment or force of the last suspense. The operation of the first is necessary to every play; the second and third are good but not indispensable accessories.\autocite[115]{freytag1895technique}
\end{quote}

In sum, Freytag's dramatic model consists of eight constructs: five parts in addition to three structural moments. Given the dramatic model in the context of our discussion, a natural question to ask is: where does temporal asymmetry arise? Similar to the discussion of equilibrium-disequilibrium-equilibrium above, the schematic description as a simplification appears symmetrical visually, such that the third part (namely, climax) stands as the midpoint of the plot. However, the symmetry exists only in its visually simplified form. If we consider the dramatic model (as form) in relation to its content (or in Schenkerian-style terminology, elaboration), we are able to discover temporal asymmetry at a more specific level of organization (or in Schenkerian-style terminology, foreground level). Now, the three structural moments (or crises in Freytag's terminology) concern more specific dramatic structures than the five parts, and the reason is intuitive: the five parts concern portions of the drama spanning large timespans, whereas the three crises are dramatic moments concerning instantaneous timepoints. Therefore, the five parts govern larger (in terms of timespans) dramatic structures than the three crises (in terms of timepoints, which can be interpreted as zero-length timespans).

\begin{figure}[!h]
   \centering
   \includegraphics[width=.8\textwidth]{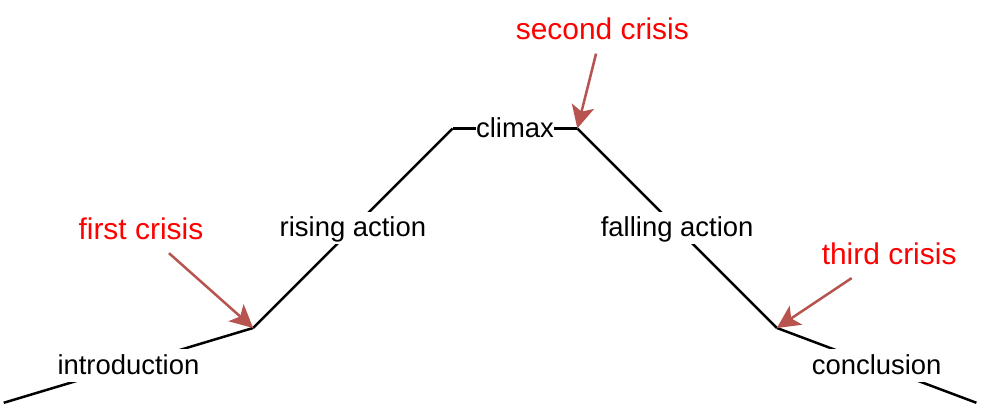}
   \caption{A reproduction of figure \ref{fig:freytag pyramid}, annotated in red to highlight the placement of the three structural moments}
   \label{fig:freytag pyramid reproduced}
\end{figure}

As our focus shifts to the three crises, we notice that their distribution within the five parts is uneven: between the beginning and the climax there is only one moment of crisis, whereas the other two moments of crisis are placed in between the climax and the ending (see figure \ref{fig:freytag pyramid reproduced}). As we examine the placement of the three crisis, we observe that the asymmetry is solely due to one missing piece: the moment between the rising action and the climax. In other words, by adding the moment between the rising action and the climax as the fourth crisis, we are able to recover a symmetrical scheme: two crises before the climax, corresponding to two crises after the climax.

\begin{figure}[!h]
   \centering
   \includegraphics[width=.8\textwidth]{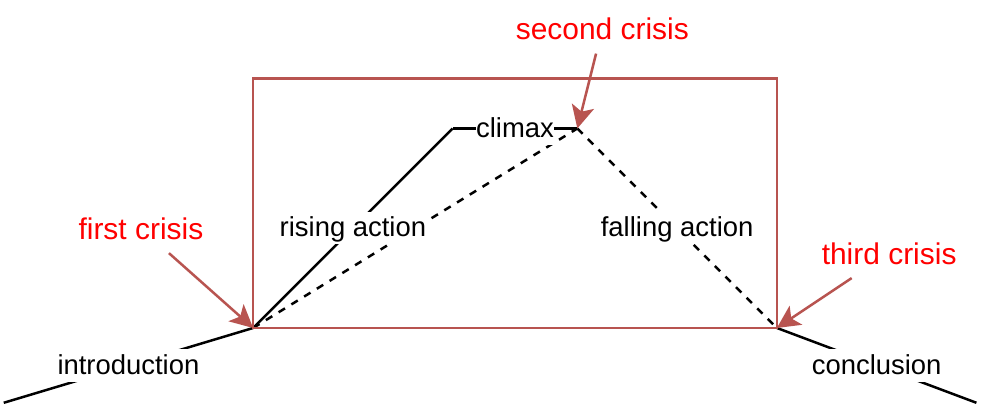}
   \caption{Substructure of a five-part drama delineated by three crises, substructure enclosed by red rectangle, whose narrative is depicted by dotted lines}
   \label{fig:pyramidal substructure}
\end{figure}

We can already observe temporal asymmetry if we consider the substructure delineated by the three crises alone. By substructure, we mean that the story begins with the first crisis and ends on the third crisis (see figure \ref{fig:pyramidal substructure}). In other words, the story is to be read without the introduction or the conclusion. Musicians can imagine analogous case where a piece is performed with the introduction and the coda (if applicable) omitted. Even though the piece is by definition no longer complete, we are still able to capture essential features of the piece in the remaining parts. Now, such substructure when considered on its own, exhibits temporal asymmetry similar to that depicted in figure \ref{fig:temporal asymmetry as displacement}: a skewed shape as a result of delaying the climax.

\begin{figure}[!h]
   \centering
   \includegraphics[width=.9\textwidth]{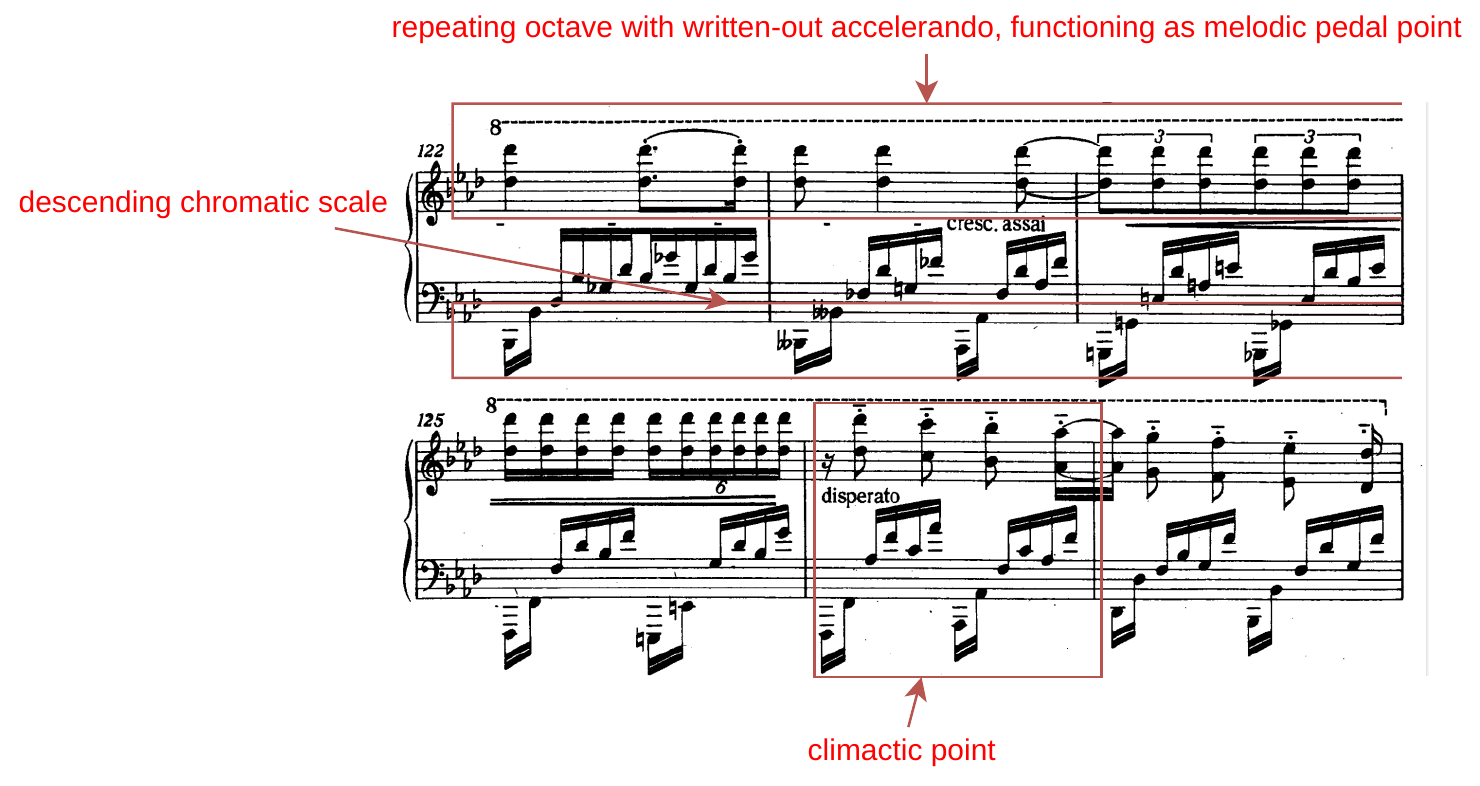}
   \caption{Liszt, Transcendental Etudes No.~10, S.139, excerpt, annotated to highlight the climactic point and its preparation}
   \label{fig:liszt etude 10 climax}
\end{figure}

Why would there be a missing crisis whose presence can conveniently achieve a symmetric description of the dramatic structure? One potential answer is an intuitive one: because the arrival of the climax admits no possibility for any scenic effect. The sheer dramatic force at the climax denies any form of flowery elaborations: it must continue the course of the rising action in the most direct and powerful manner, with whose raw dramatic power any scenic elaboration would be considered distractions. Performers are familiar with this idea that climax must be un-elaborated: in order to bring out the utmost climactic point of the piece, one must refrain from detailed nuanced interpretation in detail. For example, consider Liszt's transcendental etude no. 10. Among the 12 transcendental etudes, no. 10 stands out as it contains no programmatic title, therefore we may claim that the piece makes up a purely musical narrative. Now, its climax is located unambiguously: a series of repeating octaves with written-in accelerando and the semitone descent in the left hand musically cry out the climax on measure 126 (see figure \ref{fig:liszt etude 10 climax}), as if the repeating octaves in the right hand are performers pounding on the blackboard to tell the audience: ``wake up folks! You do not want to sleep through the upcoming passage''.

For performers, one of the effective strategies is to arrive at the climax on measure 126 in a straightforward manner, interpreting the crescendo, written-in octave repetition and descending semitone in their simplest and most literal form, thus conveying a sense of uninterrupted momentum. It would be utterly absurd to introduce nuanced timing or dynamics (for example, rubato in the right hand or a series of half-measure long crescendos in the left hand) with the hope to achieve somewhat magical effects. In other words, attempts to express any scenic effect before the climax often leads to disappointment in its literal sense: a ruined performance. Listeners are disappointed by the fact that the performance of the climax does not meet listeners' expectation, as the climax is the one of a few places one cannot use tricks such as evading (as in evading cadence) anticipated arrivals.

The three crises in the dramatic structure all convey a sense of dramatic turning, such that the contour of the plot changes. For example, the first crisis (namely, the exciting force) occurs at ``a point where, in the soul of the hero, there arises a feeling or volition which becomes the occasion of what follows; or where the counter-play resolves to use its lever to set the hero in motion''.\autocite[121]{freytag1895technique} Therefore, contour of the plot changes from an uneventful state of life to an agitated one (albeit initially small in degree). The second crisis (namely, the tragic force) presents a reversal of the plot's contour, where the ``tragic force, the banishment, begins here; what seems about to become the highest elevation of the hero, becomes by his untamable pride just the opposite; he is overthrown''.\autocite[131]{freytag1895technique} Finally, the third crisis (namely, the force of the final suspense) is another twist of the plot that reverses, albeit falsely, the contour of the plot. It is a moment in which:

\begin{quote}
Just at the time when the weight of an evil destiny has already long burdened the hero, for whom the active sympathy of the audience is hoping relief, although rational consideration makes the inherent necessity of his destruction very evident, -- in such a case, it is an old, unpretentious poetic device, to give the audience for a few moments a prospect of relief. This is done by means of a new, slight suspense; a slight hindrance, a distant possibility of a happy release, is thrown in the way of the already indicated direction of the end.\autocite[136]{freytag1895technique}
\end{quote}

As discussed above, would be inappropriate to place the fourth crisis between the rising action and the climax because it does not contain any dramatic turning points. On the contrary, the moment between the rising action and the climax requires the forceful continuation of the rising action which asserts the existing contour of plot, instead of reversing its course.

The unequal distribution of the three crises among the five parts puts a heavy emphasis on the second half of the plot since it is more eventful that the first half: containing two structural moments of crises instead of one. To this end, we obtain a kind of temporal asymmetry without referring to specific temporal proportions: even with the assumption that the climax part lies at the midpoint dividing the plot into two equal halves, we can still conclude that the second half is dramatically more important. In particular, the combination of the climax (the midpoint of the play) and the tragic force (in the second half) constitutes the dramatic highlight of the entire play:

\begin{quote}
In the case where the climax is connected with the downward movement by a tragic force, the structure of the drama presents something peculiar, through the juxtaposition of two important passages which stand in sharp contrast with each other. \autocite[130]{freytag1895technique}
\end{quote}

Consequently, the climax along with the tragic force that follows it ``gives to the drama with tragic force a magnitude and expanse of the middle part, which -- if the playful comparison of the lines may be carried out, -- changes the pyramidal form into one with a double apex''.\autocite[132]{freytag1895technique} The implied temporal asymmetry is increasingly obvious by now.

Firstly, why is there no counterpart to the combination of the climax and the tragic force? Could there be some analogous combination of the climax and something that precedes the climax? As discussed above, such grouping is nonsensical because there is no structural moment preceding the climax that is adjacent to it (the exciting force is between the introduction and the rising movement). In other words, there is no time-reversal equivalent to the climax/tragic force combination, which is exactly the condition of temporal asymmetry discussed in section \ref{sec:distinguishability}: dramatic constructs behave differently under different temporal context. In this case, the tragic force behaves differently from its counterpart (which is the hypothetical dramatic moment between the rising action and the climax) because its counterpart does not exist at all.

\begin{figure}[!h]
   \centering
   \includegraphics[width=.8\textwidth]{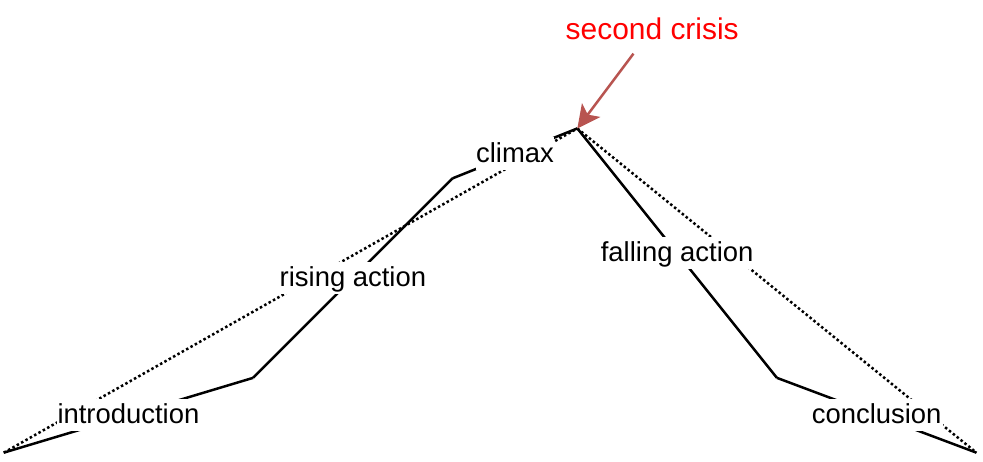}
   \caption{The pyramidal structure transformed into a double-apex structure by putting emphasis on the second crisis, as a result, the underlying structure (notated in dotted lines) is skewed, achieving similar effects to delaying the climax}
   \label{fig:pyramidal structure double apex}
\end{figure}

Secondly, in the case of ``double apex'' mentioned by Freytag, what is the implication? One interpretation is that the tragic force is of equal importance to the climax. As such, we may modify figure \ref{fig:freytag pyramid reproduced} such that it gives the tragic force its own apex as well as turning the climax into a continuously intensifying part attaining ``magnitude and expanse''\autocite[132]{freytag1895technique} (see figure \ref{fig:pyramidal structure double apex}). Consequently, the underlying shape of the dramatic structure is skewed (see dotted lines in figure \ref{fig:pyramidal structure double apex}) because we take into consideration the second crisis (i.e.~tragic force), which invariably alters the dramatic structure, achieving similar effects to delaying the climax as in figure \ref{fig:temporal asymmetry as displacement}.

\part*{Interlude}
\addcontentsline{toc}{part}{Interlude}

\chapter{Organicism}
\label{chapter:organicism}

\epigraph{Running brooks, poetry, and music reflect our exuberant life, above which the spectre of time hovers}

Music and the arrow of time meet in the middle: the organic and lively quality of life.

Readers should notice that throughout the paper, the discussion of time (temporal asymmetry in particular) is often accompanied by a certain lively quality. For example, in section \ref{sec:narrative as foundation}, we discussed the consequence of using narrative as a metaphor for music: music can be interpreted as a journey. One of the unique features of journey is that it contains a sense of proactive force: the protagonist embarking on the journey proactively experience and participate in the journey. Therefore we are entitled to call a series of event a journey instead of mere happenings: the additional property of a journey that differentiates itself from mere happenings as in planetary motion is the very proactive force, signaling the presence of life behind events.

Metaphorically, the lively quality is spectre-like in our discussion: it appears from time to time, as if it constantly lingers, ready to make its appearance. Like in a typical horror movie, the spectre appears constantly yet randomly, such that the protagonist encounters it at unexpected places: in recurring nightmares, under the table, in the closet, in the mirror, and most importantly, in the heart of the audience.

It is then natural to ask: why does organicness make its frequent appearance in our discussion of temporal asymmetry in music? The reason is that it is native to the discussion of both time and music. The term ``native'' means the following. On the one hand, without explicitly invoking the notion of temporal asymmetry, organicism is discussed in music on an everyday basis. On the other hand, in a conversation concerning the arrow of time in physics, without mentioning music at all, we are also justified to discuss the relationship between the arrow of time and life. Therefore, in order to better understand the topic of temporal asymmetry as it relates to music, it is helpful to survey a more complete picture: the pairwise relationships between music, arrow of time, and life.

Both music and the arrow of time are intimately related to the concept of life. More precisely, the arrow of time gives rise to essential features of both music and life. For music, the arrow of time governs the course of its narrative, in such a way that our psychological and emotional states fluctuate accordingly. For life, the arrow of time governs the life cycle of every organism, in such a way that it gives rise to birth, death, and resurrection. Poetically, we may equally say that the life cycle of organism is nothing but another instance of narrative, just as we previously claimed that the history of the universe is also a narrative (see page \pageref{prop:history as narrative}). A probably more interestingly and frequently discussed topic is the third pair: the relationship between music and life.

\section{Organicism in music}

Unlike the arrow of time, the term organicism has been extensively discussed in the field of music. Consequently, upon hearing the word, we construe it as a notion native to music instead of a term borrowed from other fields. It is universally acknowledged that music should be organic, yet what constitutes the organicness of music is where meaningful debates and discourses in music unfold (no pun intended).

Arguably, performers are most familiar with and sensitive to the organic quality in music due to the nature of stage performance. The act of performing on stage depends so crucially on the lively quality of music without which the performance becomes indifferent to a MIDI rendition of notes. This is not to say that composers, analysts and theorists are less aware of the organic quality of music. However, one must notice again the potential pitfalls entailed by notation: as composers, analysts and theorists must approach music \emph{through} notated symbols, one is in danger of becoming preoccupied by symbolic manipulations and failing to recognize the immediate emotional impact of music. For example, in harmonic analysis, upon repeated labeling of notated symbols using terms such as ``dominant/predominant/tonic'', we are in danger of forgetting the very emotional effects behind those terms, such that emotionally moving excerpts causing tension and release in our heart are reduced to repeated occurrences of annotations in terms of the pattern ``PD-D-T''. For performers, because there is no notation standing in between the music and the performer, the performer is in direct contact with the emotional impact of the music. It is true that performers might become desensitized of the emotional impact due to repeated practice. However, the sounding of a dominant-tonic resolution can never be reduced to anything but the very feeling of tension and release.

For performers, what exactly is considered organic in music? There is no universally acknowledged answer, not because it is a vague and ill-defined concept, but because it is all-encompassing. For performers, the organic quality is discussed on an everyday basis. We may even claim that the \emph{only} criterion for good performance is to make the music come to life. Such criterion entails all conditions required for good performance: technical proficiency, musical sincerity, naturality of phrasing and breathing, nuanced timing and dynamics, and so on. While presenting the concept of organicness as the all-encompassing quality characterizing musical beauty is certainly poetic, it nevertheless informs us little of what organicism in music is practically about. It is true that one can invoke statement in the style of theology (see section \ref{section:theology}), that the all-encompassing power of the organic quality in music is too great to be comprehended by human rationality. Therefore, the statement continues, the best way to comprehend organicness in music is through spirituality: close your eyes, and let your imagination and tender heart guide you in finding the liveliness of music. In short, organicness in music may be considered resistant to verbal description. It is certainly an artistically valid statement, as verbal description, being another form of notated symbols, is inherently limiting insofar as stage performance is concerned.

Among music scholars and composers whose analytical tradition necessarily involves notated symbols, the notion of organicness is more varying and specific. Here, we enumerate four aspects of organicism in music. Due to the vast number of interpretations, the enumeration here is deemed incomplete. However, it aims to capture properties essential for music to be considered organic, thus potentially providing a better understanding of the term:

\begin{itemize}
\item
  \textbf{Organicist condition 1 (growth)}: the process of music unfolding in time should be considered analogous to the biological growth of a living organism.
\item
  \textbf{Organicist condition 2 (purposiveness)}: the process of music unfolding in time should be goal-oriented, thereby fulfilling certain purposes.
\item
  \textbf{Organicist condition 3 (holism)}: the process of music unfolding in time should be considered an organic unity, which cannot be reduced to sum of its parts.
\item
  \textbf{Organicist condition 4 (motion)}: the process of music unfolding in time should be considered as expressing some kind of motion.
\end{itemize}

Before discussing each condition in detail, we should observe that all four conditions are in close relation to temporal asymmetry: growth, purposiveness and motion all entail a sense of directionality and tendency in time. Specifically, when we discuss growth in its biological sense, the growing object grows \emph{into} something; when a process is purposive, then it is goal-oriented to \emph{arrive at} some purpose-fulfilling events; finally, the directionality of motion is self-evident. It remains to show that the holism condition relates to temporal asymmetry, which will be discussed in section \ref{section:music as organic whole}. It is true that the four conditions are selected partly because they can echo the theme of this project. However, this does not mean that the four selected conditions are ad-hoc. Akin to the discussion of narrative: we pick tension-release as the generalized definition of narrative, firstly because it is conveniently a musician's concept. Secondly, nevertheless, it is, regardless of its etymological origin and the author's subjective bias, effective in characterizing narrative from a more universal and psychological perspective.

\subsection{Music as organic growth}
\label{section:music as growth}

The metaphor of biological growth is arguably the most intuitive interpretation of organicism in music: organicism as organism-like, with biological growth and reproduction as two evident features of organisms. Therefore, by analogy, we interpret a piece of music in terms of its narrative as a process of organic growth. The word unfolding is particularly useful metaphorically: a motive can be interpreted as an embryo or a seed, from which the leaves and flowers are unfolded (literally, as flowers un-fold when they blossom) and expanded from this single seed.

Biologically, one of the most outstanding features of organic growth is the singleness of genetic encoding. The entirety of an organism is developed based on a single DNA serving as the grand blueprint for the embryonic development. The variety of organs in an organism is nothing but different expressions of the same DNA, through the magical process called cell specialization by which undifferentiated prototypical stem cells are specialized and turned into cells needed for individual parts: heart, muscles, bones, and so on.

Consequently, for organism, we have a peculiarly counterintuitive observation: the part is at the same time contained in the whole and containing the whole. Therefore, given any cell, we can infer the following two statements: 1) it is contained in some part of the organism; 2) it contains the DNA which can generate the entire organism.

The two statements above precisely echo Schoenberg's remark on motive:

\begin{quote}
{[}The motive{]} includes elements, at least, of every subsequent musical figure, one could consider it the `smallest common multiple'. And since it is included in every subsequent figure, it could be considered the `greatest common factor'.\autocite[9]{schoenberg1970}
\end{quote}

Without invoking the biological metaphor, we might be perplexed by Schoenberg's remark as it uses vague terms such as smallest common multiple and greatest common factor whose mathematical meaning does not quite apply here. However, with the two statements given above, we can immediately see the correspondence (hence explanation):

\begin{itemize}
\item
  ``{[}The motive{]} includes elements, at least, of every subsequent musical figure'': the motive is like DNA, which can generate all subsequent cells of an organism. Therefore, the part contains the whole.
\item
  ``it is included in every subsequent figure'': the motive is like DNA, whose exact copy is found in every subsequent cell it generates. Therefore, the whole contains the part.
\end{itemize}

As Schoenberg's remark illustrates, the concept of organic growth is one of the justifications for motivic development. The motive as the atomic musical unit is ``often considered the `germ' of the idea'',\autocite[8]{schoenberg1970} from which the piece grows and develops to the extent that ``almost every figure within a piece reveals some relationship to it''.\autocite[8]{schoenberg1970}

Such idea is not unique to Schoenberg. As Cook remarks in his ``epistemologies of music theory'', the ``aesthetic model of `unity in diversity'"\autocite[83]{cook2002epistemologies} is not limited to Schoenberg's theoretical thinking:

\begin{quote}
{[}It{]} is remarkable how many twentieth-century music theorists specifically refer to {[}the aesthetic model of ``unity in diversity''{]} -- among them not only Schoenberg's followers (Keller and Walker as well as Réti) but also Schenker, as most notably expressed in the motto ``semper idem sed non eodem modo'' (always the same, but not in the same way) displayed between divisions in the second volume of \emph{Kontrapunkt} and on the title page of \emph{Der freie Satz}.\autocite[83]{cook2002epistemologies}
\end{quote}

For composers and performers alike, one of the universally acknowledged criteria for good artwork (composition and performance) is the idea of organic unity: something (which does not have to be about motives) has to somehow unify the work, giving listeners the idea that this is a holistic piece instead of a collage of excerpts. Although arguably a considerable amount of trends challenge the idea of organic unity, such as the Romantic trends which put emphasis on fragmentation and miniatures, the idea of organic unity remains largely unchallenged among performers. For performers, even when performing a collection of miniatures, the performer has to deliberately come up with some unifying theme that can present the concert as a whole. For example, the unifying theme can be the keyword ``miniature'' itself, such that when listeners ask ``why do you program your concert this way?'', the performer can use the keyword ``miniature'' as the germinating seed giving rise to the organic unity of the concert: ``because this concert is motivated by the concept of miniature as a germinating seed, from which every piece in the program is related to form an organic whole''. It would be absurd for the performer to answer that ``I program this way because I haphazardly feel like doing so'' (unless the concert is about haphazardness, e.g.~chance music): in addition to the issue of attitude, the performer fails to construe his concert as an organic whole.

In addition to Schoenberg, many authors propose similar ideas, which can be viewed as ways to paraphrase the same idea: music unfolds as embryo develops. Therefore, organicism is not strictly a particular school of thought, but recurring trend that makes its frequent appearance through history. In short, according to the ``Cambridge history of Western music theory'', organicist view can be concisely summarized as follows: music concerns itself with ``processes of germination and growth across the piece'',\autocite[933]{cambridge_history_of_theory} whose growth is like the organic growth ``from seed to harvest''.\autocite[821]{cambridge_history_of_theory}

\subsection{Music as fulfillment of purpose}
\label{section:music as fulfillment of purpose}

The idea of purposiveness is more metaphorical than organic growth. For organic growth, we are able to find evidence of growth from the material of scores and sounds using pattern matching: parallelism in pitches, rhythmic patterns and intervallic relations serves as evidence of organic growth. On the contrary, when we say that a music exhibits purpose, we do not mean that we can find purpose in music, namely, a sequence of events. But rather, we mean that we find musical events that are somehow reflecting our own purpose, which is a uniquely human attribute. In other words, the notion of purpose is more symbolic in music: when we say that the cadence represents the process of fulfilling purposes, we somehow symbolically use harmonic resolution to represent fulfillment of purposes. Therefore, what do we mean when say that a piece of music is human-like, such that it can somehow fulfill purposes?

The crux of the matter lies in the notion of goal-oriented behavior. Before continuing the discussion, we must recognize that the term ``goal-oriented'' is a concept completely native to music: no one would be surprised if such term is uttered in a music class (whereas the utterance of ``entropy'' is more surprising). For example, composers and performers alike talk about goal-oriented phrases, such that one should perform in a way that the music moves \emph{toward} the arrival point, namely, the goal of the phrase. One should be perplexed by the use of terms ``arrival point'' and ``goal'': they possess purposiveness. In particular, arrival point and goal seem to suggest that a particular moment, rather than other moments, in music is treated as destination. There is a hidden preference here. When we say that the last note of the phrase is the destination, we are preferring it over others, thus making a deliberate choice. We are less likely to say that ``the phrase should arrive at the passing tone'' because we feel that the passing tone serves as means to achieve some ends. Now, the identification of means and ends exhibits purposiveness. In short, only through purposiveness, are we entitled to utter terms such as ``goal'', ``arrival point'' and ``destination''. Without us imposing purposes, every moment in the music can technically be considered an arrival point because, by definition, our performance arrives at (i.e.~visits) every moment in the music.

As a side note, in terms of historical context, the discussion of goal-oriented behavior in music may partly have been influenced by contemporary trends in other disciplines. For example, we may speculate that, it is partly due to the influence of Hegelian teleology that music scholars in the 19th century are accustomed to the mindset of imposing purpose to whatever topic is at hand, which ranges from the structure of a single piece, to the structure of the entire music history. For example, Christensen makes the following remark about the Hegelian influence in music theory:

\begin{quote}
Fétis attempted to chronicle the evolution of harmonic thought culminating in his own formulation of tonalité. Inspired by Hegel's philosophy of history, Fétis saw music theorists as vessels of an emerging tonal consciousness scrolling across time, and he was therefore not slow to either praise or censure any given writer depending upon how closely the writer was able to give voice to this tonal spirit.\autocite[14]{cambridge_history_of_theory}
\end{quote}

One of the characteristic features of Hegelian teleology is that, akin to Hobbe's conception of the Leviathan, communities as collection of individuals are anthropomorphized to form a collective consciousness, thereby making the history of the world goal-driven: history evolves in order to fulfill some purposive \emph{teleo}, as if some personified force is behind the evolution of history. Therefore, it would be more than natural to describe a piece of music as a process fulfilling its purposive \emph{teleo} as well: the entirety of music history can be viewed as the collection of all compositional works achieving some \emph{teleo}; inversely, a single work of music can be seen as an instance of history attaining its own development and \emph{teleo}.

Meanwhile, instead of attributing the present-day discussion of goal-oriented behavior in music to the historical legacy of 19th century thought, we must recognize its ahistorical aspect. In other words, goal-oriented behavior is intrinsic to life, in particular, intelligent life. For example, in artificial intelligence, one of the objectives is to artificially replicate (intelligent) life. Therefore, researchers must capture essential features of intelligent life. One of the fundamental questions is to differentiate between intelligent agent and lifeless object. For example, what is the difference between a person and an air conditioner? Notice first that they both react to the environment, such that it seems insufficient to say that responding to environment alone is the essential feature of life. We need something more deliberate and more proactive: pursuing goals in fulfillment of purposes. In the book ``Developing intelligent agent systems'', the authors distinguish between agent and object as follows:

\begin{quote}
Another key property of agents is that they pursue goals over time, that is, they are proactive. {[}\ldots{]} Although objects can be reactive, and can be seen as having an implicit goal, they are not proactive in the sense of having multiple goals, and of these goals being explicit and persistent. Thus, proactiveness is another property that distinguishes agents from objects. \autocite[2]{padgham2005developing}
\end{quote}

As a side note, note how the use of the term ``proactive'' echoes our discussion of the proactive force essential to music (see section \ref{sec:narrative as foundation}). Through goal-oriented behavior, agents exhibit qualities resembling intelligent life. Thus, in claiming that a piece of music is organic, we partly mean that the piece, analogous to intelligent agents, fulfills goals as the music unfolds in time.

\subsection{Music as organic whole}
\label{section:music as organic whole}

Construing music as an organic whole is a restatement of the slogan ``the whole is more than the sum of its parts''.\autocite{sep-physics-holism} Equivalently, we say that all parts within a piece of music function like ``organs''\autocite[895]{cambridge_history_of_theory} in an organism, through whose cooperation and inter-relatedness the piece is to be construed as a whole in the holist sense: the whole cannot be understood by individually inspecting its parts.

Musically, we listen to a piece of music and appreciate its beauty. We ask ``where is the beauty residing''? We then take out a magnifier and look at every measure of the piece, but cannot find beauty in \emph{any} of the measures. To our dismay, we look into every single note, but cannot find beauty in \emph{any} of the notes either.

\begin{figure}[!h]
   \centering
   \includegraphics[width=.8\textwidth]{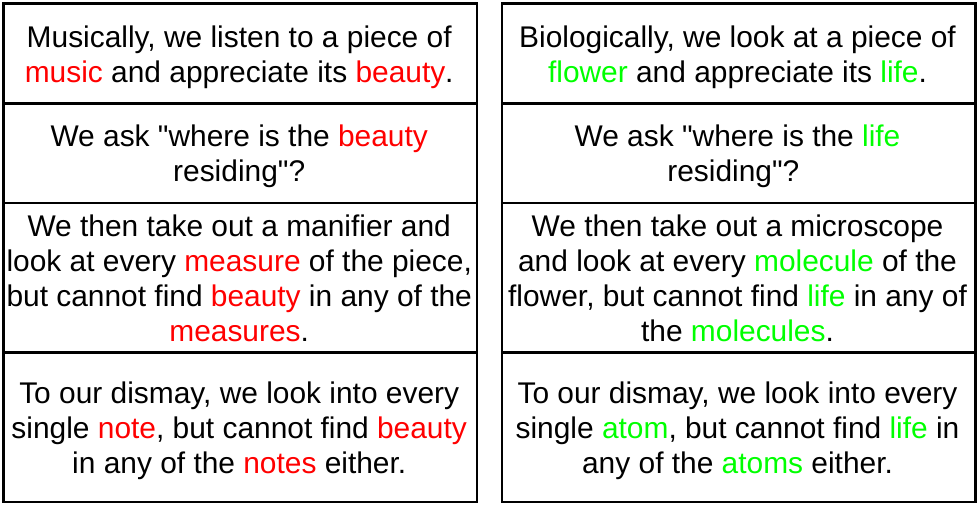}
   \caption{A side by side comparison highlighting word substitution}
   \label{fig:music holism vs biological holism}
\end{figure}

The example above concerning music can be translated to its exact analogue in biology, word for word (see figure \ref{fig:music holism vs biological holism}):

\begin{quote}
Biologically, we look at a piece of flower and appreciate its life. We ask ``where is the life residing''? We then take out a microscope (or something more powerful) and look at every molecule of the flower, but cannot find trace of life in \emph{any} of the molecules. To our dismay, we look into every single atom, but cannot find trace of life in \emph{any} of the atoms either.
\end{quote}

The crux of the matter is that we replace the term ``note'' by ``atom'', and ``beauty'' by ``life''. Moreover, we feel that such analogy is natural, in the sense that we do consider individual notes as atomic (variant of atom), and beauty as lively (variant of life).

Zuckerkandl makes similar observations concerning holism in music:

\begin{quote}
A melody is a whole-and a temporal whole, a whole whose parts are given as a sequence, as temporal succession. Such a whole can, in general, be of two kinds. Either it is a sum of parts, which are successively added to one another, an and-sum, as it is termed, the result of simple addition; or it is ``the whole which is more than the sum of its parts,'' a Gestalt. If then listeners each hear one tone, the totality of their sensations is an and-sum; if one listener hears ten tones, the totality of his sensations is a Gestalt - a melody. \autocite[229]{zuckerkandl_1973}
\end{quote}

From here we are able to see how the arrow of time makes its appearance again in the holistic property of music: it is only through the unfolding of music in time, that we are able to mentally synthesize its parts to form a whole, and construe a piece of music holistically instead of producing, in Zuckerkandl's words, an and-sum. Specifically, the holism of music arises from the difference in the epistemic access of the past and the future, between memory and anticipation. In fact, Zuckerkandl spells this point out explicitly:

\begin{quote}
Temporal Gestalten appear possible because the past instant can be held in memory; the future instant can be anticipated in expectation.\autocite[230]{zuckerkandl_1973}
\end{quote}

\subsection{Music as motion}

The dynamic quality of music is readily felt. We intuitively understand music as a dynamic process involving constant motion and energetic force. For example, one of the (if not the) most frequently uttered teachings in music lessons is to make music move forward. However, this moving forward is not the same as simply rushing. It is for such reason that performers must consciously shape the phrases, plan the dynamics, and tune nuanced timings agogically in order to move the music forward. Additionally, as discussed in section \ref{section:music as fulfillment of purpose}, the motion is goal-oriented such that it moves \emph{toward} some point of arrival: it can be the crucial moment of a phrase, the high point of a hairpin, or the climax of the entire piece. Therefore, music is about a special type of motion: musical motion.

In addition to the acoustic vibration which physically moves the air, we feel that the motion is more than that of mere sound waves. After all, the motion of sound waves is also found in noise and all non-music sound. Music has to be about some additional motion, some motion that is more vital. The term vital here is to be understood as in vitalism. Again, we infer that music is about a special type of motion called musical motion.

However, we have not defined musical motion at all. So far, we infer musical motion negatively, noticing that it cannot be simple motion because for performers motion in music is not the same as simply rushing, and for listeners motion in music is not the same as motion of sound waves. It turns out that we can only construe musical motion from an organicist view, i.e.~it is the type of motion akin to the vital energy of life. Therefore upon listening to music and its motion, we are reminded of the vital energy of life (through which we strive for fulfillment of purposes). It is for the reason that musical motion reminds us of the essence of life, so that we are entitled to say that we are moved by music.

Notice that the usage of the word ``move'' already entails musical motion discussed above. Additionally, the statement that music has the ability to move people is not simply a convenient metaphor. Instead, we may argue that its root can be traced back to the Baroque theory of the affects (equivalently, affections). One of the characteristics of the Baroque interpretation of the affects is that affects are objectified emotions in the following sense: emotions are categorized into affects such as anger, fear, sorrow, joy, etc. For example, the affect of anger describes the general state of being angry. Therefore, the affect of anger can be interpreted as an object to be manipulated and in particular, moved (in its literal sense). For example, in a Baroque suite, each movement may be dedicated to an affect, such that the movement arouses its corresponding affect. Because affects are considered as an object (i.e.~we can talk about \emph{the} affect of fear universally instead of my personal fear or your personal fear), the metaphor of motion becomes surprisingly appropriate and even literal: music as motion can move the affects as objects just as physical motion can move physical objects. Therefore, the idea of moving the affects is less metaphorical than physical. For example, in the ``A history of Western music'', the authors mentions Descartes' idea of the animal spirit:

\begin{quote}
The affections were thought of as relatively stable states of the soul, each caused by a certain combination of spirits, or ``humors,'' in the body. According to René Descartes, once these spirits were set in motion by external stimuli through the senses, they conveyed their motions to the soul, thus bringing about specific emotions.\autocite{norton_history}
\end{quote}

Consequently, music as motion, from the perspective of the Baroque theory on affects, is to be interpreted literally.

Another school of thought which puts paramount emphasis on musical motion is energeticism. That it is considered as ``another'' school of thought instead of a continuation of, for example, the Baroque theory of affects is partly due to an observation made by Rothfarb in his article ``Energetics'' surveying energeticism:

\begin{quote}
As a rubric for music-theoretical literature focused on music's dynamic qualities, ``energetics'' is unrestrictively broad in scope on the one hand, and restrictively narrow on the other. It is broad because ever since ancient times authors have identified motion as a fundamental aspect of music, and narrow because specific references to ``energy'' in music, or analogies with force, power, or similar concepts from the domain of physics, are historically limited, appearing first with regularity in the decades straddling 1900.\autocite[927]{rothfarb2002energetics}
\end{quote}

Among the proponents of energeticism in its narrow sense indicated by Rothfarb, Kurth can be considered as exemplary in illustrating the concept of energeticism. In her article about Kurth bearing the very title ``Ernst Kurth and his concept of music as motion'', Hsu observes one of Kurth's conception of motion, quoting his words:

\begin{quote}
What one calls the soul of an art, can never consist of rigidly established external forms, but can only be felt as a flowing state; it cannot be thought of as a substance to be delineated, but rather as direction, motion, striving, as impulse and will, that merely erupt in various ways into a wealth of external forms.\autocite[12]{hsu1966ernst}
\end{quote}

We should note how Kurth's energeticist perspective echoes many ideas appearing in this project (or vice versa). Firstly, as noticed in the beginning of this section, the musical motion energeticists address is not to be understood in its simple physical sense. But rather, it is necessarily accompanied by life-related attributes: volition, purposiveness, and goal-oriented behavior. Secondly, these life-related ideas make musical motion particularly relevant to the discussion of section \ref{section:music as fulfillment of purpose}, to the extent that we may even claim that music as motion is the same as music as fulfillment of purposes. Consequently, the discussion in this section is simply a restatement of section \ref{section:music as fulfillment of purpose}. Finally, at a larger scope, we should notice how music as motion shifts our attention to the psychological aspect of music, which is the central thesis of section \ref{section:what should narrative of music be about}.

As a result of the similarities between energeticist perspective on musical motion and the discussion of temporal asymmetry in this project, we may argue that this project is advocating energeticism. Specifically, the arrow of time is the manifestation of musical motion and energetics, whose volitional power gives rise to the psychological arrow of time in music.

Consequently, the temporal asymmetry caused by delaying the climax as illustrated in figure \ref{fig:temporal asymmetry as displacement} can be interpreted literally: the arrow of time exerts a musical force as the carrier of our volitional power, through whose effect the climax is moved and pushed to the right side of the score, creating a temporal displacement we call musical motion. Moreover, such description is not merely metaphorical , as the musical force exerted by the arrow of time is as real as musical motion, to which Kurth ``did not intend the word {[}motion{]} in a figurative sense, but insisted on its original and literal meaning''.\autocite[12]{hsu1966ernst}

\section{Life in relation to the arrow of time}
\label{section:life in relation to the arrow of time}

Poetically, it is through the arrow of time that we appreciate the value of life. In fact, we may consider our reflection of life to be the motivation for discussing time as a topic in itself. In other words, in order to discuss the topic of life meaningfully, we must consider it through the lens of time, in particular, the arrow of time. The Britannica article on time points out that historically, our self-reflection may be the original motivation for the topic of time:

\begin{quote}
The irreversibility and inexorability of the passage of time is borne in on human beings by the fact of death. Unlike other living creatures, they know that their lives may be cut short at any moment and that, even if they attain the full expectation of human life, their growth is bound to be followed by eventual decay and, in due time, death.\autocite{britannica_time}
\end{quote}

It turns out that this intuition that the arrow of time is destructive to life is also captured by scientific inquiries, in particular, by the concept of entropy as a measure of orderliness. Recall that throughout section \ref{section:thermodynamic arrow of time}, we focus on one of the (if not the most) commonly observed phenomena: disorderly possibilities astronomically outnumber orderly possibilities. We can certainly claim that the reason is superficial: since we define what is considered orderly or disorderly, the conclusion that the world tends to favor disorder is simply a result of our bias. However, in the discussion of life which is itself the origin of all subjective biases, we are justified to base our argument on bias. In fact, bias is not only allowed, but necessary. Therefore, even when admitting that the notion of order and disorder is merely a social construct solely defined by our biased preference, we can nevertheless assert that disorder is unfavorable to life.

In other words, life aligns itself with the orderly aspect of the universe. Because orderly possibilities are outnumbered, physical phenomena which we can call life are also relatively rare. Moreover, because life aligns itself with order, from the perspective of life, the majority of the world appears hostile: the world seems to act in a way that is opposite to what we strive for. It is exactly due to the hostility (from the perspective of living organisms such as us who read this sentence) of the world that performers must diligently practice in order to maintain the high level of artistry and technique. Poetically, the hostility of the world is the reason that humanity must proactively strive for a better world through volition and hard work.

Given the hostile environment in which life arises, a meaningful question to ask is the following: what is the role of life? This question can be treated as a scientific question as well as an artistic one. Artistically, the hero in a tragedy must overcome sorrows, trials, misfortunes and conflicts, as Fate does not seem to bless the hero. Through compassion and sympathy, as audience, we momentarily become the hero and ask: what can I do and what is my role in the midst of this unfortunate circumstance? We may argue that such question is exactly what Hamlet means when he utters the most well-known line: ``To be, or not to be, that is the question''.\autocite[309]{shakespeare2014arden} We look for purpose and value which can justify and guide our striving, \emph{given} the context of the destructive universe.

Meanwhile, scientifically, we are also curious about the role of life, namely, organisms. In particular, we are curious about the characteristics of life such that we are able to identify life in theories of physics. It would be odd to have theories describing microscopic behavior of atoms and macroscopic behavior of planetary motions, yet leaving little place for living organisms which we value as more fundamental and real than atoms and celestial bodies. As part of his poetic endeavor, physicist Schrodinger (whose equation makes its appearance on page \pageref{equations of motion}) writes about his interpretation of the role of life in his book bearing the self-evident title ``What is life''. Being neither a professional biologist nor professional poet, he characterizes life using the physicist's idea of entropy.

Schrodinger defines life as entities feeding on ``negative entropy''.\autocite[70]{schrodinger1992life} By doing so, ``living matter evades the decay to equilibrium''.\autocite[69]{schrodinger1992life} Recall that entropy can be interpreted as a measure of the degree of disorder. The term equilibrium refers to the state of maximal entropy, i.e.~maximal degree of disorder. For living organisms, the state of maximal entropy is intuitive: death. On the one hand, thermodynamics tells us that Nature favors disorder, simply because the number of disorderly states is vastly bigger than that of orderly ones. In Schrodinger's words:

\begin{quote}
An isolated system or a system in a uniform environment {[}\ldots{]} increases its entropy and more or less rapidly approaches the inert state of maximum entropy. We now recognize this fundamental law of physics to be just the natural tendency of things to approach the chaotic state (the same tendency that the books of a library or the piles of papers and manuscripts on a writing desk display) unless we obviate it.\autocite[72]{schrodinger1992life}
\end{quote}

On the other hand, life is special in that living organisms feed on negative entropy, thus maintaining (or even increasing) orderliness. In his book, Schrodinger primarily addresses the biological aspect of orderliness as avoiding death: through biological metabolism we are able to maintain orderly arrangement of molecules within our body, maintaining the state of being alive for an extended period of time.\autocite[70]{schrodinger1992life} However, we can see how life as maintaining orderliness can also be applied to artistic endeavors.

We live a biological life, whose metabolism keeps us alive to fulfill life's purpose in Schrodinger's definition: feeding on negative entropy to avoid the decay of death. Meanwhile we live an artistic life, in which we create works of art in our image: organic entities that are as lively as ourselves (we should note that this statement is particularly true in the context of organicism). The potentially blasphemous words ``in our image'' (suggestive of Genesis 1:27 (KJV) ``So God created man in his own image'') lie in the heart of artistic creation, where the lively quality residing in us is passed to the work of art, thereby bringing it to life. In other words, we try to be gods in our own artistic micro-cosmos, in ways that can be poetically described using Hamlet's utterance:

\begin{quote}
I could be bounded in a nutshell and count myself a king of infinite space \autocite[305]{shakespeare2014arden}
\end{quote}

Therefore, Schrodinger's definition of life can be extended to include artistic endeavors: we maintain and create orderliness not only biologically in our bodies, but also in the artworks we produce. In particular, we arrange objects (strokes for visual artists, notes for composers, bodily movements for performers) into orderly patterns such that they are distinguished from random and disorderly patterns. To this end, Schrodinger's characterization of life in terms of entropy and orderliness becomes the very thesis for organicism. In Liu's poetic rendition of life as low-entropy entities, we can even sense that life and art refer to the same thing:

\begin{quote}
Entropy increased in the universe, and order decreased. The process was like the boundless wings of the giant balance bird pressing down upon all of existence. But low-entropy entities were different. The low-entropy entities decreased their entropy and increased their order, like columns of phosphorescence rising over the inky-dark sea. This was meaning, the highest meaning, higher than enjoyment. To maintain this meaning, low-entropy entities had to continue to exist.\autocite[467]{liu2016death}
\end{quote}

On the one hand, through the arrow of time, life is defined in terms of entropy. On the other hand, artistic creation reflects life's striving for orderliness. At this point, organicism becomes literal: artwork and life are one, unified as in Shelley's words:

\begin{quote}
\centering
A tone

Of some world far from ours

Where music and moonlight and feeling

Are one.\autocite[106]{hodgson2021cambridge}
\end{quote}

\part{Musical consequence}

\chapter{Recurrence}

Based on the distinguishability condition (see section \ref{sec:distinguishability}), the distinguishability between the past and the future as a result of the epistemic difference between memory and prediction (or artistically, anticipation) gives rise to temporal asymmetry in music. As discussed in section \ref{sec:distinguishability}, it particularly addresses the observation that two events with similar content may be interpreted as drastically different because they are placed at different temporal locations within a piece. For example, the equilibrium-disequilibrium-equilibrium paradigm discussed in section \ref{sec:distinguishability} differentiates between the two instances of equilibrium by its temporal placement relative to the disequilibrium part of the narrative. In fact, we may claim that the highlight of the equilibrium-disequilibrium-equilibrium paradigm is precisely the interplay between two transitions: from the initial equilibrium, the story is destabilized into disequilibrium; from the state of disequilibrium, the story restores its equilibrium in the end.

We should note that the concept of recurrence is implied here. When we speak of two events similar in content that are placed at different temporal locations, the similarity in content induces recurrence. In particular, two similar events in a piece necessarily implies that the second event later in temporal location is a re-occurrence of the first event. However, it is worth mentioning that there is a certain subtlety concerning the idea of recurrence. The subtlety lies in the observation that recurrence entails two layers of meaning. On the one hand, recurrence describes content, referring to parallelism between musical events such as repetition in music. On the other hand, recurrence describes the cognitive process in which we recall a past event from memory caused by the present moment, thereby causing recurrence of the past event: it occurs in both the past and the present. In fact, every time we recall from memory, we have recurrence of the past. However, while saying that recurrence is found every time we recall from memory is true, it is nevertheless too general to be useful for analytical purposes. Therefore, in this discussion, we focus on the first layer of the meaning of recurrence which is more specific: parallelism between musical events.

Even though in our discussion, the concept of recurrence is derived as an implication of temporal asymmetry, recurrence by itself is nevertheless one of the central topics in music. We may even claim that the idea of recurrence is one of the most powerful expressive tools in music for composers and performers alike. For example, we may interpret that in Schoenberg's view, the organic structure of a composition arises from the balanced interplay between variety and unity, providing ``variety in unity''.\autocite[21]{schoenberg1970} Just as introducing new musical material (such as the secondary theme in sonata-allegro form) is the primary way to promote variety, recurrence (i.e.~repetition) of existing musical material (such as the recapitulation in sonata-allegro form as modified repetition) is the primary way to achieve unity.

In fact, we may even argue that music itself, albeit being temporal, is fundamentally cyclical. As Lasser remarks:

\begin{quote}
The musical experience, being time-based, is also cyclical in nature. We build structure in our minds by recognizing returns of events. Meter, motives, harmonic progressions, large tonal structures are all premised on the basic concept of cyclical return and derive their meaning from this very concept.\autocite[4]{lasser2008}
\end{quote}

One of Lasser's main points is to notice that a symbol or an event in a narrative ``acquires meaning only as it returns over and again throughout a story''.\autocite[4]{lasser2008} In other words, temporal ``structures can only be created through cyclic return''.\autocite[4]{lasser2008} For example, imagine how analysis of formal structures is virtually impossible without relying on recurrence: it is only through the notion of recurrence that we are entitled to discuss concepts such as recapitulation in ABA form or reprise/ritornello in rondo form. Without invoking recurrence, \emph{all} musical forms would be reduced to meaningless labeling in the following schemes: AB, ABC, ABCD, ABCDEF, \ldots. In other words, recurrence is essential for musical coherence. In fact, we may argue that the unique feature of music is that it is a self-referential and non-representational art form, has its origin in recurrence. In particular, by definition of non-representational art, musical structure relies solely on the internal relations between its parts, such that a moment in a piece of music refers to other moments in the piece instead of objects in the external world (i.e.~reality). Now, the possibility of establishing internal relations arises largely from recurrence. Through musical events that make their returns throughout the piece, we are able to understand musical structure as internally organized. For example, we are able to claim that ``this moment relates to the beginning, where I heard \emph{the same} tune''. In short, a piece of music, when considered as narrative unfolding in time, is internally organized precisely because recurrent events, borrowing Lasser's idea, ``provide inner structure which can be independent from the plot''.\autocite[4]{lasser2008}

\begin{figure}[!h]
   \centering
   \includegraphics[width=.8\textwidth]{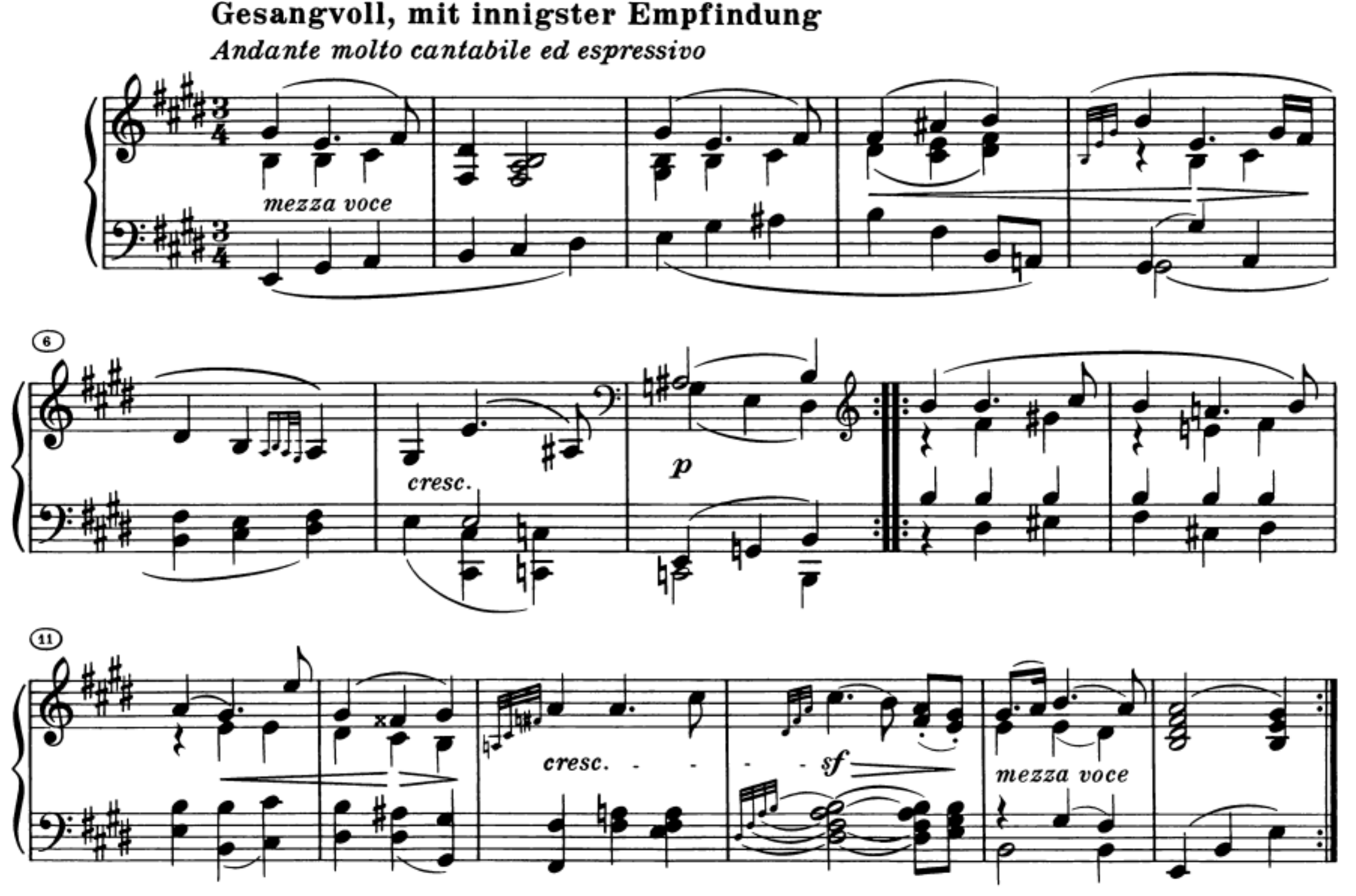}
   \caption{Beethoven, Piano Sonata No.~30, Op. 109, third movement, beginning}
   \label{fig:beethoven op109 mov3 beginning}
\end{figure}

\begin{figure}[!h]
   \centering
   \includegraphics[width=.8\textwidth]{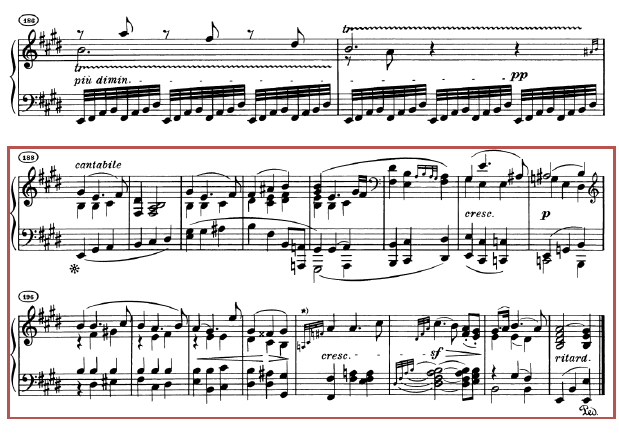}
   \caption{Beethoven, Piano Sonata No.~30, Op. 109, third movement, ending, annotated in red rectangle to identify restatement of the opening theme in figure \ref{fig:beethoven op109 mov3 beginning}}
   \label{fig:beethoven op109 mov3 ending}
\end{figure}

Meanwhile, the idea of recurrence is also crucial in bringing out expressive performance. For performers, one of the interpretive challenges is the treatment of repetition. The challenge arises due to the complicated nature of repetition. For example, consider the following dilemma. On the one hand, in general, it would be unmusical to play the repeated passage in exactly the same manner as its original occurrence. It is for this concern of variety that vocalists add improvisatory ornaments when the first section returns in a da capo aria, creating a modified return. However, on the other hand, sometimes it is equally unmusical to simply say ``do something differently with the repetition'' when it would arguably be preferable to, surprisingly, produce identical playing. For example, consider the third movement of Beethoven's piano sonata no. 30 (see figure \ref{fig:beethoven op109 mov3 beginning}). The movement is in variation form, but with one peculiar compositional feature: instead of ending the piece with the last variation, the opening theme is repeated exactly (see figure \ref{fig:beethoven op109 mov3 ending}).

In order to produce a justified performance, we need to answer the following question: what could be the reasons to justify this anomaly of exact restatement of the opening theme? To this end, we may resort to poetic interpretations: the restatement of the opening theme is symbolic of the idea of homecoming. As the music reaches the opening theme in the end, we metaphorically imagine that the weary wayfarer has finally returned home. We can further imagine that the homecoming of the wayfarer is more spiritual than physical: in fact, the wayfarer is already dead before he/she is able to make the return trip. Due to the lingering homesickness, his/her spirit revisits home before leaving this mortal world. In his/her spiritual return to home, he/she finds that everything at home appears familiar, but with a sense of quintessential quietness (which is fitting metaphor considering Beethoven's hearing issue), hinting at the unrealness of the homecoming.

How does this overly imaginative metaphor inform us about the interpretation? The key feature of the metaphor is that the spirit's homecoming should be identical to reality in details, to the point that the spirit does not even realize that he/she has died. Translating this feature to interpretive decisions, the metaphor suggests that the repetition should be carried out literally, such that the ending is an exact restatement of the beginning. In particular, the tenderness of the heart indicated by the expressive mark ``Gesangvoll, mit innigster Empfindung'' is restored in the end, and becomes even more outstanding after the exhausting journey represented by the variations in between the two utterances of the opening theme. As indicated by the transition into the theme in figure \ref{fig:beethoven op109 mov3 ending}, the restatement of the theme is prepared by a gradual decrescendo accompanied by endless trills, as if all previous variations musically evaporate without leaving any definite trace. Therefore, the final restatement of the theme as a restoration of the initial cantabile quality is particularly evident in the context of the transition preparing it. As a result, it is preferable to restore the cantabile quality of the opening theme unmodified, under a new light: homecoming. In other words, the performer may creatively render the beginning and the ending in exactly the same way, hence highlighting how the perception of the passage may be completely different as their sole difference lies in the temporal placement: the opening and the ending are equally tender-hearted, yet because of our journey (as discussed in section \ref{sec:narrative as foundation}, journey is a powerful metaphor for narrative) through the variations in between, the return of the opening theme gives us a completely different state of mind, regardless of the sameness in musical content.

As the example above illustrates, the interpretive challenge of repetition for performers is a result of the multifaceted role recurrence plays in music. Consequently, in order to arrive at a justified interpretive decision, performers ask about the function and role of repetition which is specific to each piece. Because the term repetition subsumes so many musical instances that are virtually impossible to generalize, they must be studied on an individual basis. However, at the same time, we should nevertheless realize that regardless of the specific interpretive decisions made about repetition, temporal asymmetry alone can shape recurrence so that similar events are construed differently because of their difference in temporal placement. In other words, instead of saying that temporal asymmetry complicates interpretation of repetition because performing repetition requires awareness of the temporal context in which repetition takes place, we can optimistically argue that it is due to temporal asymmetry that recurrence in music attains the manifold of expressive power: a simple repetition becomes capable of expressing a rich set of emotions.

In short, musicians take advantage of recurrence to control the flow of musical narrative. Because musical narrative is nothing but the time evolution of listeners' psychological states, musicians can thus control how listeners' emotions fluctuate in response to the performance. In other words, one of the strategies is to use recurrence as the guide for the listeners in navigating within the piece: upon hearing recurring materials, the listeners are fully aware of the context in which the they are situated.

\section{Resurrection as recurrence in action: a musical case study of Bach's St Matthew Passion (BWV 244) and Christmas Oratorio (BWV 248) in relation to a recurring chorale tune}
\label{section:resurrection as recurrence in action}

The dichotomy between birth and resurrection is a telling example of recurrence. Both words essentially describe the same physical process through which one is brought into life. However, the difference in temporal placement completely determines their interpretation and identification, to the extent that they are denoted by two distinct terms. In other words, because we feel that the meanings of birth and resurrection are sufficiently different in nature, albeit the fact that they describe the same physical process, that we need to distinguish between them in terminology. In fact, we may consider birth and resurrection as one of the canonic examples where temporal asymmetry alone determines the very definition of an event: when the process of coming into life is placed before trials and sorrows of life, it is called birth; when the same process is placed after trials and sorrows, it is called resurrection.

The theme of resurrection is explicitly dealt with in Bach's compositional output (in particular, his sacred compositions). Therefore, the discussion of resurrection in context of Bach's sacred work is in little danger of intentional fallacy, a fallacy in which we make far-fetched interpretation of the music in terms of the extramusical idea of resurrection. Specifically, resurrection is expressed through the particular theological lens: the life of Jesus with three milestone events consisting of birth, death and resurrection. Naturally, the three milestone events are captured in the liturgical calendar, whose focus revolves around the three liturgical days: Christmas signifying Jesus' birth, Good Friday signifying Jesus' crucifixion, and Easter signifying Jesus' resurrection. In relation to the topic of resurrection, we are to examine a special portion of Bach's works whose composition is dedicated to the liturgical days signifying the three key moments in Jesus' life. Given Bach's religiously pious attitude, it would be unsurprising that St Matthew Passion (BWV 244) which is dedicated to Jesus' crucifixion is of personal importance to Bach, and as a result of the composer's invested effort, is of analytical interest.

Among Bach's sacred compositions, one characteristic compositional feature is the invocation of chorales. We may arguably claim that the role of chorales for Bach is akin to the role of the chorus in Greek tragedies such as those by Aeschylus: instead of being part of the narrative, they serve as extra-narrative commentary about the narrative from the perspective of spectators (namely, audience members). For example, in the introduction to Aeschylus' ``The Oresteia'', editors observe that Aeschylus' ``unique tragic style is especially remarkable for its extensive and intensive use of the chorus''\autocite{aeschylus}, through which

\begin{quote}
a chorus of twelve (in Aeschylus' time) or fifteen (for most of the careers of Sophocles and Euripides), who would sing and dance formal songs and whose Chorus Leader would engage in dialogue with the characters or offer comment on the action.\autocite{aeschylus}
\end{quote}

Metaphorically, chorus is like the lucky audience members who are invited to the stage, whose reactions and comments to the play are also included as part of the program. Similarly, Bach's use of chorale can be interpreted as commentary from the congregation. As Hill remarks in his article:

\begin{quote}
The chorales in general are considered to represent Bach's idea of how the congregation should respond to the Passion narrative. {[}\ldots{]} Bach put words if not into the mouths, then into the minds, of his congregation, and many commentators have recognized that the chorales are meant to represent their thoughts. Spitta wrote, ``the greater number of chorales are set in a severe style, and accentuate the congregational feeling in all its modesty and force''; Schweitzer observed, ``the feelings of the Christian spectators are expressed in chorale verses''; Geiringer believed that the chorales express the ``reaction of the congregation''; and Pelikan described one particular chorale as ``the response of the church and of the pious soul''.\autocite[519]{hill1996time}
\end{quote}

Therefore, we may interpret Bach's use of chorale to be an organizational tool to help listeners navigate within the piece, offering hints for interpretation. For example, a hypothetical listener may be lost in the midst of the complex narrative trajectory of St Matthew passion, to the extent that he/she may start wondering: ``where am I situated in the plot, and am I supposed to feel happy or sad at this very moment?''. Chorales come to the rescue by offering suggested reactions to the plot, based on which the hypothetical listener can develop his/her own understanding of the plot accordingly. Metaphorically, Bach presents the passion with chorales as annotations that serve as an analysis of itself, thus offering insights for our understanding of the work.

Additionally, we should note how Bach's use of chorales is relevant to our discussion as it relates to both the topic of narrative as the time evolution of psychological states (i.e.~topic of chapter \ref{chapter:temporal arts as narrative}) and recurrence (i.e.~topic of this chapter).

For the topic of narrative as the time evolution of psychological states, we observe that chorales as commentary to the plot, naturally puts an emphasis on the psychological aspect of narrative. Therefore, based on our claim in section \ref{section:tension-release as psychological narrative} that narrative must be based on psychological states such as tension-release, the collection of chorales captures essential aspects of the biblical narrative. In fact, because of our shift of focus to the psychological aspect of narrative, chorales become more important than other movements in the work. One may argue that such conclusion is counterintuitive, precisely because chorales function as commentary: they do not describe the dramatic action of the plot (i.e.~the content of the story, involving characters and their interactions) which is what enables the development of the plot. Instead, chorales function in a way similar to the concept of operatic aria: the dramatic time is suspended, leaving the performance to focus on the timeless contemplation of the plot that just took place. To counter such potential objection, we notice that chorale movements are crucial not only analytically, but also in terms of the performance effect: they are moments that have direct impact on the listeners emotional states since the chorale tunes are not expressed through plot but tunes familiar to the listeners. In fact, based on the Lutheran tradition, the chorales may be performed by performers and listeners alike, resulting in the Lutheran idea of ``increasing worshipers' participation through music''.\autocite{norton_history} Therefore, the direct impact is evident, as suddenly the performance is expanded in scope: the performers now consist of the totality of people in the performance venue. We may even poetically say that, the transition into the chorales turns a Baroque ensemble into a late-Romantic ensemble similar to that in performing Mahler's symphony of a thousand.

For the topic of recurrence, because chorales offer listeners guidance in navigating within the piece, we can consider them as structural pillars in the following sense: chorales mark key moments within the piece. Such observation is based on the assumption that we make commentary only about important events, and being musical commentary, chorales also correspond to key events within the piece. Therefore, we are able to analyze Bach's passions and oratorios through reduction, by focusing solely on the sequence of chorales as if they constitute a suite-like composition on their own. Specifically, in relation to the topic of recurrence, we focus on a special case where chorale tunes repeat within the work, thus informing us about the overall structure of the work through recurrence.

\subsection{The puzzle of St Matthew's Passion}

\begin{figure}[!h]
   \centering
   \includegraphics[width=.8\textwidth]{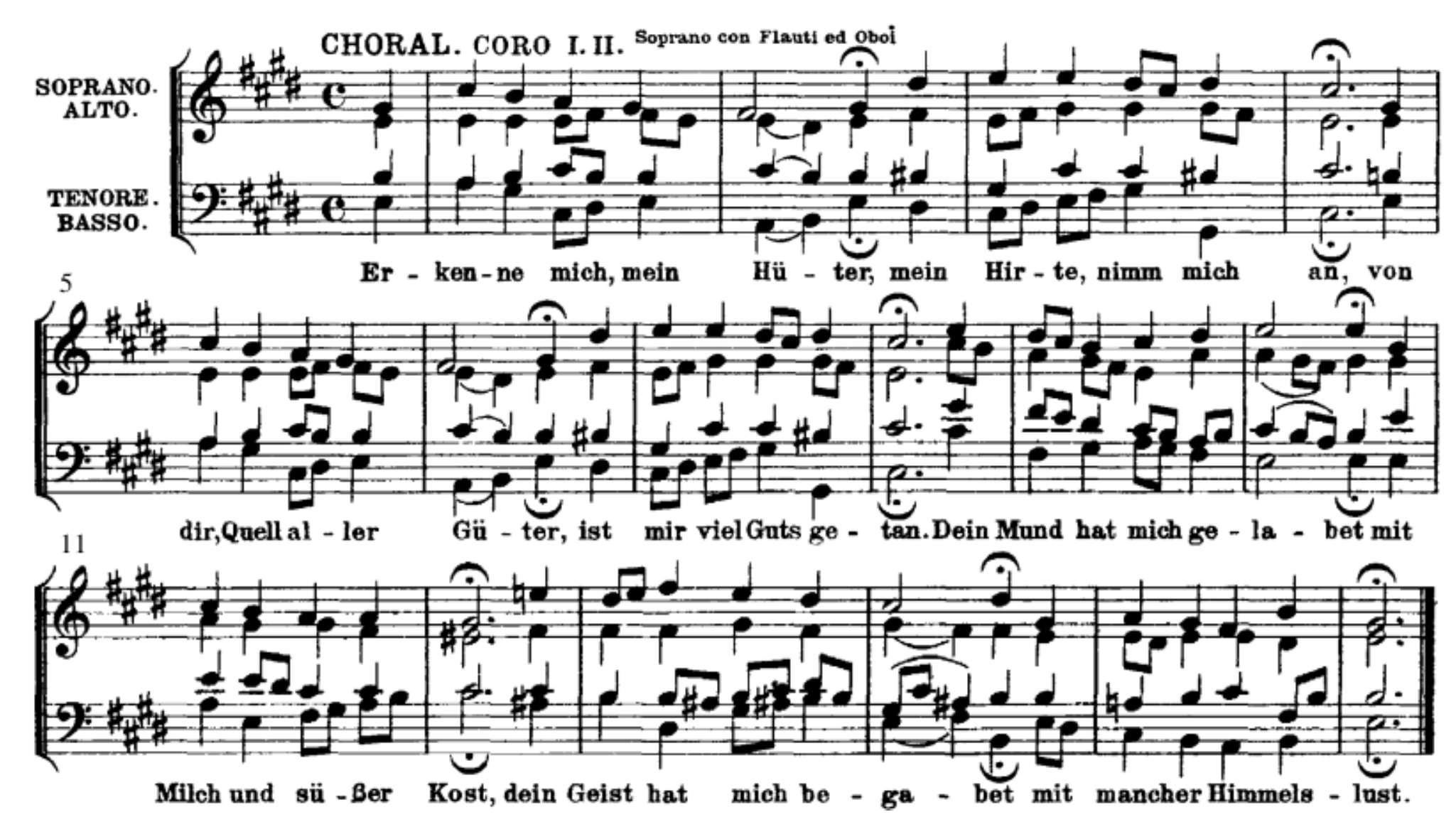}
   \caption{Bach, St Matthew Passion, BWV 244, movement 15 (NBA), incipit ``Erkenne mich, mein Hüter'' (vocal score reduction)}
   \label{fig:bach bwv244 mov15}
\end{figure}

\begin{figure}[!h]
   \centering
   \includegraphics[width=.8\textwidth]{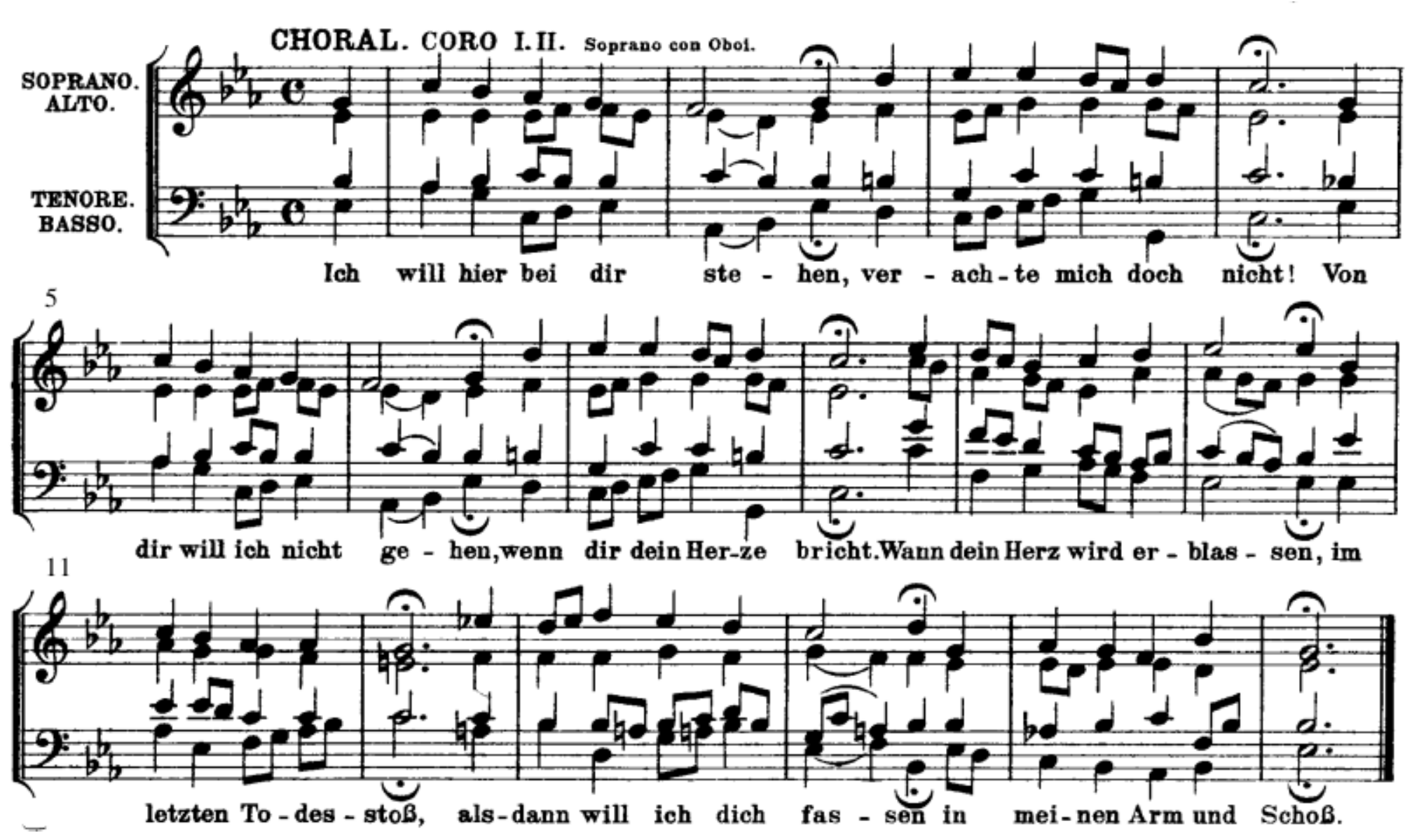}
   \caption{Bach, St Matthew Passion, BWV 244, movement 17 (NBA), incipit ``Ich will hier bei dir stehen'' (vocal score reduction)}
   \label{fig:bach bwv244 mov17}
\end{figure}

\begin{figure}[!h]
   \centering
   \includegraphics[width=.8\textwidth]{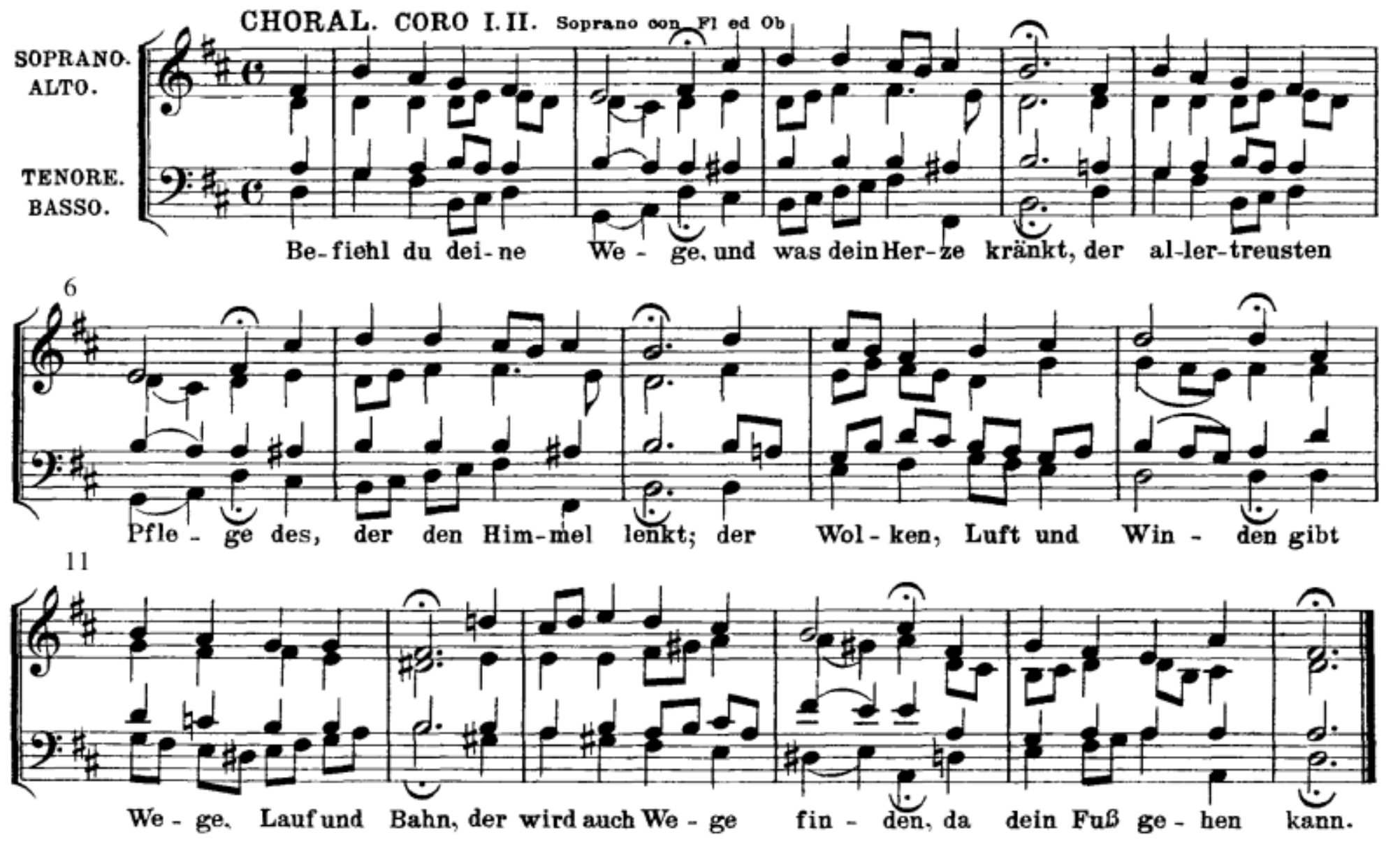}
   \caption{Bach, St Matthew Passion, BWV 244, movement 44 (NBA), incipit ``Befiehl du deine Wege'' (vocal score reduction)}
   \label{fig:bach bwv244 mov44}
\end{figure}

\begin{figure}[!h]
   \centering
   \includegraphics[width=.8\textwidth]{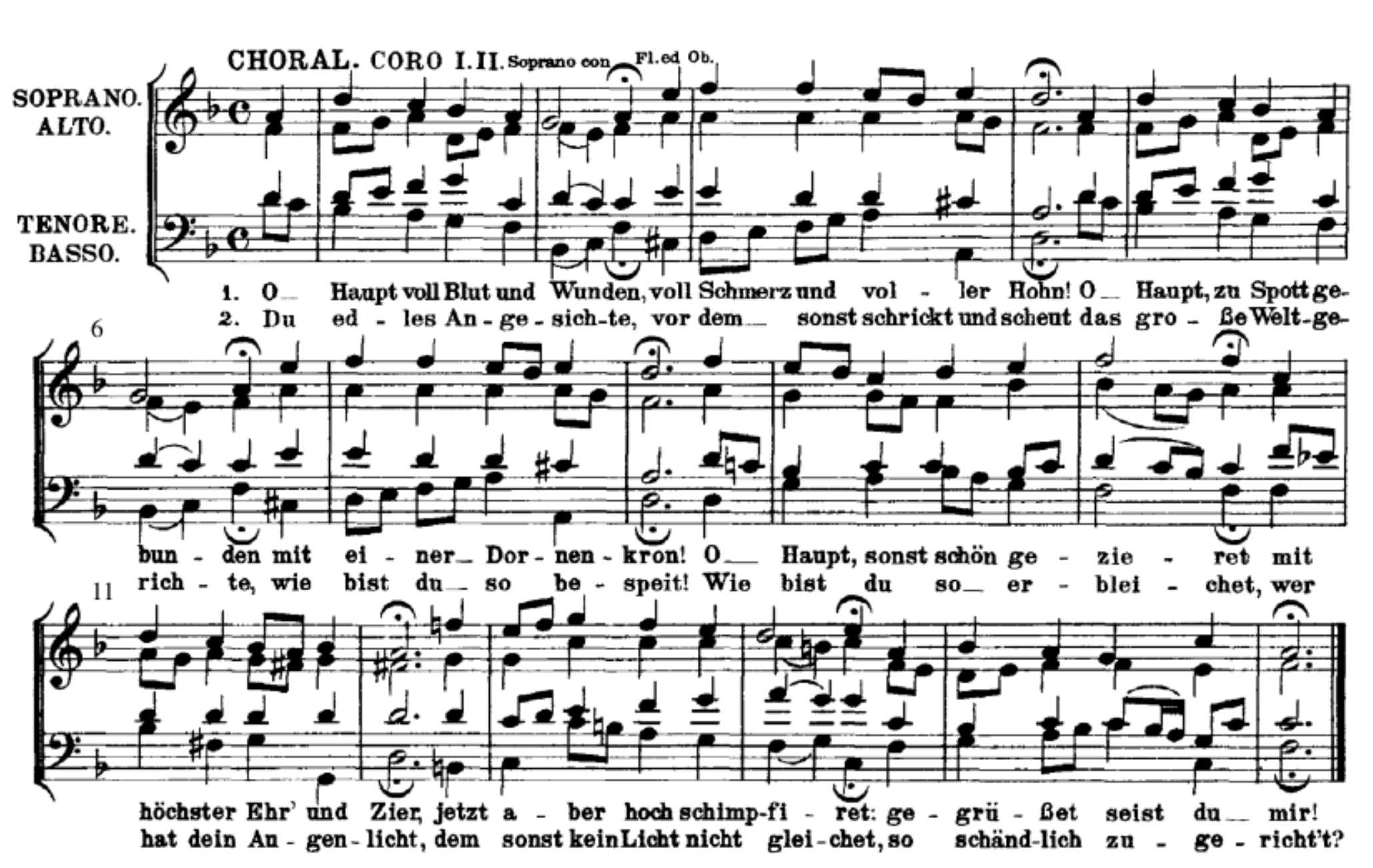}
   \caption{Bach, St Matthew Passion, BWV 244, movement 54 (NBA), incipit ``O Haupt voll Blut und Wunden'' (vocal score reduction)}
   \label{fig:bach bwv244 mov54}
\end{figure}

\begin{figure}[!h]
   \centering
   \includegraphics[width=.8\textwidth]{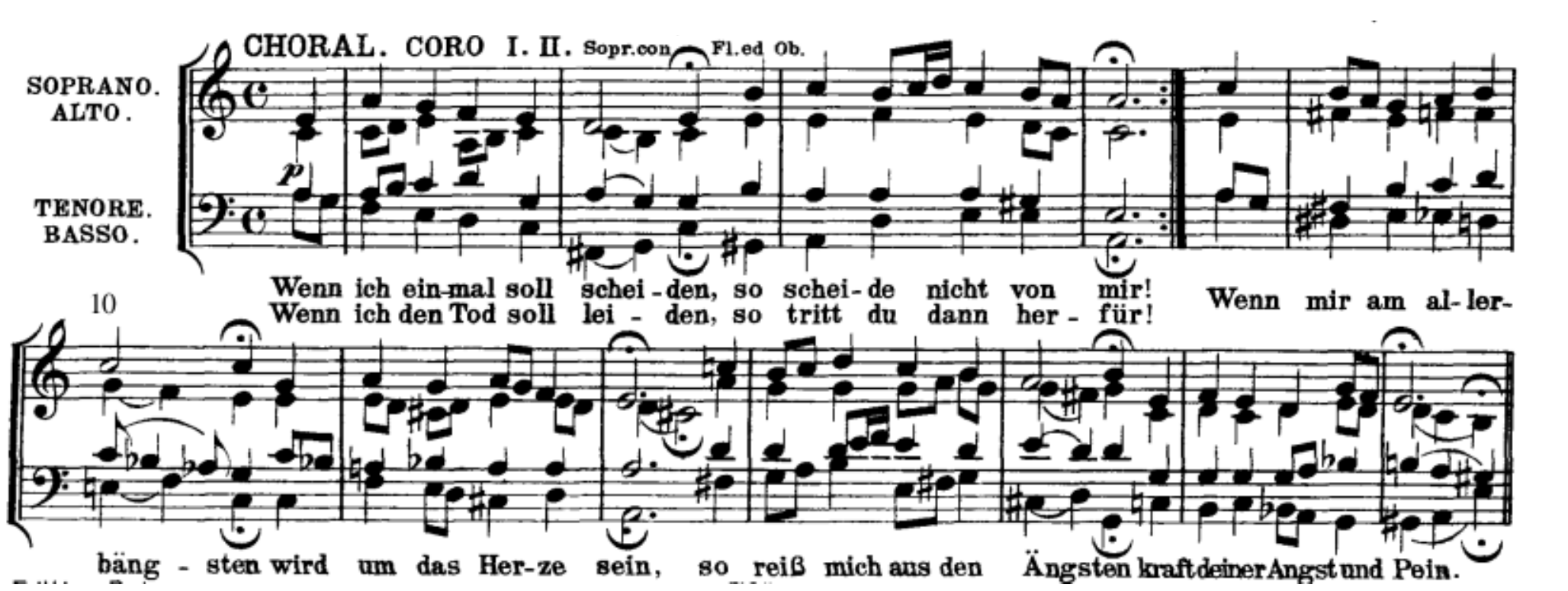}
   \caption{Bach, St Matthew Passion, BWV 244, movement 62 (NBA), incipit ``Wenn ich einmal soll scheiden'' (vocal score reduction)}
   \label{fig:bach bwv244 mov62}
\end{figure}

One of the characteristic features of St Matthew's Passion is a recurring chorale that is particularly outstanding in the midst of its constantly developing narrative. The chorale tune ``O Haupt voll Blut und Wunden'' appears five times throughout the passion at movements 15 (see figure \ref{fig:bach bwv244 mov15}), 17 (see figure \ref{fig:bach bwv244 mov17}), 44 (see figure \ref{fig:bach bwv244 mov44}), 54 (see figure \ref{fig:bach bwv244 mov54}), and 62 (see figure \ref{fig:bach bwv244 mov62}).

Before making any reference to the text and biblical narrative, the five occurrences of the chorale tune are remarkable due to the sheer number of repetitions. Even without any analytical approaches, listeners are more likely to remember the five occurrences then any other movement from the piece simply because of the power of repetition: repetition as reinforcement of memory.

Meanwhile, Hill makes the following observation concerning the five occurrences of the chorale tune:

\begin{quote}
Bach inserts the melody known as ``O Haupt voll Blut Wunden'' into the Passion at five very important points in the plot: the scene on the Mount of Olives (twice), the trial before Pilate, the scourging, and immediately following Jesus's death. As Eric Chafe puts it, this group of settings is ``one of the best-known elements in the St.~Matthew Passion and the only one that the majority of listeners will instinctively recognize as a series.'' As such a series, the appearances of this melody guide the congregation's thoughts through the work, culminating in the central tragedy of the plot, the death of Jesus.\autocite[520]{hill1996time}
\end{quote}

As discussed above, chorales in general serve as structural pillars guiding one's understanding of the whole piece. The five occurrences of the same chorale tune, as a special case where five chorales are related by the special relation called equality, shows even more power of guidance: the hypothetical listener can be completely ignorant of the plot, and simply by comparing the difference in harmonization between the five occurrences, he/she can grasp the dynamic (as in fluctuation of musical tension, not volume) of the entire passion. In other words, the five occurrences can be considered as an abstract (as in academic papers) summarizing the entire passion, through listening to which one can grasp the structural outline. Additionally, speculating on the composer's intention with the assumption of good faith (e.g.~the composer makes compositional choices for meaningful musical reasons, instead of reasons related to meeting the publisher's deadline), the five occurrences of the same tune must be due to compositional choice such that Bach deliberately chooses to occupy the valuable moments of chorales with five settings of the same tune. Therefore, for analytical purposes, we may consider the series of the five occurrences of the chorale tune as a composition on its own, serving as an analytical reduction of the passion's overall structure.

Construing the five occurrences as a single composition, we may observe one anomaly: the piece is incomplete in the sense that the ending does not convey a sense of resolution or finality. Instead, the ending is destabilizing, creating a sense of open-ended question. As listeners, we are naturally sensitive (even without receiving training in music theory and analysis) to anomalies in the sense that listeners are sensitive to places where established repetitive patterns break. Such capability may be attributed to evolution where prehistoric humans need to develop the ability to spot anomalies in order to survive. For example, in the wild, humans need to be able to spot poisonous snakes which are often disguised under the established visual patterns consisting of plants. Now, as listeners and analysts, what we do is in fact similar to prehistoric humans: compare the five occurrences and see where established patterns break.

\begin{figure}[!h]
   \centering
   \includegraphics[width=.8\textwidth]{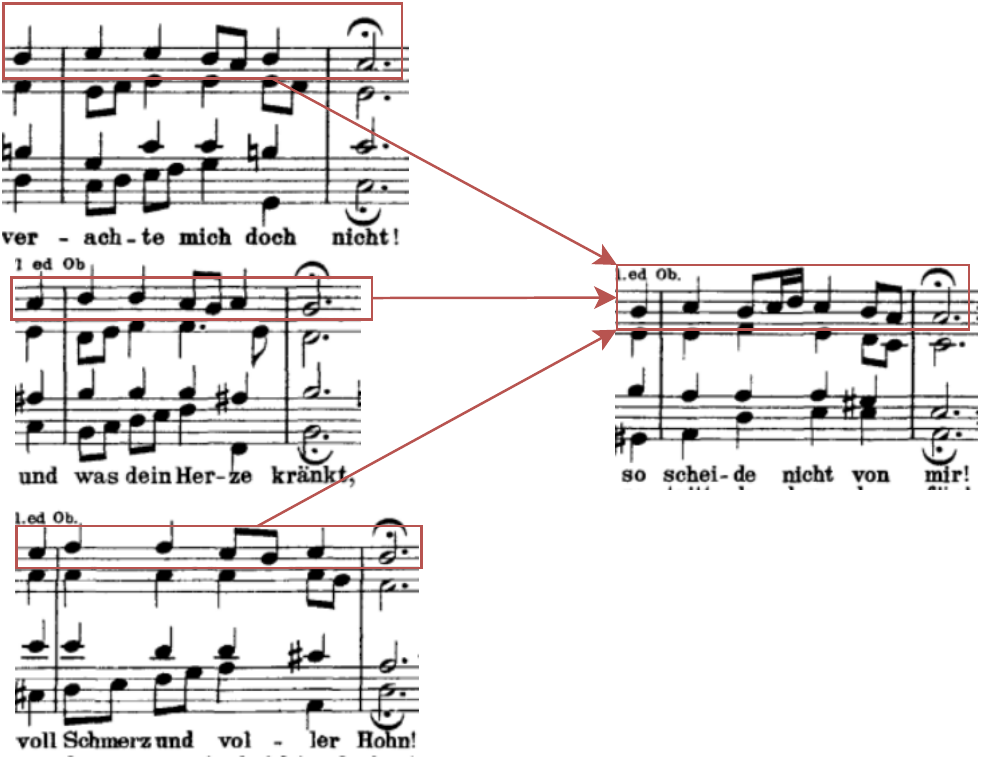}
   \caption{Comparison between the four occurrences (the first occurrence is omitted as it is a transposed duplicate of the second), annotated to indicate difference in melodic contour (left: first three occurrences, right: last occurrence)}
   \label{fig:bach bwv244 melodic difference}
\end{figure}

By comparing the five occurrences side by side, we may observe that the last occurrence, namely, movement 62 (see figure \ref{fig:bach bwv244 mov62}) stands out as the most inflected occurrence because of its melodic contour as well as chromaticism in harmonization. Melodically, one of the most noticeable pattern breaking occurs at measure 3, where the identical melodic contour found in the first four occurrences is modified in the last occurrence, creating audible difference to the listeners (see figure \ref{fig:bach bwv244 melodic difference}).

\begin{figure}[!h]
   \centering
   \includegraphics[width=.8\textwidth]{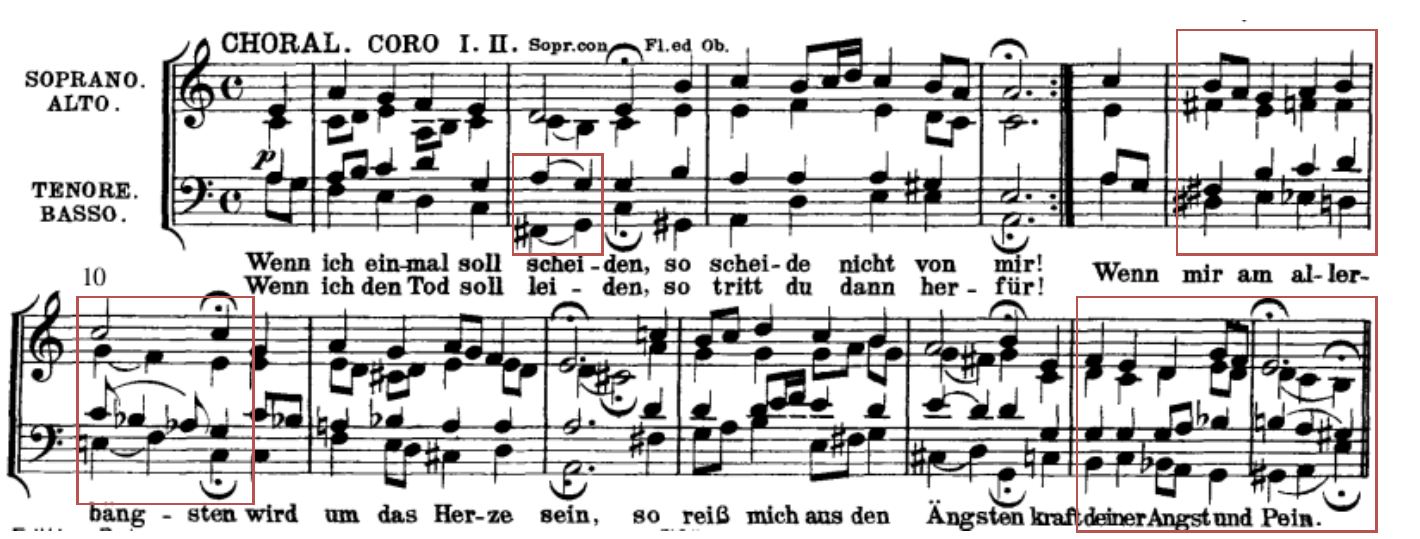}
   \caption{Bach, St Matthew Passion, BWV 244, movement 62 (NBA), incipit ``Wenn ich einmal soll scheiden'' (vocal score reduction), annotated to indicate harmonic anomalies}
   \label{fig:bach bwv244 mov62 harmonic anomalies}
\end{figure}

Meanwhile, harmonically, the heavy use of chromaticism in the last occurrence, when compared to the previous four occurrences, is readily observable. In fact, the heavy use of chromaticism coupled with the text may be considered as an instance of word-painting, where Bach employs the Renaissance modal tradition of using chromaticism to invoke the imagery of pain and agony. To name a few places of startling chromaticism, consider figure \ref{fig:bach bwv244 mov62 harmonic anomalies}. The first red rectangle indicates chromatic inflection of the bass. One may argue that the downbeat of measure 2 is simply an applied chord, namely, V/V. However, such harmonic interpretation is unsuitable if we consider the previous occurrences of the same measure: if we were to keep the same harmonization of measure 2, then the bass in this case should be F instead of F-sharp. Therefore, from a pattern-breaking perspective, the bass consists of chromatically inflecting note F into F-sharp. As a consequence of the chromatic inflection, the bass motion leading into the downbeat of measure 2 becomes a tritone, whose dissonance is symbolic of a twisted emotional state. The second and third red rectangle in figure \ref{fig:bach bwv244 mov62 harmonic anomalies} highlights Bach's heavy use of chromaticism, to the extent that we may claim that Bach maximizes the chromaticism in horizontal motion at the expense of contrapuntal convention. For example, in measure 10, the alto and tenor create an exceptionally noticeable parallel fifth even when the parallel fifth is technically avoided by the use of syncopation, thus drawing the listener's attention to the syncopated tenor voice. The syncopated tenor voice then arrives at the third beat of measure 10 in a descending motion falling from flat-\(\hat 6\) to \(\hat 5\), which metaphorically creates a strong downward pulling force.

The harmonic anomalies through chromaticism suggest that there must be motivations strong enough to justify Bach's use of chromaticism even at the expense of of contrapuntal convention and harmonious sonority. The potential motivation becomes evident once we consider the last occurrence of the chorale tune in relation to the plot: it is set to refer to the moment in the plot which ``immediately follows Jesus' death''.\autocite[532]{hill1996time} Consequently, from the perspective of word-painting, Bach deliberately invokes discomforting chromaticism in order to convey a sense of agony. Additionally, from the perspective of comparison between the five occurrences of the chorale tune, chromaticism unique to the last occurrence signals that it must be considered separately from all previous occurrences. In other words, Bach uses chromaticism in order to single out the last occurrence through anomaly.

The ending indicated by the last red rectangle in figure \ref{fig:bach bwv244 mov62 harmonic anomalies} is particularly outstanding in its anomaly. In addition to chromaticism in the word-painting sense, i.e.~as chromatic inflections, the final cadence creates ambiguity in its harmonic interpretation. Specifically, when interpreted in the context of A minor key, which is suggested by the key signature, we have a V-I-V progression, creating a half cadence. Alternatively, the sonority, especially with the sustained treble voice, is characteristic of plagal cadence, in which case the key is to be interpreted as E minor (with the major chord interpreted as a picardy third). However, both interpretations are problematic. If the cadence is a half cadence in A minor, then the ending is incomplete as it does not end on the tonic. If instead, the cadence is a plagal cadence in E minor, then the chorale setting involves a modulation from A minor to E minor, which also lacks finality. We may argue that such ambiguity is exactly what Bach intends if we consider the incipit ``Wenn Ich Einmal Soll Scheiden'' (When I must depart one day). If we take the interpretation of ending in E minor, then the modulation from A minor to E minor is symbolic of the term ``depart''. Consequently, the open-endedness of the last occurrence is a deliberate choice, signaling the idea that Jesus' crucifixion is not the end of the story.

A puzzle arises from observing the last occurrence of the chorale tune ``O Haupt voll Blut Wunden'': since the ending suggests that the story is incomplete, where are we to find its continuation? As Hill observes:

\begin{quote}
The last cadence of ``Wenn ich einmal soll scheiden'' is a loaded gesture, a sign whose interpretation is at best equivocal. Clearly, it is a sign that indicates a different tradition is being invoked in the last setting of this melody than the four settings that preceded it. But what else does the sign tell us? Does it signify closure or continuation? If closure, Jesus is dead and salvation is questionable; if continuation, Jesus will rise on Easter Sunday and salvation is certain for those who truly believe. Bach makes the sign very hard to read, however, by the way he sets it into relief against the three harmonies that open the last phrase, which point to an ending in C major, and the way he drags the harmonies into a completely different system through the interjection of the Bb. In short, Bach gives us a sign whose significance is obscured in situ.\autocite[542]{hill1996time}
\end{quote}

In fact, such puzzle can be extended to the entire passion as its lack of finality may pose challenge to our understanding of its theological implications. On the one hand,

\begin{quote}
it is very curious that not a single one of the arias or choruses that follow Jesus's death deal with his resurrection; every one instead focuses only on his death. {[}On the other hand, a{]} careful reading of Martin Luther's ``Meditation on Christ's Passion'' makes the downplaying of the resurrection in the St.~Matthew Passion highly problematic, for Luther believed that the resurrection, not the crucifixion, provided salvation.\autocite[514]{hill1996time}
\end{quote}

In short, the puzzle can be stated as follows: how are we to understand Jesus' crucifixion whose meaning depends so heavily on his resurrection, when St Matthew passion deliberately avoids the theme of resurrection altogether?

One potential solution is to accept the incompleteness of St Matthew passion, and to construe it as part of a larger narrative. In other words, the passion conveys a sense of open-endedness precisely because the story continues. The incompleteness of St Matthew passion is no longer a defect or puzzle, but a carefully-designed feature: through the absence of resurrection in the plot, we are in an overwhelmingly anxious mood to anticipate its continuation where we are to find resurrection as the resolution. Surprisingly, the chorale tune ``O Haupt voll Blut Wunden'' turns out to be tremendously helpful in locating the continuation of St Matthew passion.

\subsection{The Christmas oratorio as continuation of St Matthew passion}

At first it appears counterintuitive that the Christmas oratorio can be considered as the continuation of St Matthew passion because of the liturgical calendar. The temporal ordering of the liturgical calendar is unambiguous: Christmas takes place first, which is followed by Good Friday; after Good Friday, Easter takes place. After all, how could one possibly misunderstand the direction of time through which life events are ordered? How can we embark on a journey in which we arrive before departure? However, treating the Christmas oratorio as continuation of St Matthew passion is suggesting that Christmas occurs after Good Friday. We should note that from a theological perspective, assuming a Christian view, one can indeed make a case for the view that Christmas happens after Good Friday. Christmas occurring after Good Friday implies that Jesus is born after he is crucified. At this point, we should note that considering the implied message that Jesus is born after crucifixion, the situation is not too puzzling anymore. After all, assuming a Christian view, Jesus \emph{is} born after crucifixion, and it is nothing other than the story of his resurrection. The only remaining question then is: if Christmas is placed after Good Friday, then where do we put Easter?

Instead of invoking any sort of theological argument, our discussion takes musical observations as points of departure. Therefore, it should be made clear that, interpreting the Christmas oratorio as the continuation of St Matthew passion should be limited to music in scope. By the term continuation, we simply refer to a possible musical interpretation where the Christmas oratorio can be considered a purely musical continuation, as if in designing a concert program, it is musically fitting to place St Matthew passion in the first half and Christmas oratorio in the second half. In short, this discussion bears no theological implication, and religiously involved readers should not attempt to extend this discussion to any theological contexts. In fact, in order to focus on musical observations, we may safely assume that our hypothetical analyst carrying out this analysis of St Matthew passion and the Christmas oratorio is completely ignorant of Christianity. In particular, to our hypothetical analyst's knowledge, Christmas is about Santa with deer floating around to distribute gifts, and Good Friday is good as in good mood.

\begin{figure}[!h]
   \centering
   \includegraphics[width=.8\textwidth]{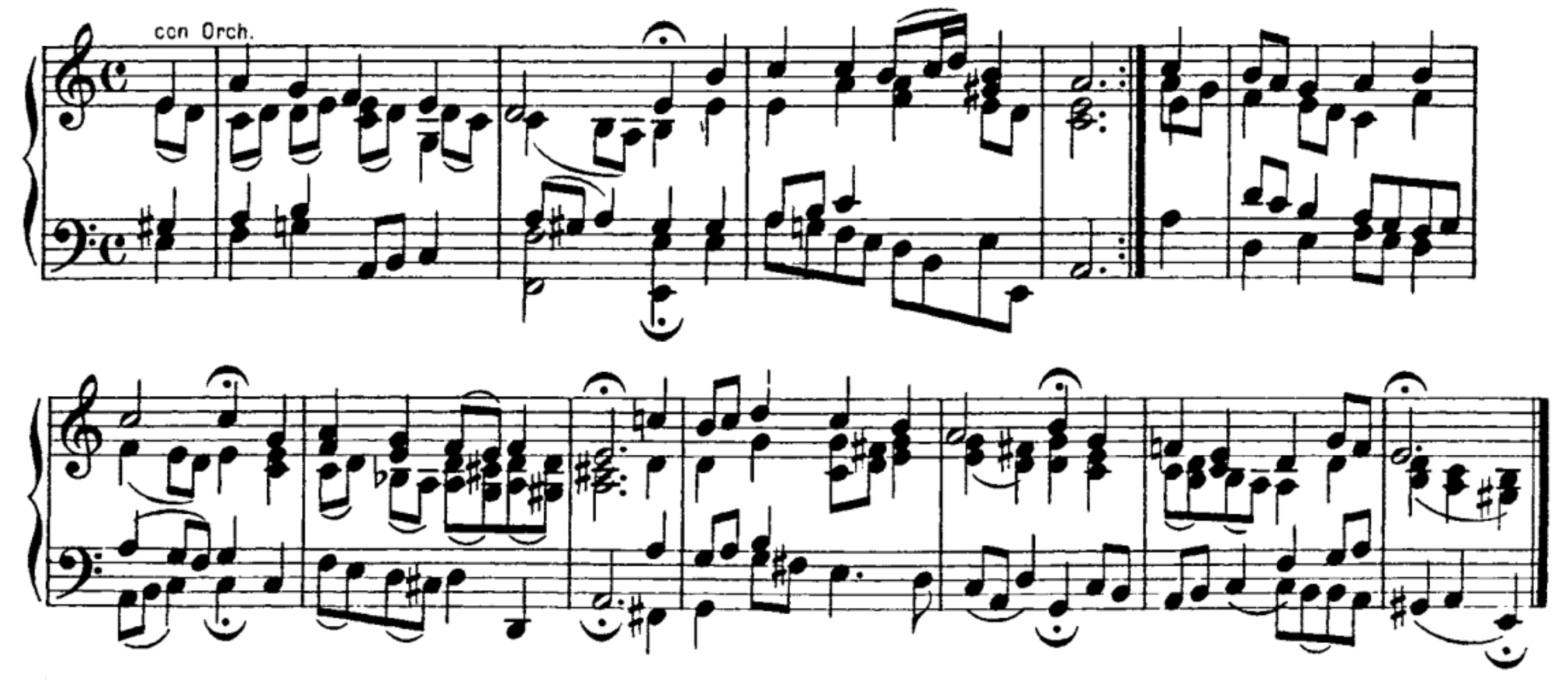}
   \caption{Bach, the Christmas Oratorio, BWV 248, movement 5 (vocal score reduction)}
   \label{fig:bach bwv248 mov5}
\end{figure}

\begin{figure}[!h]
   \centering
   \includegraphics[width=.8\textwidth]{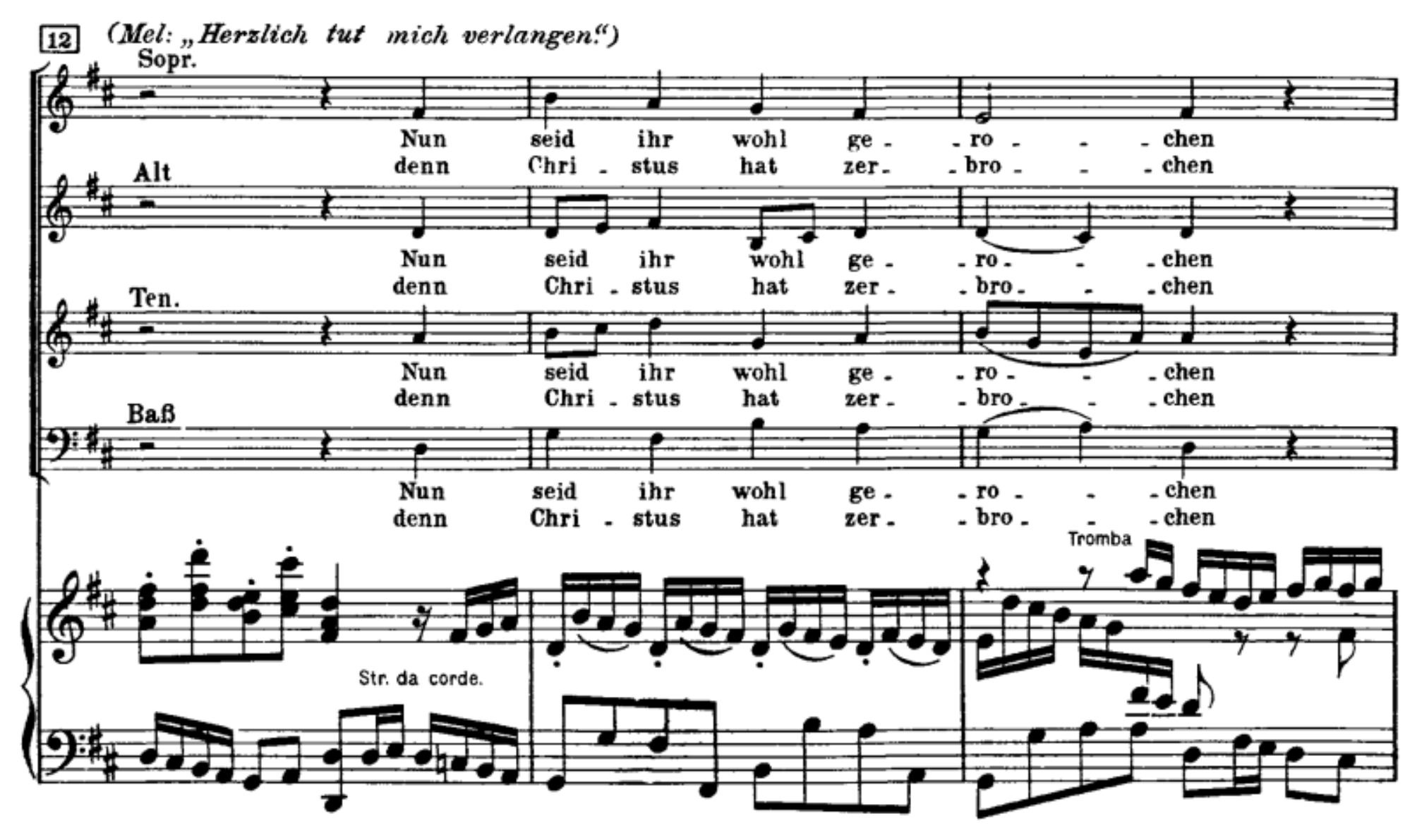}
   \caption{Bach, the Christmas Oratorio, BWV 248, movement 64, excerpt indicating vocal entrance after extensive instrumental introduction (vocal score reduction)}
   \label{fig:bach bwv248 mov64}
\end{figure}

The five occurrences of ``O Haupt voll Blut und Wunden'' in St Matthew passion alone are musically intriguing. It seems that Bach's interest in the very chorale tune motivates him to incorporate the tune several more times. In another major sacred work, the Christmas oratorio (BWV 248), Bach incorporates the tune of ``O Haupt voll Blut und Wunden'' again, at two of the most crucial places of the entire work: the opening chorale (see figure \ref{fig:bach bwv248 mov5}), and the finale (see figure \ref{fig:bach bwv248 mov64}).

Without comparison to St Matthew passion, the analysis of the Christmas oratorio itself shows that the chorale tune is structurally important. As Durr comments:

\begin{quote}
{[}The{]} first and last chorales of the entire work, nos. 5 and 64, were sung to the same melody may have been intended by Bach as a form of thematic unification.\autocite[105]{durr2005cantatas}
\end{quote}

Given the chorale tune placed at the two temporal ends of the entire work which is gigantic in scale, we can express its structural importance in threefold layers.

The first layer of importance concerns the nature of placement. We should note that the structural importance of beginning and ending is universal to all works of music. In other words, the structural importance of beginning and ending is purely syntactical, without considering particular semantics. The reason for their importance may have root in psychology, as they greatly influence our perception of the entire piece. For example, for performers, the beginning and ending of an performance often require the most practice. They are not necessarily places that are musically or technically challenging. However, because they are the two places that definitively shape listeners' reception of the performance, performers need to invest considerable amount of practice to ensure that the beginning and ending of a performance are satisfactory. Now for the Christmas oratorio, even though technically the chorale tune does not appear in the very beginning. Instead, it appears as the fifth movement, before which the oratorio presents a chorus opening, two recitatives, and an aria. However, it is the first chorale of the entire oratorio. Given the performance tradition that listeners also participate in singing the chorale, we may arguably claim that its placement as the fifth movement does not hinder its role as the beginning of the oratorio.

The second layer of importance concerns the recurring contents. Given the syntactical importance of the beginning and ending, the structural importance is strengthened once we assign the same musical material to the beginning and ending. Upon placing the same musical material at the two temporal ends of a work, we obtain structural unity (or in Durr's comment, ``thematic unification''\autocite[105]{durr2005cantatas}) through recurrence. In particular, through reminiscence and recalling, listeners are able to relate the two temporal ends (which are the most important parts according to first layer of importance discussed above) in relation to musical events in between them. Additionally, one should take into consideration the scale of this oratorio: composed in six parts, the oratorio, to some extent, can be considered as a series of six cantatas. One of the implications is about performance: instead of being performed in a single occasion, the piece is often performed in a series of performances. For example, in Bach's time, the Christmas oratorio was intended to be performed ``over six different services from Christmas to Epiphany'',\autocite[44]{durr2005cantatas} corresponding to six major feast days. Counterintuitively, the multi-performance span of the work does not diminish the value of thematic unity brought by recurrence. In fact, the value of thematic unity is strengthened. In particular, upon hearing the vocal entrance of the finale, listeners immediately get a sense of long-range recalling: ``wait, we have heard of this melody a few weeks ago!''. Such recalling is more inevitable given the tradition that the chorale is actively performed by the listeners as well.

\begin{figure}[!h]
   \centering
   \includegraphics[width=.8\textwidth]{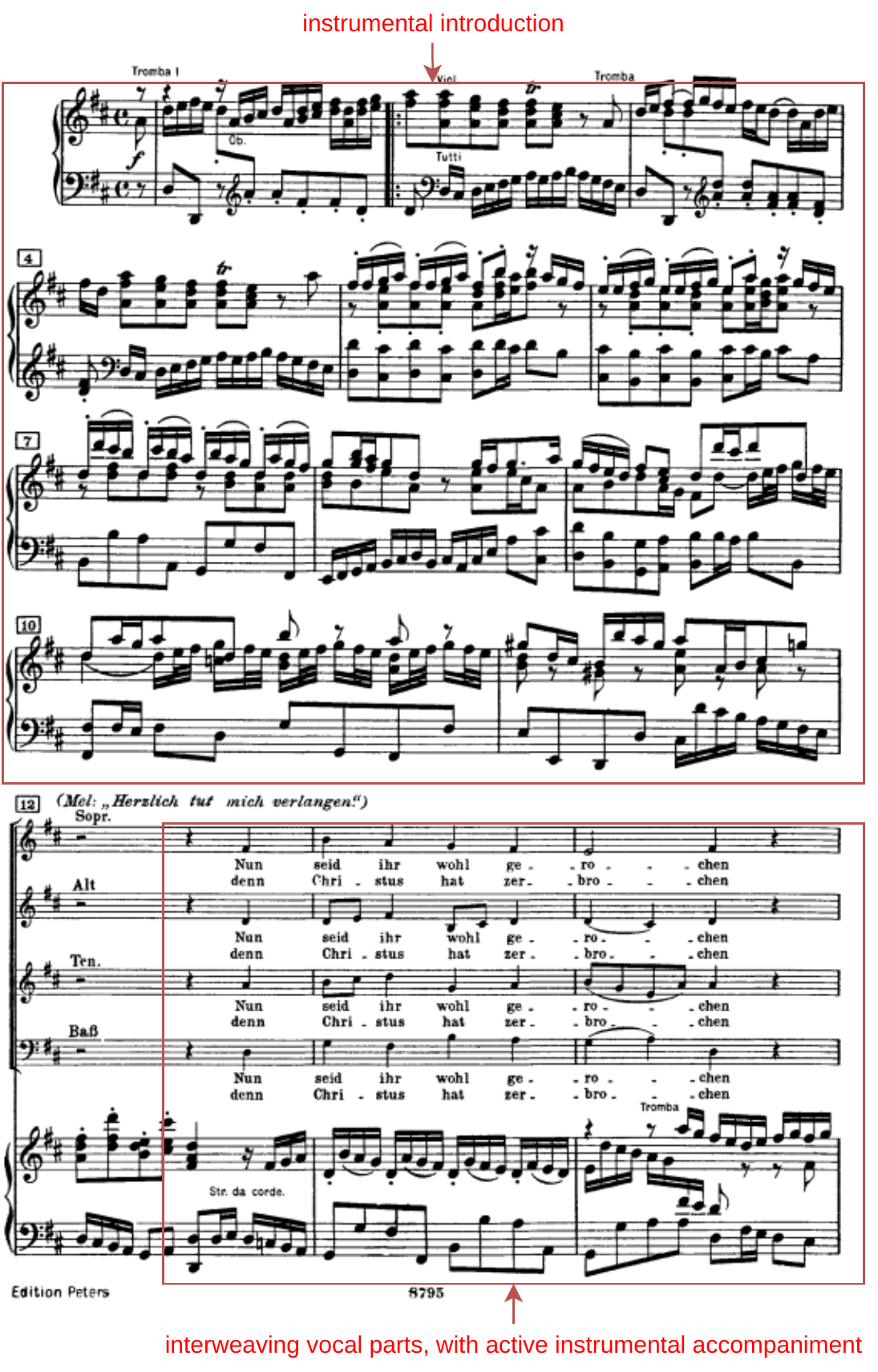}
   \caption{Bach, the Christmas Oratorio, BWV 248, movement 64, opening, annotated to indicate interactions between instrumental and vocal parts (vocal score reduction)}
   \label{fig:bach bwv248 mov64 annotated}
\end{figure}

The third layer of importance concerns the finale as a full-fledged elaboration of the beginning chorale tune. In other words, the finale may be considered as an orchestral arrangement of the chorale tune. Metaphorically, the finale can be considered as an antiphonal dialogue between the orchestra and the choir, with instrumental and vocal parts alternating (see figure \ref{fig:bach bwv248 mov64 annotated}). In particular, the chorale tune is interwoven with active instrumental accompaniments.

\begin{figure}[!h]
   \centering
   \includegraphics[width=.8\textwidth]{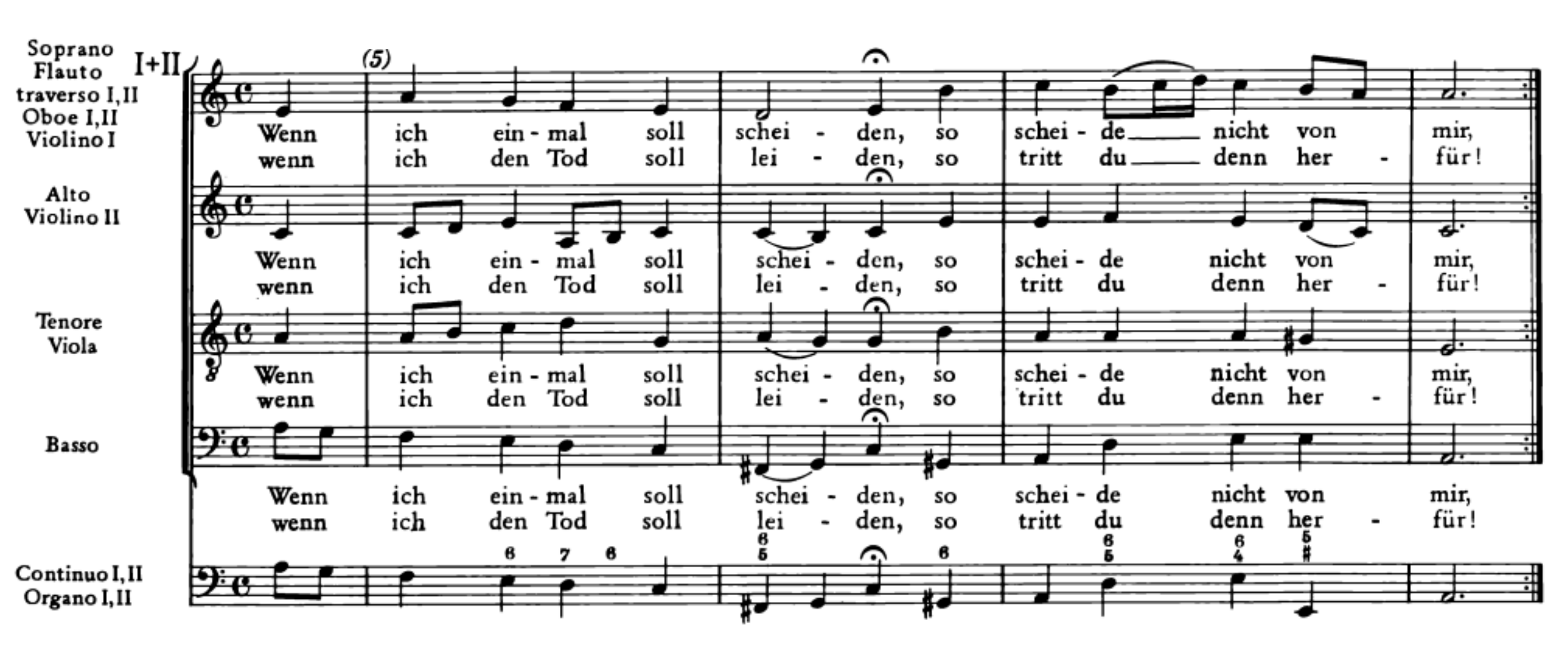}
   \caption{Bach, St Matthew Passion, BWV 244, movement 62 (NBA), incipit ``Wenn ich einmal soll scheiden'', beginning (full score)}
   \label{fig:bwv244 mov62 scoring}
\end{figure}

We should note that such elaborate treatment of this chorale tune is unprecedented in the context of our discussion concerning its seven occurrences (five occurrences in St Matthew passion and two in the Christmas oratorio). In the other six occurrences, the role the instrumental parts is minimal: to provide harmonic support through basso continuo and melodic duplication. Without loss of generality, consider movement 62 in St Matthew passion in terms of its scoring (see figure \ref{fig:bwv244 mov62 scoring}). Because the role of instrumental parts is minimal, the scoring is particularly simple, as there is no need to write out the instrumental parts.

In comparison, the role of instrumental parts is more active in the finale of the Christmas oratorio. Extending the dialogue metaphor above, the instrumental parts are now on an equal footing with the vocal parts, thus creating musical interactions (instead of mere harmonic support as in basso continuo) between the instrumental accompaniment and the vocal chorale.

How should we construe the relation between the opening chorale and the finale, in light of the active role of instrumental parts in the finale? We may invoke the exact idea of organicism discussed in chapter \ref{chapter:organicism}: the opening chorale is a germinating seed, through whose organic growth, it grows into a living organism, in which the dancing instrumental parts are its flesh and bones. The poetically organicist description continues:

\begin{quote}
In the finale, the instrumental parts are emancipated to attain musical autonomy. Now, the instrumental parts, being autonomous, have life of their own. Consequently, they can join the congregation in celebrating the ecstasy of eternal life.
\end{quote}

After asserting the structural importance of the chorale tune ``O Haupt voll Blut und Wunden'' in the Christmas oratorio, we now turn our discussion to comparison between the oratorio and St Matthew passion in relation to this chorale tune. Durr warns us about the potential pitfalls of relating the two works based on the common chorale tune:

\begin{quote}
Less likely is the theory that with this melody he wished to anticipate Christ's Passion. For the melody Herzlich tut mich verlangen {[}whose melody is identical to that of ``O Haupt voll Blut und Wunden''{]} was, at that time, not so closely connected with Paul Gerhardt's Passion hymn O Haupt voll Blut und Wunden in the consciousness of the congregation that this allusion would have been readily understood. Moreover, the tune was so commonly used in Leipzig for the hymn Wie soll ich dich empfangen that none of the listeners would have guessed that a special reference to the Passion lay behind it.\autocite[105]{durr2005cantatas}
\end{quote}

However, we may challenge his remark by noticing the salient resemblance between the opening chorale of the Christmas oratorio and the last (i.e.~fifth) occurrence of the chorale tune in St Matthew passion. In other words, the association is not based solely on the chorale \emph{melody} but its detailed \emph{harmonization} as well. In previous discussion, we claimed that the last occurrence of the chorale tune in St Matthew passion is unique among its occurrences. Therefore, here we focus on observations that enable us to group the last occurrence of the chorale tune in St Matthew passion and the opening chorale of the Christmas oratorio in one group through similarity, and distinguishing them from the other 4 occurrences in St Matthew passion through difference.

\begin{figure}[!h]
   \centering
   \includegraphics[width=.8\textwidth]{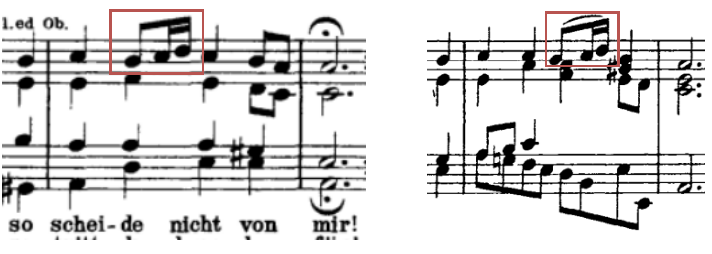}
   \caption{Side by side comparison focusing on measure 3, annotated to indicate melodic contour, left: movement 62 of St Matthew passion, right: movement 5 of the Christmas oratorio}
   \label{fig:comparison bwv244 bwv248 measure3}
\end{figure}

By comparing figure \ref{fig:bach bwv244 mov62} and \ref{fig:bach bwv248 mov5} side by side, we observe that melodically one of the most salient similarities is the melodic contour in measure 3 characterized by sixteenth-note pattern (see figure \ref{fig:comparison bwv244 bwv248 measure3}). It is true that the two melodic contours are not identical, in the sense that the sixteenth-note pattern is placed on the second beat for the example in St Matthew passion and the third beat for the example in the Christmas oratorio. However, when comparing the melodic contour containing the sixteenth-note pattern found in the two examples in figure \ref{fig:comparison bwv244 bwv248 measure3} to the other occurrences (see figure \ref{fig:bach bwv244 melodic difference}), we find that the similarity characterized by the sixteenth-note pattern outweighs the difference.

\begin{figure}[!h]
   \centering
   \includegraphics[width=.5\textwidth]{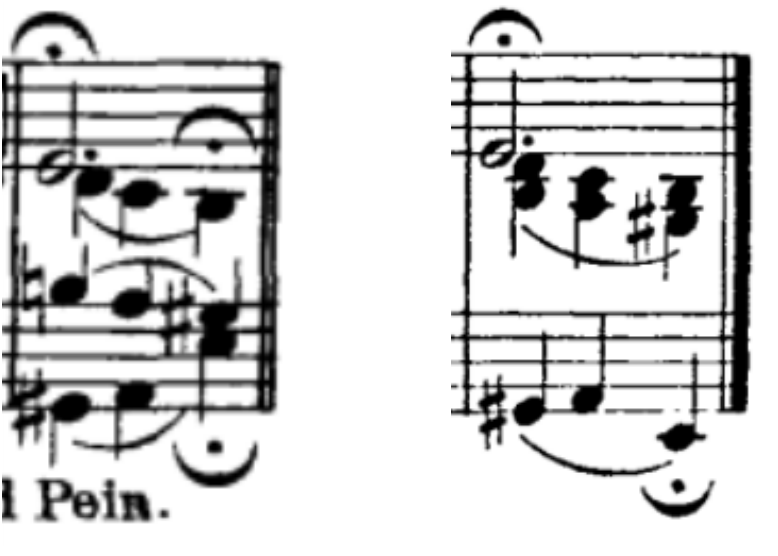}
   \caption{Side by side comparison focusing on the final cadence, left: movement 62 of St Matthew passion, right: movement 5 of the Christmas oratorio}
   \label{fig:comparison bwv244 bwv248 cadence}
\end{figure}

Hamonically, one of the most evident similarities is the key signature: both are in A minor. Most importantly, in addition to the shared key signature, they share the same harmonic ambiguity in the final cadence (see figure \ref{fig:comparison bwv244 bwv248 cadence}): a cadence whose ambiguous harmonic identity directly contributes to the open-endedness of St Matthew passion. In particular, as discussed above, it can either be a half cadence in the key of A minor, or a plagal cadence in the key of E minor. In fact, the harmonic similarity (or more precisely, identical harmonization) between their endings is the observation that enables us to claim that the two harmonizations must have been related intentionally by the composer. It would be very unlikely that the two harmonizations coincidentally share the same key signature in addition to the same harmonic ambiguity found in the ending.

Additionally, from the perspective of compositional chronology, Christmas oratorio was composed a few years after St Matthew passion, therefore we may argue that there is little danger of intentional fallacy in speculating that Bach explicitly had St Matthew passion in mind when incorporating the chorale tune ``O Haupt voll Blut und Wunden'' in composing the Christmas oratorio.

Combining the musical similarity as well as speculation based on compositional chronology, we arrive at a seemingly peculiar conclusion: the Christmas oratorio can be viewed as continuation of St Matthew passion. In particular, the Christmas oratorio picks up exactly where St Matthew passion leaves off: the chorale harmonization characterized by the final plagal cadence.

With such interpretation, the puzzle of St Matthew passion is resolved: the incompleteness at the end of the passion continues into the Christmas oratorio, with whose unfolding, the incompleteness is finally completed and resolved in the finale, as the chorale tune has grown organically into the celebration of eternal life.

\subsection{Summary}

The musical case study discussed in this section illustrates the power of temporal asymmetry in relation to the notion of recurrence. In particular, our conclusion is that the Christmas oratorio admits a twofold interpretation:

\begin{itemize}
\item
  When interpreted as Christmas oratorio by itself, it symbolizes Jesus' birth entailed by the meaning of Christmas
\item
  When interpreted as continuation of St Matthew passion, it symbolizes Jesus' resurrection
\end{itemize}

The crux of the matter lies, unsurprisingly, at the arrow of time: the arrow of time alone has the ability to change how events are defined. Therefore, if we assign the Christmas oratorio different temporal placements in relation to St Matthew passion, its meaning shifts accordingly. Specifically, if the Christmas oratorio precedes St Matthew passion, as indicated by the liturgical calendar, then we have the theme of birth (as in newborn). Meanwhile, if the Christmas follows St Matthew passion, as indicated by our analysis, then we have the theme of resurrection.

\chapter{Delaying the climax}

The climax of a piece is often delayed due to the sheer force of our artistic passion. The statement starts as a metaphor. However, just as the arrow of time is as real as it is metaphorical, delaying the climax of a piece as a result of the arrow of time should also receive serious consideration.

As our discussion in chapter \ref{chapter:temporal arts as narrative} concludes, all temporal arts can be interpreted as instances of narrative. Additionally, core concepts of narrative should be universal to all temporal arts. Therefore, instead of considering concepts such as characters which are non-existent in music insofar as music is considered a non-representational art form, we focus on the psychological aspect of narrative whose description is more universal. In particular, we consider the description of narrative in terms of tension-release. Now, the climax refers to nothing other than the place in a narrative where the tension is maximal. Therefore, the discussion of climax, as a concept universal to all temporal arts, is meaningful in a musical discourse.

Whenever we speak of delaying the climax, we mean a dynamic process (i.e.~delaying) in which the climax is originally supposed to be somewhere in the middle. However, due to the mysterious artistic force, the climax is figuratively pushed to some later point in the piece (see figure \ref{fig:temporal asymmetry as displacement}), resulting in its delay. Therefore, what we are interested in is not simply the idea that the climactic point is not placed in the middle of a piece, but a complete description of the motion of the climactic point with which it gradually moves from its initial position to its final position.

In particular, the dynamic process that delays the climax can often be described as an expansion process by which a formal structure is expanded. In its general statement, the expansion delays the climax by expanding pre-climax parts, thus resulting in delayed climax.

The dynamic process of delaying the climactic point admits a threefold interpretation, focusing on performance perspective, analytical perspective, and historical perspective, respectively.

\section{Performance perspective: Rachmaninoff and his idea of unique climax}

As performers, we interpret the statement (that by the force of our artistic passion, the climax of a piece is often delayed) seriously because it has both real-world and aesthetic consequences: it can practically guide our playing to produce aesthetically appealing performances. When we speak of a thrilling and breath-taking performance, one way in which we can describe the performance is that the music is able to hold its tension. As a result of holding tension, listener's attention is also held captive, to the extent that the listener listens attentively without the slightest feeling of sleepiness. Consequently, we can naturally ask the question: what \emph{exactly} do we mean by holding the musical tension? Moreover, what is musical tension?

\begin{figure}[!h]
   \centering
   \includegraphics[width=.8\textwidth]{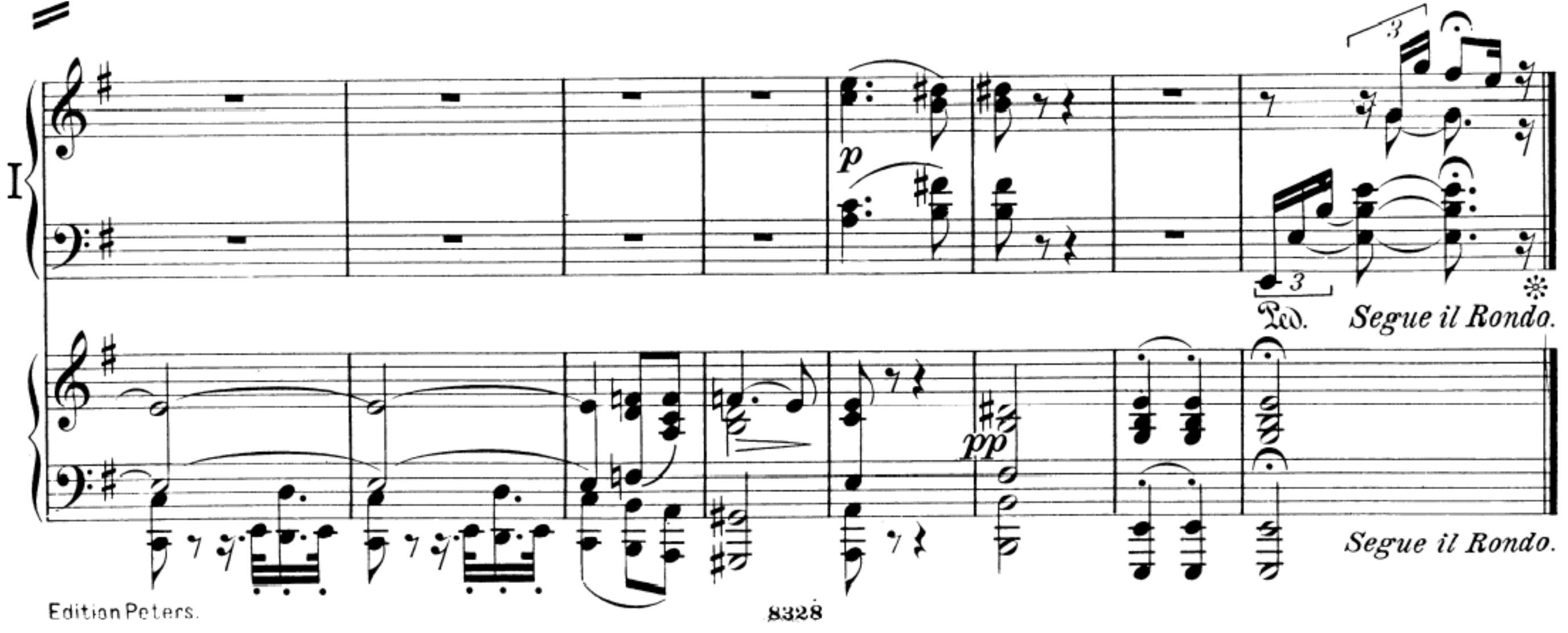}
   \caption{Beethoven, Piano Concerto No.~4, Op. 58, second movement, ending}
   \label{fig:beethoven op58}
\end{figure}

The subtlety of artistic expression implies that formulating a simple answer to the questions above is virtually impossible. However, we do know part of the answer: we know what musical tension is \emph{not} about. For example, tension has little to do with physical volume. In fact, one of the common mistakes students in instrumental lessons commit is to associate musical tension with fast speed and loud volume: whenever instructors say ``hold the musical tension'', then the playing tends to either speed up or increases volume. Meanwhile, experienced performers are able to maintain musical tension without necessarily producing loud volume. In fact, we may even hold our breath in startling performances of pianissimo passages or even silence. For example, consider the Beethoven's fourth piano concerto. The ending of the second movement (see figure \ref{fig:beethoven op58}) is breathtaking precisely because of its mystic quality, giving a sense of ethereal silence. However, such silence conveys no finality. Instead, the ascending gesture played by the piano in the last measure breaks the conclusiveness of the orchestral part. Specifically, the ascending gesture stirs up curiosity as well as hope, leaving the listeners in the psychological state of active anticipation.

While analysts may be interested in the tonal trajectory in which the E-minor final chord of the second movement prepares the C major tonality in the next movement, pianists are often more interested in performance aspects with the goal of psychologically manipulating listeners' mentality (in a positive way). In this example, one detail is particularly noteworthy: the final sixteenth-note rest. Besides the stylistic feature that Beethoven is particularly exact with his notational intention such that rests must be understood and executed with rhythmic precision, performers in general take advantage of rests as a performance strategy to create musical tension. The tension here in fact can be translated to the listeners' muscle tension: when the sixteenth-note rest becomes the silent carrier of temporal suspense, good-hearted listeners become physically unable to move, afraid of producing noise that undermines such temporal suspense. Therefore, in instrumental lessons, instructors may metaphorically teach students about the musical tension at this very moment of sixteenth-note rest:

\begin{quote}
Imagine that listeners in the audience have candies at hand. Now your goal is to hold the musical tension especially at this sixteenth-note rest, to the extent that no one in the audience wishes to unwrap his/her candies, making the annoying noise.
\end{quote}

As seen in section \ref{section:tension-release as psychological narrative}, the difficulty in capturing musical tension verbally is partly because it does not exist in music, but in us. In other words, musical tension, as part of the tension-release model, describes our psychological response to musical events. Therefore, looking for tension in music scores almost always results in approximations: there is no quantity called musical tension per se inside music scores. As a result, it is virtually impossible to generally characterize musical tension by inspecting musical events alone. Instead, performers must study works on an individual basis to understand the specific meaning of musical tension in relation to the psychological response to individual musical events.

Therefore, instead of characterizing musical tension using musical parameters (or more precisely, acoustic parameters) such as dynamics, rhythm and timbre, we resort to descriptions of our subjective experience. Instead of asking what musical tension \emph{is}, we ask about what it \emph{does} musically. When performers are capable of holding musical tension, they convey a sense of proficiency of understanding and controlling the holistic structure of the piece, i.e.~the big-picture. In other words, the performance should appear organized (in addition to spontaneous) to the listeners through the performer's holistic grasp and control of the overall musical structure.

Specifically, one of the strategies to assert control of the overall musical structure is to sustain and control musical tension in order to present climactic moments that are carefully designed and prepared. In order to present climactic moments in ways that do not appear haphazard, one of the strategies is to insert a uniqueness condition: each piece should contain one and only one climax.

The uniqueness condition for climax is a useful performance strategy whose historical origin is hard to trace. Nevertheless, historically, we may observe that Rachmaninoff expresses this very idea as a pianist. As Dubal notices in his book ``The art of the piano'':

\begin{quote}
For Rachmaninoff, the musical syntax had to be precise and clear, everything building toward a specific ``point,'', as he called it, of climax. It was his main job as an interpreter to make that ``point'' clear to his audience. If he missed it, he considered the performance a failure, and he was known to be disconsolate after such concerts.\autocite[286]{dubal2004art}
\end{quote}

In fact, this remark might be the quintessential characterization of Rachmaninoff as a pianist. In the Grove article about Rachmaninoff, the section about Rachmaninoff as a performer is relatively brief in comparison to other sections in the article. However, within the relatively brief section, a considerable amount of discussion is dedicated to his idea of unique climax:

\begin{quote}
Whatever music he was playing, his performances were always carefully planned, being based on the theory that each piece has a `culminating point'. `This culmination', as he told the poet Marietta Shaginian, `may be at the end or in the middle, it may be loud or soft; but the performer must know how to approach it with absolute calculation, absolute precision, because, if it slips by, then the whole construction crumbles, and the piece becomes disjointed and scrappy and does not convey to the listener what must be conveyed' (quoted in Apetian, 1957).\autocite{RachmaninoffRakhmaninovRachmaninovSerge}
\end{quote}

From Rachmaninoff's words, we are reminded of the observations discussed previously: the climax is not about musical parameters appearing on score. Therefore, any attempts to characterize climax in terms of dynamics, rhythm and timbre are likely to fail, as climactic points in different pieces may exhibit completely different (even opposing) musical behaviors. Moreover, the idea of designing (or more precisely, albeit less musically, engineering) a unique climax for the entire piece conveys an organicist sense of purposiveness which is the very thesis of section \ref{section:music as fulfillment of purpose}. At this point, we already perceive a kind of directionality reminiscent of the arrow of time: by setting a moment in a piece as its unique climax, our goal as performers is to build our performance \emph{toward} (notice the implied directionality) the climax.

We should find Rachmaninoff's notion of climax similar to the notion of dramatic climax discussed in section \ref{section:unique orientability}. However, the similarity should not appear surprising because one of the central claims of chapter \ref{chapter:temporal arts as narrative} is to construe all temporal arts as instances of narrative. Therefore, Rachmaninoff's remark on the holistic structure of the performance can be considered in the light of dramatic structure discussed in section \ref{section:unique orientability}: a music performance can be itself considered a five-act tragedy, whose structure highlights the climax in relation to its preparation (i.e.~rising action) and its resolution (i.e.~falling action).

Given the unique climax in performing a piece we are then able to discuss how it is delayed. To this end, we resort to an analytical perspective through which performance is often informed. Specifically, we should note that even though the climax differs drastically in its musical behavior across different pieces, the choice of climax is nevertheless not arbitrary. In other words, performers cannot haphazardly roll a dice to determine which measure will be considered the climax because the performance interpretation must be largely informed by the analytical perspective of the piece. The correlation (albeit flexible, giving performers opportunities to express individuality) between compositional decision and performance decision is precisely the reason performers should be aware of analysis, theory, and composition (and vice versa, composers should be aware of performance practice and idioms).

\section{Analytical perspective: Schenkerian analysis}

Schenkerian analysis is arguably one of the most illustrative analytical approaches to give us insights about temporal asymmetry in relation to the placement of structural events in a piece. The reason for its illustrative power is straightforward: by relying on graphical representations of analysis (in the form of Schenkerian graphs), we can visualize temporal asymmetry by identifying pictorial asymmetry, offering a more intuitive grasp of temporal structure through visualization.

\begin{figure}[!h]
   \centering
   \includegraphics[width=.8\textwidth]{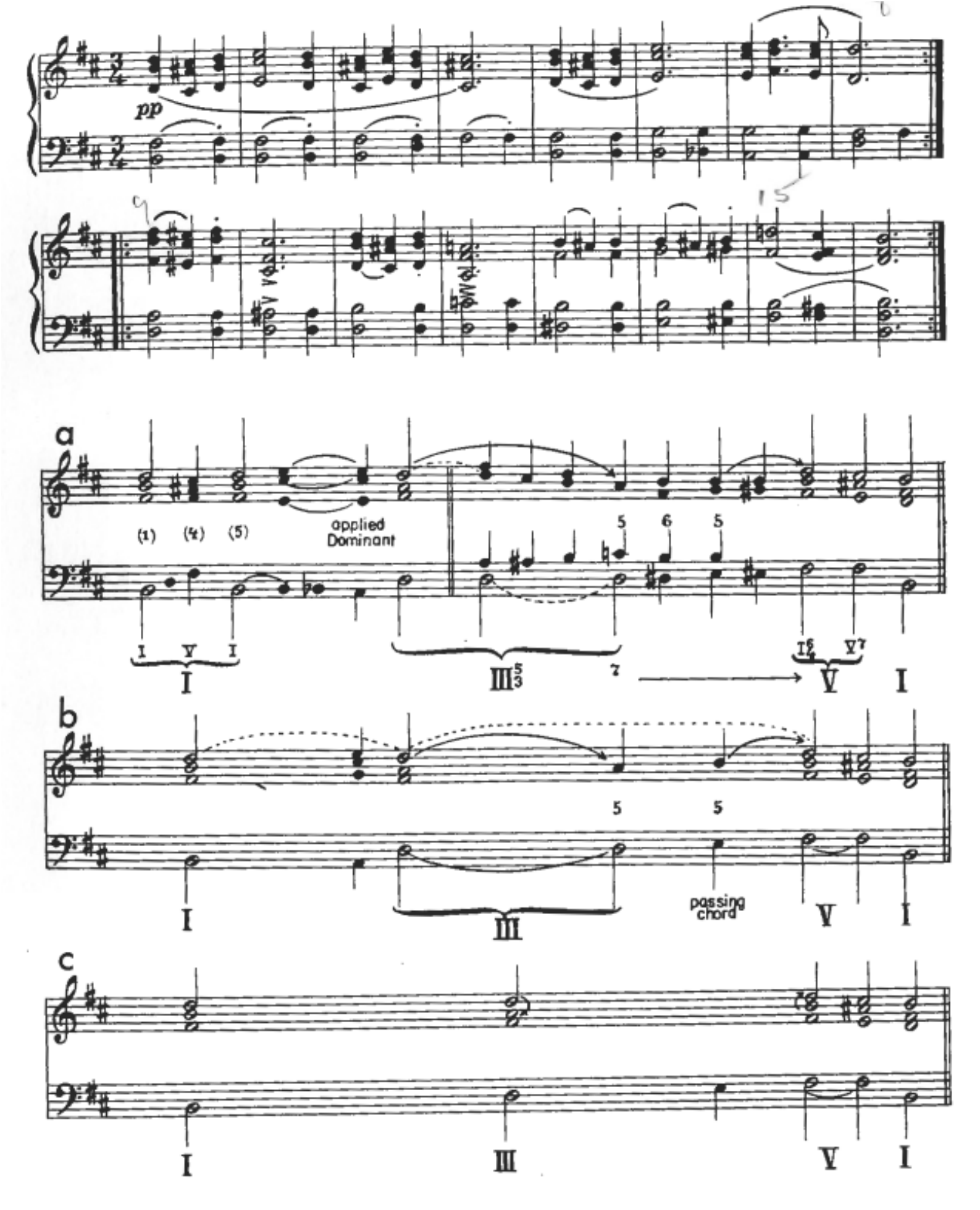}
   \caption{Schenkerian graphs analyzing Schubert's Waltz, Op. 18, No.~10, produced by Salzer}
   \label{fig:salzer schenkerian graphs}
\end{figure}

In particular, one of the peculiar characteristics of Schenkerian graphs is that, as the analysis dives into the background layer, notes start to cluster near the end. Without loss of generality, consider Salzer's Schenkerian graph produced to analyze Schubert's waltz, op. 18, no.~10 (see figure \ref{fig:salzer schenkerian graphs}).\autocite{salzer_1952_2} In the figure, the original excerpt is presented on the top (which can be viewed as the most foreground layer containing all surface details). Then, the graphs labeled using ``a'', ``b'' and ``c'' as a succession can be considered an analytical process in which we progressively dive from a foreground layer (graph labeled by ``a'') into a background layer (graph labeled by ``c''). Now, if we observe graph ``c'', we notice that chords remaining in the background layer correspond to, thanks to Salzer's visual alignment of graphs, mostly the second half of the excerpt. Specifically, the first half of the excerpt is uneventful in the sense that it is represented by a single harmony. On the contrary, the ending of the excerpt is eventful since the last three chords of the excerpt are all admitted into the background layer.

The implication of this characteristic in Schenkerian analysis is the following: structurally important chords are mostly distributed in the second half of a piece. We should note that the term ``structurally important'' attains well-defined meaning: they are chords that survive into the background layer. In figure \ref{fig:salzer schenkerian graphs}, one observes that the succession of graphs ``a'', ``b'' and ``c'' effectively constitutes a reductional process, through which structurally less important chords are filtered out. Consequently, as we arrive at the background layer, only structurally important chords remain.

Before proceeding with our discussion, we need to briefly cover analytical approaches essential to Schenkerian analysis. As an analytical tool, Schenkerian analysis can be considered as a way of capturing and encoding the hierarchical structure of music that is intuitively understood by musicians. As Drabkin describes in his summary of Schenkerian analysis:

\begin{quote}
If one were to attempt to reduce Schenker's understanding of music to a single concept, ``hierarchy'' would perhaps be the best choice. For Schenker, music -- great music -- is tonal, and hence a composition is governed ultimately by its principal chord, the tonic triad; all other harmonic functions are subordinate to the tonic, and analysis must always make a distinction between essential and passing harmonies. Similarly, the notes of a melody can be described as either essential or transitional. Moreover, the notion of essential versus passing, of harmonic versus non-harmonic, applies not only to the surface of the music but informs the deeper levels, too: a harmony might be essential at one level but transitional at another, a passing note at one level might be the start of an important ``linear progression'' at another.\autocite[816]{drabkin_2002}
\end{quote}

For example of how musical hierarchy is understood without explicitly invoking any music theory, in teaching beginner level piano, instructors emphasize the importance of grouping running notes: running sixteenth notes should be grouped in units of four notes, such that accents occur per quarter note. Instructors may bring up the topic of grouping in the context of technique, such that students can play running notes with less muscle stiffness. However, grouping is also meaningful musically, as the playing is more appealing because the phrase now sounds less chaotic: instead of a collection of running notes, we are able to construe the phrase as overlaying running sixteenth notes (as elaboration) on top of a melody in quarter notes. In fact, such simplified example of grouping sixteenth notes can be considered a toy model to illustrate the analytical approach of Schenkerian approach. In Schenkerian terms, the running sixteenth notes make up the foreground layer, representing a shallow (as evident from the word ``surface'' as in musical surface) structural level of music. Meanwhile, the melody in quarter notes makes up a layer that is more background than the layer containing sixteenth notes. We can keep constructing layers that is increasingly more background until we arrive at a layer beyond which we cannot proceed deeper, in which case Schenker uses the term ``Ursatz'' to denote the most background layer.

As observed in figure \ref{fig:salzer schenkerian graphs}, we discover that asymmetrical distribution of structurally important chords as we move into the background layer toward the Ursatz. Given the observation, we can then naturally ask about the origin of the asymmetrical distribution: what is the musical cause responsible for the asymmetrical distribution of structurally important chords appearing in the background layer? In Schenkerian analysis, the answer is given by the notion of prolongation. However, in order to present the topic of prolongation in relation to the asymmetrical structure of the Ursatz, we need to introduce a shift in mindset.

\subsection{Top-down versus bottom-up approaches}

One of the most fascinating features of Schenkerian analysis is that it admits a pair of complementary interpretations: bottom-up and top-down. In this paper, we adopt the convention that the term ``bottom" denotes the shallowest structural level, and the term ``top" denotes the deepest structural level. The difference between the two interpretations amounts to reading the graphs in figure \ref{fig:salzer schenkerian graphs} in different directions: one can start from the top of the page, reading forward; meanwhile, one can start from the bottom of the page, reading backward. Conventionally from an analytical perspective, Schenkerian graphs are discussed from a bottom-up fashion. We are given a piece of music which is by definition the most foreground layer (i.e.~the bottom of hierarchical layers). As analysts, our goal is to analyze the given piece of music through reduction: through successive removal of foreground elements, we derive background layers that are simpler in the sense of containing fewer notes: borrowing Drabkin's terminology from the quote above,\autocite[816]{drabkin_2002} only the ``essential'' elements remain while ``passing/transitional'' ones are filtered out along the process of producing graphs for the background layer.

Meanwhile, Schenkerian analysis admits a top-down approach that is fundamentally more organicist. Instead of taking an analyst approach, we can instead take a creative approach where we are initially given the background layer. Then our goal is to produce foreground layer, embarking on a process that is essentially the time-reverse of the bottom-up approach discussed above: through successive addition of foreground elements, we derive a more elaborate foreground layer. In fact, word choice reveals different perspective. In Salzer's conception for example, instead of calling foreground elements passing or transitional, he extensively uses the term ``elaboration'' to denote foreground elements. As suggested by the very word ``elaboration'', the implied process must proceed from the background layer to a foreground one, in order to enable us to use the term ``elaboration'' as in elaboration \emph{of} something. It is with the mindset of this top-down approach that the Schenkerian idea of prolongation makes sense at all. According to the Grove article on the term ``Prolongation'', prolongation is ``the generation of the harmonic and contrapuntal substance of a piece by a linear elaboration of its fundamental structure (Ursatz)''.\autocite{grove_rolongation} Now, similar to the idea of elaboration, we need to move in the direction from the background to the foreground in order to understand prolongation as in prolongation \emph{of} the background layer.

It is worth emphasizing the organicist implication in the top-down approach. The process of obtaining foreground layers as elaboration of background Ursatz can be interpreted as a process of organic growth exactly as it is discussed in section \ref{section:music as growth}. In fact, we may arguably claim that Schenker's understanding of musical structure is based on this organicist top-down approach more heavily than the reductional bottom-up approach which is more analytically useful. In other words, it is true that the bottom-up and top-down approaches are equally justified insights concerning musical structure, Schenker nevertheless prefers the top-down approach. In particular, we may claim that Schenker construes Schenkerian graphs from a top-down perspective where the graphs are to be considered as an organic growth of the Ursatz. Schenker spells out such organicist view concerning the Ursatz multiple times. For example, in his discussion of the Urlinie (being the upper voice of Ursatz, can be considered as part of the Ursatz), he takes the background layer as point of departure through whose generative process we obtain foreground elaborations:

\begin{quote}
The Urlinie bears in itself the seeds of all the forces that shape tonal life. With the cooperation of the harmonic degrees, the Urlinie indicates the paths to all elaboration and so also to the composition of the outer voices, in whose intervals the marriage of strict and free composition is so wonderfully and mysteriously consummated.\autocite[21]{drabkin2004tonwille}
\end{quote}

Moreover, in discussing the idea of diminution as a form of prolongation, Schenker again explicitly invokes an organicist metaphor:

\begin{quote}
Diminution relates to the Urlinie as ﬂesh in the bloom of life relates to a man's skeleton.\autocite[53]{drabkin2004tonwille}
\end{quote}

\subsection{Organicist consequence: temporal asymmetry as a result of prolongation}
\label{section:temporal asymmetry as a result of prolongation}
Using the top-down approach, Schenkerian analysis may be viewed as a description of a generative process in which the Ursatz organically grows into the foreground events we hear as listeners in concert. The generative process is to be understood as an active process of expansion. Music begins with a simple germinating seed called the Ursatz. Then, anthropomorphically, music attains its own will, through whose volition it grows according to generative rules called elaboration and prolongation.

Consequently, our observation that most of the chords in the background layer are distributed in the second half of a piece can be translated as follows: prolongation favors the beginning of Ursatz. In other words, elaboration occurs most intensively in the first half of a piece. Consequently, we invariably obtain temporal asymmetry in the form of delaying the climax: regardless of where the climax is originally placed, as a result of the expansive process, it is delayed as elaborations necessarily occupy durational extension. In light of music as an active process of expansion, we enumerate a few instances.

\subsection{Abstract schema for formal expansion}

The smallest meaningful formal unit can be labeled as ``AB'', representing a two-part division splitting the formal unit. Temporal asymmetry forces expansion of this formal unit to focus on elaboration and prolongation of part A.

\subsubsection{Expansion through left-replication: AB to AAB}
\label{section:left-replication}

One of the viable strategies of expanding the formal unit AB is induced by the \textbf{left-replication} operation. The operation replicates the leftmost component (namely, A) of the formal unit, resulting in a new formal unit: AAB. In short, the left-replication operation transforms AB into AAB.

\subsubsection{Example: bar form}
\label{section:bar form}

Just as ABA form is often associated with sonata form (with the exposition, the development, and the recapitulation commonly labeled by A, B, and A at the design level), AAB form is often associated with the term ``bar form''. According to the Grove article on this very term, it is considered ``one of the most common of all musical form genres''.\autocite{grove_barform} Given its real-world prevalence and importance, we are able to show that left-replication is effective in explaining a considerable portion of musical phenomena.

In the context of our discussion, bar form can be considered the canonic real-world application of left-replication, where the resulting form is explicitly spelled out and labeled as AAB. Moreover, left-replication offers additional insights: from the perspective of left-replication, we no longer understand the label AAB as mere labeling of segments of a piece. Instead, we understand AAB as a result of an active expansion (or borrowing terms from previous discussion, organic growth) of AB.

Furthermore, in light of left-replication, we can better appreciate the uniqueness of AAB form: its time-reverse counterpart, namely, ABB, does not exist as a widely-recognized form. From the perspective of left-replication, the lack of a time-reverse counterpart to AAB form translates exactly to the statement that there is no right-replication that can transform AB into ABB.

\subsubsection{Example: phrase structure}
\label{section:phrase structure}

One of the peculiar properties of bar form discussed in section \ref{section:bar form} is that it is not limited to description of musical forms. It can describe musical events at arbitrary organizational levels. As McClatchie quotes Lorenz's remark on the concept of bar form:

\begin{quote}
Just as the change of feeling can occur after two normal strophes of average length, so it can follow in even the tiniest musical shoots. . . . It can also just as well follow a gigantic, cumulative double intensification of great expanse. The essence of the Bar does not reside in the actual length, but in the distribution of its powers. Whether the Bar occupies 3 measures or 1000 is irrelevant; it is always a regular Bar if the aforementioned essence is fulfilled: a double appearance as against a single balancing occurrence of equal weight.\autocite{mcclatchie1998analyzing}
\end{quote}

In other words, we can consider other seemingly unrelated musical structures at various organizational levels in terms of bar form. In this section, we focus on phrase structure as an example of local musical structures. Based on Lorenz's remark on bar form, it should be possible to analyze phrase structure in terms of bar form, or in our generalized notion, left-replication.

It turns out that a particular type of phrase structure called the sentence can be considered exactly as phrase-level bar form. For a standard eight-measure phrase, a sentence is a type of phrase structure in which the first half (i.e.~the first four measures) consists of a repeated musical unit, while the second half consists of a holistic musical gesture as the continuation of the first half. The result of such division scheme is a phrase structure in the form (in the case of an eight-measure phrase) of 2+2+4. As Laitz remarks, one of the defining feature of sentence structure is its short-short-long pattern:

\begin{quote}
Whenever this rhetorical formal device of short-short-long (usually in the proportion 1:1:2) occurs within music, it is described as an instance of sentence structure.\autocite[310]{laitz2003complete}
\end{quote}

As a side note, if we compare Laitz's characterization of sentence structure with that of bar form by Lorenz, we can notice how both characterizations describe musical organizations which are essentially independent of specific levels of organization. In other words, just as bar form is not limited to describing formal structures, sentence structure is not limited to describing phrase-level structures. Therefore, we may equivalently claim that musical form with a structure of AAB can be considered exactly as form-level sentence structure.

\begin{figure}[!h]
   \centering
   \includegraphics[width=.8\textwidth]{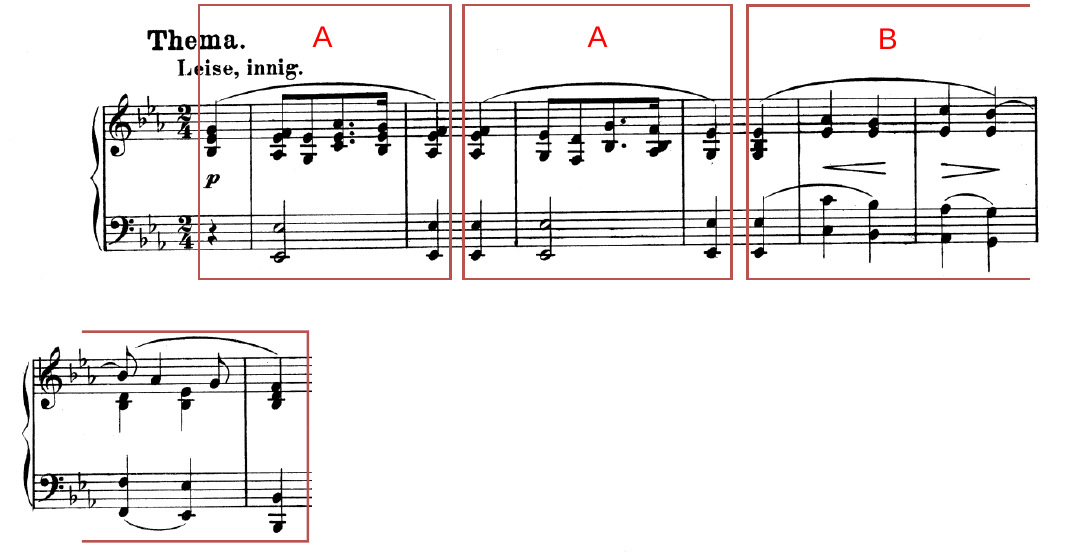}
   \caption{Schumann, Theme and Variations in E-flat major for piano, WoO 24, opening theme, annotated to indicate AAB structure}
   \label{fig:schumann ghost variation}
\end{figure}

In order to see how the 2+2+4 characterization of sentence structure can be considered equivalent to the AAB characterization of bar form, we need to explain how the 1:1:2 durational proportion suggests a grouping resembling the AAB scheme of bar form. In particular, we consider sentence structure with its musical content. One of the implications of the 1:1:2 durational proportion in sentence structure is the parallelism in content. In particular, the first half represented by the 1:1 durational proportion often accompanies parallelism in content as well. Therefore, the sentence structure is often characterized in terms of its content division: the 1:1:2 durational proportion corresponds to AAB indicating parallelism in musical content. For example, consider the opening theme of Schumann's ghost variations (see figure \ref{fig:schumann ghost variation}), the labeling of AAB has a twofold meaning: A and B represent short and long in durational values, respective; A and B represent parallelism in content such that AA denotes repetition (in this case, transposed repetition).

\subsubsection{Example: Schenkerian interruption}
\label{section:schenkerian interruption}

\begin{figure}[!h]
   \centering
   \includegraphics[width=.5\textwidth]{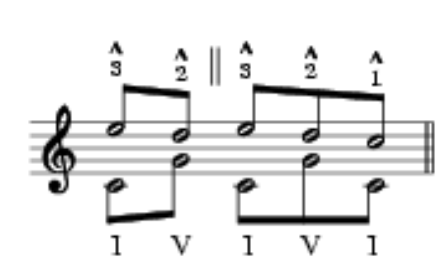}
   \caption{The notion of interruption in Schenkerian analysis, quoting from the Grove article on ``Interruption'': ``The symbol for an interruption is a double stroke on the same line as the capped arabic numerals representing the melodic scale steps of the fundamental line (Urlinie). {[}\ldots{]} The dominant that immediately precedes the interruption is called the Divider.''}
   \label{fig:schenkerian interruption}
\end{figure}

A special type of Schenkerian prolongation is to prolong the Ursatz itself instead of particular elements of the Ursatz, resulting in the notion of interruption (see figure \ref{fig:schenkerian interruption}).\autocite{grove_interruption} The term interruption attains its meaning figuratively: the Urlinie represented by \(\hat3-\hat2-\hat1\) is interrupted so that it has to be reset and somehow start again, thus requiring ``a return to the starting-point of the fundamental structure''.\autocite{grove_interruption}

\begin{figure}[!h]
   \centering
   \includegraphics[width=.5\textwidth]{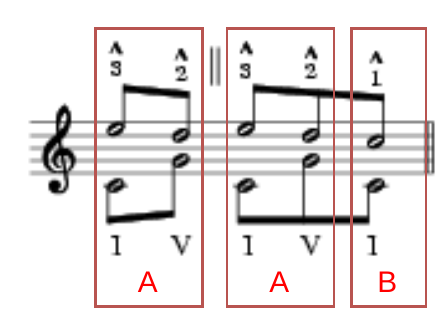}
   \caption{The notion of interruption in Schenkerian analysis, annotated to illustrate the left-replication operation which transforms AB into AAB}
   \label{fig:schenkerian interruption annotated}
\end{figure}

Alternatively, instead of a figurative interpretation where interruption interrupts a progression, the interruption can be interpreted as an example of left-replication, where the Ursatz originally labeled as AB is expanded by replicating the A section, resulting in an AAB structure (see figure \ref{fig:schenkerian interruption annotated}).

\subsubsection{Example: repetition of exposition in sonata-allegro form}

One of the most conventional ways of labeling sonata-allegro form is ABA. However, such labeling potentially falls victim to the pitfall of notation. Specifically, analysts often overlook one of the most important features of sonata form: the repetition of exposition. On the contrary, performers are consciously aware of its importance because one has to consciously decide whether to play the repetition, and if so, how to keep listeners engaged. Repetition of exposition is one of the pitfalls of notation again. Analysis is heavily based on notated score. However, as repetition sign in notated score takes up virtually no space, it is often neglected.

In the book ``Elements of sonata theory'', authors make similar warnings about downplaying the role of repeated exposition:

\begin{quote}
{[}Repeat{]} signs should not be taken for granted, passed over lightly in analysis, or omitted in performance. Repeat signs are never insignificant. Block-repetitions are an integral component of the style, and composers can work with this defining convention in a variety of ways.\autocite[21]{hepokoski2006elements}
\end{quote}

Consequently, in describing sonata form, an alternative description is to analyze with repetitions written-out, in which case the ABA form is expanded to AABA form. Such process of expansion is an example of left-replication: the leftmost component is replicated in expanding the form.

\begin{figure}[!h]
   \centering
   \includegraphics[width=.8\textwidth]{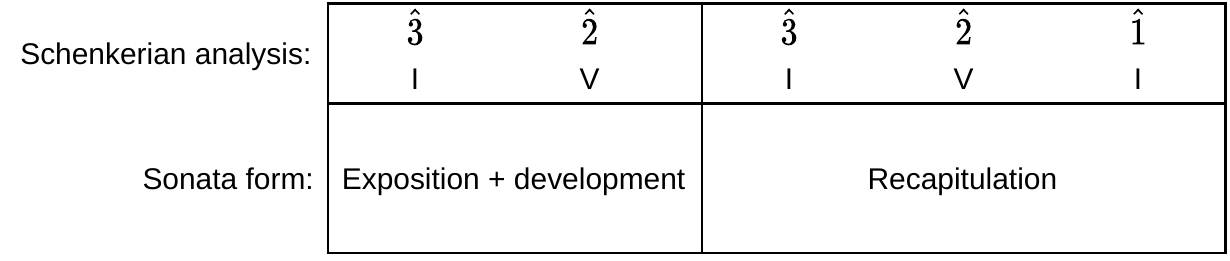}
   \caption{A schematic diagram illustrating the correspondence between Schenkerian interruption and sonata form}
   \label{fig:diagram interruption sonata form}
\end{figure}

It is worth noting the relationship between sonata form and Schenkerian interruption discussed in section \ref{section:schenkerian interruption}. In particular, Schenkerian interruption is often invoked to explain sonata form: the recapitulation in sonata form corresponds to the moment of interruption (see figure \ref{fig:diagram interruption sonata form}). In other words, ``the return to the starting-point of the fundamental structure corresponds to the beginning of the recapitulation in the musical foreground''.\autocite{grove_interruption}

\begin{figure}[!h]
   \centering
   \includegraphics[width=\textwidth]{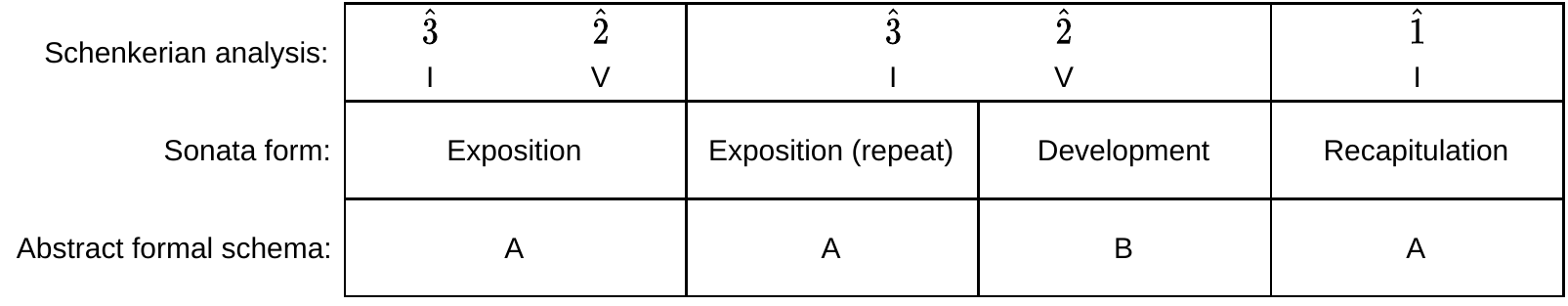}
   \caption{A schematic diagram offering an alternative interpretation (compare to figure \ref{fig:diagram interruption sonata form}) of formal structure as a consequence of incorporating exposition repeat into formal analysis}
   \label{fig:diagram interruption sonata form AABA}
\end{figure}

Alternatively, we may describe the relationship from the perspective of left-replication. Both the Schenkerian interruption and sonata form are examples of left-replication where an AB section is expanded to AAB. Additionally, we should realize that in this case, we obtain a different way of understanding sonata form in terms of Schenkerian interruption. Instead of relating the moment of interruption to the recapitulation as in figure \ref{fig:diagram interruption sonata form}, we treat repeat of exposition as structurally important to be included in the analysis, resulting in the correspondence between the moment of interruption and the moment of exposition repeat (see figure \ref{fig:diagram interruption sonata form AABA}).

\section{Historical perspective: maximalism}

Delaying the climax of a piece can also be viewed as a historical process, in which the placement of the climax gradually shifts to a later moment in the piece as history progresses in time. In particular, the historical evolution of musical form is often related to the formal expansion in scale. Peculiarly, the formal expansion does not preserve durational proportion between formal parts. Just like the left-replication operation discussed in section \ref{section:left-replication} and effect of prolongation discussed in section \ref{section:temporal asymmetry as a result of prolongation}: formal expansion favors the beginning more than the ending. Therefore, delaying the climax is one of the symptoms of formal expansion, as most of the expanded material lies before or at the climactic point, but not after. Now, formal expansion as a dynamic process in which the climactic point is delayed attains a historical perspective: the mysterious artistic force mentioned in the beginning of this chapter is to be understood as a historical force.

\subsection{Historical description}

One of the developmental trends in music history, especially during the late-Romanticism period, is termed by Taruskin as ``maximalism''. According to him, the term maximalism is defined concisely as ``the radical intensification of means toward traditional expressive ends''.\autocite{taruskin2006music_early_20th_century} Even though for Taruskin, the term maximalism refers to a very specific stylistic trend within a very specific time period, namely, within ``the period 1890--1914'',\autocite{taruskin2006music_early_20th_century} we can nevertheless see how the idea of maximalism is applicable to music history in general precisely because virtually everything can be maximized: maximalism is about radical intensification of means, but what these means refer to is up to our own interpretation. However, for the purpose of our discussion, we focus on a very specific aspect of maximalism: the temporal aspect of which delayed climax is a result.

Given maximalism as intensification of means toward ends, Taruskin then elucidates means and ends in detail:

\begin{quote}
What were the traditional ends given radically intensified or maximalized expression? Pound has already mentioned emotional expression, one of the prerequisites of romantic art. Another, from the very beginning of romanticism, was a sense of religious awe in the presence of the sublime. A third, sometimes an ally of the other two but potentially a subversive diversion (hence the most essentially ``modernist'') was sensuality.\autocite{taruskin2006music_early_20th_century}
\end{quote}

From Taruskin's characterization of expressive ends, we should recognize its emphasis on the psychological aspect of music. The emphasis should remind us of the discussion in section \ref{section:tension-release as psychological narrative}, whose very thesis is that narrative universal to all temporal arts must be based on our subjective experience of the artwork, in particular, tension-release. Therefore, we may notice that the three aspects of traditional ends (namely, emotional expression, religious awe in the presence of the sublime, and sensuality) are all descriptions of our emotional reactions to artworks, which are nothing but manifestations of tension-release. In fact, because of the emphasis on the psychological aspect in characterizing expressive ends concerning maximalism, the topic of maximalism benefits from the universality in describing our subjective experience discussed in section \ref{section:tension-release as psychological narrative}: because our emotional states are universal in the sense of being style-agnostic and insensitive to time periods, the topic of maximalism can be addressed outside the time period it originally intends to represent. For example, trying to talk about maximalism in Renaissance music appears less anachronistic than talking about futurism in Baroque music, precisely because the themes of maximalism (i.e.~emotional expression, religious awe in the presence of the sublime, and sensuality) are timeless.

On the other hand, means in the context of maximalism are more specific. According to Taruskin:

\begin{quote}
What were the intensified means? One involved the two dimensions in which musical works exist, the temporal and the sonorous {[}\ldots{]} Turning musical works into awe-inspiring mountains---by extending their length, amplifying their volume, and complicating their texture---became an obsession. Another way of amplifying the sense of musical space {[}\ldots{]} was to increase the range and maneuverability of ``tonal navigation,'' that is, the range of key relationships. Yet another area {[}\ldots{]} was the sheer level of tolerable (or at least tolerated) dissonance, and even more important, the postponement of its resolution.\autocite{taruskin2006music_early_20th_century}
\end{quote}

Even though specific compositional devices vary in achieving intensified means, we nevertheless find that intensified means enumerated by Taruskin share a common ground: there is a sense of expansion resulting from maximizing whatever musical parameters of interest. One of the most direct means is to expand in temporal scale. As observable from Taruskin's enumeration of intensified means, expansion in temporal scale makes its appearance twice: the first appearance occurs when Taruskin mentions ``the two dimensions in which musical works exist, the temporal and the sonorous'',\autocite{taruskin2006music_early_20th_century} and the second occurs when he spells out ``the postponement of {[}the dissonance's{]} resolution''.\autocite{taruskin2006music_early_20th_century} The twice appearance of expansion in temporal scale suggests that it is one of the indispensable elements of maximalism, at least in Taruskin's description.

It should not be surprising that expansion in temporal scale should be an (if not the most) important theme of maximalism. After all, maximalism is nothing other than the radical intensification of emotional intensity (whether it is passion, awe, or whatever else). Now, manipulating the temporal dimension of music is exceptionally effective in emotional intensification precisely because one's emotional state in response to music is primarily shaped by the interplay between memory and anticipation. Given a piece of music, its temporal dimension is the determining factor of memory and anticipation because they are possible only in relation to the concept of past and future.

A particular strategy to expand in temporal scale is to focus on the preparation of the climax. As discussed in the beginning of this chapter, the climax is well-defined and meaningful in a musical discourse because it is conceptually based on tension-release, whose focus on the psychological aspect of narrative gives rise to a description universal to all temporal arts. Consequently, the preparation of the climax is also well-defined and meaningful in a musical discourse.

In analytical descriptions of sonata form, the preparation of the climax, i.e.~formal component that leads into the climactic moment, attains a specific name: the retransition. Therefore, it would be unsurprising that in discussing maximalism using the example of Mahler's second symphony, Taruskin pays special attention to the retransition:

\begin{quote}
The best place to look to observe Mahler's response to it is the moment that traditionally carried the highest charge in a symphonic first movement: the ``retransition'' to the recapitulation, where sufficient ``dominant tension'' had to be generated to motivate a ``double return'' commensurate in strength to the length and range of the preceding development.\autocite{taruskin2006music_early_20th_century}
\end{quote}

\begin{figure}[!h]
   \centering
   \includegraphics[width=.8\textwidth]{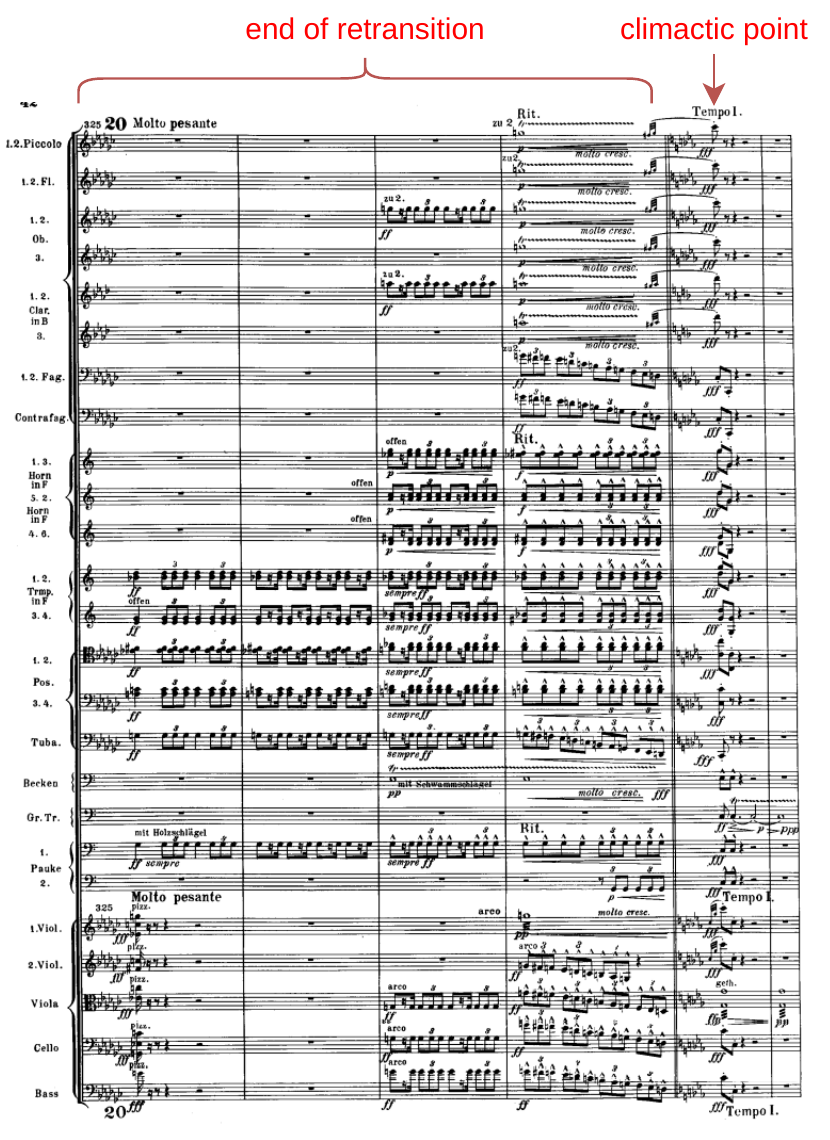}
   \caption{Mahler, Symphony No.~2, first movement, retransition before the recapitulation, annotated to highlight the repeating dissonant chords as expanded retransition delaying the arrival of the climactic point}
   \label{fig:mahler 2 retransition}
\end{figure}

In Taruskin's discussion, he focuses on the aspect of harmony to explain maximalism through intensified dissonance. However, it is equally noteworthy to observe the temporal effect of this section of repeated dissonant chord: repetition as sustaining (or more precisely, accumulating) tension (see figure \ref{fig:mahler 2 retransition}). In particular, we should note how the repeated dissonant chords as additionally inserted (i.e.~they could have been repeated fewer times if not for the sake of maximalism) musical materials, essentially stretches the retransition, thus prolonging the development section within the sonata form.

\subsubsection{Evolution of form as a process of expansion}

In fact, the retransition in the Mahler excerpt above as an example of maximalism can be put in a larger historical trend. As discussed above, maximalism is, to some extent, universal that it is not limited to the specific time period of 1890-1914 as Taruskin indicates. Instead, it can be considered as a continuation or extension of an ongoing historical trend: evolution of musical form as a process of expansion. In other words, the historical tendency to intensify emotional expression has always been one of the historical trends. Meanwhile, maximalism in the narrow sense defined by Taruskin is simply an intensification of this very trend, whose degree of intensity is alarming enough for us to spell out the specialized term ``maximalism''.

Therefore, delaying the climax, or more precisely, formal expansion whose result is delayed climax, can be considered as a theme applicable to the evolution of form in general. With this interpretation, maximalism in the sense defined by Taruskin is simply the climactic point of the evolution of form, when the evolution here is itself interpreted as a historical narrative.

As aforementioned, a particular strategy of expansion is to prolong the preparation of the climax. For sonata form, this strategy translates to the expansion of the development section. As Hepokoski and Darcy observe in their book ``Elements of sonata theory'':

\begin{quote}
The development is variable in length, although in the period 1760 -- 90 one would normally expect it to occupy a smaller space than that established by the exposition. Longer, more elaborate developments in the 1780s, 1790s, and later decades are monumentalized statements that invite special attention.\autocite[19]{hepokoski2006elements}
\end{quote}

Besides expansion of the development section, there are other ways to delay the climax by expanding pre-climax parts. For example, the development of repetition scheme can be interpreted as an expansion of pre-climax parts:

\begin{quote}
Within eighteenth-century sonatas and symphonies one may find both parts repeated (\textbar\textbar: exposition :\textbar\textbar: development -- recapitulation :\textbar\textbar). This is the most formal and earliest norm. Many late-century first movements, especially those after about 1760, repeat only the first part (the exposition), although in works prior to 1790 one need not be surprised to see the second part also repeated. After that date, repeating the second part is an uncommon gesture that invites analytical interpretation. It is also possible to find both parts unrepeated. This occurs in lighter works {[}\ldots{]} Nonrepeated expositions within first movements do sometimes occur in more broadly scaled and ambitious works after 1780, but when they do {[}\ldots{]} they are exceptional and need to be considered as consciously expressive choices.\autocite[20]{hepokoski2006elements}
\end{quote}

We can summarized the above observation as follows: the repetition of exposition is more resilient to omission over time. In particular, the repetition of the second part as a norm is discarded first. Meanwhile, the omission of repeating the first part is often associated with works smaller in scale. Therefore, for large-scale works, the most common repetition scheme is to repeat the first part but not the second. The result is an expansion of pre-climax parts. Before the climax (i.e.~the retransition leading into the recapitulation), we encounter a repeated exposition. Meanwhile, after the climax, because of the lack of repetition, we only encounter an unrepeated recapitulation.

\subsection{Organicist interpretation}
\label{section:organicist interpretation}

The evolution of form as a process of expansion also admits an organicist interpretation analogous to that in section \ref{section:temporal asymmetry as a result of prolongation}. In particular, the evolution of form is to be understood as an organic growth (with the term ``organic growth'' discussed in section \ref{section:music as growth}) of form. For example, we may thus construe the evolution of sonata form as a process of growth. The simple binary form is the embryonic stage of sonata form. Then, through the will of music history, sonata form expands in scale, passing through stages of expanding the exposition, development and other parts as its organs. Finally, it reaches its adulthood: the monumental sonata form associated with maximalism.

Moreover, history of music is anthropomorphized to attain purpose (with the term ``purpose'' discussed in section \ref{section:music as fulfillment of purpose}), through whose volition musical forms grow and expand. In fact, we may claim that history is the field where the writing style invoking anthropomorphism which lies at the heart of organicism has its etymological origin. In particular, the purposiveness crucial to an organicist understanding of music has its etymological root in an teleological understanding of history: the will of history is the driving force behind all musical development.

With such organicist interpretation, we are able to understand maximalism under a new light. Taruskin already hints upon the connection between maximalism and degeneracy. For example, he quotes Pound's comment in introducing maximalism:

\begin{quote}
This is the whole flaw of ``emotional'' music. It is like a drug: you must have more drug, and more noise each time, or this effect, this impression which works from the outside, in from the nerves and sensorium upon the self---is no use, its effect is constantly weaker and weaker.\autocite{taruskin2006music_early_20th_century}
\end{quote}

Moreover, he quotes Clara Schumann's reaction to Wagner's Tristan und Isolde, which is considered an example of maximalism as well:

\begin{quote}
It is not emotion that the opera portrays, it is a disease, and they tear their hearts out of their bodies, while the music expresses it all in the most nauseous manner.\autocite{taruskin2009oxford_19century}
\end{quote}

With an organicist view, we can better understand why people often associate maximalism with degeneracy. When maximalism is described using organicist terms, organic growth is turned into pathological mutation, and volition is turned into greed.

Consequently, we arrive at a poetic conclusion. The evolution of musical form is understood as a process of organic growth through expansion. One of the results of expansion is delaying the climax. The gradually delayed climax gives us an increasingly clear image of the arrow of time, as we can see its force and effect on moving the climax. Now we naturally wonder: what does the arrow of time do? Maximalism points us to an answer: the destructive tendency of our universe. Meanwhile, physicists arrive at the same answer.

\part*{Postlude}
\addcontentsline{toc}{part}{Postlude}

\chapter{Arrow of time in relation to organicism}

\epigraph{The spectre of time hovers, whose arrow tries to cast death and destruction. With cheerful laughter, we escaped. The running brooks carry us home, where nine-tailed foxes sing joyfully.}

If we are to trace the origin of temporal asymmetry, then it is nowhere to be found but in us: intelligent agents who are able to possess as well as reflect on life. Evidently the statement is an exaggeration since the cosmological arrow of time, i.e.~the expansion of the universe, is hardly about us. However, the crux of the matter is as follows. On the one hand, our possession of life, i.e.~the characteristics that enable us to identify life-bearing entities, is determined by temporal asymmetry in the form of entropy. As discussed in section \ref{section:life in relation to the arrow of time}, life can be characterized as entities who strive to evade the destructive tendency of the universe by actively maintaining a low-entropy state. On the other hand, as we reflect on the organic quality of music in terms of the expansion of musical forms, we see the arrow of time at work: the climax is delayed as a result of the organic growth of formal structures. Meanwhile, as maximalism discussed in section \ref{section:organicist interpretation} warns us, the end of such growth is the same as the effect of time's arrow: eventual destruction at world's end.

As discussed in section \ref{section:temporal asymmetry in narrative}:

\begin{quote}
The history of the universe is nothing but another form of story-telling.
\end{quote}

Poetically, we may then interpret the evolution of the universe as the unfolding of an five-act tragedy. It is a tragedy because we readily see the destructive force that underlies the story's unfolding. Through the tragic force the protagonist is brought to inevitable destruction. Similarly, through the arrow of time, the universe is brought to the eventual state of maximal entropy. Finally, maximalism in music informs us that the sheer musical tension created by delaying the climax exhausts our emotional capacity, resulting in an emotional breakdown.

How are we to handle this tragic force of destruction? One of the most organicist answers is found in the very value of tragedy itself: catharsis. According to Aristotle, the value of tragedy is to purify the soul ``through pity and fear''\autocite{aristotle_poetics} invoked by the tragic force. Consequently, the destructive tendency of the universe is poetically resolved:

\begin{quote}
The organic growth of musical form delays the climax, resulting in an catharsis at the retransition before the recapitulation. The sheer power of elevated musical tension results in our emotional breakdown. The breakdown purifies our soul, such that when we reach the recapitulation, we are back home (tonally and spiritually).
\end{quote}

\part*{Appendices}
\addcontentsline{toc}{part}{Appendices}
\appendix

\chapter{The evolution of equations of motion}
\label{appendix:evolution of equations of motion}

To describe the motion of an object, we associate each moment in time with a position in space. Therefore, a complete description of an object's motion is ``a `map' (or `function') from time to space'': \autocite[3]{lawvere2009conceptual} we are mapping each moment in time to the position of the object at that moment. However, the rules abiding by which each moment in time is associated with a position in space are not arbitrary. In mechanics, the central task is to ask: what are the physical laws according to which objects move (or more generally, behave)? The description of these physical laws is encoded in the form of equations of motion. Through equations of motion, we are able to predict an object's motion by knowledge of the map from time to space. Therefore, the development of mechanics is nothing other than the process of discovering and interpreting various equations of motion. It follows that by examining equations of motion and their interpretations, we are able to gain insights into the historical development of mechanics.

At this point, given the importance of equations of motion in all formulations of mechanics, we may already recognize the ubiquity of time, as equations of motion (regardless of their content) uniquely specify the motion as a map from time to space. Therefore, we are always reminded that we are dealing with maps from time to space. Now, it suffices to consider three examples in mechanics to elucidate what the ubiquity of time entails: Newtonian mechanics, Lagrangian formulation, and quantum mechanics.

In Newtonian mechanics, motion is encoded in a straightforward way. Without loss of generality, we may consider the motion of a single particle. The position of a particle is represented by a point in a three-dimensional Euclidean space, namely \((x,y,z)\) in Cartesian coordinates. The representation is straightforward because the mathematical point \((x,y,z)\) refers to the real physical entity of the particle. Similarly, the concepts of velocity and acceleration are also represented by points (or more accurately, vectors) in a three-dimensional Euclidean space. They are also intuitive because a point/vector in three-dimensional Euclidean space captures the notions of direction and magnitude, which are needed for an intuitive understanding of velocity and acceleration in everyday life. Finally, the concept of mass is also immediately intelligible. Now, Newton's second law gives the equation of motion: \(F = ma\). The law is intuitive because every component in the equation corresponds to some meaningful physical quantity that has a counterpart in reality. This is exactly why we can state Newton's second law verbally: the force is equal to mass times acceleration.

In comparison, Lagrangian mechanics is less intuitive. Lagrangian mechanics is often considered to be a reformulation of Newtonian mechanics in the following sense. Firstly, it deals with the same set of physical problems and yields the same results as Newtonian mechanics. Secondly, however, it differs from Newtonian mechanics in terms of how physical problems are presented, resulting in a drastically different set of perspectives and methods. In Lagrangian mechanics, we start with representing particles using generalized coordinates in configuration space. The terminology already suggests how counterintuitive the approach is. To see an example, we consider a collection of 10 particles. In the Newtonian approach, we represent their positions using 10 points in a three-dimensional Euclidean space, such that we have \((x_1,y_1,z_1),...,(x_{10},y_{10},z_{10})\). However, in the Lagrangian approach, their positions are encoded by a single point (i.e.~generalized coordinate) in a 30-dimensional Euclidean space (i.e.~configuration space): \((x_1,y_1,z_1,...,x_{10},y_{10},z_{10})\). One may argue that the difference is merely about grouping and arrangement of the 30 variables needed to specify the system's state. However, one must realize how we lose the intuitive grounding in the Lagrangian approach. A point in three-dimensional space refers to real physical entities, namely, the position of a particle in reality. The same is true for a collection of 10 points: their representations have physical meaning. However, interpreting them using a single point in a 30-dimensional space attains no physical counterpart in reality: it resides in an imagined space of mathematical beauty and elegance, at the expense of losing the intuitive grasp.

Given the generalized coordinates and configuration space, one then constructs a quantity called the Lagrangian, equipped with an equation called the Euler-Lagrange equation. Without going into mathematical details, we may describe the general Lagrangian approach as follows: the equations of motion are given by ``optimizing'' the Lagrangian such that it satisfies the Euler-Lagrange equation. What do we mean by the term optimizing the Lagrangian? We go back to the example of 10 particles. As discussed, motion is nothing but a map from time to space. Therefore, for motion of 10 particles, we can represent them using 10 points in three-dimensional space, so we have 10 maps from time to three-dimensional space, with each map encoding the motion of a particle. Alternatively, we can picture a single moving point in 30-dimensional space, so the motion of 10 particles is captured by a single map from time to 30-dimensional space. Note that the two views are equivalent, i.e.~they encode exactly the same information. Therefore, if we know how 10 points in three-dimensional space move, then we know how the corresponding single point in 30-dimensional space moves, and vice versa. Now, the question is: according to what laws does the point in 30-dimensional space move? What governs the motion (note that the term is now metaphorical) of this imaginary point? The answer is that the laws are given by the Euler-Lagrange equation, or more generally, the principle of least action. In particular, the point in 30-dimensional space moves in a way that its motion, as a map from time to the 30-dimensional space, makes the Lagrangian quantity (which depends on this map) ``optimized'' in such a way that the Euler-Lagrange equation is satisfied.

One can indeed recover Newton's second law in Lagrangian mechanics as a special case. However, we are no longer able to express it verbally. If we take the Lagrangian quantity to be the peculiar choice of kinetic energy minus potential energy, then the Euler-Lagrange equation is naturally reduced to the simple equation \(F = ma\), thus recovering Newton's second law. In such a way, if one is to describe Newton's second law from a Lagrangian perspective verbally, the description would have to start with the quantity of kinetic energy minus potential energy, which again has no intuitive physical meaning. Upon contemplation, one should realize how startling the Lagrangian approach is: the entire process of solving for physical motion is counterintuitive because virtually nothing except the end result refers to physically meaningful entities. Instead, it focuses on conceptual constructs such as generalized coordinates, configuration spaces, Lagrangians, and the Euler-Lagrange equation. Meanwhile, nature seems to favor these conceptual constructs as they prioritize the beautiful and the sublime of our world over reality. Physicist Weyl once remarked that

\begin{quote}
My work always tried to unite the true with the beautiful; but when I had to choose one or the other, I usually chose the beautiful.\autocite[217]{birds_and_frogs}
\end{quote}

Adding to his remark, we are entitled to poetically say: physics makes more sense if we represent the world with beauty instead of reality in mind. As our discussion of Lagrangian mechanics illustrates, nature imposes physical laws which admit descriptions that are particularly elegant.

The counterintuitive approach found in Lagrangian formulation equally applies to quantum mechanics, if not more. The methodological approach of quantum mechanics is similar to the Lagrangian formulation, in that we construct some mathematical object (the Lagrangian for Lagrangian formulation, and wavefunctions for quantum mechanics) to describe the state of a physical system. Given the state-encoding mathematical object, it is then equipped with rules (Euler-Lagrange equation and Schrodinger's equation, respectively) that specify the time evolution of the state. The rules are presented as a set of mathematical problems (in the form of what is called differential equations) whose solutions enable us to predict the time evolution. Without going into details, we readily see that the world is modeled in an increasingly abstract and counterintuitive fashion. For example, wavefunctions, as the fundamental building block of quantum mechanics, attain no direct physical meaning. They are simply ways to encode the information needed to describe the quantum state. It is through manipulations of wavefunctions that we are able to extract from them useful results that have physical meaning.

The three examples above give a potential way to interpret the historical development of mechanics: the discipline focuses on increasingly abstract representation of the physical world. It is interesting to notice a seeming paradox: on the one hand, theories gradually deviate from the way we intuitively understand reality because they involve formalisms that are increasingly abstract. However, on the other hand, physicists find that abstract representations of the physical world often give a more fundamental description of reality. The meaning of the term ``fundamental'' is twofold here. Firstly, fundamental means usefulness and beauty, in the sense that abstract representations often are more powerful in solving problems in physics. In other words, they often provide a unified theory to some extent, which is what enables physicist D'Alembert (yes you are not mistaken, D'Alembert is also the person who wrote a survey of Rameau's harmonic theory) to artistically claim that

\begin{quote}
To someone who could grasp the universe from a unified standpoint, the entire creation would appear as a unique truth and necessity. \autocite[2600]{gaither2012}
\end{quote}

The second meaning of fundamental is somewhat philosophical. It involves the ontological argument that these abstract mathematical entities (take wavefunctions for example) are not ``mere tool for prediction''. \autocite{sep-qm-copenhagen} Instead, they are the first principles, i.e.~fundamental premises, of the world through which the reality as it appears to us is derived. In other words, abstract mathematical representations are the irreducible definition of the physical world. Therefore, we should not ask what the physical meaning of wavefunction is. But instead, we should take wavefunction as the premise of our physical world, and ask: what do we mean by words we use everyday such as force, mass, and acceleration? Do we really know their physical meanings? Or more poetically, is the word ``physical meaning'' well-defined at all? Such an ontological view is an example of interpretation of physical formalisms, which frequently causes heated debates among physicists and philosophers alike.

\pagebreak
\singlespace
\printbibliography

\end{document}